%% file: main.tex
\documentclass[aps,prc,superscriptaddress,longbibliography, showpacs,floatfix,letterpaper,preprintnumbers,nofootinbib,notitlepage]{revtex4-1}
\usepackage[letterpaper, total={7in,9.5in}]{geometry}
\usepackage{subcaption, caption}
\usepackage[colorlinks=true, linkcolor=blue, citecolor=blue, urlcolor=blue]{hyperref}
\usepackage{amssymb}
\usepackage[title]{appendix}
\usepackage{color}
\usepackage{graphicx}
\usepackage{dcolumn}% Align table columns on decimal point
\usepackage{bm}% bold math
\usepackage{amsmath,mathrsfs}
\usepackage{braket}
\usepackage{xfrac}
\usepackage{soul}
\usepackage{lineno}
\usepackage{slashed}
\usepackage{appendix}
\usepackage{tabularx,enumitem,paralist}
\newcolumntype{R}{>{\raggedleft\arraybackslash}X}%
\newcolumntype{I}[1]{>{\begin{compactitem}}p{#1}<{\end{compactitem}}}

\begin{document}

\title{Glauber quark and gluon contributions to quark energy loss at next-to-leading order and next-to-leading twist}

\author{Amit Kumar}
\email[Corresponding author: ]{amit.kumar@uregina.ca}
\affiliation{Department of Physics, University of Regina, Regina, Saskatchewan S4S 0A2, Canada}

\author{Gojko Vujanovic}
\email[Corresponding author: ]{gojko.vujanovic@uregina.ca}
\affiliation{Department of Physics, University of Regina, Regina, Saskatchewan S4S 0A2, Canada}

\date{\today}

%%%%%%%%%%%%%%%%%%%%%%%%%%%%%%%%%%%%%%%%%%%%%%%%%%%%%%%%%%%%%%%%%%%%%%%%%%%%%%%%%%%%%%%%%%%%%%%%
%%%%%%%%%%%%%%%%%%%%%%%%%%%%%%%%%%%%%%%%%%%%%%%%%%%%%%%%%%%%%%%%%%%%%%%%%%%%%%%%%%%%%%%%%%%%%%%%
%%%%%%%%%%%%%%%%%%%%%%%%%%%%%%%%%%%%%%%%%%%%%%%%%%%%%%%%%%%%%%%%%%%%%%%%%%%%%%%%%%%%%%%%%%%%%%%%
\begin{abstract}
The higher-twist formalism is used at $O(\alpha^2_s)$ to compute all possible medium-induced single-scattering emission kernels for an incoming highly energetic and virtual quark traversing the nuclear environment. The effects of the heavy-quark mass scale are taken into account [\href{https://journals.aps.org/prc/abstract/10.1103/PhysRevC.94.054902}{Phys. Rev. C 94, 054902 (2016)}] both in the initial state as well as in the final state, along with interactions involving both in-medium Glauber gluons and quarks [\href{http://dx.doi.org/10.1016/j.nuclphysa.2007.06.009}{Nucl. Phys. A 793, 128 (2007)}], as well as coherence effects [\href{https://journals.aps.org/prc/abstract/10.1103/PhysRevC.105.024908}{Phys. Rev. C 105, 024908 (2022)}]. As this study is a continuation of our work on medium-induced photon production [\href{https://journals.aps.org/prc/abstract/10.1103/bmmf-9zv5}{Phys. Rev. C 112, 025204 (2025)}], the general factorization procedure for $e$-$A$ deep-inelastic scattering is still used. An incoming quark energy loss in the nuclear medium yields four possible scattering kernels $\mathcal{K}_i$ with the following final states: (i) $q+g$; (ii) $g+g$; (iii) $q+\bar{q}'$, where the quark $q$ may have a flavor different from the antiquark $\bar{q}'$; and (iv) $q+q'$, where, again, $q$ may have a flavor different from $q'$. The collisional kernels include full phase factors from all nonvanishing diagrams and complete first-order derivative in the longitudinal direction ($k^-$) as well as second-order derivative in the transverse momentum ($k_{\perp}$) gradient expansion. Furthermore, in-medium parton distribution functions and the related jet transport coefficients have a hard transverse-momentum dependence (of the emitted quark or gluon) present within the phase factor.
\end{abstract}
%%%%%%%%%%%%%%%%%%%%%%%%%%%%%%%%%%%%%%%%%%%%%%%%%%%%%%%%%%%%%%%%%%%%%%%%%%%%%%%%%%%%%%%%%%%%%%%%
%%%%%%%%%%%%%%%%%%%%%%%%%%%%%%%%%%%%%%%%%%%%%%%%%%%%%%%%%%%%%%%%%%%%%%%%%%%%%%%%%%%%%%%%%%%%%%%%
%%%%%%%%%%%%%%%%%%%%%%%%%%%%%%%%%%%%%%%%%%%%%%%%%%%%%%%%%%%%%%%%%%%%%%%%%%%%%%%%%%%%%%%%%%%%%%%%

\maketitle

%%%%%%%%%%%%%%%%%%%%%%%%%%%%%%%%%%%%%%%%%%%%%%%%%%%%%%%%%%%%%%%%%%%%%%%%%%%%%%%%%%%%%%%%%%%%%%%%
%%%%%%%%%%%%%%%%%%%%%%%%%%%%%%%%%%%%%%%%%%%%%%%%%%%%%%%%%%%%%%%%%%%%%%%%%%%%%%%%%%%%%%%%%%%%%%%%
%%%%%%%%%%%%%%%%%%%%%%%%%%%%%%%%%%%%%%%%%%%%%%%%%%%%%%%%%%%%%%%%%%%%%%%%%%%%%%%%%%%%%%%%%%%%%%%%
\section{Introduction}
%%%%%%%%%%%%%%%%%%%%%%%%%%%%%%%%%%%%%%%%%%%%%%%%%%%%%%%%%%%%%%%%%%%%%%%%%%%%%%%%%%%%%%%%%%%%%%%%
%%%%%%%%%%%%%%%%%%%%%%%%%%%%%%%%%%%%%%%%%%%%%%%%%%%%%%%%%%%%%%%%%%%%%%%%%%%%%%%%%%%%%%%%%%%%%%%%
%%%%%%%%%%%%%%%%%%%%%%%%%%%%%%%%%%%%%%%%%%%%%%%%%%%%%%%%%%%%%%%%%%%%%%%%%%%%%%%%%%%%%%%%%%%%%%%%
Ultrarelativistic heavy-ions collisions carried out at the Relativistic Heavy-Ion Collider (RHIC) and the Large Hadron Collider (LHC) produce a deconfined state of quarks and gluons, called quark-gluon plasma (QGP). One of the primary goals of these collisions is to constrain properties of QGP, through, e.g., the modifications it imparts on high-energy quark- and gluon-initiated jets. Jet evolution in the QGP is a multiscale process, and different physics are involved at different virtuality scales. While highly energetic, nearly on-shell jet partons are described through effective Boltzmann transport, which includes the hard thermal loop formalism supplemented by Landau-Pomeranchuk-Migdal (LPM) resummation \cite{Arnold:2000dr,Arnold:2002zm,Arnold:2003zc}, the goal of this contribution is to explore the evolution of jet partons using the higher-twist (HT) formalism \cite{Guo:1998rd,Guo:2001tz,Qiu:2004da,Kang:2015mta,Kang:2014hha,Kang:2016ron} in the single-scattering limit \cite{Guo:2000nz,Wang:2001ifa,Majumder:2009ge} for high energy and virtual partons.\footnote{While modifications to the radiation spectrum induced by single scattering can happen at any virtuality, as virtuality is lowered multiple scatterings become important \cite{Majumder:2007hx,Majumder:2009ge}. Thus, for sufficiently short-lived partons, single scattering is enough to describe their in-medium radiation, as done here.}  In addition to these theoretical developments, there has also been a surge in various Monte Carlo implementations of these approaches, such as MATTER~\cite{Majumder:2013re,Cao:2017qpx} at high virtuality, with LBT \cite{Cao:2016gvr,JETSCAPE:2017eso} and MARTINI \cite{Schenke:2009gb} at lower virtuality, while the JETSCAPE framework \cite{Putschke:2019yrg,JETSCAPE:2019udz} has enabled a more holistic simulation to be devised, better covering the virtuality-dependent dynamics of jet energy loss in a nuclear environment. The JETSCAPE framework also incorporates event-by-event simulations of the nuclear medium, including hydrodynamical simulation of the QGP, along with pre-hydrodynamical simulations and hadronic transport, allowing sophisticated Bayesian model-to-data comparisons to be conducted \cite{JETSCAPE:2024cqe}. To push Bayesian analysis further, jet-medium interactions should be explored in a more discerning way by separating the gluonic from the fermionic contributions to jet-medium transport coefficients (or accounting for both of them), as is done herein. The higher-twist formalism separates the perturbative (jet) and non-perturbative (nuclear medium) parts of jet-medium interactions.   

Modern multistage simulations of the nuclear medium in heavy-ion collisions relies on a combination of gluon-dominated initial state, where color-glass condensate (CGC) is an appropriate model to describe initial-state dynamics~\cite{Schenke:2012wb,McDonald:2023qwc,Heffernan:2023gye,Heffernan:2023utr}, followed by hydrodynamical attractor-inspired simulations that bridge the gap to fluid dynamical evolution of the QGP~\cite{Kurkela:2018vqr,Kurkela:2018wud}. Such an approach has been recently used to study photon production \cite{Gale:2021emg} and this calculation provides the necessary ingredients to enable future studies of jets inside a multistage evolution of the hot medium. It has long been believed that the HT formalism \cite{Ellis:1982wd, Qiu:1990xxa, Qiu:1990xy} and CGC \cite{McLerran:1993ni, Ayala:1995kg, Gelis:2010nm, Kovchegov:2012mbw} approaches differ fundamentally due to two main factors \cite{Fu:2023jqv}. First, HT formalism is based on generalized QCD collinear factorization, in which medium properties are encoded in multiparton correlation functions. However, CGC formalism is based on transverse momentum dependent factorization at small $x$, where the evolution of the distribution function is given by Jalilian-Marian-Iancu-McLerran-Weigert-Leonidov-Kovner and Balitsky-Kovchegov nonlinear equations \cite{Balitsky:1995ub, Kovchegov:1999ua, Jalilian-Marian:1997qno, Jalilian-Marian:1997jhx, Jalilian-Marian:1997ubg, Kovner:2000pt, Iancu:2000hn, Iancu:2001ad, Ferreiro:2001qy}. The second difference emanates from the treatment of multiple scattering; CGC treats single and multiple scattering on the same footing within the semi-classical approximation. Using direct photon production in {\it p-A} collisions, the authors of Refs.~\cite{Fu:2023jqv,Fu:2024sba} show that the HT formalism is in fact consistent with the  CGC approach at finite $x$, when including sub-eikonal terms and LPM phase interference (twist-4) terms arising from two in-medium gluon scatterings. This calculation demonstrates a deeper understanding of the CGC and HT formalisms, showing how both approaches yield the same, unique, result for a given set of observables. Owing to the discovery presented in Refs. \cite{Fu:2023jqv,Fu:2024sba}, the results presented in this paper for the medium modifications of the hard parton in \textit{e-A} collisions, obtained within the HT formalism, should be compatible with the CGC formalism; both equivalently describing the high-energy nuclear medium at early times.

\begin{figure}[!h]
    \centering 
    \begin{subfigure}[t]{0.2\textwidth}
        \includegraphics[height=0.75in]{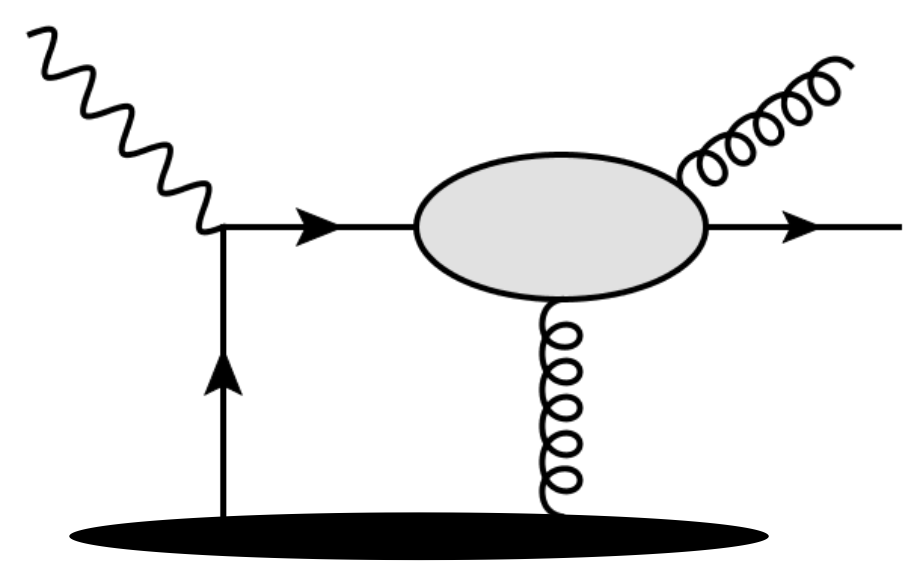}
        \caption{ A gluon and quark in the final state. }
    \end{subfigure}%
    \begin{subfigure}[t]{0.2\textwidth}
        \includegraphics[height=0.75in]{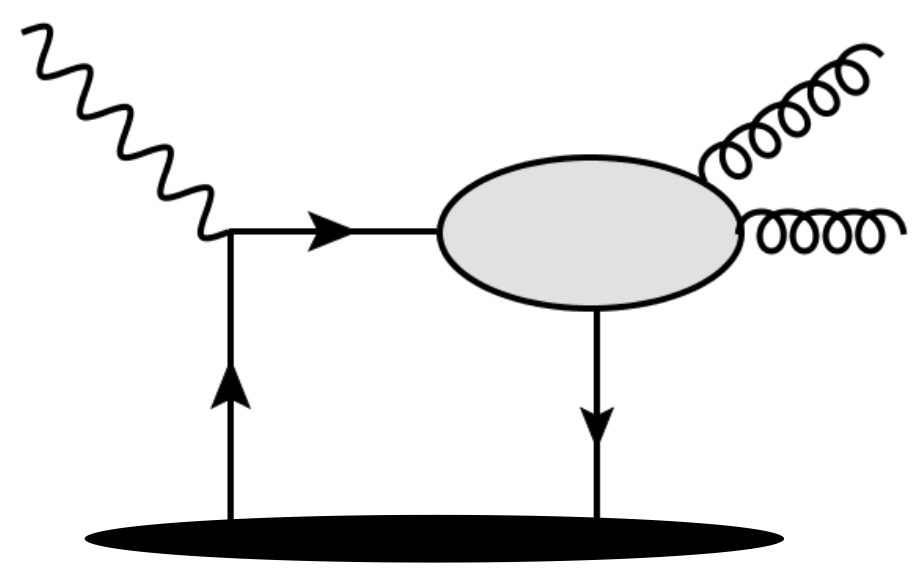}
        \caption{Two gluons in the final state.}
    \end{subfigure}
    \begin{subfigure}[t]{0.2\textwidth}
        \includegraphics[height=0.75in]{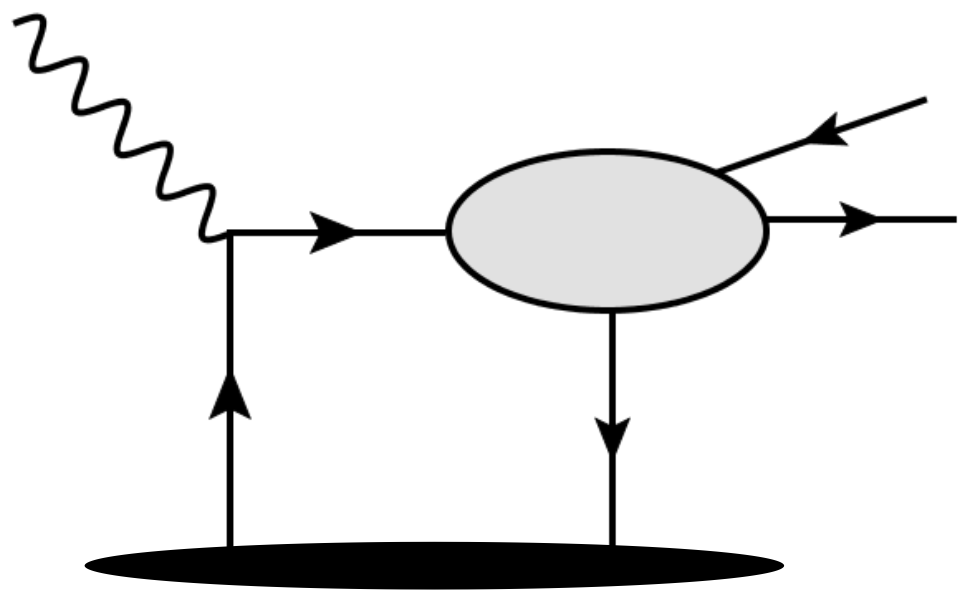}
        \caption{Quark anti-quark in the final state.}
    \end{subfigure}
    \begin{subfigure}[t]{0.2\textwidth}
        \includegraphics[height=0.75in]{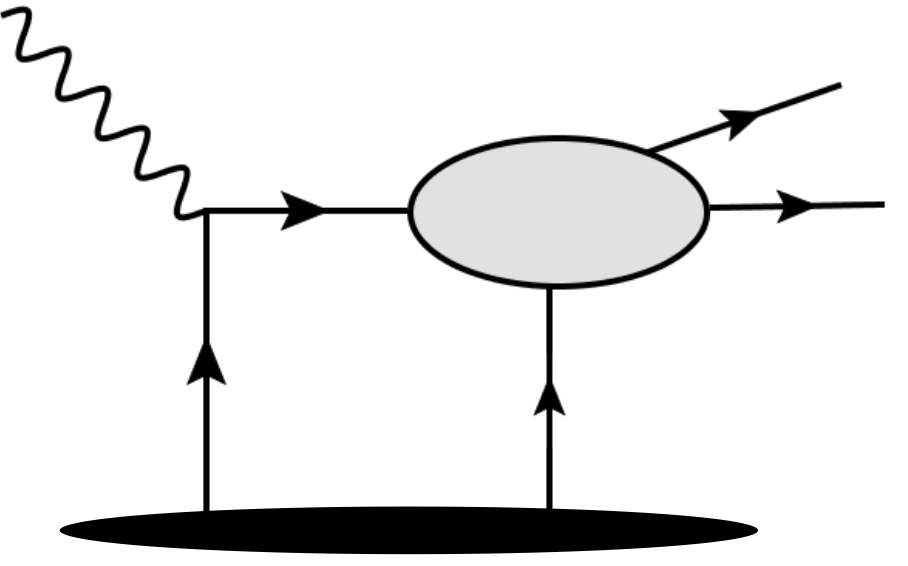}
        \caption{Two quarks in the final state.}
    \end{subfigure}
\caption{Scattering diagrams with one emission and one scattering for a quark initiating jet.}
\label{fig:HT_gluon}
\end{figure}

The production investigated herein is induced by processes illustrated in Fig.~\ref{fig:HT_gluon}, where the HT formalism in the single-scattering-induced radiation limit is used to calculate their production kernels $\mathcal{K}$. The first kernel $\mathcal{K}_1$ captures processes involving a Glauber gluon~\cite{Idilbi:2008vm} exchange with the nuclear medium, as depicted in the blob diagram Fig.~\ref{fig:HT_gluon}(a). The second kernel $\mathcal{K}_2$ involves an annihilation process between the jet quark and the in-medium antiquark (or vice versa) thus giving rise to quark-to-gluon conversion depicted in Fig.~\ref{fig:HT_gluon}(b). A subset of diagrams in Fig.~\ref{fig:HT_gluon} were already explored in our previous calculation \cite{Kumar:2025egh}. However, interactions in the medium can also involve virtual gluons giving either a quark-quark final state, encoded in the third kernel $\mathcal{K}_3$, or a quark-antiquark final state encapsulated within the fourth kernel $\mathcal{K}_4$, as shown in Figs.~\ref{fig:HT_gluon}(c) and~\ref{fig:HT_gluon}(d), respectively. The diagrams in Fig.~\ref{fig:HT_gluon} will be referred to as Kumar-Vujanovic (KV) kernels. As was the case in our previous study \cite{Kumar:2025egh}, there will be Glauber gluon~\cite{Idilbi:2008vm} and quark contributions to in-medium (i) jet transverse momentum broadening, as well as (ii) longitudinal drag, taking into account the heavy-quark mass scale~\cite{Abir:2015hta}, coherence effects \cite{Sirimanna:2021sqx}, together with fermion-to-boson conversion processes \cite{Schafer:2007xh}. 

This paper presents modifications of the hard parton traversing a large nucleus computed within a generalized factorization procedure used for e-A deep-inelastic scattering. Within the higher-twist formalism, perturbative and non-perturbative portions in the scattering kernel are explicitly separated, thus providing a clear setup to investigate how energetic quarks propagate through any nuclear environment. A direct application of this approach is to study medium-induced energy loss quantitatively, via deep inelastic scattering (DIS) in the upcoming Electron-Ion Collider (EIC) or through jets in heavy-ion collisions, by constraining nuclear jet transport coefficients, such as transverse momentum broadening and longitudinal momentum drag, within hot or cold nuclear matter. 

In contrast to the physics to be explored by the EIC, in relativistic heavy-ion collisions the composition of the hot nuclear medium formed at early times is well described by glasma dymamics \cite{Schenke:2012wb, McDonald:2023qwc,Heffernan:2023utr, Heffernan:2023gye}. 
At later stages, the particle composition of the hot nuclear medium changes, as quark-antiquark pairs are dynamically generated, finally leading to hydrodynamical behavior. This process is also called hydrodynamization. In the hydrodynamic regime, the lattice QCD equation is used \cite{Heffernan:2023gye}, and quarks and gluons are assumed to have reached their thermodynamic occupation numbers. Given that jets are sensitive to the entire dynamical evolution of the nuclear medium, they probe hydrodynamization dynamics of quark flavors. A prehydroynamical model that accounts for flavor hydrodynamization, such as that of Ref.~\cite{Gale:2021emg}, can be implemented inside of a realistic simulation of heavy-ion collisions using the modular JETSCAPE framework, enabling the next generation of jet-medium Bayesian analysis.

This paper is organized into several sections. Section~\ref{sec:DIS} describes the separation of the leptonic tensors and the hadronic tensor in DIS, including the various scattering kernels explored here. Section~\ref{sec:gluon_quark} provides details on the calculation of one diagram that contributes to the $q\to q+g$ scattering kernel, with the aim of precisely introducing the power-counting and approximation scheme used in this work. All other diagrams contributing to the various collisional kernels are presented in the appendixes. Section~\ref{sec:full_kernel} gives the results for the four scattering kernels in Fig.~\ref{fig:HT_gluon}, their collinear expansion along with a discussion, is presented in Sec.~\ref{sec:collinear_exp}, while a summary and an outlook are given in Sec.~\ref{sec:summary_outlook}. 

%%%%%%%%%%%%%%%%%%%%%%%%%%%%%%%%%%%%%%%%%%%%%%%%%%%%%%%%%%%%%%%%%%%%%%%%%%%%%%%%%%%%%%%%%%%%%%%%
%%%%%%%%%%%%%%%%%%%%%%%%%%%%%%%%%%%%%%%%%%%%%%%%%%%%%%%%%%%%%%%%%%%%%%%%%%%%%%%%%%%%%%%%%%%%%%%%
%%%%%%%%%%%%%%%%%%%%%%%%%%%%%%%%%%%%%%%%%%%%%%%%%%%%%%%%%%%%%%%%%%%%%%%%%%%%%%%%%%%%%%%%%%%%%%%%
\section{Hadronic Tensor in Deep Inelastic Scattering}\label{sec:DIS}
\begin{figure}[!h]
\includegraphics[width=0.32\textwidth]{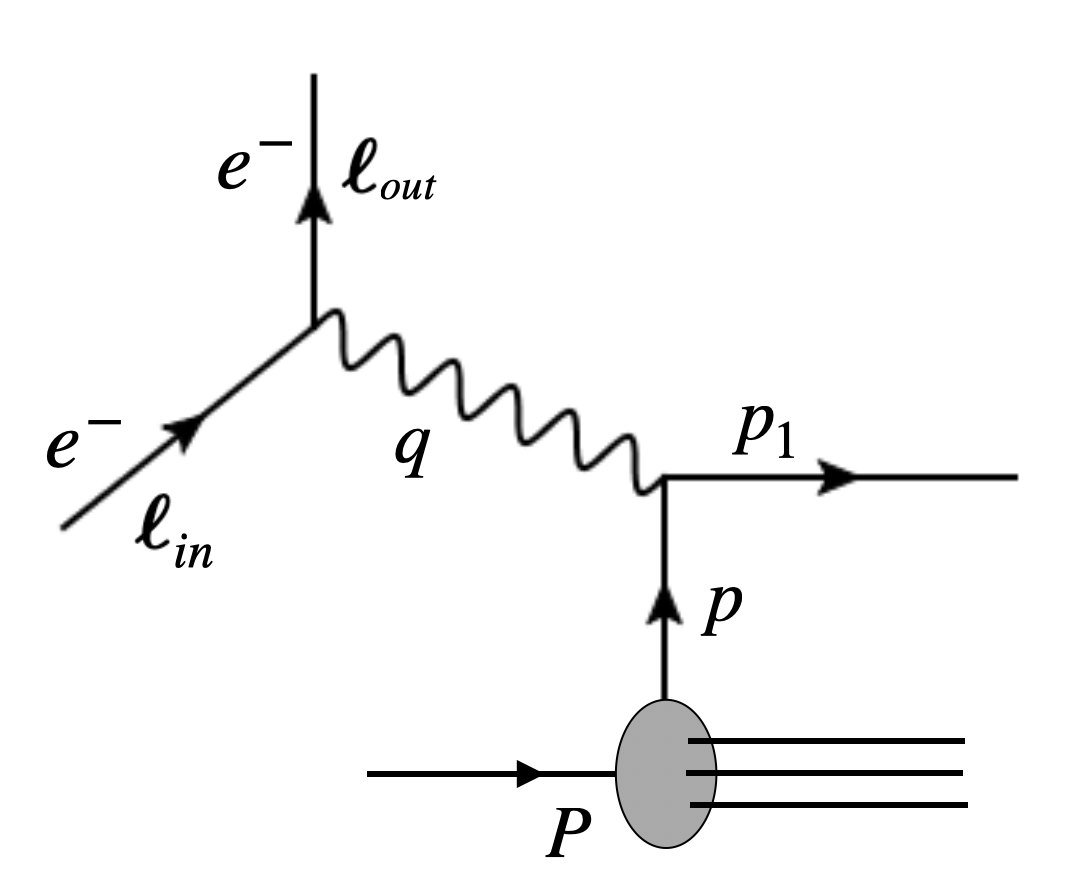}
\caption{A schematic diagram of deep-inelastic scattering between an electron and a nucleon inside the nucleus. The virtual photon carries momentum $q$, whereas the struck quark carries momentum $p$. The nucleus momentum is $\mathcal{P}=AP$, where $P$ is the momentum of the nucleon.}
\label{fig:DIS_eA_collision}
\end{figure}

%%%%%%%%%%%%%%%%%%%%%%%%%%%%%%%%%%%%%%%%%%%%%%%%%%%%%%%%%%%%%%%%%%%%%%%%%%%%%%%%%%%%%%%%%%%%%%%%
%%%%%%%%%%%%%%%%%%%%%%%%%%%%%%%%%%%%%%%%%%%%%%%%%%%%%%%%%%%%%%%%%%%%%%%%%%%%%%%%%%%%%%%%%%%%%%%%
%%%%%%%%%%%%%%%%%%%%%%%%%%%%%%%%%%%%%%%%%%%%%%%%%%%%%%%%%%%%%%%%%%%%%%%%%%%%%%%%%%%%%%%%%%%%%%%%
DIS reactions can be used to study (i) how jets exchange their energy and momentum with a nuclear environment and (ii) how the jet's particle-composition changes in the nuclear medium. The fundamental DIS-type reaction explored herein is as follows:
\begin{equation}
e^-(\ell_{\rm in})+A(\mathcal{P}) \to e^-(\ell_{\rm out})+X. 
\label{eq:DIS_reaction}
\end{equation}
where an incoming electron with momentum $\ell_{\rm in}$ collides with a nucleus $A$ with momentum $\mathcal{P}$, giving an outgoing electron with momentum $\ell_{\rm out}$ and a final state $X$. This reaction is illustrated in Fig.~\ref{fig:DIS_eA_collision}, where $X$ refers to the state on the far right-hand side of that figure. As depicted, Fig.~\ref{fig:DIS_eA_collision} is a leading order process, where all possible states $X$ are implicitly accounted for.

At next-to-leading order (NLO), the substructure of $X$ is explored more explicitly by allowing the highly off-shell quark, which has momentum $p_1$ in Fig.~\ref{fig:DIS_eA_collision}, to produce identified particles in the (perturbative) final state depicted in Fig.~\ref{fig:HT_gluon}. Figure~\ref{fig:HT_gluon} (a) gives a medium-modified gluon radiation spectrum, which is changed due to coherent single-scattering interactions with a Glauber gluon~\cite{Idilbi:2008vm}, and is the only vacuum process, of the form $q\to q+g$, that is modified in the medium. Interactions with Glauber quarks in nuclear matter shown in Figs.~\ref{fig:HT_gluon}(b)--\ref{fig:HT_gluon}(d) thus open novel channels that are not available in vacuum. The processes in Fig.~\ref{fig:HT_gluon}(b) correspond to $q\to g+g$ (through in-medium annihilation with a Glauber quark, of course), while Fig.~\ref{fig:HT_gluon}(c) corresponds either to $q\to q+\bar{q}'$ or $q\to q'+\bar{q}'$, where $q'$ may be a different quark flavor than $q$. Finally, Fig.~\ref{fig:HT_gluon}(d) corresponds to $q\to q+q'$, where $q'$ may be a different quark flavor from $q$ once more. 

The general expression for the cross section of the reaction in Eq.~\ref{eq:DIS_reaction} is 
\begin{eqnarray}
E_{\ell_{\rm out}}\frac{d^3 \sigma}{d^3\ell_{\rm out} } &=& \frac{\alpha^2_{\rm em}}{2\pi s}\frac{L_{\mu\nu} W^{\mu\nu}}{Q^4},\nonumber\\
L^{\mu\nu}&=&\frac{1}{2}{\rm Tr}\left[\slashed{\ell}_{\rm in}\gamma^\mu\slashed{\ell}_{\rm out}\gamma^\nu \right],
\label{eq:crossection_eX}
\end{eqnarray}
where $\alpha_{\rm em}=e^2/(4\pi)$ is the fine-structure constant, $s=\left(\mathcal{P}+\ell_{\rm in}\right)^2$ is the usual Mandelstam variable, $\slashed{\ell}=\ell_\mu\gamma^\mu$ with $\gamma^\mu$ are the usual Dirac matrices, while $q^\mu= \ell^\mu_{\rm out}-\ell^\mu_{\rm in}$ and gives $q^\mu q_\mu=-Q^2$. The hadronic tensor $W^{\mu\nu}$ in Eq.~\ref{eq:crossection_eX} contains the QCD portion of the interaction and is the main interest here. It is defined as a complete matrix element given by \cite{Collins:2011zzd} 
\begin{eqnarray}
W^{\mu\nu} &=& \sum_{X} \delta^{(4)}(p_{X} - \mathcal{P} -q) \langle AP, S| j^{\mu}(0)  |X \rangle \langle X | j^{\nu}(0)| AP, S\rangle, \nonumber\\
j^{\mu}(x) &=& \sum_{f} e_{f} \bar{\psi}_{f}(x) \gamma^{\mu} \psi_{f}(x),
\end{eqnarray}
where the sum over $X$ denotes the usual Lorentz-invariant sum (and integral) over all hadronic states, and $S$ is the spin state of the target. The fractional charge $e_f=2/3$ for the up, charm, and top quarks, while being $e_f=-1/3$ for the down, strange, and bottom quarks. Note that the electric unit charge $e$ has been accounted for in the $\alpha^2_{\rm em}$ term on the right-hand side of Eq.~\ref{eq:crossection_eX}, therefore neither $L_{\mu\nu}$ nor $W^{\mu\nu}$ contain any $\alpha_{\rm em}$. The light-cone coordinates are chosen (cf. Ref.~\cite{Kumar:2025egh} for details) where 
\begin{eqnarray}
q^\mu =\left[q^+,q^-,{\bf q}_\perp={\bf 0}_\perp\right] = \left[\frac{-Q^2}{2q^-},q^-,0,0\right],
\end{eqnarray}
such that the $q^+$ and $q^-$ components of the incoming virtual momentum are large, i.e., $O(1)$, giving $q^\mu\sim[O(1),O(1),{\bf 0}_\perp]Q$. Note that the hard scale is given by $Q=\sqrt{-q^2} \gg \Lambda_{\rm QCD}$. 

In this paper, we focus on the situation where the highly virtual photon carrying four-momentum $q^\mu$ strikes a nucleon traveling in the positive $z$ direction. In this setup, the struck quark has a very small $p^-\sim \lambda^2 Q$ momentum component --- where the dimensionless parameter $\lambda$ is a small quantity $\lambda^2 \llless 1$ --- while the large momentum component is $p^+\sim Q$, resulting in the four-momentum scaling: $p^\mu\sim\left[O(1),O\left(\lambda^2\right),{\bf 0}_\perp\right]Q$. The momentum components of the quark after the scattering are organized as $p^\mu_1\sim\left[O\left(\lambda^2\right),O(1),{\bf 0}_\perp\right]Q$.\footnote{Note that the mass $M$ of the struck quark can be sizable $\frac{M}{Q}\sim O(\lambda)$, and hence $p^\mu_1=\left[\frac{M^2-Q^2+2\left(p^{+}q^{-}-\frac{M^2}{2p^+}\frac{Q^2}{2q^-}\right)}{2p^{-}_1},p^-_1,{\bf 0}_\perp\right]$.} Thus, $\lambda$ is used as a small scale to establish a perturbative series expansion.

The real final-state parton radiation spectrum contained within $W^{\mu\nu}$ is divided as follows:
\begin{eqnarray}
\frac{dW^{\mu\nu}}{dy}
&=& \frac{dW^{\mu\nu}_0}{dy}+\sum^{4}_{i=1} \frac{dW^{\mu\nu}_i}{dy},\nonumber
\end{eqnarray}
where $\frac{dW^{\mu\nu}_0}{dy}$ is the vacuum contribution to gluon radiation from the jet, while the in-medium correction is enclosed is $\sum_i \frac{dW^{\mu\nu}_i}{dy}$ and is depicted in Fig.~\ref{fig:HT_gluon}. Separating the vacuum $W^{\mu\nu}_0$ from the in-medium $W^{\mu\nu}_i$ contributions gives the following:
\begin{eqnarray}
\frac{dW^{\mu\nu}_0}{dy}&=& \sum_{f}\int dx F^{A}_{_f}(x) \mathcal{H}^{\mu\nu}_0 \mathcal{K}_0, \nonumber\\
\frac{dW^{\mu\nu}_i}{dy}&=& \sum_{f} \int dx F^{A}_{_f}(x) \mathcal{H}^{\mu\nu}_0 \mathcal{K}_i,\nonumber\\
\mathcal{H}^{\mu\nu}_0&=&\frac{e^2_f}{2}(2\pi)\delta\left[\left(q+xP\right)^2\right]{\rm Tr}\left[\slashed{P}\gamma^\mu\left(\slashed{q}+x\slashed{P}\right)\gamma^\nu\right],
\label{eq:W_total}
\end{eqnarray}
where the tensor $\mathcal{H}^{\mu\nu}_0$ encodes the perturbative partonic contributions, with no emission and no in-medium scattering. The radiated gluon momentum fraction $y$ is given as $y=\ell^{-}_{2}/q^-$, where the radiated final state gluon momemtum ($\ell_2$) is traveling in the negative $z$ direction and is collinear to the direction of the final-state quark. The parton distribution function (PDF), present in both vacuum and in-medium portions of $W^{\mu\nu}$, is defined as
\begin{eqnarray}
F^{A}_{f}(x)=A\int \frac{dy^-}{2\pi} e^{-ixP^+y^-}\frac{1}{2} \left\langle P \left\vert \bar{\psi}_{_f} \left(y^-\right)\gamma^+\psi_{_f}(0) \right\vert P \right\rangle,
\end{eqnarray}
and encodes the probability of finding a quark of flavor $f$, with momentum fraction $x$, in the nucleus $A$. The momentum fraction $x$ carried by the struck quark is $x=p^+/P^+$, where $p^+$ is the first component of the quark's light-cone momentum, while $P^+$ is the corresponding momentum component of the nucleon in the nucleus. The expectation value $\langle P \vert \bar{\psi}_{f}\left(y^-\right)\gamma^+\psi_{f}(0) \vert P \rangle$ is a two-point fermionic correlator with a light-cone separation $y^-$ along negative $z$ direction.\footnote{The quark spin and color averaging factors $1/(2N_{c})$ are absorbed in the PDF.} The main objective of this contribution is to compute the kernels $\mathcal{K}_i$, where $\mathcal{K}_0$ describes the vacuum kernel, while the different $\mathcal{K}_{i=1,2,3,4}$ encode in-medium interactions described in Figs.~\ref{fig:HT_gluon}(a)--\ref{fig:HT_gluon}(d), respectively. Separate discussions of the hadronic tensor for the vacuum and medium-modified radiation are thus presented.

%%%%%%%%%%%%%%%%%%%%%%%%%%%%%%%%%%%%%%%%%%%%%%%%%%%%%%%%%%%%%%%%%%%%%%%%%%%%%%%%%%%%%%%%%%%%%%%%%%%%%%%%%%%%
%%%%%%%%%%%%%%%%%%%%%%%%%%%%%%%%%%%%%%%%%%%%%%%%%%%%%%%%%%%%%%%%%%%%%%%%%%%%%%%%%%%%%%%%%%%%%%%%%%%%%%%%%%%%
\subsection{Single gluon emission from the hard quark without in-medium scattering: The vacuum contribution}
%%%%%%%%%%%%%%%%%%%%%%%%%%%%%%%%%%%%%%%%%%%%%%%%%%%%%%%%%%%%%%%%%%%%%%%%%%%%%%%%%%%%%%%%%%%%%%%%%%%%%%%%%%%%
%%%%%%%%%%%%%%%%%%%%%%%%%%%%%%%%%%%%%%%%%%%%%%%%%%%%%%%%%%%%%%%%%%%%%%%%%%%%%%%%%%%%%%%%%%%%%%%%%%%%%%%%%%%%
\begin{figure}[!h]
    \centering
    \includegraphics[width=0.9\textwidth]{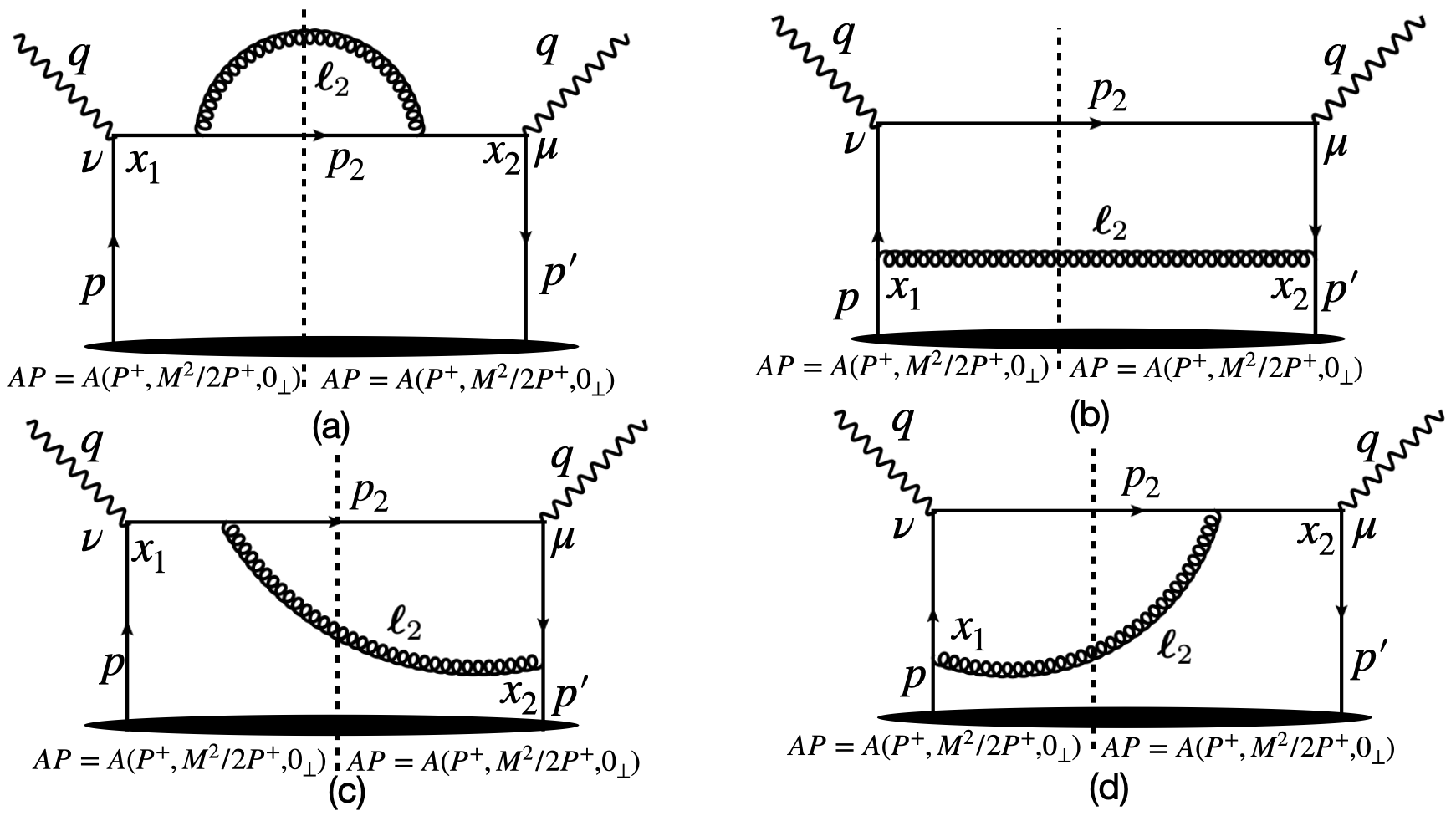}
    \caption{Forward scattering diagrams of leading order gluon production from the quark. The cut-line (i.e., dashed line) represents the final state.}
    \label{fig:W_0_vacuum}
\end{figure}

Four diagrams contribute to single-gluon emission in vacuum. In the light-cone gauge $(A^-=0)$, Fig.~\ref{fig:W_0_vacuum}(a) represents the dominant, leading log, contribution, with the remaining three diagrams being subleading.\footnote{The subleading nature of diagrams in Fig.~\ref{fig:W_0_vacuum}(b--d) is analogous to photon production from an incoming virtual quark \cite{Kumar:2025egh}.} The hadronic tensor for the diagram shown in Fig.~\ref{fig:W_0_vacuum}(a) is given as
\begin{eqnarray}
\begin{split}
W^{\mu\nu}_{0,a} & =\sum_f 2 \left[-g^{\mu\nu}_{\perp\perp}\right] e^2_f  \int d (\Delta X^{-}) e^{iq^{+}\Delta X^{-}} \left\langle AP \left| \bar{\psi}_{_f}\left(\Delta X^-\right) \frac{\gamma^{+}}{4} \psi_{_f}(0)\right| AP\right\rangle\\
  & \times g^2_s C_F C_A \int  \frac{dy}{2\pi}\frac{d^2 \ell_{2\perp}}{(2\pi)^2} e^{ -i\left\{ \frac{\pmb{\ell}^2_{2\perp} + yM^2}{2y\left(1-y\right)q^-}\right\} \Delta X^{-} } \left[\frac{1+ \left(1-y\right)^2}{y}\right]   \frac{[\pmb{\ell}^2_{2\perp} +M^2 y^4 \kappa ]}{ \left[  \pmb{\ell}^2_{2\perp}  + y^2  M^2 \right]^2},
\end{split}\label{eq:W_0_vacuum_a}
\end{eqnarray}
where $M$ is the mass of the struck quark, $C_A=N_c=3$,  $C_F=(N^2_c-1)/(2N_c)=4/3$,  while
\begin{eqnarray}
\kappa=\frac{1}{1+\left(1-y\right)^2},
\label{eq:kappa}
\end{eqnarray}
also included in Appendix~\ref{append:list_of_function} for the reader's convenience. The emitted gluon, carrying momentum $\ell_2$, is traveling in the negative $z$ direction, having momentum fraction $y=\ell^{-}_2/q^-$; in fact, $\ell^\mu_2=\left[\frac{\ell^{2}_{2\perp}}{2yq^-},yq^-,\pmb{\ell}_{2\perp}\right]$. Note that in Eq.~(\ref{eq:crossection_eX}) and Eq.~(\ref{eq:W_0_vacuum_a}), one factor of $e^2$ has already been accounted for in the hadronic tensor, giving $\alpha^2_{em}$ in Eq.~(\ref{eq:crossection_eX}). 

Any calculation of $W^{\mu\nu}$ proceeds by first obtaining the full $T$-matrix amplitude $T^{\mu\nu}$ of a given process before extracting the forward scattering amplitude (Eq. (7.49) in Ref.~\cite{Peskin:1995ev}) using
\begin{eqnarray}
W^{\mu\nu}&=&{\rm Disc}\left[T^{\mu\nu}\right].
\label{eq:T_disc}
\end{eqnarray}
The power counting of the small scale $\lambda$ within $\ell_2$ reveals $\ell^\mu_2\sim\left[O(\lambda^2),O(1),O(\lambda),O(\lambda) \right]Q$. The outgoing quark's momentum is given by $p^\mu_2=\left[\frac{\ell^2_{2\perp}+M^2}{2\left(1-y\right)q^-},\left(1-y\right)q^-,-\pmb{\ell}_{2\perp}\right]$, which scales as $p^\mu_2\sim [O(\lambda^2),O(1),O(\lambda),O(\lambda)]Q$ and admits $p^2_2=M^2\geq 0$. Having established the vacuum result, including the relevant $\lambda$ power counting, interactions with the medium are next considered. 
%%%%%%%%%%%%%%%%%%%%%%%%%%%%%%%%%%%%%%%%%%%%%%%%%%%%%%%%%%%%%%%%%%%%%%%%%%%%%%%%%%%%%%%%%%%%%%%%
%%%%%%%%%%%%%%%%%%%%%%%%%%%%%%%%%%%%%%%%%%%%%%%%%%%%%%%%%%%%%%%%%%%%%%%%%%%%%%%%%%%%%%%%%%%%%%%%
\subsection{Classification of single-scattering-induced parton radiation diagrams}
%%%%%%%%%%%%%%%%%%%%%%%%%%%%%%%%%%%%%%%%%%%%%%%%%%%%%%%%%%%%%%%%%%%%%%%%%%%%%%%%%%%%%%%%%%%%%%%%
%%%%%%%%%%%%%%%%%%%%%%%%%%%%%%%%%%%%%%%%%%%%%%%%%%%%%%%%%%%%%%%%%%%%%%%%%%%%%%%%%%%%%%%%%%%%%%%%
In this section, we consider DIS between the virtual photon and the nucleus in which the struck quark, after hard scattering, undergoes various in-medium scatterings with the nuclear environment, allowing for multiple final states. The in-medium QCD scattering kernels for a quark contributing at $O(\alpha^2_s)$ are classified using the identity of final-state particles. The first kind of kernel ($\mathcal{K}_{1}$) contains a real gluon and a quark in the final state shown in Fig.~\ref{fig:HT_gluon}(a), and the gluon spectrum is modified in the nuclear medium via single Glauber-gluon~\cite{Idilbi:2008vm} scattering herein. A total of 11 possible diagrams contribute to $\mathcal{K}_1$, which are shown in Fig.~\ref{fig:kernel-1_all}. Section \ref{sec:gluon_quark} presents the details of the calculation giving $\mathcal{K}_1$, for one of the diagrams. The remaining diagrams are in Appendix \ref{append:kernel-1}.   

The second kind of kernel $(\mathcal{K}_{2})$ represents the two real gluons emission process $q\to g+g$ where a Glauber quark is used to annihilate the incoming highly virtual quark, as depicted in Fig.~\ref{fig:HT_gluon} (b). There are a total of five possible central cut diagrams contributing to this kernel, shown in Fig.~\ref{fig:kernel-2_all}, are discussed in Appendix \ref{append:kernel-2}. 

The third kind of kernel $(\mathcal{K}_{3})$ represents both the $q\to q+\bar{q}'$ process where a Glauber antiquark is the scattering partner of the highly virtual incoming quark, as well as the $q\to q'+\bar{q}'$ where a Glauber antiquark is used to annihilate the incoming highly virtual quark. Both $q\to q+\bar{q}'$ and $q\to q'+\bar{q}'$ possibilities are depicted in Fig.~\ref{fig:kernel-3_all}, with four contributing central cut diagrams discussed in Appendix \ref{append:kernel-3}. 

The last kind of kernel $(\mathcal{K}_{4})$ represents both the $q\to q+q'$ process where a Glauber quark is the scattering partner of the highly virtual incoming quark. The two possible diagrams are depicted in Fig.~\ref{fig:kernel-4_all}, while their hadronic tensor is calculated in Appendix \ref{append:kernel-4}. 

Of course, there is nothing special about our assumption of an incoming highly virtual quark. Our results also hold for an incoming highly virtual antiquark instead, with the corresponding change conjugation applied to the Glauber quark. 

%%%%%%%%%%%%%%%%%%%%%%%%%%%%%%%%%%%%%%%%%%%%%%%%%%%%%%%%%%%%%%%%%%%%%%%%%%%%%%%%%%%%%%%%%%%%%%%%
%%%%%%%%%%%%%%%%%%%%%%%%%%%%%%%%%%%%%%%%%%%%%%%%%%%%%%%%%%%%%%%%%%%%%%%%%%%%%%%%%%%%%%%%%%%%%%%%
%%%%%%%%%%%%%%%%%%%%%%%%%%%%%%%%%%%%%%%%%%%%%%%%%%%%%%%%%%%%%%%%%%%%%%%%%%%%%%%%%%%%%%%%%%%%%%%%
\section{Single-Scattering induced emission: the one gluon and one quark final state}
\label{sec:gluon_quark}
%%%%%%%%%%%%%%%%%%%%%%%%%%%%%%%%%%%%%%%%%%%%%%%%%%%%%%%%%%%%%%%%%%%%%%%%%%%%%%%%%%%%%%%%%%%%%%%%
%%%%%%%%%%%%%%%%%%%%%%%%%%%%%%%%%%%%%%%%%%%%%%%%%%%%%%%%%%%%%%%%%%%%%%%%%%%%%%%%%%%%%%%%%%%%%%%%
%%%%%%%%%%%%%%%%%%%%%%%%%%%%%%%%%%%%%%%%%%%%%%%%%%%%%%%%%%%%%%%%%%%%%%%%%%%%%%%%%%%%%%%%%%%%%%%%
\begin{figure}[ht!]
    \centering
    \includegraphics[width=0.80\textwidth]{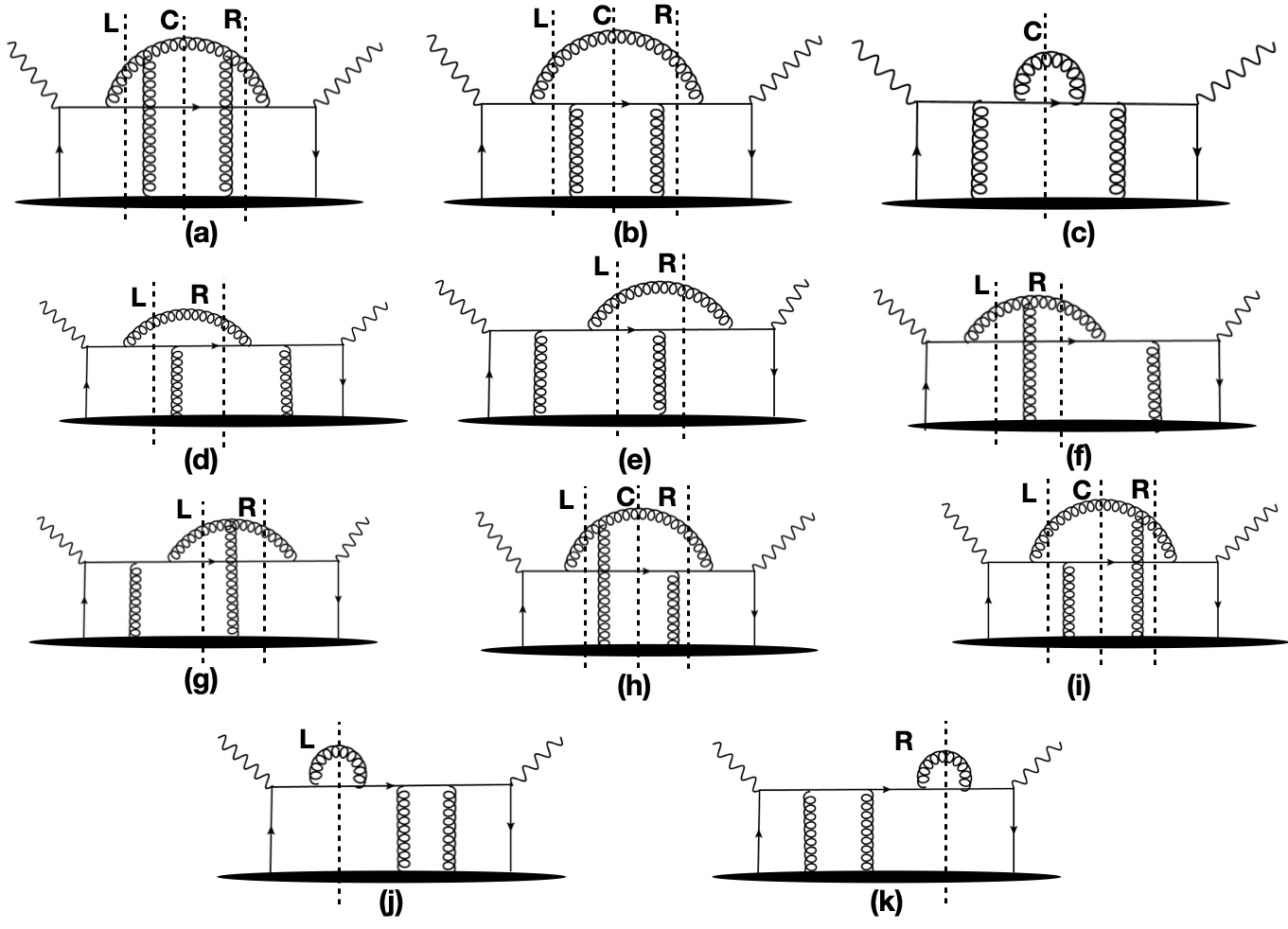}
    \caption{Diagrams for scattering kernel-1. }
    \label{fig:kernel-1_all}
\end{figure}
This section presents the calculation of all possible diagrams in which the hard quark produced in the primary hard scattering undergoes a single-gluon emission and single Glauber-gluon~\cite{Idilbi:2008vm} exchange with the nuclear medium. The diagrams are referred to as kernel-1 and are depicted in Fig.~\ref{fig:kernel-1_all}. There are a total of 23 diagrams. The calculation is performed in light-cone gauge $n\cdot A=A^-=0$, where light-cone vector $n=[1,0,0_{\perp}]$. The polarization tensor of the gluon propagator is given as
\begin{eqnarray}
d^{(X)}_{\mu\nu}=-g_{\mu\nu}+\frac{X_{\mu}n_{\nu}+n_{\mu}X_{\nu}}{n\cdot X},
\label{eq:d_X_sigma}
\end{eqnarray}
where $X$ is the gluon's momentum.

Presented below is the detailed calculation of the central-cut diagram in Fig.~\ref{fig:kernel-1_all}(a). The remaining diagrams are calculated in Appendix \ref{append:kernel-1}.

\begin{figure}[ht!]
    \centering
    \includegraphics[width=0.45\textwidth]{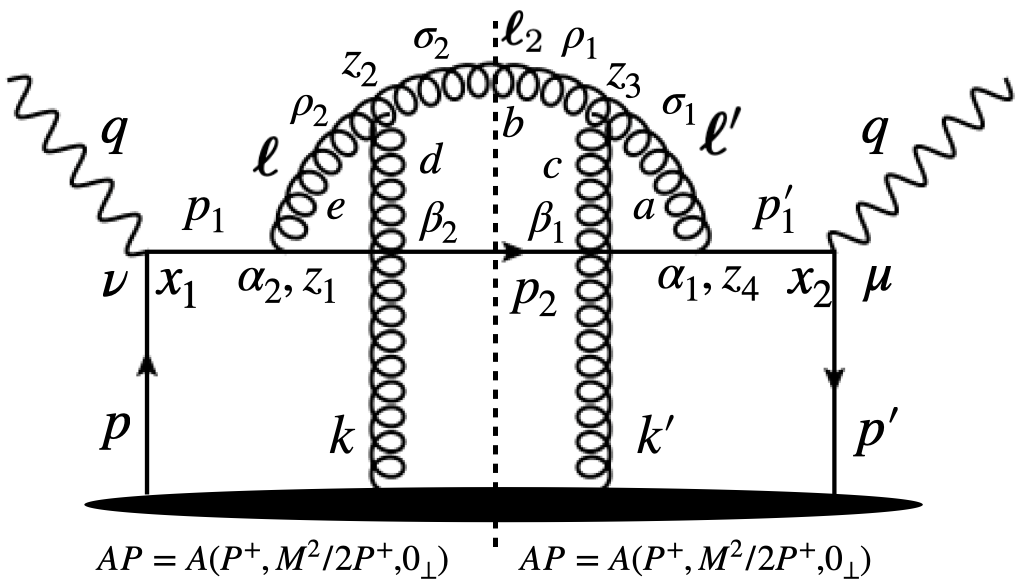}
    \caption{A forward scattering diagram in kernel-1. }
    \label{fig:kernel-1_DGG_centercut}
\end{figure}

The amplitude for the central-cut diagram in Fig.~\ref{fig:kernel-1_DGG_centercut} is given as
\begin{eqnarray}
\begin{split}
T^{\mu \nu}_{1,c} &= \sum_f e^2_f g^4_s \int d^4 x_1 d^4 x_2 d^4 z_1 d^4 z_2 d^4 z_3 d^4 z_4 \int \frac{d^{4} p_{1}}{(2 \pi)^{4}}  \frac{d^{4} p'_{1}}{(2 \pi)^{4}}  \frac{d^{4} \ell}{(2 \pi)^{4}}  \frac{d^{4} \ell'}{(2 \pi)^{4}}  \frac{d^{4} \ell_{2}}{(2 \pi)^{4}}  \frac{d^{4} p_{2}}{(2 \pi)^{4}}  \\  
& \times e^{ix_2\left(q-p'_{1}\right) }  e^{ix_1\left(p_{1}-q\right) }  e^{iz_{1}(\ell + p_{2} - p_{1})} e^{iz_{2}(\ell_2-\ell)}  e^{iz_{3}(\ell' -\ell_{2})} e^{iz_{4}(p'_{1} -\ell' - p_{2})} \\
& \times \left\langle AP\left| \mathrm{Tr} \left[  \psi_{_f}(x_1) \bar{\psi}_{_f}(x_2) \gamma^{\mu} \frac{ i\left(\slashed{p}'_{1} +M\right)}{\left[\left(p'_{1}\right)^2 -M^2 + i \epsilon\right]} it^a\gamma^{\alpha_1} 
\frac{i\left(\slashed{p}_{2} +M\right)}{\left[p^{2}_{2} -M^2 + i \epsilon\right]} it^e\gamma^{\alpha_2} \frac{i\left(\slashed{p}_{1}+M\right)}{\left[p^{2}_{1} -M^2 + i \epsilon\right]} \gamma^{\nu} \right] \right| AP \right\rangle \\
& \times if^{abc} g^{\sigma_1 \rho_1}\left(-\ell' - \ell_2\right)^{\beta_1} A^{c}_{\beta_{1}} if^{bed} g^{\sigma_2 \rho_2} \left(-\ell_2 -\ell\right)^{\beta_2} A^{d}_{\beta_{2}} \frac{id^{(\ell)}_{\alpha_2 \rho_2}}{\left(\ell^2+i\epsilon\right)} \frac{id^{(\ell_2)}_{\sigma_2 \rho_1}}{\left(\ell^2_2+i\epsilon\right)} \frac{id^{(\ell')}_{\sigma_1 \alpha_1}}{\left(\ell'^2+i\epsilon\right)},  
\end{split} \label{eq:K1_T1_center1}
\end{eqnarray}
where the trace Tr[...] is over the Dirac matrices and the Gell-Mann SU(3) color matrices. The notation $\sum_f$ represents the sum over quark flavors. We apply Cutkosky's~\cite{Cutkosky:1960sp,Peskin:1995ev} procedure to evaluate the discontinuity along the cut-line and obtain the hadronic tensor. The discontinuity associated with the gluon propagator and quark propagator is given as
\begin{eqnarray}
\mathrm{Disc} \left[ \frac{1}{\ell^2_2 +i\epsilon}\right] &=& -2\pi i \delta\left(\ell^2_2\right),\nonumber\\
\mathrm{Disc}\left[\frac{1}{p^2_2-M^2+i\epsilon}\right] &=& -2\pi i\delta(p^2_2 -M^2).
\end{eqnarray}
The resulting hadronic tensor is given by
\begin{eqnarray}
\begin{split}
W^{\mu \nu}_{1,c} &= \sum_f e^2_f g^4_s \int d^4 x_1 d^4 x_2 d^4 z_1 d^4 z_2 d^4 z_3 d^4 z_4 \int \frac{d^{4} p_{1}}{(2 \pi)^{4}}  \frac{d^{4} p'_{1}}{(2 \pi)^{4}}  \frac{d^{4} \ell}{(2 \pi)^{4}}  \frac{d^{4} \ell'}{(2 \pi)^{4}}  \frac{d^{4} \ell_{2}}{(2 \pi)^{4}}  \frac{d^{4} p_{2}}{(2 \pi)^{4}}  \\  
& \times e^{ix_2\left(q-p'_{1}\right) }  e^{ix_1\left(p_{1}-q\right) }  e^{iz_{1}(\ell + p_{2} - p_{1})} e^{iz_{2}(\ell_2-\ell)}  e^{iz_{3}(\ell' -\ell_{2})} e^{iz_{4}(p'_{1} -\ell' - p_{2})} \\
& \times \left\langle AP\left| \mathrm{Tr} \left[  \psi_{_f}(x_1) \bar{\psi}_{_f}(x_2) \gamma^{\mu} \frac{ (-i)\left(\slashed{p}'_{1}+M\right)}{\left[\left(p'_{1}\right)^{2} -M^2 - i \epsilon\right]} (-i)t^a\gamma^{\alpha_1} \left(\slashed{p}_{2} +M\right) it^e\gamma^{\alpha_2} \frac{i\left(\slashed{p}_{1}+M\right)}{\left[p^{2}_{1} -M^2 + i\epsilon\right]} \gamma^{\nu} \right] \right| AP \right\rangle \\
& \times if^{abc} g^{\sigma_1 \rho_1}\left(\ell' + \ell_2\right)^{\beta_1} A^{c}_{\beta_1} if^{bed} g^{\sigma_2 \rho_2} \left(\ell_2 +\ell\right)^{\beta_2} A^{d}_{\beta_2} \frac{id^{(\ell)}_{\alpha_2 \rho_2}}{\left(\ell^2+i\epsilon\right)} d^{(\ell_2)}_{\sigma_2 \rho_1} \frac{(-i)d^{(\ell')}_{\sigma_1 \alpha_1}}{\left[\left(\ell'\right)^2-i\epsilon\right]} (2\pi) \delta\left(p^{2}_{2} -M^2\right) (2\pi) \delta\left(\ell^2_2\right).
\end{split} \label{eq:K1_Wmunu1_center1}
\end{eqnarray}
To separate the perturbative and nonperturbative portions of this calculation, a power-counting scheme is established. The incoming quark before primary scattering is moving in a positive direction, i.e., $p=\left[p^+, M^2/2p^{+},\pmb{0}_{\perp}\right]$, and thus $p^\mu\sim\left[O(1),O\left(\lambda\right),\pmb{0}_\perp\right]Q$. The same $\lambda$ scales also hold for $p'$ since $p'^+/Q\sim 1$ and $p'=\left[p'^{+}, M^{2}/2p'^{+}, \pmb{0}_{\perp}\right]$. Isolating the leading nonperturbative component, which is $\psi(x_1) \bar{\psi}(x_2)$ for the first scattering correlator herein, allows to write $\psi(x_1)\bar{\psi}(x_2)$ in terms of a scalar function $T(x_1,x_2)$:
\begin{equation}
\begin{split}
& \psi(x_1) \bar{\psi}(x_2)=\slashed{p}T(x_1,x_2)=p^+\gamma^- T(x_1,x_2)\Longrightarrow {\rm Tr}[\gamma^+\psi(x_1)\bar{\psi}(x_2)]=p^+{\rm Tr}[\gamma^+\gamma^-] T(x_1,x_2) \\ & \Longrightarrow \psi(x_1)\bar{\psi}(x_2)=\gamma^-{\rm Tr}\left[\bar{\psi}(x_2)\frac{\gamma^+}{4}\psi(x_1)\right].
\end{split}
\label{eq:psi_psi_bar}
\end{equation}
In the light-cone gauge $A^{-}=0$, the Glauber gluon~\cite{Idilbi:2008vm} emanating from the medium has $A^+ \ggg A_\perp$, and thus $(\ell' + \ell_2)^{\beta_1} A^{c}_{\beta_1}(z_3)\approx (\ell'^{-} + \ell_2^{-}) A^{c+}(z_3)$ and $(\ell_2+\ell)^{\beta_2} A^{d}_{\beta_2}(z_2)\approx (\ell^{-}_2+\ell^{-}) A^{d+}(z_2)$. In addition, we assume that the hard quark produced from the primary hard scattering with the nucleon struck by the virtual photon will undergo further rescatterings while traversing the remaining $A-1$ nucleons. As in $Q\ggg 1$, any scatterings after the first one are assumed to be independent, and thus the correlators are factorized,
\begin{equation}
\left\langle AP \left| {\rm Tr}\left[\bar{\psi}_{_f}(x_2)\frac{\gamma^+}{4}\psi_{_f}(x_1)\right]A^{c+}(z_3)A^{d+}(z_2)\right| AP \right\rangle \approx
\left\langle P \left\vert {\rm Tr}\left[\bar{\psi}_{_f}\left(x_2\right)\frac{\gamma^+}{4}\psi_{_f}\left(x_1\right)\right]\right\vert P \right\rangle \left\langle P_{A-1} \left| A^{c+}\left(z_3\right)A^{d+}\left(z_2\right)\right|P_{A-1}\right\rangle,
\label{eq:transport_coeff}
\end{equation}
where the first term will be absorbed in the definition of the nuclear parton distribution function, while the second term will be included in the scattering kernel. 
The color trace can be simplified as
\begin{equation}
i^2 f^{bed} f^{abc} {\rm Tr}[t^{a}t^{e}] = - \frac{f^{bad} f^{abc}}{2} =      \frac{f^{dab} f^{cab}}{2} = C_{A}\frac{\delta^{dc}}{2} = C_{A} {\rm Tr}\left[t^d t^c\right], 
\end{equation}
where $C_A=N_c$. The resulting $W^{\mu\nu}$ is given as
\begin{eqnarray}
\begin{split}
W^{\mu \nu}_{1,c} &= C_{A} \sum_f e^2_f g^4_s \int d^4 x_1 d^4 x_2 d^4 z_1 d^4 z_2 d^4 z_3 d^4 z_4 \int \frac{d^{4} p_{1}}{(2 \pi)^{4}}  \frac{d^{4} p'_{1}}{(2 \pi)^{4}}  \frac{d^{4} \ell}{(2 \pi)^{4}}  \frac{d^{4} \ell'}{(2 \pi)^{4}}  \frac{d^{4} \ell_{2}}{(2 \pi)^{4}}  \frac{d^{4} p_{2}}{(2 \pi)^{4}}  \left\langle P \left| \bar{\psi}_{_f}(x_2) \frac{\gamma^{+}}{4} \psi_{_f}(x_1)\right| P\right\rangle \\  
& \times e^{ix_2\left(q-p'_{1}\right) }  e^{ix_1\left(p_{1}-q\right) }  e^{iz_{1}(\ell + p_{2} - p_{1})} e^{iz_{2}(\ell_2-\ell)}  e^{iz_{3}(\ell' -\ell_{2})} e^{iz_{4}(p'_{1} -\ell' - p_{2})} 
\left\langle P_{A-1} \left\vert {\rm Tr}\left[A^{+}\left(z_3\right)A^+\left(z_2\right)\right] \right\vert P_{A-1} \right\rangle \\
& \times \mathrm{Tr} \left[  \gamma^{-} \gamma^{\mu} \frac{ \left(\slashed{p}'_{1}+M\right)}{\left[\left(p'_{1}\right)^{2} -M^2 - i \epsilon\right]} \gamma^{\alpha_1} \left(\slashed{p}_{2} +M\right) \gamma^{\alpha_2} \frac{\left(\slashed{p}_{1}+M\right)}{\left[p^{2}_{1} -M^2 + i\epsilon\right]} \gamma^{\nu} \right]  \\
& \times  g^{\sigma_1 \rho_1}\left(\ell'^{-} + \ell_2^{-}\right)  g^{\sigma_2 \rho_2} \left(\ell^{-}_2 +\ell^{-}\right)  \frac{d^{(\ell)}_{\alpha_2 \rho_2}}{\left(\ell^2+i\epsilon\right)} d^{(\ell_2)}_{\sigma_2 \rho_1} \frac{d^{(\ell')}_{\sigma_1 \alpha_1}}{\left[\left(\ell'\right)^2-i\epsilon\right]} (2\pi) \delta\left(p^{2}_{2} -M^2\right) (2\pi) \delta\left(\ell^2_2\right).
\end{split}
%\label{eq:}
\end{eqnarray}

After performing the change of variable $p'_1=q+p'$ and $p_1=q+p$, which stems from energy and momentum conservation (see Fig.~\ref{fig:kernel-1_DGG_centercut}), the integration measure $d^4 p'_1$ transforms as $d^4 p'_1\to d^4 p'$, while $d^4 p_1\to d^4 p$.
We also perform the integrals over $d^4z_1$ and $d^4z_4$ gives  
$(2 \pi)^{4} \delta^{(4)}\left(-q-p+\ell+p_{2}\right) (2 \pi)^{4} \delta^{(4)}\left(q+p'-\ell'-p_{2}\right)$,
which allows for the $d^4 \ell$ and $d^4 \ell'$ integration. The $W^{\mu \nu}_{1,c}$ becomes
\begin{eqnarray}
\begin{split}
W^{\mu \nu}_{1,c} &= C_{A} \sum_f e^2_f g^4_s \int d^4 x_1 d^4 x_2  d^4 z_2 d^4 z_3  \int \frac{d^{4} p}{(2 \pi)^{4}}  \frac{d^{4} p'}{(2 \pi)^{4}}   \frac{d^{4} \ell_{2}}{(2 \pi)^{4}}  \frac{d^{4} p_{2}}{(2 \pi)^{4}}  \left\langle P \left| \bar{\psi}_{_f}(x_2) \frac{\gamma^{+}}{4} \psi_{_f}(x_1)\right| P\right\rangle \\  
& \times e^{-ix_2p' }  e^{ix_1p }   e^{iz_{2}(\ell_2 +p_2-q-p)}  e^{iz_{3}(q+p'-p_2 -\ell_{2})}  \left\langle P_{A-1} \left\vert {\rm Tr}\left[A^{+}\left(z_3\right)A^+\left(z_2\right)\right] \right\vert P_{A-1} \right\rangle \\
& \times \mathrm{Tr} \left[  \gamma^{-} \gamma^{\mu} \frac{ \left(\slashed{q}+\slashed{p}'+M\right)}{\left[\left(q+p'\right)^{2} -M^2 - i \epsilon\right]} \gamma^{\alpha_1} \left(\slashed{p}_{2} +M\right) \gamma^{\alpha_2} \frac{\left(\slashed{q}+\slashed{p}+M\right)}{\left[\left(q+p\right)^{2} -M^2 + i\epsilon\right]} \gamma^{\nu} \right]  \\
& \times  g^{\sigma_1 \rho_1} g^{\sigma_2 \rho_2}  \left(q^{-} -p^{-}_{2} + \ell_2^{-}\right)^2     \frac{d^{\left(q+p-p_2\right)}_{\alpha_2 \rho_2}}{\left[\left(q+p-p_2\right)^2+i\epsilon\right]} d^{\left(\ell_2\right)}_{\sigma_2 \rho_1} \frac{d^{\left(q+p'-p_2\right)}_{\sigma_1 \alpha_1}}{\left[\left(q+p'-p_2\right)^2-i\epsilon\right]} (2\pi) \delta\left(p^{2}_{2} -M^2\right) (2\pi) \delta\left(\ell^2_2\right).
\end{split}
%\label{eq:}
\end{eqnarray}
Introducing new variables for distances
\begin{eqnarray}
\Delta x &=& x_2 - x_1,\quad x = \frac{ x_2 + x_1 }{2}, \nonumber\\
\Delta z &=& z_3 - z_2,\quad z = \frac{ z_3 + z_2 }{2},
\label{eq:X_Delta_X_zeta}
\end{eqnarray}
leaves the integration measure unchanged, namely $d^4 x_2 d^4 x_1 = d^4 x d^4 (\Delta x)$ and $d^4 z_3 d^4 z_2 = d^4 z d^4 (\Delta z)$. Moreover, that the two-point correlator $\langle P| \bar{\psi}_{_f}(x+\Delta x/2) \gamma^{+} \psi_{_f}(x-\Delta x/2) |P\rangle$ is invariant under translation by four-vector $x$. This is primarily true owing to the fact that the incoming state and the outgoing state are identical. Thus, the hadronic tensor becomes
\begin{eqnarray}
\begin{split}
W^{\mu \nu}_{1,c} &= C_{A} \sum_f e^2_f g^4_s \int d^4 x d^4 (\Delta x)  d^4 z d^4 (\Delta z)  \int \frac{d^{4} p}{(2 \pi)^{4}}  \frac{d^{4} p'}{(2 \pi)^{4}}   \frac{d^{4} \ell_{2}}{(2 \pi)^{4}}  \frac{d^{4} p_{2}}{(2 \pi)^{4}}  \left\langle P \left| \bar{\psi}_{_f}\left(\Delta x\right) \frac{\gamma^{+}}{4} \psi_{_f}\left(0\right)\right| P\right\rangle \\  
& \times e^{-i\left(x+\frac{\Delta x}{2}\right)p' }  e^{i\left(x-\frac{\Delta x}{2}\right)p }   e^{i\left(z-\frac{\Delta z}{2}\right)\left(\ell_2 +p_2-q-p\right)}  e^{i\left(z+\frac{\Delta z}{2}\right)\left(q+p'-p_2 -\ell_{2}\right)}  \left\langle P_{A-1} \left|{\rm Tr}\left[A^{+}\left(z+\frac{\Delta z}{2}\right)A^+\left(z-\frac{\Delta z}{2}\right)\right] \right| P_{A-1} \right\rangle \\
& \times \mathrm{Tr} \left[  \gamma^{-} \gamma^{\mu} \frac{ \left(\slashed{q}+\slashed{p}'+M\right)}{\left[\left(q+p'\right)^{2} -M^2 - i \epsilon\right]} \gamma^{\alpha_1} \left(\slashed{p}_{2} +M\right) \gamma^{\alpha_2} \frac{\left(\slashed{q}+\slashed{p}+M\right)}{\left[\left(q+p\right)^{2} -M^2 + i\epsilon\right]} \gamma^{\nu} \right]  \\
& \times  g^{\sigma_1 \rho_1} g^{\sigma_2 \rho_2}  \left(q^{-} -p^{-}_{2} + \ell_2^{-}\right)^2     \frac{d^{\left(q+p-p_2\right)}_{\alpha_2 \rho_2}}{\left[\left(q+p-p_2\right)^2+i\epsilon\right]} d^{\left(\ell_2\right)}_{\sigma_2 \rho_1} \frac{d^{\left(q+p'-p_2\right)}_{\sigma_1 \alpha_1}}{\left[\left(q+p'-p_2\right)^2-i\epsilon\right]} (2\pi) \delta\left(p^{2}_{2} -M^2\right) (2\pi) \delta\left(\ell^2_2\right).
\end{split}
\label{eq:K1_Wmunu_x_dx}
\end{eqnarray}
Equation~(\ref{eq:K1_Wmunu_x_dx}) becomes singular when the denominator of the propagator for $p_1$, $\ell$, $\ell'$ and $p'_1$ vanishes. Computing this integral is easiest in the complex plane of $p^{+}$ and $p'^{+}$, where both $p^{+}$ and $p'^{+}$ have two simple poles.\footnote{One of the propagators takes the form 
\begin{eqnarray}
\left[\left(q+p\right)^{2} - M^{2} + i \epsilon\right]^{-1}&=&\left[2\left(q^{+}+p^+\right)\left(q^{-}+p^-\right)-\left\vert\pmb{q}_\perp+\pmb{p}_\perp\right\vert^2-M^2+i\epsilon\right]^{-1} 
\approx \left[2q^{-}\left(q^{+}+p^+\right)\left[1+O\left(\lambda^2\right)\right]-M^2+i\epsilon\right]^{-1} \nonumber \\ 
\Rightarrow \left[\left(q+p\right)^{2} - M^{2} + i \epsilon\right]^{-1} &\approx& \frac{1}{2q^-}\left[q^{+}+p^{+}-\frac{M^2}{2q^-}+i\epsilon\right]^{-1}, 
\end{eqnarray}
where the established power counting $p^-/q^-\sim \lambda^2$ together with $\pmb{p}_\perp=\pmb{0}_\perp$ was used to simplify the full propagator to the expression above. A similar procedure is used for $\left[\left(q+p-p_{2}\right)^2+i\epsilon\right]$.
} The contour integration for $p^{+}$ can be carried out as
\begin{equation}
\begin{split}
C_{1} & = \oint \frac{dp^{+}}{(2\pi)} \frac{e^{ip^{+}(x^{-}_{1}-z^{-}_{2})}}{\left[\left(q+p\right)^{2} - M^{2} +i\epsilon\right]\left[\left(q+p-p_{2}\right)^{2} + i\epsilon\right]} \\
      & = \oint \frac{dp^{+}}{(2\pi)} \frac{e^{ip^{+}(x^{-}_{1} - z^{-}_{2}) }}{2q^{-}\left[q^{+} + p^{+} - \frac{M^2}{2q^{-}} + i \epsilon\right] 2\left(q^{-} -p^{-}_{2}\right)\left[q^{+} + p^{+} -p^{+}_{2} - \frac{\pmb{p}^{2}_{2\perp}}{2(q^{-}-p^{-}_{2})} + i\epsilon\right]}   \\
      & = \frac{(2\pi i)}{2\pi} \frac{\theta(x^{-}_{1} - z^{-}_{2})}{4q^{-}(q^{-}-p^{-}_{2})} \left[\frac{e^{i\left(-q^{+} + \frac{M^2}{2q^-}\right)\left(x^{-}_{1}-z^{-}_2\right)}}{\left\{\frac{M^2}{2q^-} -p^+_2 -\frac{\pmb{p}^2_{2\perp}}{2(q^{-}-p^{-}_{2})}\right\}}+
      \frac{e^{i\left(-q^{+}+p^+_2 + \frac{\pmb{p}^2_{2\perp}}{2\left(q^{-}-p^-_2\right)}\right)\left(x^{-}_{1}-z^-_2\right)}}{ \left\{p^+_2 +\frac{\pmb{p}^2_{2\perp}}{2\left(q^{-}-p^-_2\right)}-\frac{M^2}{2q^-}\right\}}\right] \\
      & = \frac{(2\pi i)}{2\pi} \frac{\theta(x^{-}_{1} - z^{-}_{2})}{4q^{-}(q^{-}-p^{-}_{2})}  e^{i\left(-q^{+}+\frac{M^2}{2q^-}\right)\left(x^{-}_{1}-z^-_2\right)}\left[ \frac{ -1 + e^{i \mathcal{G}^{(p_2)}_M(x^{-}_{1} - z^{-}_2)} }{ \mathcal{G}^{(p_2)}_M}  \right],
\end{split}
\label{eq:C_1_in_W1}
\end{equation}
where
\begin{equation}
\mathcal{G}^{(p_2)}_M = p^{+}_{2} + \frac{\pmb{p}^{2}_{2\perp}}{2(q^{-}-p^{-}_{2})} - \frac{M^2}{2q^{-}}, %= \frac{\pmb{\ell}^2_{2\perp} + y^2M^2}{2y(1-y)q^-}.
\end{equation}
which is also present in Appendix~\ref{append:list_of_function} for convenience. The contour integration for $p'^+$ proceeds analogously, giving
\begin{equation}
\begin{split}
C_{2} & = \oint \frac{dp'^{+}}{(2\pi)} \frac{e^{-ip'^{+}(x^{-}_{2}-z^{-}_{3})}}{\left[\left(q+p'\right)^{2} - M^{2} - i \epsilon\right]\left[\left(q+p'-p_{2}\right)^{2}-i\epsilon \right]} \\
      & = \oint \frac{dp'^{+}}{(2\pi)} \frac{e^{-ip'^{+}(x^{-}_{2} - z^{-}_{3}) }}{2q^{-}\left[q^{+} + p'^{+} - \frac{M^2}{2q^{-}} - i \epsilon\right] 2(q^{-} -p^{-}_{2} )\left[q^{+} + p'^{+} -p^{+}_{2} - \frac{\pmb{p}^{2}_{2\perp}}{2(q^{-}-p^{-}_{2})} - i\epsilon\right]}  \\
      & = \frac{(-2\pi i)}{2\pi} \frac{\theta\left(x^{-}_{2} - z^{-}_{3}\right)}{4q^{-}(q^{-}-p^{-}_{2})} e^{-i\left(-q^{+} + \frac{M^2}{2q^-}\right)\left(x^{-}_{2}-z^{-}_{3}\right)}\left[\frac{ -1 + e^{-i\mathcal{G}^{(p_2)}_M\left(x^{-}_{2}-z^{-}_3\right)}}{\mathcal{G}^{(p_2)}_M}\right].
\end{split}
%\label{eq:}
\end{equation}

As the final expression for $C_1$ and $C_2$ is independent of $p$ and $p'$, respectively, the dependence on these variables in Eq.~(\ref{eq:K1_Wmunu_x_dx}) remains within $e^{ip(x_{1}-z_2)}$ and $e^{ip'(z_3-x_2)}$ as well as the trace over $\gamma$-matrices. While our $\lambda$-power counting scheme constrains the size of momentum variables, the same cannot be said about position variables. Thus, the $e^{ip(x_{1}-z_2)}$ and $e^{ip'(z_3-x_2)}$ phase factors must remain intact. As the trace in Eq.~(\ref{eq:K1_Wmunu_x_dx}) only contributes at $O(\lambda^2)$ in $p$ and $p'$, the only nontrivial contribution remaining to the $p$ and $p'$ integrals stems solely from $e^{ip(x_1-z_2)}$ and $e^{ip'(z_3-x_2)}$ phase factors. To perform the remaining integrals for $p$ and $p'$, the following substitutions are used $p=\left[p^+,p^{-},\pmb{0}_\perp\right]=\left[p^+,\frac{M^2}{2p^{+}}+ \delta p^-, \pmb{0}_{\perp}\right]$ and $p'=\left[p'^{+},p'^{-},\pmb{0}_\perp\right] = \left[p'^{+},\frac{M^2}{2p'^{+}} +\delta p'^{-},\pmb{0}_{\perp}\right]$, where $\delta p^-\sim O\left(\lambda^2\right)$ and $\delta p'^-\sim O\left(\lambda^2\right)$. Thus, the integrals over $dp^- d^2 p_{\perp} d p'^{-} d^2p'_{\perp}$ simply become integrals over $d(\delta p^-) d^2 p_{\perp} d(\delta p'^{-}) d^2p'_{\perp}$ yielding 
\begin{eqnarray}
(2\pi)^6 \delta\left(x^{+}-\frac{\Delta x^{+}}{2} - z^{+} + \frac{\Delta z^{+}}{2} \right) \delta^{(2)}\left(\pmb{x}_{\perp} - \frac{\Delta \pmb{x}_{\perp}}{2} - \pmb{z}_{\perp} + \frac{\Delta\pmb{z}_{\perp}}{2} \right)\nonumber\\ 
\times\delta\left(-x^{+}-\frac{\Delta x^{+}}{2} + z^{+} + \frac{\Delta z^{+}}{2} \right) \delta^{(2)}\left(-\pmb{x}_{\perp} - \frac{\Delta \pmb{x}_{\perp}}{2} + \pmb{z}_{\perp} + \frac{\Delta\pmb{z}_{\perp}}{2} \right).
\label{eq:spacetime_delta}
\end{eqnarray}
Performing the integral over spacetime variables $(x^{+}, \pmb{x}_{\perp})$ and $(\Delta x^+,\Delta\pmb{x}_{\perp})$ using $\delta$-functions in Eq.~(\ref{eq:spacetime_delta}) yields
\begin{eqnarray}
\begin{split}
W^{\mu \nu}_{1,c} &= C_{A} \sum_f e^2_f g^4_s \int d x^{-} d (\Delta x^{-})  d^4 z d^4 (\Delta z)  \int   \frac{d^{4} \ell_{2}}{(2 \pi)^{4}}  \frac{d^{4} p_{2}}{(2 \pi)^{4}}  \left\langle P \left| \bar{\psi}_{_f}\left(\Delta x^{-}, \Delta z^{+}, \Delta \pmb{z}_{\perp} \right) \frac{\gamma^{+}}{4} \psi_{_f}\left(0\right)\right| P\right\rangle \\  
& \times e^{i\left(q^{+}-\frac{M^2}{2q^{-}}\right)\left(\Delta x^{-} -\Delta z^{-} \right) }     e^{i\Delta z (q-p_2 -\ell_{2})} 
\left[ -1 + e^{i\mathcal{G}^{(p_2)}_{M}\left(x^{-}-z^{-} -\frac{\Delta x^-}{2} + \frac{\Delta z^-}{2}\right)} \right] \left[-1 + e^{-i\mathcal{G}^{(p_2)}_{M}\left(x^{-}-z^{-} +\frac{\Delta x^-}{2} - \frac{\Delta z^-}{2}\right)} \right]
\\
& \times \theta\left(x^{-}-z^{-} -\frac{\Delta x^-}{2} + \frac{\Delta z^-}{2}\right)\theta\left(x^{-}-z^{-} +\frac{\Delta x^-}{2} - \frac{\Delta z^-}{2}\right)\frac{1}{\left[4q^{-}\left(q^{-}-p^{-}_{2}\right)\right]^2} \left[\mathcal{G}^{(p_2)}_{M}\right]^{-2}\\
& \times \mathrm{Tr} \left[ \gamma^{-} \gamma^{\mu}  \left(\slashed{q}+\slashed{p}'+M\right) \gamma^{\alpha_1} \left(\slashed{p}_{2} +M\right) \gamma^{\alpha_2} \left(\slashed{q}+\slashed{p}+M\right) \gamma^{\nu} \right]  \left\langle P_{A-1} \left| {\rm Tr}\left[A^{+}\left(z + \Delta z/2\right)A^+\left(z-\Delta z/2\right)\right]\right| P_{A-1}\right\rangle \\
& \times  g^{\sigma_1 \rho_1} g^{\sigma_2 \rho_2}  \left(q^{-} -p^{-}_{2} + \ell_2^{-}\right)^2 d^{(q+p-p_2)}_{\alpha_2 \rho_2} d^{(\ell_2)}_{\sigma_2 \rho_1} d^{(q+p'-p_2)}_{\sigma_1 \alpha_1} (2\pi) \delta\left(p^{2}_{2} -M^2\right) (2\pi) \delta\left(\ell^2_2\right).
\end{split}
\label{eq:k1_intermediate_1}
\end{eqnarray}

The next step is to perform $d\ell^{+}_2$ and $dp^+_2$ using the $\delta$-functions $\delta(\ell^2_2)$ and $\delta(p^2_2 -M^2)$ and $\lambda$-power counting. Indeed
\begin{eqnarray}
\delta\left(\ell^2_2\right)\nonumber&=&\delta\left(2\ell^+_2\ell^-_2 -\pmb{\ell}^2_{2\perp}\right)=\frac{1}{2\ell^-_2} \delta\left(\ell^+_2-\frac{\pmb{\ell}^2_{2\perp}}{2\ell^-_2}\right),\nonumber\\
\delta(p^2_2 -M^2)&=&\delta\left(2p^+_2 p^-_2 -\pmb{p}^2_{2\perp}-M^2\right)=\frac{1}{2p^-_2}\delta\left(p^+_2 -\frac{\pmb{p}^2_{2\perp} +M^2}{2p^-_2} \right).
\label{eq:kernel1_deltafunction_cut-line}
\end{eqnarray}
Defining the momentum fraction $y$ as $\ell^{-}_{2}=yq^{-}$, allows us to rewrite $d\ell^{-}_{2} = q^{-} dy$. Furthermore, energy and momentum conservation in Fig.~\ref{fig:kernel-1_DGG_centercut} implies that 
\begin{equation}
q+p=p_1=\ell_2+\ell=\ell_2+(p_2-k) \Longleftrightarrow  q+p-\ell_2-p_2+k=0.
\label{eq:lpq_4_mom_cons}
\end{equation}
While the $\delta$-functions can be used to perform the $\ell^+_2$ and $p^+_2$ integrals, the $p^-_2$ integral can be performed using $\lambda$-power counting. Indeed, as $k^\mu\sim\left[O\left(\lambda^2\right),O\left(\lambda^2\right),O(\lambda),O(\lambda)\right]Q$, while $\ell^\mu_2\sim\left[O\left(\lambda^2\right),O(1),O(\lambda),O(\lambda) \right]Q$ and $p^\mu_2\sim \left[O\left(\lambda^2\right),O(1),O(\lambda),O(\lambda)\right]Q$, using energy and momentum conservation implies
\begin{eqnarray}
0&=&q^{-}+p^{-}-\ell^{-}_{2}-p^{-}_{2} + k^{-}, \nonumber\\
0&=&q^-+O\left(\lambda^2\right)-\ell^-_2-p^-_2+O(\lambda),
\label{eq:kernel1-l2_p2_k}
\end{eqnarray}
and thus the following change of variable $p^-_2=q^{-} - \ell^{-}_{2} + k^{-} + \delta p^{-}_{2}$, where $\delta p^-_2\sim O\left(\lambda^2\right)$ is a small quantity, induces a change in the integration measure $dp^{-}_{2} = d\left(\delta p^{-}_{2}\right)$. Thus, the integration over $dp^{-}_{2}$ yields a $\delta(\Delta z^{+})$, as the only function in Eq.~(\ref{eq:k1_intermediate_1}) that is not small is $e^{-ip^{-}_{2}(\Delta z^+)}$, since $\Delta z^+$ is not subjet to the power counting in $\lambda$. Any other dependence on $p^-_2$ seen in Eq.~(\ref{eq:k1_intermediate_1}) can simply be set to $q^{-}-\ell^-_2+k^-$. Defining 
\begin{eqnarray}
    \eta = \frac{k^-}{yq^-}
\end{eqnarray}
(see also Appendix~\ref{append:list_of_function}) and applying the following transformation $\pmb{p}_{2\perp} + \pmb{\ell}_{2\perp} = \pmb{k}_{\perp}$, allows us to express $d^{2}p_{2\perp} \to d^{2}k_{2\perp}$, for a fixed $\ell_2$, thus giving
\begin{eqnarray}
\begin{split}
W^{\mu \nu}_{1,c} &= C_{A} \sum_f e^2_f g^4_s \int d x^{-} d (\Delta x^{-})  d^4 z d^4 (\Delta z)  \int   \frac{dy d^2  \ell_{2\perp}}{(2 \pi)^{3}}  \frac{d^{2} k_{\perp}}{(2 \pi)^{2}} \delta\left(\Delta z^+\right) e^{-i\Delta z^+k^-}\left\langle P \left| \bar{\psi}_{_f}\left(\Delta x^{-}, \Delta z^{+}, \Delta \pmb{z}_{\perp} \right) \frac{\gamma^{+}}{4} \psi_{_f}\left(0\right)\right| P\right\rangle \\  
& \times e^{i\left(q^{+}-\frac{M^2}{2q^{-}}\right)\Delta x^{-}}  
\left[ -1 + e^{i\mathcal{G}^{(p_2)}_{M}\left(x^{-}-z^{-} -\frac{\Delta x^-}{2} + \frac{\Delta z^-}{2}\right)} \right] 
\left[-1 + e^{-i\mathcal{G}^{(p_2)}_{M}\left(x^{-}-z^{-} +\frac{\Delta x^-}{2} - \frac{\Delta z^-}{2}\right)} \right] e^{-i\Delta z^{-}\mathcal{H}^{(\ell_2,p_2)}_M} e^{i\pmb{k}_{\perp}\cdot \pmb{\Delta z}_{\perp}} 
\\
& \times \theta\left(x^{-}-z^{-} -\frac{\Delta x^-}{2} + \frac{\Delta z^-}{2}\right) \theta\left(x^{-}-z^{-} +\frac{\Delta x^-}{2} - \frac{\Delta z^-}{2}\right) \frac{1}{\left[4q^{-}\left(q^{-}-p^{-}_{2}\right)\right]^2} \left[\mathcal{G}^{(p_2)}_{M}\right]^{-2} \frac{q^-}{2yq^{-} 2 \left(1-y+\eta y\right)q^-}\\
& \times \mathrm{Tr} \left[\gamma^{-} \gamma^{\mu} \left(\slashed{q}+\slashed{p}'+M\right) \gamma^{\alpha_1} \left(\slashed{p}_{2} +M\right) \gamma^{\alpha_2} \left(\slashed{q}+\slashed{p}+M\right) \gamma^{\nu} \right]\left\langle P_{A-1}\left|{\rm Tr}\left[A^{+}\left(z+\Delta z/2\right)A^+\left(z-\Delta z/2\right)\right] \right| P_{A-1}\right\rangle \\
& \times  g^{\sigma_1 \rho_1} g^{\sigma_2 \rho_2}  \left(2yq^{-} - k^{-}\right)^2 d^{\left(q+p-p_2\right)}_{\alpha_2 \rho_2} d^{\left(\ell_2\right)}_{\sigma_2 \rho_1} d^{\left(q+p'-p_2\right)}_{\sigma_1 \alpha_1}.
\end{split}
\label{eq:k1_y_l2_kperp_2}
\end{eqnarray}
Note that the two-point gauge field operator $ \langle P_{A-1}|{\rm Tr}[A^+(z+\Delta z/2) A^+(z-\Delta z/2)] |P_{A-1}\rangle$  is invariant under translation by four-vector $z$. This is primarily true owing to the fact that the incoming state $|P_{A-1}\rangle$ and the outgoing state $\langle P_{A-1}|$ are identical. Therefore, any $z$ dependence seen in the operator expectation value is not physical. The phases that depend on the relative distances $\Delta x^-=x^{-}_{2}-x^{-}_{1}$, such as $e^{i\left(q^{+} - \frac{M^2}{2q^-}\right)\Delta x^{-}}$, are absorbed in the definition of the quark PDF, while the phases $e^{-i\Delta z^{-}\mathcal{H}^{(\ell_2,p_2)}_M} e^{i\pmb{k}_{\perp}\cdot \pmb{\Delta z}_{\perp}}$ are included within the distribution function of the nuclear medium.

Next, we consider the following quantity that appears  in the second line of Eq.~(\ref{eq:k1_y_l2_kperp_2}):
\begin{eqnarray}
    \begin{split}
    \mathcal{R}  & = \left[ -1 + e^{i\mathcal{G}^{\left(p_2\right)}_{M}\left(x^{-}-z^{-} -\frac{\Delta x^-}{2} + \frac{\Delta z^-}{2}\right)} \right] \left[-1 + e^{-i\mathcal{G}^{\left(p_2\right)}_{M}\left(x^{-}-z^{-} +\frac{\Delta x^-}{2} - \frac{\Delta z^-}{2}\right)} \right] \\
    & \Longrightarrow \mathcal{R}  \approx  \left[ 2 -2 \cos\left\{\mathcal{G}^{\left(p_2\right)}_{M}\left(x^{-}-z^{-}\right)\right\}\right].
\end{split}
\label{eq:2_2cos_k1}
\end{eqnarray}
The relation in Eq.~(\ref{eq:2_2cos_k1}) assumes that the phase factors spanning the size of a nucleon are negligible compared to those spanning the large nucleus. Indeed, the distances $\Delta x^{-}$ and $\Delta z^{-}$ solely span the size of the nucleon, whereas the distance $x^{-}-z^{-}$ spans the size of the nucleus. In Fig.~\ref{fig:kernel-1_DGG_centercut}, it is assumed that the primary and secondary scatterings both happen in different nucleons inside the nucleus, and the two nucleons participating in the scattering are identical to the two nucleons in the complex-conjugate side. Keeping terms in $\mathcal{R}$ arising from $\Delta x^{-}$ and $\Delta z^{-}$ leads to subleading corrections that are ignored here. This entails that $x^{-}_{2} - z^{-}_{3} \approx x^{-}_{1} - z^{-}_{2}$, where $x^{-}_{1} - z^{-}_{2}$ represents the distance between first scattering and second scattering on the amplitude side, while $x^{-}_{2} - z^{-}_{3}$ represents the same distance on the complex-conjugate side. It is a standard practice~\cite{Deng:2009ncl,Sirimanna:2021sqx} to ignore the fluctuations from $\Delta x^{-}$ and $\Delta z^{-}$ in the length integration $\zeta^-$ but to keep these fluctuations in the definition of the initial state parton distribution function of the nucleon and in the in-medium transport coefficient. As in previously published papers related to the higher-twist formalism~\cite{Deng:2009ncl,Sirimanna:2021sqx}, we make these assumptions explicit, thus reducing the length integration term $\mathcal{R}$ to $\left[2 -2 \cos\left\{\mathcal{G}^{\left(p_2\right)}_{M}\zeta^{-}\right\}\right]$. As $\theta(x^{-}_{1} - z^{-}_{2})$ suggests that $x^{-}_{1} - z^{-}_{2}>0$, while $\theta(x^{-}_{2} - z^{-}_{3})$ implies $x^{-}_{2} - z^{-}_{3}>0$, we define a new length integration variable 
\begin{eqnarray}
\zeta^{-} &=& x^{-} - z^{-} \approx x^{-}_{2} - z^{-}_{3} \approx x^{-}_{1} - z^{-}_{2} .
\label{eq:zeta_minus_def}
\end{eqnarray}

Introducing $\zeta^-$ in $W^{\mu\nu}$ allows to perform the integrals over $d^4 z$,\footnote{The integral over $d^4 z$ gives an overall normalization factor, which is absorbed in the redefinition of the operator product expectation value.} and $\Delta z^+$, giving
\begin{eqnarray}
\begin{split}
W^{\mu \nu}_{1,c} &= C_{A} \sum_f e^2_f g^4_s \int d \zeta^{-} d (\Delta x^{-}) d (\Delta z^{-}) d^2 (\Delta z_{\perp}) \int \frac{dy d^2  \ell_{2\perp}}{(2 \pi)^{3}}  \frac{d^{2} k_{\perp}}{(2 \pi)^{2}}  
e^{i\left(q^{+}-\frac{M^2}{2q^{-}}\right)\Delta x^{-}   } 
\left\langle P \left| \bar{\psi}_{_f}\left(\Delta x^{-}\right) \frac{\gamma^{+}}{4} \psi_{_f}\left(0\right)\right| P\right\rangle \\  
& \times \theta\left(\zeta^-\right)  \frac{\left[ 2 -2 \cos\left\{\mathcal{G}^{(p_2)}_{M} \zeta^-\right\} \right]}{\left[4q^{-}\left(1-\eta\right)yq^{-}\right]^2} \left[\mathcal{G}^{(p_2)}_{M}\right]^{-2} \frac{q^-}{2yq^{-} 2 (1-y+\eta y)q^-} e^{-i\Delta z^{-}\mathcal{H}^{(\ell_2,p_2)}_M} e^{i\pmb{k}_{\perp}\cdot \pmb{\Delta z}_{\perp}} \\
& \times \mathrm{Tr} \left[  \gamma^{-} \gamma^{\mu}  \left(\slashed{q}+\slashed{p}'+M\right) \gamma^{\alpha_1} \left(\slashed{p}_{2} +M\right) \gamma^{\alpha_2} \left(\slashed{q}+\slashed{p}+M\right) \gamma^{\nu} \right] \left\langle P_{A-1} \left\vert {\rm Tr}\left[A^{+}\left(\zeta^-, \Delta z^{-}, \Delta z_{\perp}\right)A^+\left(\zeta^-,0\right)\right] \right\vert P_{A-1} \right\rangle \\
& \times g^{\sigma_1 \rho_1} g^{\sigma_2 \rho_2}  \left(2yq^{-} - \eta y q^{-}\right)^2 d^{(q+p-p_2)}_{\alpha_2 \rho_2} d^{(\ell_2)}_{\sigma_2 \rho_1} d^{(q+p'-p_2)}_{\sigma_1 \alpha_1}.
\end{split}
\label{eq:k1_y_l2_kperp_zeta_length}
\end{eqnarray}
The trace in Eq.~(\ref{eq:k1_y_l2_kperp_zeta_length}) can be simplified to get
\begin{equation}
\begin{split}
    & \mathrm{Tr} \left[  \gamma^{-} \gamma^{\mu}  \left(\slashed{q}+\slashed{p}'+M\right) \gamma^{\alpha_1} \left(\slashed{p}_{2} +M\right) \gamma^{\alpha_2} \left(\slashed{q}+\slashed{p}+M\right) \gamma^{\nu} \right] g^{\sigma_1 \rho_1} g^{\sigma_2 \rho_2} d^{\left(q+p-p_2\right)}_{\alpha_2 \rho_2} d^{\left(\ell_2\right)}_{\sigma_2 \rho_1} d^{\left(q+p'-p_2\right)}_{\sigma_1 \alpha_1} \\
  %& = \\
  & = \frac{8q^{-} \left[ -g^{\mu\nu}_{\perp\perp} \right]  }{\left(1-\eta\right)^2 y \left(1-y+\eta y\right)} \left[ \frac{1+ \left(1-y\right)^2}{y}\right] \left[ \left(\pmb{\ell}_{2\perp} -\pmb{k}_{\perp}\right)^2 + \kappa  y^4 M^2  \right],
\end{split}
\label{eq:K1_trace_dgg_1}
\end{equation}
where $\kappa$ is defined in Eq.~(\ref{eq:kappa}). Using the expression in Eq.~(\ref{eq:K1_trace_dgg_1}), the hadronic tensor becomes
\begin{eqnarray}
\begin{split}
W^{\mu\nu}_{1,c} & = C_{A} \sum_f 2 \left[-g^{\mu\nu}_{\perp\perp}\right]  e^2_f g^4_{s} \int d (\Delta x^{-}) e^{i\Delta x^{-}\left(q^{+}-\frac{M^2}{2q^-}\right)} \left\langle P \left| \bar{\psi}_{_f}(\Delta x^-) \frac{\gamma^{+}}{4} \psi_{_f}(0)\right| P\right\rangle\\
  & \times  \int d (\Delta z^{-})d^2 \Delta z_{\perp} \frac{dy}{2\pi}\frac{d^2 \ell_{2\perp}}{(2\pi)^2} \frac{d^2 k_{\perp}}{(2\pi)^2} \left[\frac{1+ \left(1-y\right)^2}{y}\right] 
  \left[ \frac{\left(1-\frac{\eta}{2}\right)^2}{\left(1-\eta\right)^2}\right]
  e^{-i\Delta z^{-}\mathcal{H}^{(\ell_2,p_2)}_M} e^{i\pmb{k}_{\perp}\cdot \Delta\pmb{z}_{\perp} } \\ 
  & \times \int d \zeta^- \theta\left(\zeta^-\right) \frac{\left[\left(\pmb{\ell}_{2\perp} - \pmb{k}_{\perp}\right)^2+\kappa y^4 M^2 \right]\left[2 - 2\cos\left\{\mathcal{G}^{(p_2)}_{M}\zeta^-\right\}\right]}{\left[\left(\pmb{\ell}_{2\perp} - \pmb{k}_{\perp}\right)^2  + y^2\left(1-\eta\right)^2  M^2 \right]^2}\\
  & \times \left\langle P_{A-1} \left\vert {\rm Tr}\left[A^{+}\left(\zeta^-, \Delta z^-, \Delta z_{\perp}\right)A^+\left(\zeta^-,0\right)\right]\right\vert P_{A-1}\right\rangle,
\end{split}
\label{eq:K1_W_final_dia1}
\end{eqnarray}
where, %$\kappa$ is defined in Eq.~(\ref{eq:kappa}), while,
for completeness, 
\begin{eqnarray}
 \mathcal{G}^{(p_2)}_M &=& p^{+}_{2} + \frac{\pmb{p}^{2}_{2\perp}}{2(q^{-}-p^{-}_{2})} - \frac{M^2}{2q^{-}} = \frac{\left(\pmb{\ell}_{2\perp}-\pmb{k}_{\perp}\right)^2 + y^2\left(1-\eta\right)^2M^2}{2y\left(1-y+\eta y\right)\left(1-\eta\right)q^-},\nonumber\\
\mathcal{H}^{(\ell_2,p_2)}_M &=& \ell^{+}_{2} + p^{+}_{2} - \frac{M^2}{2q^-} =  \frac{ \pmb{\ell}^{2}_{2\perp} -yM^2}{2yq^{-}} +\frac{\left(\pmb{\ell}_{2\perp} - \pmb{k}_{\perp}\right)^2+M^2}{2q^{-}\left(1-y+\eta y\right)}.
\label{eq:H_l2-p2-M}
\end{eqnarray} 
Note that momentum variables, such as $\mathcal{G}^{(p_2)}_M$ and $\mathcal{H}^{(\ell_2,p_2)}_M $ are in Appendix~\ref{append:list_of_function} for the reader's convenience. 
%\newpage
%%%%%%%%%%%%%%%%%%%%%%%%%%%%%%%%%%%%%%%%%%%%%%%%%%%%%%%%%%%%%%%%%%%%%%%%%%%%%%%%%%%%%%%%%%%%%%%%
%%%%%%%%%%%%%%%%%%%%%%%%%%%%%%%%%%%%%%%%%%%%%%%%%%%%%%%%%%%%%%%%%%%%%%%%%%%%%%%%%%%%%%%%%%%%%%%%
%%%%%%%%%%%%%%%%%%%%%%%%%%%%%%%%%%%%%%%%%%%%%%%%%%%%%%%%%%%%%%%%%%%%%%%%%%%%%%%%%%%%%%%%%%%%%%%%
\section{Full scattering kernel at next-to-leading order and next-to-leading twist}
\label{sec:full_kernel}
%%%%%%%%%%%%%%%%%%%%%%%%%%%%%%%%%%%%%%%%%%%%%%%%%%%%%%%%%%%%%%%%%%%%%%%%%%%%%%%%%%%%%%%%%%%%%%%%
%%%%%%%%%%%%%%%%%%%%%%%%%%%%%%%%%%%%%%%%%%%%%%%%%%%%%%%%%%%%%%%%%%%%%%%%%%%%%%%%%%%%%%%%%%%%%%%%
%%%%%%%%%%%%%%%%%%%%%%%%%%%%%%%%%%%%%%%%%%%%%%%%%%%%%%%%%%%%%%%%%%%%%%%%%%%%%%%%%%%%%%%%%%%%%%%%
In the preceding sections, the goal was to present in detail the steps involved in the derivation of a hadronic tensor for a given diagram and to highlight the power-counting strategy employed herein. In this section, for each kernel, all diagrams are combined and a full scattering kernel is provided for the gluonic and fermionic emissions from quark-initiated jet-medium interactions in each category: $q\to q+g$, $q\to g+g$, $q\to q+\bar{q}'$, and $q\to q'+\bar q'$. The $q\to q+g$ case is analyzed first, which involves a Glauber gluon~\cite{Idilbi:2008vm} exchange with the medium, while all other diagrams subsequently explored involve a Glauber quark exchange with the medium. Kernels involving Glauber quarks are especially interesting, as quark degrees of freedom develop dynamically as the heavy-ion collision transitions from a glasma-like initial state to the QGP. So, kernels $\mathcal{K}_2$ through $\mathcal{K}_4$ are sensitive to flavor hydrodynamization dynamics, thus complementing our photon study in Ref.~\cite{Kumar:2025egh}. The present calculation accounts for heavy-quark mass scales, full phase factors, and fermion-to-boson conversion processes. Monte Carlo simulations of jet-medium interactions involving highly virtual partons, including Bayesian analyses constraining $\hat{q}$ \cite{JETSCAPE:2024cqe}, have an unexplored theoretical systematic uncertainty due to the incomplete accounting of the Glauber-quark contribution to parton energy loss. This knowledge gap is addressed in the subsections below, from a theoretical perspective. 

Note that some of the diagrams in kernel-1 through kernel-4 are structurally identical to those in Ref.~\cite{Kumar:2025egh}, with the only difference being the replacement of the final-state photon line by a gluon line, together with the corresponding substitution $e^{2}e^{2}_{f} \rightarrow g^2_s$ and the appropriate SU(3) Casimir factors. As the goal herein is to provide self-contained and transparent discussion of all diagrams, these derivations are reproduced in appendixes.

%%%%%%%%%%%%%%%%%%%%%%%%%%%%%%%%%%%%%%%%%%%%%%%%%%%%%%%%%%%%%%%%%%%%%%%%%%%%%%%%%%%%%%%%%%%%%%%%
%%%%%%%%%%%%%%%%%%%%%%%%%%%%%%%%%%%%%%%%%%%%%%%%%%%%%%%%%%%%%%%%%%%%%%%%%%%%%%%%%%%%%%%%%%%%%%%%
\subsection{Single-scattering induced emission kernel: One gluon and one quark final state}
\label{subsec:quark_gluon}
%%%%%%%%%%%%%%%%%%%%%%%%%%%%%%%%%%%%%%%%%%%%%%%%%%%%%%%%%%%%%%%%%%%%%%%%%%%%%%%%%%%%%%%%%%%%%%%%
%%%%%%%%%%%%%%%%%%%%%%%%%%%%%%%%%%%%%%%%%%%%%%%%%%%%%%%%%%%%%%%%%%%%%%%%%%%%%%%%%%%%%%%%%%%%%%%%
For kernel-1, a total of 23 diagrams were identified (Fig.~\ref{fig:kernel-1_all}), including the left-cut and right-cut diagrams. These are presented in Appendix~\ref{append:kernel-1}. In order to add these diagrams, we institute $\Delta x^-=x^{-}_2 -x^{-}_1$, $\Delta z^-=z^{-}_3 -z^{-}_2$, and $\zeta^- = x^{-}_1-z^{-}_{2} = x^{-}_2-z^{-}_{3}$. The phase factors (i.e. complex exponentials) that depend on $\Delta x^-$ are absorbed in the definition of the nucleon parton distribution function, whereas those that depend on the relative distance $\Delta z^- = z^{-}_{3} -z^{-}_2$ are absorbed in the definition of the gluon/quark distribution in the medium. Following these definitions, the diagrams within each kernel are summed.

Diagrams in Fig.~\ref{fig:kernel-1_all} are a common setup \cite{Guo:2000nz,Aurenche:2008hm,Abir:2015hta,Sirimanna:2021sqx} for quark energy loss in the nuclear medium, using the HT formalism. The first HT calculation  \cite{Guo:2000nz}, considered a subset of diagrams in Fig.~\ref{fig:kernel-1_all} for a light quark propagating through the nuclear medium. The setup in Ref.~\cite{Guo:2000nz} was later extended to include corrections from the heavy-quark mass scale \cite{Abir:2015hta}. Finally, Ref.~\cite{Sirimanna:2021sqx} takes into account all diagrams in Fig.~\ref{fig:kernel-1_all} to explain how a light quark propagates through the nuclear medium, while our result extends those of in Ref.~\cite{Sirimanna:2021sqx} to include heavy-quark mass scales. The hadronic tensor for an incoming  quark $Q$ of mass $M$ and flavor $f$ reads
\begin{equation}
W^{\mu\nu}_{1,\rm full} = \sum_{f} 2 \left[-g^{\mu\nu}_{\perp\perp}\right]  e^2_f \int d (\Delta x^{-}) e^{i\Delta x^{-}\left(q^{+}-\frac{M^2}{2q^-}\right)} \left\langle P \left| \bar{\psi}_{_f}(\Delta x^-) \frac{\gamma^{+}}{4} \psi_{_f}(0)\right| P\right\rangle\times \mathcal{K}^{\left(Q;Q,g\right)}_1 ,
\end{equation}
% u,c,top
where $e_f=2/3$ for up-type quarks ($u,c, t$) and $e_f=-1/3$ for down-type quarks ($d,s,b$). The effective medium-modified scattering kernel for type-1 processes $\mathcal{K}^{\left(Q;Q,g\right)}_{1}$ is
\begin{equation}
\begin{split}
\mathcal{K}^{\left(Q;Q,g\right)}_1 &= g^4_s \int d (\Delta z^{-})d^2 \Delta z_{\perp} \frac{dy}{2\pi}\frac{d^2 \ell_{2\perp}}{(2\pi)^2} \frac{d^2 k_{\perp}}{(2\pi)^2} e^{-i\left(\Delta z^{-}\right)\mathcal{H}^{(\ell_2,p_2)}_M} e^{i\pmb{k}_{\perp}\cdot \Delta\pmb{z}_{\perp} } \\ 
  & \times \int d \zeta^- \theta\left(\zeta^-\right) \mathcal{S}^{(Q;Q,g)}_1  \left\langle P_{A-1} \left\vert {\rm Tr}\left[A^{+}\left(\zeta^-, \Delta z^-, \Delta z_{\perp}\right)A^+\left(\zeta^-,0\right)\right]\right\vert P_{A-1}\right\rangle,
\end{split}
\end{equation}
with
\begin{equation}
\begin{split}
\mathcal{S}^{\left(Q;Q,g\right)}_1&=C_{A}\left[\frac{1+ \left(1-y\right)^2}{y}\right]\\
&\times\left\{\frac{\left(1-\frac{\eta}{2}\right)^2}{\left(1-\eta\right)^2}\frac{\left[\left(\pmb{\ell}_{2\perp} - \pmb{k}_{\perp}\right)^2+ \kappa y^4 M^2 \right]\left[2 - 2\cos\left\{\mathcal{G}^{(p_2)}_{M}\zeta^-\right\}\right]}{\left[\left(\pmb{\ell}_{2\perp} - \pmb{k}_{\perp}\right)^2  + y^2\left(1-\eta\right)^2  M^2 \right]^2} + \frac{n\left(1+\frac{\eta}{2}\right)^2}{\left(1+\eta\right)}\frac{\left[\pmb{\ell}^2_{2\perp} + \kappa y^4 M^2\right]}{\left[\pmb{\ell}^2_{2\perp} + y^2 M^2 \right]^2}\left[\cos\left\{\mathcal{G}^{\left(\ell_2\right)}_{M}\zeta^{-}\right\}-1\right]\right\}\\
  &+C_F\left[\frac{1+\left(1-y\right)^2}{y}\right]\frac{\left[\pmb{\ell}^2_{2\perp}+\kappa y^4M^2\right]} {\left[\pmb{\ell}^2_{2\perp}+y^2M^2\right]^2}\left[1-\cos\left\{G^{(\ell_2)}_M\zeta^-\right\}\right](2-n)\\
  &+C_F\left[\frac{\left(1+\eta y\right)^2+\left(1-y+\eta y\right)^2}{y}\right]\left[\frac{\left\{\left(1+\eta y\right)\pmb{\ell}_{2\perp}-y\pmb{k}_{\perp}\right\}^2+\kappa y^4 M^2}{J^2_1}\right]\\
  &+\left[\frac{C_A}{2}-C_F\right]\left[\frac{1+\left(1-y\right)^2+\eta y\left(2-y\right)}{y}\right]\left[\frac{\left(1+\eta y\right)\pmb{\ell}^2_{2\perp}-y\pmb{\ell}_{2\perp}\cdot\pmb{k}_{\perp}+\kappa y^4 M^2}{\left[\pmb{\ell}^2_{2\perp}+y^2M^2\right]J_1}\right]\left[2-2\cos\left\{\mathcal{G}^{(\ell_2)}_M\zeta^-\right\}\right]\\
  &-\frac{C_A}{2}\left[\frac{1+\left(1-y\right)^2}{y}\right]\left[\frac{1-\frac{\eta}{2}}{1-\eta}\right]\left[\frac{\left(\pmb{\ell}_{2\perp}-\pmb{k}_{\perp}\right)\cdot\left(\pmb{\ell}_{2\perp}-y\pmb{k}_{\perp}\right)+\kappa y^4 M^2}{\left[\left(\pmb{\ell}_{2\perp}-\pmb{k}_{\perp}\right)^2+y^2\left(1-\eta\right)^2 M^2\right]J_1}\right]\left[2 - 2\cos\left\{\mathcal{G}^{(p_2)}_{M}\zeta^-\right\}\right]\\
  &-\frac{C_A}{2}\left[\frac{1+\left(1-y\right)^2}{y}\right]\left[\frac{\left(\pmb{\ell}_{2\perp}-\pmb{k}_{\perp}\right)\cdot\pmb{\ell}_{2\perp}+\kappa y^4 M^2}{\left[\left(\pmb{\ell}_{2\perp}-\pmb{k}_{\perp}\right)^2+y^2\left(1-\eta\right)^2 M^2\right]\left[\pmb{\ell}^2_{2\perp}+y^2 M^2\right]}\right]\left[1-\frac{\eta}{2}\right]\\
  &\times\left\{\frac{\left[2-2\cos\left\{\mathcal{G}^{(p_2)}_M\zeta^-\right\}-2\cos\left\{\mathcal{G}^{(\ell_2)}_M\zeta^-\right\}+2\cos\left\{\Delta\mathcal{G}_M\zeta^-\right\}\right]}{1-\eta}+n\left[ 4 \cos\left\{\mathcal{G}^{(\ell_2)}_{M}\zeta^-\right\}  - 2\cos\left\{\Delta\mathcal{G}_{M}\zeta^-\right\} \right]\right\}\\
  &-C_F\left[\frac{1+\left(1-y\right)^2}{y}\right]\frac{n\left[\pmb{\ell}^2_{2\perp} + \kappa y^4 M^2\right]}{\left[\pmb{\ell}^2_{2\perp} + y^2 M^2\right]^2}\cos\left\{\mathcal{G}^{(\ell_2)}_{M} \zeta^- \right\},
\end{split}
\label{eq:S_1}
\end{equation}
where $\Delta\mathcal{G}_M=\left(\mathcal{G}^{(p_2)}_{M}-\mathcal{G}^{(\ell_2)}_{M}\right)$. Equation~(\ref{eq:S_1}) is now examined in detail. The first two lines stem from Fig.~\ref{fig:kernel-1_all}(a) combining all three cuts.\footnote{The first term on the second line is from the central cut [Eq.~(\ref{eq:K1_W_final_dia1})] of Fig.~\ref{fig:kernel-1_all}(a), while adding the corresponding left and right cuts [Eq.~(\ref{eq:K1_W_final_add_left-rightcut})] gives the second term of the second line.} The third line originates from Fig.~\ref{fig:kernel-1_all}(b) again combining all three cuts.\footnote{ The term associated with the factor of 2 in $2-n$ corresponds to central-cut [Eq.~(\ref{eq:K1_bc_a_central_fi})] in Fig.~\ref{fig:kernel-1_all}(b) and the term with factor of $n$ is associated with its left- and right-cut [Eq.~(\ref{eq:K1_Wmunu_bc_a_add_left_rightcut_final})].} The fourth line comes from the central cut of Fig.~\ref{fig:kernel-1_all}(c) [Eq.~(\ref{eq:K1_bc_b_fi_wmunu})]. The fifth line combines the right-cut of Fig.~\ref{fig:kernel-1_all}(d) and the left-cut of Fig.~\ref{fig:kernel-1_all}(e) [Eq.~(\ref{eq:K1_de_nonvan})].\footnote{Note that combining left-cut of Fig.~\ref{fig:kernel-1_all}(d) and the right-cut of Fig.~\ref{fig:kernel-1_all}(e) gives zero [Eq.~(\ref{eq:K1_de_vanishing})].} The sixth line combines the right-cut of Fig.~\ref{fig:kernel-1_all}(f) and the left-cut of Fig.~\ref{fig:kernel-1_all}(g) [Eq.~(\ref{eq:K1_fg_non-vanishing})].\footnote{Again, combining left-cut of Fig.~\ref{fig:kernel-1_all}(f) and the right-cut of Fig.~\ref{fig:kernel-1_all}(g) gives zero [Eq.~(\ref{eq:K1_fg_vanishing})].} The seventh and eight line combine Fig.~\ref{fig:kernel-1_all}(h) and Fig.~\ref{fig:kernel-1_all}(i), where adding the two central cuts [Eq.~(\ref{eq:K1_hi_centercut_final})] gives the interference $\left[2-2\cos\left\{\mathcal{G}^{(p_2)}_M\zeta^-\right\}-2\cos\left\{\mathcal{G}^{(\ell_2)}_M\zeta^-\right\}+2\cos\left\{\Delta\mathcal{G}_M\zeta^-\right\}\right]$, while the left-cut of Fig.~\ref{fig:kernel-1_all}(h) with the right-cut of Fig.~\ref{fig:kernel-1_all}(h) [Eq.~(\ref{eq:K1_d_rightcut_final_x1_x2_origin})] gives the interference $\left[ 4 \cos\left\{\mathcal{G}^{(\ell_2)}_{M}\zeta^-\right\}  - 2\cos\left\{\Delta\mathcal{G}_{M}\zeta^-\right\} \right]$.\footnote{One cannot add the left- and right-cuts of Fig.~\ref{fig:kernel-1_all}(i) to those of Fig.~\ref{fig:kernel-1_all}(h), as that would be double-counting.} Finally, the last line stems from adding Fig.~\ref{fig:kernel-1_all}(j) and (k) [Eq.~(\ref{eq:kernel1_lastdiagram_ab_wmunu_sum})].

In Eq.~(\ref{eq:S_1}), the factor $n$ is included for noncentral cut diagrams. Since the noncentral cut diagrams give rise to two gluons on the same side of the cut-line, it imposes an additional phase-space constraint as $\theta(\Delta z^{+})$ due to the time ordering of the two gluons originating from the plasma. When evaluating the hadronic tensor, the effective integral over $d(\Delta z^{+})$ becomes
\begin{eqnarray}
    \int d(\Delta z^+) \delta(\Delta z^+)  \theta(\Delta z^{+}) = \theta\left(\Delta z^+=0\right) = n.
    \label{eq:maintext_factor_n}
\end{eqnarray} 
The definition of the Heaviside function at zero, encapsulated in $n$, depends on convention.\footnote{According to Ref.~\cite{weinberger1995first}, $n=1$; Ref.~\cite{bender2013advanced} uses $n=1/2$, while other references, such as Ref.~\cite{strauss1992partial}, leave $n$ unspecified.} A detailed discussion about the origin of $n$ can be found in Appendix A of Ref.~\cite{Sirimanna:2021sqx}, which puts the bound as $0\leq n <1/2$.

%%%%%%%%%%%%%%%%%%%%%%%%%%%%%%%%%%%%%%%%%%%%%%%%%%%%%%%%%%%%%%%%%%%%%%%%%%%%%%%%%%%%%%%%%%%%%%%%
%%%%%%%%%%%%%%%%%%%%%%%%%%%%%%%%%%%%%%%%%%%%%%%%%%%%%%%%%%%%%%%%%%%%%%%%%%%%%%%%%%%%%%%%%%%%%%%%
\subsection{Single-scattering induced emission kernel: Two gluons in the final state}
\label{subsec:gluon_gluon}
%%%%%%%%%%%%%%%%%%%%%%%%%%%%%%%%%%%%%%%%%%%%%%%%%%%%%%%%%%%%%%%%%%%%%%%%%%%%%%%%%%%%%%%%%%%%%%%%
%%%%%%%%%%%%%%%%%%%%%%%%%%%%%%%%%%%%%%%%%%%%%%%%%%%%%%%%%%%%%%%%%%%%%%%%%%%%%%%%%%%%%%%%%%%%%%%%
%
\begin{figure}[ht!]
    \centering
    \includegraphics[width=0.60\textwidth]{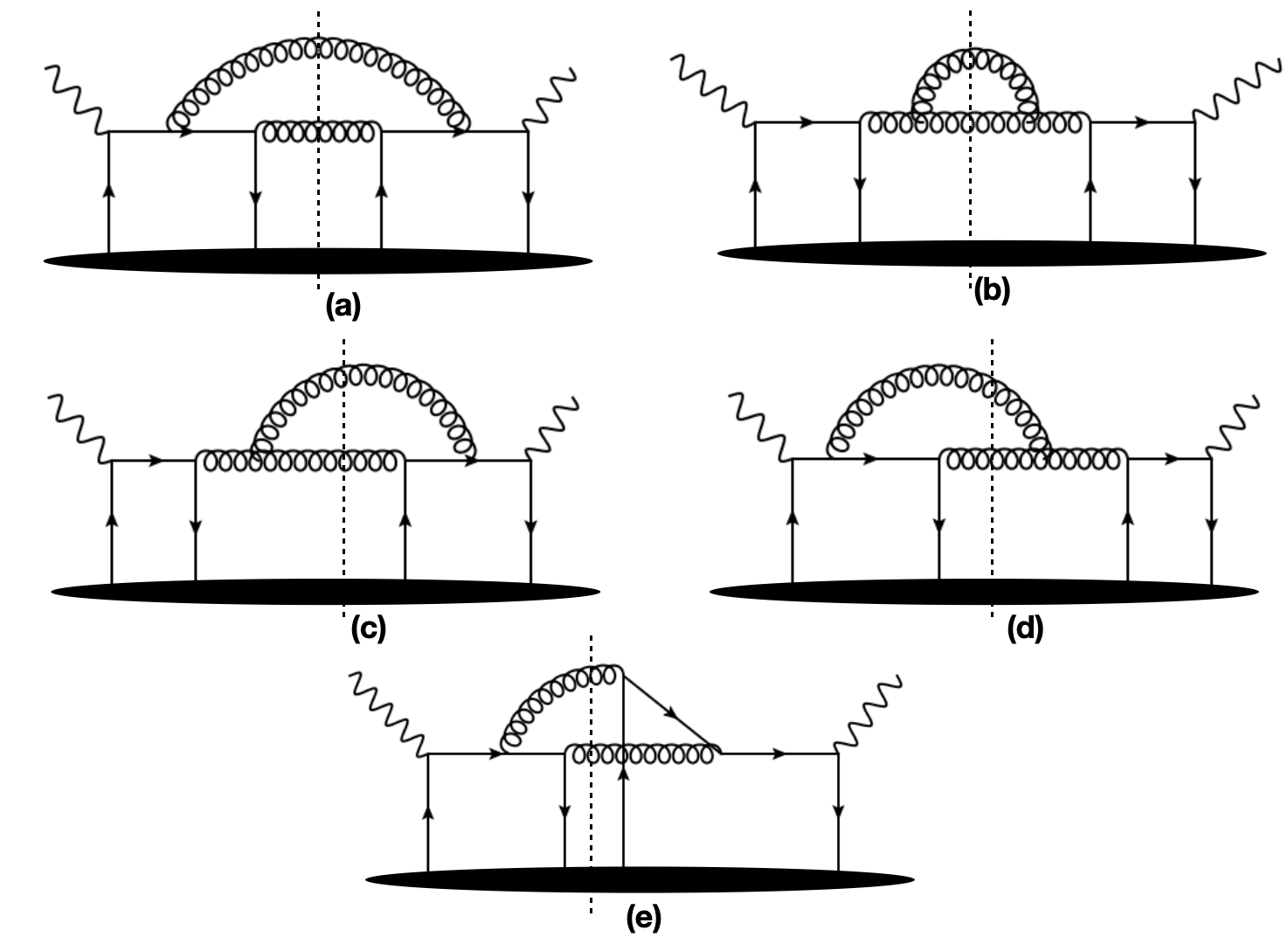}
    \caption{Diagrams for scattering kernel-2. }
    \label{fig:kernel-2_all}
\end{figure}
Interactions between a highly virtual incoming quark and the Glauber quark in the nuclear medium have not received the same amount of attention %\cite{Schafer:2007xh}
as those involving in-medium Glauber gluons~\cite{Idilbi:2008vm} discussed above. Thus, Sec. \ref{subsec:gluon_gluon} as well as Secs.~\ref{subsec:quark_antiquark} and \ref{subsec:quark_quark} present a detailed account of interactions between a highly virtual incoming quark and in-medium Glauber quarks. In kernel-2, considering first the channel with virtual quark and Glauber antiquark annihilation into two gluons, labeled as $q\to g+g$ are shown in Fig.~\ref{fig:kernel-2_all}. A complete calculation of these diagrams is presented in Appendix \ref{append:kernel-2}. Adding the hadronic tensors of all diagrams (Fig.~\ref{fig:kernel-2_all}) gives the following form of the hadronic tensor:
\begin{equation}
\begin{split}
W^{\mu\nu}_{2,\rm full}=\sum_{f} 2 [-g^{\mu\nu}_{\perp\perp}] e^2_f \int d (\Delta x^{-}) e^{iq^{+}(\Delta x^{-} )} \left\langle P \left| \bar{\psi}_{_f}( \Delta x^{-}) \frac{\gamma^{+}}{4} \psi_{_f}(0)\right| P\right\rangle \times \mathcal{K}^{\left(q;g,g\right)}_2,
\end{split}
\end{equation}
where $e_f=2/3$ for an up quark and $e_f=-1/3$ for down and strange quarks. The effective medium-modified scattering kernel for type-2 processes $\mathcal{K}^{\left(q;g,g\right)}_{2}$ is given as
\begin{equation}
\begin{split}
\mathcal{K}^{\left(q;g,g\right)}_{2}&=g^4_s \int d (\Delta z^{-})d^2 \Delta z_{\perp} \frac{dy}{2\pi}\frac{d^2 \ell_{2\perp}}{(2\pi)^2} \frac{d^2 k_{\perp}}{(2\pi)^2} e^{-i\left(\Delta z^{-}\right)\mathcal{H}^{(\ell_2,p_2)}_0} e^{i\pmb{k}_{\perp}\cdot \Delta\pmb{z}_{\perp} } \\ 
  & \times \int d \zeta^- \theta\left(\zeta^-\right) \mathcal{S}^{(q;g,g)}_2 \left\langle P_{A-1} \left|\bar{\psi}_{_{f}}(\zeta^-, 0) \frac{\gamma^+}{4} \psi_{_{f}}(\zeta^-,\Delta z^-, \Delta z_{\perp})\right| P_{A-1} \right\rangle,
\end{split}
\end{equation}
while
\begin{eqnarray}
\begin{split}
\mathcal{S}^{(q;g,g)}_2&=C_A C^2_F\left\{\left[\frac{1+\left(1-y\right)^2}{y}\right]\frac{\left[2-2\cos\left\{\mathcal{G}^{(\ell_2)}_{0} \zeta^- \right\}\right]}{q^-\left(1-y+\eta y\right)\pmb{\ell}^2_{2\perp}}+\left[\frac{1+y^2+\eta y^2\left(\eta-2\right)}{1-y+\eta y}\right]\frac{\left[2-2\cos\left\{\mathcal{G}^{(p_2)}_{0} \zeta^- \right\}\right]}{yq^-\left(\pmb{\ell}_{2\perp}-\pmb{k}_\perp\right)^2}\right\}\\
&+2C^2_AC_F\left[\frac{y\left(1+\eta y\right)^2}{1-y+\eta y}+\frac{(1+2\eta y)(1-y+\eta y)}{y}+y\left(1+\eta^2\right)\left(1-y+\eta y\right)\right]\frac{1}{\left(1+\eta y\right)^2 q^-\left[\left(1+\eta y\right)\pmb{\ell}_{2\perp}-y\pmb{k}_\perp\right]^2}\\
&-\frac{C^2_A C_F}{2}\left[\frac{1+\left(1-y+\eta y\right)^3}{y\left(1-y+\eta y\right)}\right]\left[\frac{\left(1+\eta y\right)\pmb{\ell}^2_{2\perp}-y\,\pmb{\ell}_{2\perp}\cdot\pmb{k}_\perp}{\pmb{\ell}^2_{2\perp}\left[\left(1+\eta y\right)\pmb{\ell}_{2\perp}-y\pmb{k}_{\perp}\right]^2}\right]\frac{1}{\left(1+\eta y\right)q^-}\left[2-2\cos\left\{\mathcal{G}^{(\ell_2)}_{0} \zeta^- \right\}\right]\\
&-\frac{C^2_A C_F}{2}\left[\frac{1+y^3}{y\left(1-y+\eta y\right)}\right]\frac{\left[\left(1+\eta y\right)\pmb{\ell}_{2\perp}-y\pmb{k}_\perp\right]\cdot\left[\pmb{\ell}_{2\perp}-\pmb{k}_\perp\right]}{\left(\pmb{\ell}_{2\perp}-\pmb{k}_{\perp}\right)^2\left[\left(1+\eta y\right)\pmb{\ell}_{2\perp}-y\pmb{k}_{\perp}\right]^2}\frac{1}{\left(1+\eta y\right)q^-}\left[2-2\cos\left\{\mathcal{G}^{(p_2)}_{0} \zeta^- \right\}\right]\\
&-\frac{\left(\frac{C_A}{2}-C_F\right)C_A C_F}{q^-\left(1-y+\eta y\right)}\left[\frac{1-y+2\eta y}{y}\right]\frac{\left[-\pmb{\ell}^2_{2\perp}+\pmb{\ell}_{2\perp}\cdot\pmb{k}_\perp\right]}{\left(\pmb{\ell}_{2\perp}-\pmb{k}_\perp\right)^2\pmb{\ell}^2_{2\perp}}\left[2-2\cos\left\{\mathcal{G}^{(\ell_2)}_{0} \zeta^- \right\}-2\cos\left\{\mathcal{G}^{(p_2)}_{0} \zeta^-\right\}+2\cos\left\{\Delta\mathcal{G}_0\zeta^-\right\}\right],
\end{split}
\label{eq:s2_full}
\end{eqnarray}
where $\Delta\mathcal{G}_0=\mathcal{G}^{(p_2)}_0-\mathcal{G}^{\left(\ell_2\right)}_0$. In Eq.~(\ref{eq:s2_full}), the first line corresponds to the process in Fig.~\ref{fig:kernel-2_all}(a) [Eqs.~(\ref{eq:kernel2-wmunu-final_a}) and (\ref{eq:kernel2-wmunu-final_a_l2p2Interchanged})], the second line stems from the process Fig.~\ref{fig:kernel-2_all}(b) [Eq.~(\ref{eq:kernel2_wfinal_b})], the third line and fourth lines are associated with interference processes shown in Figs.~\ref{fig:kernel-2_all}(c) and \ref{fig:kernel-2_all}(d) [Eqs.~(\ref{eq:kernel2_wfinal_c_d}) and (\ref{eq:kernel2_wfinal_cd_l2p2_interchange})] where the momenta $\pmb{\ell}_{2\perp}$ and $\pmb{p}_{2\perp}$ of the two identical final state gluons are interchanged, while the fifth line corresponds to the process in Fig.~\ref{fig:kernel-2_all}(e) [Eq.~(\ref{eq:kernel2_wfinal_e_both_added})].

Note that each term in Eq.~(\ref{eq:s2_full}) carries a suppression factor of $1/q^{-}$ compared to the terms in the scattering kernel-1 [Eq.~(\ref{eq:S_1})]. This indicates that the medium-induced quark-to-gluon conversion processes of kernel-2
are suppressed by the incoming energy of the quark $q^-$.
%%%%%%%%%%%%%%%%%%%%%%%%%%%%%%%%%%%%%%%%%%%%%%%%%%%%%%%%%%%%%%%%%%%%%%%%%%%%%%%%%%%%%%%%%%%%%%%%
%%%%%%%%%%%%%%%%%%%%%%%%%%%%%%%%%%%%%%%%%%%%%%%%%%%%%%%%%%%%%%%%%%%%%%%%%%%%%%%%%%%%%%%%%%%%%%%%
\subsection{Single-scattering induced emission kernel: One quark and one antiquark in the final state}
\label{subsec:quark_antiquark}
%%%%%%%%%%%%%%%%%%%%%%%%%%%%%%%%%%%%%%%%%%%%%%%%%%%%%%%%%%%%%%%%%%%%%%%%%%%%%%%%%%%%%%%%%%%%%%%%
%%%%%%%%%%%%%%%%%%%%%%%%%%%%%%%%%%%%%%%%%%%%%%%%%%%%%%%%%%%%%%%%%%%%%%%%%%%%%%%%%%%%%%%%%%%%%%%%
\begin{figure}[ht!]
    \centering
    \includegraphics[width=0.60\textwidth]{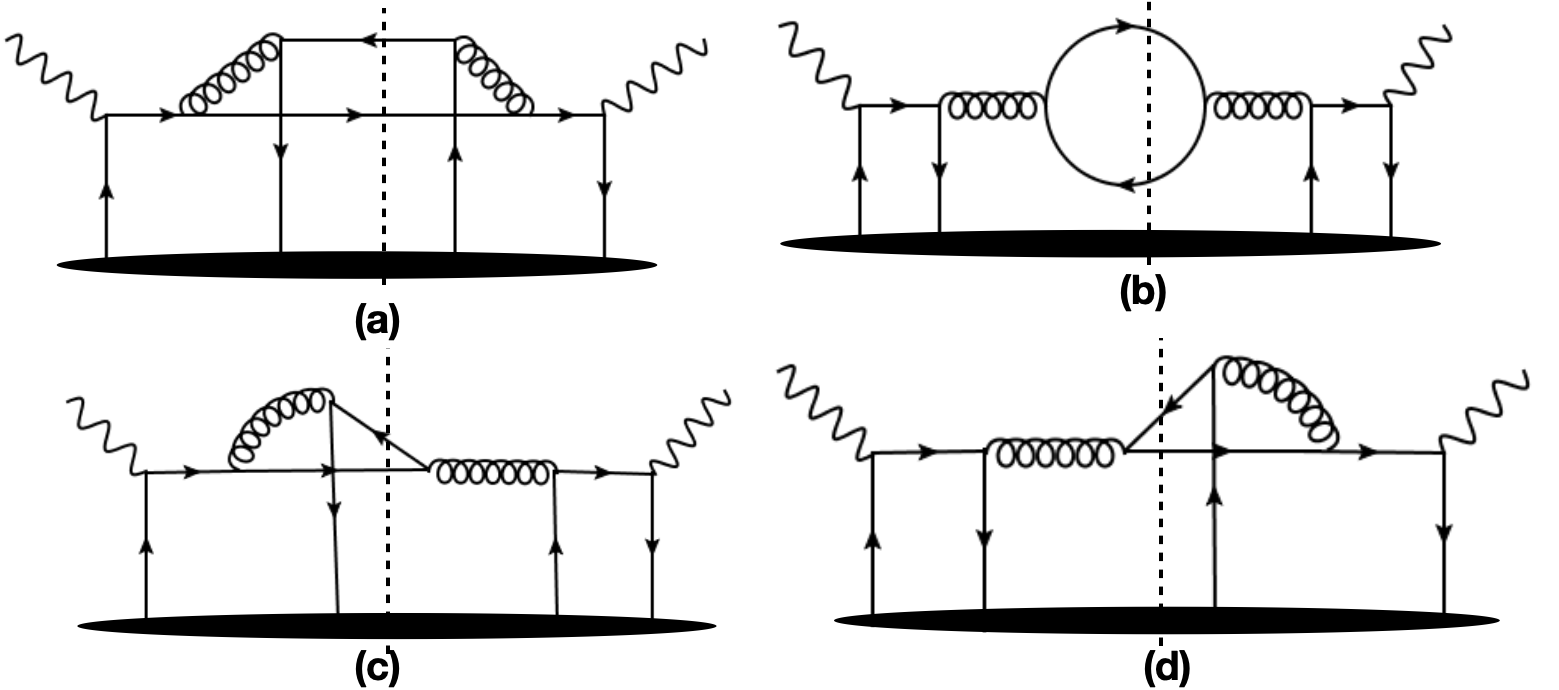}
    \caption{Diagrams for scattering kernel-3. }
    \label{fig:kernel-3_all}
\end{figure}
The single-scattering induced emission diagrams contributing to kernel-3 are shown in Fig.~\ref{fig:kernel-3_all}. The diagrams involve a Glauber quark exchange with the medium and consist of a quark and antiquark final state. A detailed calculation of these diagrams is presented in Appendix \ref{append:kernel-3}. Based on the quark flavor of the final state, the diagrams can be divided into two categories: (i) $q\to q+\bar{q}'$ and (ii) $q\to q'+\bar{q}'$, where $q'$ may be different from $q$. We first consider the situation of a heavy quark $(Q)$ transition $Q\to Q+\bar{q}'$ in Fig.~\ref{fig:kernel-3_all}(a), which gives the hadronic tensor [Eq.~(\ref{eq:app_k3_qqbar_scat_wmnunu_final})]:
\begin{equation}
W^{\mu\nu}_{3,\rm full} =  2 [-g^{\mu\nu}_{\perp\perp}] e^2_f \int d (\Delta x^{-}) e^{iq^{+}(\Delta x^{-} )} e^{-i[M^{2}/(2q^{-})](\Delta x^{-})} \left\langle P \left| \bar{\psi}_{_f}( \Delta x^{-}) \frac{\gamma^{+}}{4} \psi_{_f}(0)\right| P\right\rangle \times \mathcal{K}^{\left(Q;Q,\bar{q}'\right)}_3, 
\label{eq:W3_QQq'_final}
\end{equation}
where $e_f=2/3$ for a charm or top quark and $e_f=-1/3$ for a bottom quark. $\mathcal{K}^{\left(Q;Q,\bar{q}'\right)}_{3}$ is the effective medium-modified scattering kernel for the type-3 process [Fig.~\ref{fig:kernel-3_all}(a)], defined as
\begin{equation}
\begin{split}
\mathcal{K}^{\left(Q;Q,\bar{q}'\right)}_3&=g^4_s \sum_{f'\in \{\bar u,\bar d, \bar s\}  } \int d (\Delta z^{-}) d^2 \Delta z_{\perp} \frac{dy}{2\pi} \frac{d^2 \ell_{2\perp}}{(2\pi)^2} \frac{d^2 k_{\perp}}{(2\pi)^2} e^{-i(\Delta z^{-})\mathcal{H}^{(\ell_2,p_2)}_{M} } e^{i \pmb{k}_{\perp} \cdot \Delta\pmb{z}_{\perp}}\\
&\times\int d\zeta^{-}\theta\left(\zeta^{-}\right) \mathcal{S}^{\left(Q;Q,\bar{q}'\right)}_3 \left\langle P_{A-1} \left| \bar{\psi}_{_{f'}}(\zeta^-,0) \frac{\gamma^+}{4} \psi_{_{f'}}(\zeta^-, \Delta z^-, \Delta z_{\perp})\right| P_{A-1} \right\rangle,
\label{eq:K_3_Q_Q_q-prime}
\end{split}
\end{equation}
with
\begin{equation}
\mathcal{S}^{\left(Q;Q,\bar{q}'\right)}_3=\left[\frac{C_F C_A}{2}\right] \left[\frac{1 +\left(1-y\right)^2}{y}\right] \frac{1}{yq^-} \frac{\left[\left(\pmb{\ell}_{2\perp} - \pmb{k}_{\perp}\right)^2  + \kappa y^4 M^2 \right]}{\left[  \left(\pmb{\ell}_{2\perp} -\pmb{k}_{\perp}\right)^2 + y^2\left(1-\eta\right)^2 M^2 \right]^2 } \left[ 2 -2 \cos\left\{\mathcal{G}^{(p_2)}_{M}\zeta^-\right\} \right],
\label{eq:S_3_Q_Q_q-prime}
\end{equation}
where $M$ is the mass of the quark with flavor $f$. Note that Eq.~(\ref{eq:W3_QQq'_final}) follows from Eq.~(\ref{eq:app_k3_qqbar_scat_wmnunu_final}). The second option in Fig.~\ref{fig:kernel-3_all}(a) is one in which the light quark $q$ has a different flavor than the light antiquark $\bar q'$, and in that case one simply sets $M=0$ in Eq.~(\ref{eq:K_3_Q_Q_q-prime}) and Eq.~(\ref{eq:S_3_Q_Q_q-prime}).

The other process contributing to kernel-3 is $q\to q'+\bar{q}'$ in Fig.~\ref{fig:kernel-3_all}(b), where the quark $q'$, spanning both light and heavy varieties, has a different flavor from the light quark $q$. For that case, a fixed light quark $q$ with flavor $f$ gives the hadronic tensor [Eq.~(\ref{eq:K3_final_g_qqbar_wmunu})]:
\begin{equation}
\begin{split}
W^{\mu\nu}_{3,\rm final} & = 2 \left[ -g^{\mu\nu}_{\perp \perp} \right] e^2_{f} \int d (\Delta x^{-})   e^{iq^{+}(\Delta x^{-} )}    
  \left\langle P \left| \bar{\psi}_{_f}( \Delta x^{-}) \frac{\gamma^{+}}{4} \psi_{f}(0)\right| P\right\rangle \times \mathcal{K}^{\left(q;q',\bar{q}'\right)}_3,
\end{split} 
\end{equation}
with
\begin{equation}
\begin{split}
\mathcal{K}^{\left(q;q',\bar{q}'\right)}_3 & = g^4_s \sum_{\substack{f'\neq f\\f'\in \{u,d,s,c,b,t\}}} \int d (\Delta z^{-})d^2 \Delta z_{\perp} \frac{dy}{2\pi} \frac{d^2 \ell_{2\perp}}{(2\pi)^2} \frac{d^2 k_{\perp}}{(2\pi)^2}  e^{-i \mathcal{H}^{(\ell_2,p_2)}_{1} (\Delta z^{-})} e^{-i \pmb{k}_{\perp}\cdot \Delta\pmb{z}_{\perp}}   \\
  & \times \int  d \zeta^{-} \theta(\zeta^{-}) \mathcal{S}^{\left(q;q',\bar{q}'\right)}_3 \left\langle P_{A-1} \left| \bar{\psi}_{_{f}}(\zeta^-,0) \frac{\gamma^+}{4}  \psi_{_{f}}(\zeta^{-},\Delta z^{-}, \Delta \pmb{z}_{\perp}) \right| P_{A-1} \right\rangle, 
\end{split} 
\label{eq:S3_channel2}
\end{equation}
where, $\mathcal{H}^{(\ell_2,p_2)}_{1}$ is defined in Eq.~(\ref{eq:H1L2P2_append}) and is dependent on $f'$, while
\begin{equation}
\begin{split}
\mathcal{S}^{\left(q;q',\bar{q}'\right)}_3=\left[\frac{C_{F}C_A}{2} \right]\frac{1}{q^{-}} \frac{\left[y^2+\left(1-y+\eta y\right)^2\right]}{\left(1+\eta y\right)^2 \left[\left\{\left(1+\eta y\right)\pmb{\ell}_{2\perp} - y \pmb{k}_{\perp}\right\}^2 + M^2_{f'} \left(1+\eta y\right)^2\right]}. 
\end{split} 
\end{equation}
The last contribution to kernel-3 comes from the process $q\to q+\bar q$, where the incoming and outgoing quarks are of the same light flavor. In that scenario, all diagrams in Fig.~\ref{fig:kernel-3_all} contribute, i.e. Eq.~(\ref{eq:app_k3_qqbar_scat_wmnunu_final}), Eq.~(\ref{eq:K3_final_g_qqbar_wmunu}), and Eq.~(\ref{eq:k3_ab_intf_q_g_qqbar_qg_wmunu_added}), giving
\begin{equation}
\begin{split}
W^{\mu\nu}_{3,\rm final} & = 2 \left[ -g^{\mu\nu}_{\perp \perp} \right] e^2_{f} \int d (\Delta x^{-})   e^{iq^{+}(\Delta x^{-} )}    
  \left\langle P \left| \bar{\psi}_{_f}( \Delta x^{-}) \frac{\gamma^{+}}{4} \psi_{f}(0)\right| P\right\rangle \times \mathcal{K}^{\left(q;q,\bar{q}\right)}_3,
\end{split} 
\end{equation}
with
\begin{equation}
\begin{split}
\mathcal{K}^{\left(q;q,\bar{q}\right)}_3 &=g^4_s \int d (\Delta z^{-}) d^2 \Delta z_{\perp} \frac{dy}{2\pi} \frac{d^2 \ell_{2\perp}}{(2\pi)^2} \frac{d^2 k_{\perp}}{(2\pi)^2} e^{-i(\Delta z^{-})\mathcal{H}^{(\ell_2,p_2)}_{0} } e^{i \pmb{k}_{\perp} \cdot \Delta\pmb{z}_{\perp}}\\
&\times\int d\zeta^{-}\theta\left(\zeta^{-}\right) \mathcal{S}^{\left(q;q,\bar{q}\right)}_3 \left\langle P_{A-1} \left| \bar{\psi}_{_{f}}(\zeta^-,0) \frac{\gamma^+}{4} \psi_{_{f}}(\zeta^-, \Delta z^-, \Delta z_{\perp})\right| P_{A-1} \right\rangle,  
\end{split} 
\end{equation}
and
\begin{eqnarray}
\begin{split}
\mathcal{S}^{\left(q;q,\bar{q}\right)}_3&=\left[\frac{C_F C_A}{2}\right] \left[ \frac{ 1 + \left(1-y\right)^2 }{y}\right] \frac{1}{yq^-} \frac{1}{\left(\pmb{\ell}_{2\perp} -\pmb{k}_{\perp}\right)^2} \left[ 2 -2 \cos\left\{\mathcal{G}^{(p_2)}_{0}\zeta^-\right\} \right]\\
&+\left[ \frac{C_F C_A}{2} \right]\frac{1}{q^{-}}\frac{\left[y^2+\left(1-y+\eta y\right)^2\right]}{\left(1+\eta y\right)^2\left[ \left\{\left(1+\eta y\right)\pmb{\ell}_{2\perp} - y \pmb{k}_{\perp}\right\}^2 +\left(1+\eta y\right)^2\right]}\\
&-\left[ C_{F}C_A \left(C_F - \frac{C_A}{2}\right)\right] \frac{1}{yq^{-}} \left[ \frac{1-y+\eta y}{(1+\eta y)(1-\eta)}\right] \frac{J_2}{\left(\pmb{\ell}_{2\perp} - \pmb{k}_\perp\right)^2 \left[\left(1+\eta y\right)\pmb{\ell}_{2\perp} - y\pmb{k}_{\perp}\right]^2}  \left[ 2 -2 \cos\left\{\mathcal{G}^{(p_2)}_{0}\zeta^-\right\}\right],
\end{split} 
\label{eq:S3_channel3}
\end{eqnarray}
where $J_2$ and $\mathcal{G}^{(p_2)}_{0}$ are defined in Appendix \ref{append:list_of_function}. The expression of the scattering kernel [Eq.~(\ref{eq:S_3_Q_Q_q-prime}), Eq.~(\ref{eq:S3_channel2}), and Eq.~(\ref{eq:S3_channel3})] contains a factor of $1/q^-$, and therefore, like kernel-2, the processes in kernel-3 are suppressed by a factor of the incoming energy of the quark $q^-$. 

%%%%%%%%%%%%%%%%%%%%%%%%%%%%%%%%%%%%%%%%%%%%%%%%%%%%%%%%%%%%%%%%%%%%%%%%%%%%%%%%%%%%%%%%%%%%%%%%
%%%%%%%%%%%%%%%%%%%%%%%%%%%%%%%%%%%%%%%%%%%%%%%%%%%%%%%%%%%%%%%%%%%%%%%%%%%%%%%%%%%%%%%%%%%%%%%%
\subsection{Single-scattering induced emission kernel: Two quarks in the final state}
\label{subsec:quark_quark}
%%%%%%%%%%%%%%%%%%%%%%%%%%%%%%%%%%%%%%%%%%%%%%%%%%%%%%%%%%%%%%%%%%%%%%%%%%%%%%%%%%%%%%%%%%%%%%%%
%%%%%%%%%%%%%%%%%%%%%%%%%%%%%%%%%%%%%%%%%%%%%%%%%%%%%%%%%%%%%%%%%%%%%%%%%%%%%%%%%%%%%%%%%%%%%%%%
\begin{figure}[ht!]
    \centering
    \includegraphics[width=0.60\textwidth]{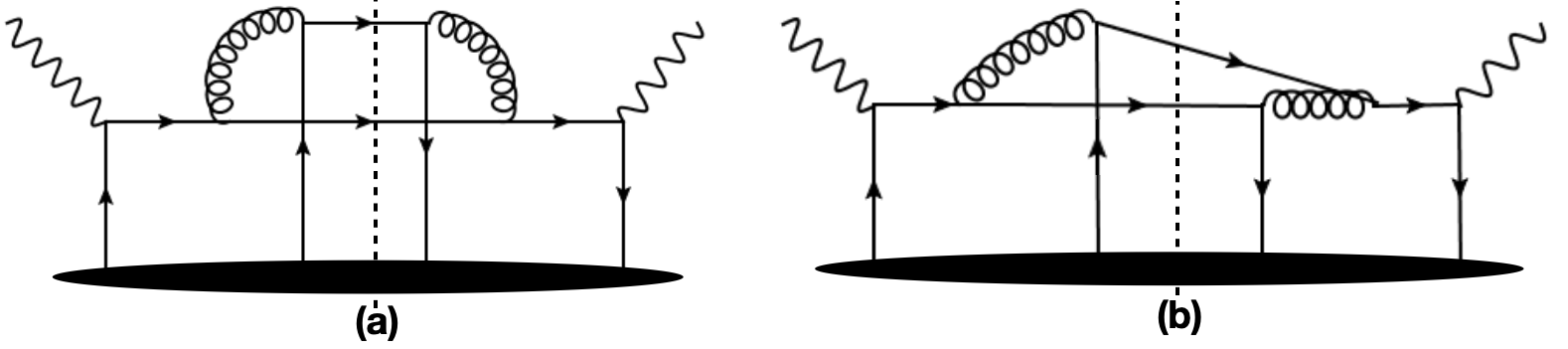}
    \caption{Diagrams for scattering kernel-4. }
    \label{fig:kernel-4_all}
\end{figure}
The single-scattering induced emission diagrams with two quarks in the final state are presented in Fig.~\ref{fig:kernel-4_all}. The evaluation of these diagrams is discussed in Appendix \ref{append:kernel-4}.
The case with two quarks in the final state is further subdivided into three categories: (i) a final state with a heavy quark $Q$ and a light quark of flavor $q'$, (ii) a final state with a light quark $q$ and another light quark of a different flavor $q'$, and (iii) final state with two light quarks of the same flavor $q$. The Feynmann diagram in Fig.~\ref{fig:kernel-4_all}(b) contributes solely to final state of type (iii). For a given heavy quark $Q$ participating in the process $Q\to Q+q'$, the hadronic tensor is [Eq.~(\ref{eq:K4_hq_light_q_scatt})]:
\begin{equation}
\begin{split}
W^{\mu\nu}_{4,\rm full} & = 2 [-g^{\mu\nu}_{\perp \perp }] e^2_f \int d (\Delta x^{-})   e^{iq^{+}(\Delta x^{-} )}  e^{-i[M^{2}/(2q^{-})](\Delta x^{-} )} \left\langle P \left| \bar{\psi}_{_f}( \Delta x^{-}) \frac{\gamma^{+}}{4} \psi_{_f}(0)\right| P\right\rangle \times \mathcal{K}^{\left(Q;Q,q'\right)}_{4},
\end{split} 
\end{equation}
where $e_f=2/3$ for a charm or top quark and $e_f=-1/3$ for a bottom quark. $\mathcal{K}^{\left(Q;Q,q'\right)}_{4}$ is the effective medium-modified scattering kernel for the type-4 process [Fig.~\ref{fig:kernel-4_all}(a)], defined as
\begin{equation}
\begin{split}
\mathcal{K}^{\left(Q;Q,q'\right)}_{4} & =  g^4_s \sum_{f'\in\{u,d,s\}} \int d (\Delta z^{-}) d^2 \Delta z_{\perp} \frac{dy}{2\pi} \frac{d^2 \ell_{2\perp}}{(2\pi)^2} \frac{d^2 k_{\perp}}{(2\pi)^2} e^{-i (\Delta z^{-})\mathcal{H}^{(\ell_2 , p_2)}_{M} }     e^{i \pmb{k}_{\perp} \cdot \Delta\pmb{z}_{\perp}}  \\
  & \times \int d\zeta^{-} \theta\left(\zeta^-\right) \mathcal{S}^{\left(Q;Q,q'\right)}_{4} \left\langle P_{A-1} \left| \bar{\psi}_{_{f'}}(\zeta^-,\Delta z^-, \Delta z_{\perp}) \frac{\gamma^+}{4} \psi_{_{f'}}(\zeta^-, 0)\right| P_{A-1} \right\rangle, 
\end{split} 
\label{eq:K_4_Q_Q_q-prime}
\end{equation}
where $f'$ spans the three light flavors, while
\begin{equation}
\begin{split}
\mathcal{S}^{\left(Q;Q,q'\right)}_{4} & =  \left[ \frac{C_F C_A}{2} \right] \left[ \frac{ 1 + \left(1-y\right)^2 }{y}\right]  \frac{1}{yq^-}  \frac{ \left[\left(\pmb{\ell}_{2\perp} - \pmb{k}_{\perp}\right)^2 +\kappa y^4 M^2 \right]}{\left[\left(\pmb{\ell}_{2\perp} -\pmb{k}_{\perp}\right)^2 + y^2\left(1-\eta\right)^2 M^2\right]^2}\left[ 2 -2 \cos\left\{\mathcal{G}^{(p_2)}_{M}\zeta^-\right\} \right].
\end{split} 
\label{eq:S_4_Q_Q_q-prime}
\end{equation}
For an incoming light-flavor quark $q$ participating in the process $q\to q+q'$, with $q'$ being of different flavor from $q$, one simply sets $M=0$ in Eq.~(\ref{eq:K_4_Q_Q_q-prime}) and Eq.~(\ref{eq:S_4_Q_Q_q-prime}). 

Finally, both diagrams in Fig.~\ref{fig:kernel-4_all} contribute to the process $q\to q +q$, where all light flavors $q$ are the same. The hadronic tensor and scattering kernel, combining Eq.~(\ref{eq:K4_hq_light_q_scatt}), Eq.~(\ref{eq:K4_light_heavy_flip}), and Eq.~(\ref{eq:K4_b_both_interf_wmunu_added}), gives:
\begin{eqnarray}
W^{\mu\nu}_{4,c}&=&2[-g^{\mu\nu}_{\perp \perp }] e^2_f \int d (\Delta x^{-}) e^{iq^{+}(\Delta x^{-} )} \left\langle P \left| \bar{\psi}_{_f}( \Delta x^{-}) \frac{\gamma^{+}}{4} \psi_{_f}(0)\right| P\right\rangle\times \mathcal{K}^{\left(q;q,q\right)}_{4}\\
\mathcal{K}^{\left(q;q,q\right)}_{4}&=&g^4_s \int d(\Delta z^{-}) d^2 \Delta z_{\perp} \frac{dy}{2\pi} \frac{d^2 \ell_{2\perp}}{(2\pi)^2} \frac{d^2 k_{\perp}}{(2\pi)^2} e^{-i(\Delta z^{-})\mathcal{H}^{(\ell_2,p_2)}_{0} } e^{i\pmb{k}_{\perp}\cdot \Delta\pmb{z}_{\perp}},\nonumber\\
&\times&\int d\zeta^-\theta\left(\zeta^-\right) \mathcal{S}^{\left(q;q,q\right)}_{4}\left\langle P_{A-1} \left| \bar{\psi}_{_{f}}(\zeta^-,\Delta z^-, \Delta z_{\perp}) \frac{\gamma^+}{4} \psi_{_{f}}(\zeta^-, 0)\right| P_{A-1} \right\rangle,
\end{eqnarray}
with
\begin{eqnarray}
\mathcal{S}^{\left(q;q,q\right)}_{4}&=& \left[\frac{C_F C_A}{2}\right] \left[\frac{1 +\left(1-y\right)^2 }{y}\right] \frac{1}{yq^-} \frac{1}{\left(\pmb{\ell}_{2\perp} -\pmb{k}_{\perp}\right)^2} \left[ 2 -2 \cos\left\{\mathcal{G}^{(p_2)}_{0}\zeta^-\right\} \right]\nonumber\\
&+&\left[\frac{C_F C_A}{2}\right] \left[ \frac{ 1 + y^2 }{1-y}\right] \frac{1}{q^-} \frac{1}{\left(1-y+\eta y\right)\pmb{\ell}^2_{2\perp}}  \left[ 2 -2 \cos\left\{\mathcal{G}^{(\ell_2)}_{0}\zeta^-\right\} \right]\nonumber\\
&+&\frac{C_F C_A\left(C_F - \frac{C_A}{2} \right)}{(1-y)yq^-} \frac{\left[ -\pmb{\ell}^{2}_{2\perp} +  \pmb{\ell}_{2\perp}\cdot \pmb{k}_{\perp} \right]}{\left(1-\eta\right)\left(\pmb{\ell}_{2\perp}-\pmb{k}_{2\perp}\right)^2 \pmb{\ell}^2_{2\perp}}
\left[2-2\cos\left\{\mathcal{G}^{(p_2)}_{0}\zeta^-\right\}-2\cos\left\{\mathcal{G}^{(\ell_2)}_{0}\zeta^-\right\}+2\cos\left\{\Delta\mathcal{G}_0\zeta^-\right\}\right],\nonumber\\
\end{eqnarray}
where $\Delta\mathcal{G}_0=\left(\mathcal{G}^{(p_2)}_{0}-\mathcal{G}^{(\ell_2)}_{0}\right)$.

Like kernel-2 and kernel-3, the scattering kernel-4 contains an additional factor of $1/q^-$, and therefore the processes in kernel-4 are also suppressed by the incoming energy of the quark $q^-$. However, note that suppression by $1/q^-$ does not imply that kernel-2 through kernel-4 should not be present inside a jet Monte Carlo simulation. 
While $q^-$ is a dynamical quantity and a large quantity for the first split, it decreases as the shower is generated with subsequent splits. Moreover, as the system created in relativistic heavy-ion collisions transitions from the early-time glasma to the QGP, quark occupation number in the medium increases, and hence, the jet energy loss starts to become sensitive to kernel-2 through kernel-4. 
%are acutely sensitive to them.    

The transport coefficient $\hat{q}$ has been estimated for a quark that traverses QGP and cold nuclear matter. The phenomenological studies carried out by JET~\cite{JET:2013cls} and JETSCAPE~\cite{JETSCAPE:2024cqe} Collaborations indicate that $\hat{q}$ is 1$-$2 GeV$^2$/fm at temperatures $T\in[300,400]$ MeV in RHIC and LHC collisions.
There also exist estimates of $\hat{q}$ based on jet quenching in DIS for cold nuclei. Using an ideal quark gas equation of state with three valence quarks at normal nucleon density in a large nucleus,  $\hat{q}$ has been estimated to be 0.02$-$0.06 GeV$^2$/fm~\cite{Deng:2009ncl,JET:2013cls}. A more recent estimate~\cite{Ru:2019qvz} based on the global fit to experimental data for semi-inclusive \textit{e-A} DIS, Drell-Yan (DY) dilepton, and heavy quarkonium production in proton-nucleus (\textit{p-A}) collisions also yields a similar value of $\hat{q}$ in cold nuclear matter. These estimates indicate that the transport coefficient $\hat{q}$ is an order of magnitude higher in hot QGP compared to the large nucleus. Assuming a similar behavior is true for the jet transport coefficient involving the fermionic correlator (see next section), one could assert that the new contributions presented in kernel-2 through kernel-4 would play a more significant role for jets traversing through QGP compared to jets in cold nuclear matter, such as in DIS.

%%%%%%%%%%%%%%%%%%%%%%%%%%%%%%%%%%%%%%%%%%%%%%%%%%%%%%%%%%%%%%%%%%%%%%%%%%%%%%%%%%%%%%%%%%%%%%%%%%%%%%%%%%%%%%%%%%%%%%%%%%%
%%%%%%%%%%%%%%%%%%%%%%%%%%%%%%%%%%%%%%%%%%%%%%%%%%%%%%%%%%%%%%%%%%%%%%%%%%%%%%%%%%%%%%%%%%%%%%%%%%%%%%%%%%%%%%%%%%%%%%%%%%%
%%%%%%%%%%%%%%%%%%%%%%%%%%%%%%%%%%%%%%%%%%%%%%%%%%%%%%%%%%%%%%%%%%%%%%%%%%%%%%%%%%%%%%%%%%%%%%%%%%%%%%%%%%%%%%%%%%%%%%%%%%%
\section{Collinear expansion and jet transport coefficients at next-to-leading order and next-to-leading twist}
\label{sec:collinear_exp}
%%%%%%%%%%%%%%%%%%%%%%%%%%%%%%%%%%%%%%%%%%%%%%%%%%%%%%%%%%%%%%%%%%%%%%%%%%%%%%%%%%%%%%%%%%%%%%%%%%%%%%%%%%%%%%%%%%%%%%%%%%%
%%%%%%%%%%%%%%%%%%%%%%%%%%%%%%%%%%%%%%%%%%%%%%%%%%%%%%%%%%%%%%%%%%%%%%%%%%%%%%%%%%%%%%%%%%%%%%%%%%%%%%%%%%%%%%%%%%%%%%%%%%%
%%%%%%%%%%%%%%%%%%%%%%%%%%%%%%%%%%%%%%%%%%%%%%%%%%%%%%%%%%%%%%%%%%%%%%%%%%%%%%%%%%%%%%%%%%%%%%%%%%%%%%%%%%%%%%%%%%%%%%%%%%%
Having presented the full scattering kernel for a hard quark traversing the nuclear medium, a Taylor expansion of four scattering kernels $\mathcal{S}_i$ around $\pmb{k}_\perp=\pmb{0}_\perp$ and $k^-=0$ is now presented, as has been done in prior higher-twist calculations \cite{Abir:2015hta,Sirimanna:2021sqx}. In general, the expansion takes the form
\begin{eqnarray}
    \mathcal{S}^{(a;b,c)}_{i}\left(\pmb{k}_{\perp},k^-\right) & = &\mathcal{S}^{(a;b,c)}_{i}\left(\pmb{k}_{\perp}=\pmb{0}_\perp,k^-=0\right) + \left. \frac{\partial \mathcal{S}^{(a;b,c)}_{i}}{ \partial k^{\rho}_{\perp}} \right|_{k=0}  k^{\rho}_{\perp}  + \frac{1}{2!} \left. \frac{\partial^2 \mathcal{S}^{(a;b,c)}_{i}}{ \partial k^{\rho}_{\perp} \partial k^{\sigma}_{\perp}} \right| _{k=0} k^{\rho}_{\perp} \otimes k^{\sigma}_{\perp} + \cdots \nonumber \\
    & + & \left. \frac{\partial \mathcal{S}^{(a;b,c)}_{i}}{\partial k^{-}} \right|_{k=0}  k^{-} + \frac{1}{2!}\left. \frac{\partial^2 \mathcal{S}^{(a;b,c)}_{i}}{\partial {k^{-}}^2} \right|_{k=0} k^{-}\otimes k^{-} + \cdots, 
\label{eq:H-taylor-expansion}
\end{eqnarray}
where $\vert_{k=0}$ is shorthand notation for all components of $k$ being evaluated to zero, $i$ spans the four kinds of kernels discussed herein, and $(a;b,c)$ represents the species involved in the $1\to 2$ process. For example, for a heavy quark radiating a gluon, $Q\to Q+g$ is labeled $(a;b,c)=(Q;Q,g)$. Assuming a homogeneous and isotropic nuclear medium leads to a vanishing second term in Eq.~(\ref{eq:H-taylor-expansion}), after integration over $k_\perp$ is performed. Homogeneity and isotropy further entails that the third term in Eq.~(\ref{eq:H-taylor-expansion}) is nonzero only when $\rho=\sigma$.

Applying collinear expansion, the effective medium-modified scattering kernel $\mathcal{K}^{(a;b,c)}_{i}$ can be written as
\begin{eqnarray}
%\begin{equation}
%\begin{split}
 \mathcal{K}^{(a;b,c)}_{i}  = g^{2}_{s} \int \frac{dy}{2\pi}\frac{d^2 \ell_{2\perp}}{(2\pi)^2} d\zeta^{-} \left[ \mathcal{R}^{(a;b,c)}_{(i;0)}\hat{\mathcal{O}}_{(i;0)} + \left(\frac{1}{2!}\mathcal{R}^{(a;b,c)}_{(i;T,2)}\hat{\mathcal{O}}_{(i;T,2)} + \frac{1}{4!}\mathcal{R}^{(a;b,c)}_{(i;T,4)}\hat{\mathcal{O}}_{(i;T,4)} + \cdots\right) \right.\nonumber\\ 
 \left. +\left(\mathcal{R}^{(a;b,c)}_{(i;L,1)}\hat{\mathcal{O}}_{(i;L,1)} + \frac{1}{2!}\mathcal{R}^{(a;b,c)}_{(i;L,2)}\hat{\mathcal{O}}_{(i;L,2)} + \cdots\right)\right],
%\end{split}
%\end{equation}
\label{eq:kernel1-coll-expansion}
\end{eqnarray}
where $\mathcal{R}^{(a;b,c)}_{(i;0)}$ is the zeroth-order term in the Taylor expansion of the $i$th collisional kernel, with $\mathcal{R}^{(a;b,c)}_{(i;L,j)}$ representing the $j$th-order derivative of $\mathcal{S}^{(a;b,c)}_i$ along the $k^-$ direction, and $\mathcal{R}^{(a;b,c)}_{(i;T,j)}$ denotes the $j$th-order derivative of $\mathcal{S}^{(a;b,c)}_i$ along $k_\perp$ direction. Note that $\mathcal{R}$'s depend solely on the momentum fraction $y$, $\zeta^-$, and $\pmb{\ell}^2_{2\perp}$. 

The operator expectation values $\hat{\mathcal{O}}_{(i;0)}$, $\hat{\mathcal{O}}_{(i;T,j)}$, and $\hat{\mathcal{O}}_{(i;L,i)}$ represent two-point gluonic or fermionic jet-medium correlations (also called jet-medium transport coefficients). The factors of $k^\rho_\perp\otimes k^\sigma_\perp$ and $k^-\otimes k^-$ in the Taylor series expansion are converted into spacetime derivatives acting on the gluonic fields $A^{\mu}$ or fermionic quark fields $\psi$ (see below) and are thereby absorbed in the definition of jet-medium transport coefficients. The expectation value $\hat{\mathcal{O}}$ of the two-point gluonic correlator will be labeled $\hat{\mathcal A}$, while $\hat{\mathcal F}$ represents the expectation value of the two-point fermionic correlator. The superscripts ${(a;b,c)}$ are only really needed for the envelopes $\mathcal{R}^{(a;b,c)}_{(i)}$ to identify the specific processes within a kernel. 
%For the expectation value of the two-point correlator $\hat{\mathcal{O}}$, once the fermionic or gluonic variety of the two-point correlator is known, the superscript simply discerns whether a heavy quark or a light quark is involved, which is already encapsulated in $\mathcal{R}^{(a;b,c)}_{(i)}$. Thus, there will be no superscripts in $\hat{\mathcal{A}}$ and $\hat{\mathcal{F}}$ correlators, only subscripts. 

The in-medium, stochastic, gluon-induced momentum broadening in the transverse direction of a quark jet is encapsulated in the operator $\hat{\mathcal{O}}_{(1;T,2)}$, labeled as $\hat{\mathcal{A}}_{(T,2)}$ from now on, and more commonly known as the jet-medium transport coefficient $\hat{q}$~\cite{Majumder:2007hx}. Similarly, $\hat{\mathcal{O}}_{(1;L,1)}$, from now on labeled as $\hat{\mathcal{A}}_{(L,1)}$, is known as $\hat e$, quantified the longitudinal momentum drag imparted by the nuclear medium on the jet parton.\footnote{The longitudinal momentum broadening $\hat{e}_2$ would be given by $\hat{\mathcal{O}}_{(1;L,2)}=\hat{\mathcal{A}}_{(L,2)}$, which the reader can straightforwardly derive using the process shown here.} There exists lattice determinations of the jet transport coefficient $\hat{q}$ containing the gluonic correlator $\hat{\mathcal{A}}$ \cite{Kumar:2020wvb,Weber:2023aea,Panero:2013pla,Majumder:2012sh,Kumar:2020pdl,Kumar:2019aop,Kumar:2018cgf}, however, the estimates of jet transport coefficient containing fermionic correlator $\hat{\mathcal{F}}$ is still unknown.

Only kernel-1 involves the gluonic two-point correlator, which are given by:
%Note, $\mathcal{R}^{(1)}_{i}$'s are independent of the momentum $k$ (and therefore independent of $\eta$), thus, only depend on the momentum fraction $y$, $\zeta^-$, $\pmb{\ell}^2_{2\perp}$, and quark mass $M$.
%
\begin{eqnarray}
 \hat{\mathcal{O}}_{(1;0)}=\hat{ \mathcal{A}}_{0} & = &   g^{2}_{s}  \int d (\Delta z^{-})d^2 \Delta z_{\perp}  \frac{d^2 k_{\perp}}{(2\pi)^2}  e^{-i\Delta z^{-}\mathcal{H}^{(\ell_2,p_2)}_M} e^{i\pmb{k}_{\perp}\cdot \Delta\pmb{z}_{\perp} }  \nonumber  \\
& & \times  \theta(\zeta^-) \left\langle P_{A-1} \left| {\rm Tr}\left[A^{+}(\zeta^-, \Delta z^-, \Delta z_{\perp})A^+(\zeta^-, 0)\right] \right| P_{A-1} \right\rangle, \\
 \hat{\mathcal{O}}_{(1;L,1)}=\hat{ \mathcal{A}}_{(L,1)} & = &   g^{2}_{s}  \int d (\Delta z^{-})d^2 \Delta z_{\perp}  \frac{d^2 k_{\perp}}{(2\pi)^2}  e^{-i\Delta z^{-}\mathcal{H}^{(\ell_2,p_2)}_M} e^{i\pmb{k}_{\perp}\cdot \Delta\pmb{z}_{\perp} }  \nonumber \\
& & \times  \theta(\zeta^-) \left\langle P_{A-1} \left| {\rm Tr}\left[i \left\{\partial^-  A^{+}(\zeta^-, \Delta z^-, \Delta z_{\perp})\right\}A^+(\zeta^-, 0)\right] \right| P_{A-1} \right\rangle, \\
\hat{\mathcal{O}}_{(1;T,2)}=\hat{ \mathcal{A}}_{(T,2)} & = &  g^{2}_{s}  \int d (\Delta z^{-})d^2 \Delta z_{\perp}  \frac{d^2 k_{\perp}}{(2\pi)^2}  e^{-i\Delta z^{-}\mathcal{H}^{(\ell_2,p_2)}_M} e^{i\pmb{k}_{\perp}\cdot \Delta\pmb{z}_{\perp} }   \nonumber \\
& & \times  \theta(\zeta^-) \left\langle P_{A-1} \left| {\rm Tr}\left[\left\{\partial_{\perp} A^{+}(\zeta^-, \Delta z^-, \Delta z_{\perp})\right\} \left\{\partial_{\perp}A^+(\zeta^-,0)\right\}\right]  \right| P_{A-1} \right\rangle.
\end{eqnarray}

Note that $\hat{\mathcal{A}}_{(L,1)}$ and $\hat{\mathcal{A}}_{T,2}$ depend explicitly on $\pmb{\ell}_{2\perp}$ via the function $\mathcal{H}^{(\ell_2,p_2)}_M$, thus these are transverse-momentum-dependent gluon parton distribution functions \cite{Kumar:2025rsa}. Furthermore, $\hat{\mathcal{A}}_{0}$ gives rise to a gauge correction to the nuclear PDF in the limit $\pmb{\ell}_{\perp} \rightarrow \pmb{0}_\perp$ and $\pmb{k}_{\perp} \rightarrow \pmb{0}_\perp$~\cite{Wang:2001ifa}, and is therefore not a jet-medium transport coefficient.  

On the other hand, kernel-2 through kernel-4 all include the fermionic two-point correlators. For kernel-2 and kernel-3, the two-point functions are
\begin{eqnarray}
 \hat{\mathcal{O}}_{(i;0)}=\hat{ \mathcal{F}}_{(0)} & = & g^{2}_{s}  \int d (\Delta z^{-})d^2 \Delta z_{\perp}  \frac{d^2 k_{\perp}}{(2\pi)^2}  e^{-i\Delta z^{-}\mathcal{H}^{(\ell_2,p_2)}_M} e^{i\pmb{k}_{\perp}\cdot \Delta\pmb{z}_{\perp} }  \nonumber \\
& &\times \theta(\zeta^-) \left\langle P_{A-1}\left|\bar{\psi}_{_f}\left(\zeta^-,0\right)\frac{\gamma^+}{4}\psi_{_f}\left(\zeta^-, \Delta z^-, \Delta\pmb{z}_{\perp}\right) \right|P_{A-1}\right\rangle, \label{eq:F_hat_0_kernel_2_3} \\
\hat{ \mathcal{O}}_{(i;L,1)}=\hat{ \mathcal{F}}_{(L,1)} & = & g^{2}_{s}  \int d (\Delta z^{-})d^2 \Delta z_{\perp}  \frac{d^2 k_{\perp}}{(2\pi)^2}  e^{-i\Delta z^{-}\mathcal{H}^{(\ell_2,p_2)}_M} e^{i\pmb{k}_{\perp}\cdot \Delta\pmb{z}_{\perp} }  \nonumber \\
& & \times \theta(\zeta^-) \left\langle P_{A-1}\left|i\left\{\partial^-\bar{\psi}_{_f}\left(\zeta^-,0\right)\right\}\frac{\gamma^+}{4}\psi_{_f}\left(\zeta^-, \Delta z^-, \Delta\pmb{z}_{\perp}\right) \right|P_{A-1}\right\rangle, \label{eq:F_hat_L1L1_kernel_2_3} \\
\hat{ \mathcal{O}}_{(i;T,2)}=\hat{ \mathcal{F}}_{(T,2)} & =& g^{2}_{s}  \int d (\Delta z^{-})d^2 \Delta z_{\perp}  \frac{d^2 k_{\perp}}{(2\pi)^2}  e^{-i\Delta z^{-}\mathcal{H}^{(\ell_2,p_2)}_M} e^{i\pmb{k}_{\perp}\cdot \Delta\pmb{z}_{\perp} }  \nonumber \\
& & \times \theta(\zeta^-) \left\langle P_{A-1}\left|\left\{\partial_{\perp}\bar{\psi}_{_f}\left(\zeta^-,0\right)\right\}\frac{\gamma^+}{4}\left\{\partial_{\perp}\psi_{_f}\left(\zeta^-, \Delta z^-, \Delta\pmb{z}_{\perp}\right)\right\} \right|P_{A-1}\right\rangle,
\label{eq:F_hat_T2T2_kernel_2_3}
\end{eqnarray}
where the index $i\in\{2,3\}$ was suppressed in $\hat{\mathcal{F}}$, given that kernel-2 and kernel-3 involve identical jet-medium transport coefficients.
The $\hat{\mathcal{F}}_{(0)}, \hat{\mathcal{F}}_{(L,1)}$, and $\hat{\mathcal{F}}_{(T,2)}$ distribution functions depend explicitly on $\pmb{\ell}_{2\perp}$ via the function $\mathcal{H}^{(\ell_2,p_2)}_M$, where for light flavors $\mathcal{H}^{(\ell_2,p_2)}_M\to\mathcal{H}^{(\ell_2,p_2)}_0$, and are transverse-momentum-dependent quark parton distribution functions. Note that the zeroth-order term in Eq.~(\ref{eq:F_hat_0_kernel_2_3}) does not represent a gauge term and cannot be absorbed in the nuclear PDF; instead, it gives rise to the leading term contributing to the parton energy loss. 

For kernel-4, the fermionic two-point function is as follows:
\begin{eqnarray}
 \hat{\mathcal{O}}_{(4;0)}=\hat{ \mathcal{F}}_{(4;0)} & = & g^{2}_{s}  \int d (\Delta z^{-})d^2 \Delta z_{\perp}  \frac{d^2 k_{\perp}}{(2\pi)^2}  e^{-i\Delta z^{-}\mathcal{H}^{(\ell_2,p_2)}_M} e^{i\pmb{k}_{\perp}\cdot \Delta\pmb{z}_{\perp} }  \nonumber \\
& &\times \theta(\zeta^-) \left\langle P_{A-1}\left|\bar{\psi}_{_f}(\zeta^-, \Delta z^-, \Delta\pmb{z}_{\perp})\frac{\gamma^+}{4}\psi_{_f}\left(\zeta^-,0\right) \right|P_{A-1}\right\rangle, \\
\hat{ \mathcal{O}}_{(4;L,1)}=\hat{ \mathcal{F}}_{(4;L,1)} & = & g^{2}_{s}  \int d (\Delta z^{-})d^2 \Delta z_{\perp}  \frac{d^2 k_{\perp}}{(2\pi)^2}  e^{-i\Delta z^{-}\mathcal{H}^{(\ell_2,p_2)}_M} e^{i\pmb{k}_{\perp}\cdot \Delta\pmb{z}_{\perp} }  \nonumber \\
& & \times \theta(\zeta^-) \left\langle P_{A-1}\left|i\left\{\partial^-\bar{\psi}_{_f}(\zeta^-, \Delta z^-, \Delta\pmb{z}_{\perp})\right\}\frac{\gamma^+}{4}\psi_{_f}\left(\zeta^-,0\right) \right|P_{A-1}\right\rangle, \\
\hat{ \mathcal{O}}_{(4;T,2)}=\hat{ \mathcal{F}}_{(4;T,2)} & =& g^{2}_{s}  \int d (\Delta z^{-})d^2 \Delta z_{\perp}  \frac{d^2 k_{\perp}}{(2\pi)^2}  e^{-i\Delta z^{-}\mathcal{H}^{(\ell_2,p_2)}_M} e^{i\pmb{k}_{\perp}\cdot \Delta\pmb{z}_{\perp} }  \nonumber \\
& & \times \theta(\zeta^-) \left\langle P_{A-1}\left|\left\{\partial_{\perp}\bar{\psi}_{_f}(\zeta^-, \Delta z^-, \Delta\pmb{z}_{\perp})\right\}\frac{\gamma^+}{4}\left\{\partial_{\perp}\psi_{_f}\left(\zeta^-,0\right)\right\} \right|P_{A-1}\right\rangle,
\end{eqnarray}
where the arguments of $\psi$ and $\bar{\psi}$ are reversed compared to the second and third kernels, thus an additional subscript is present. 
%%%%%%%%%%%%%%%%%%%%%%%%%%%%%%%%%%%%%%%%%%%%%%%%%%%%%%%%%%%%%%%%%%%%%%%%%%%%%%%%%%%%%%%%%%%%%%%%
%%%%%%%%%%%%%%%%%%%%%%%%%%%%%%%%%%%%%%%%%%%%%%%%%%%%%%%%%%%%%%%%%%%%%%%%%%%%%%%%%%%%%%%%%%%%%%%%
\subsection{Expansion of the one quark and gluon final state}
\label{subsec:quark_gluon_expanded}
%%%%%%%%%%%%%%%%%%%%%%%%%%%%%%%%%%%%%%%%%%%%%%%%%%%%%%%%%%%%%%%%%%%%%%%%%%%%%%%%%%%%%%%%%%%%%%%%
%%%%%%%%%%%%%%%%%%%%%%%%%%%%%%%%%%%%%%%%%%%%%%%%%%%%%%%%%%%%%%%%%%%%%%%%%%%%%%%%%%%%%%%%%%%%%%%%
For the case where a heavy quark $Q$ undergoes $Q\to Q+g$ illustrated in Fig.~\ref{fig:kernel-1_all}
%%%%%%%%%%%%%%%%%%%%%%%%%%%%%%%%%%%%%%%%%%%%%%%%%%%%%%%%%%%%%%%%%%%%%%%%%%%%%%%%%%%%%%%%%%%%%%%%%%%%%%%%
% Below is version 1 of the result, which will now be simplified
%%%%%%%%%%%%%%%%%%%%%%%%%%%%%%%%%%%%%%%%%%%%%%%%%%%%%%%%%%%%%%%%%%%%%%%%%%%%%%%%%%%%%%%%%%%%%%%%%%%%%%%%
\begin{equation}
\begin{split}
\mathcal{R}^{\left(Q;Q,g\right)}_{\left(1;T,2\right)}&=C_A \left[\frac{1 +\left(1-y\right)^2}{y}\right] \mathfrak{a}+C_F\left[\frac{1+\left(1-y\right)^2}{y}\right]\mathfrak{b}+C_F\left[\frac{1+\left(1-y\right)^2}{y}\right]\mathfrak{c}\\
&+\left[\frac{C_A}{2}-C_F\right]\left[\frac{1+\left(1-y\right)^2}{y}\right]\mathfrak{d}-\frac{C_A}{2}\left[\frac{1+\left(1-y\right)^2}{y}\right]\mathfrak{e}-\frac{C_A}{2}\left[\frac{1+\left(1-y\right)^2}{y}\right]\mathfrak{f}\\
&-C_F\left[\frac{1+\left(1-y\right)^2}{y}\right]\mathfrak{g},
\label{eq:R_1T2}
\end{split}
\end{equation}
where
\begin{eqnarray}
\mathfrak{a}&=&\frac{4}{\left(1+\chi\right)^2}\left[\left[1-2\left(\frac{3+\kappa y^2\chi}{1+\chi}\right)+6\frac{\left(1+\kappa y^2\chi\right)}{\left(1+\chi\right)^2}\right]\frac{\left[2-2\cos\left\{\mathcal{G}^{(\ell_2)}_M\zeta^-\right\}\right]}{\pmb{\ell}^4_{2\perp}}\right.\nonumber\\
&&\left.\quad\quad\quad\quad\quad+2\beta\left[3+\kappa y^2\chi-4\left(\frac{1+\kappa y^2 \chi}{1+\chi}\right)\right]\frac{\sin\left\{\mathcal{G}^{(\ell_2)}_M\zeta^-\right\}}{\pmb{\ell}^2_{2\perp}}\right.\nonumber\\
&&\left.\quad\quad\quad\quad\quad+2\beta^2\left(1+\kappa y^2\chi\right)\cos\left\{\mathcal{G}^{(\ell_2)}_M\zeta^-\right\}\right],\\
\mathfrak{b}&=&0,\\
\mathfrak{c}&=&\frac{4}{\left(1+\chi\right)^2}\left[1-2\left(\frac{3+\kappa y^2\chi}{1+\chi}\right)+6\left(\frac{1+\kappa y^2\chi}{\left(1+\chi\right)^2}\right)\right]\frac{y^2}{\pmb{\ell}^4_{2\perp}},\\
\mathfrak{d}&=&\frac{4y^2}{\left(1+\chi\right)^3}\left[2\left(\frac{1+\kappa y^2\chi}{1+\chi}\right)-\left(2+\kappa y^2 \chi\right)\right]\frac{\left[2-2\cos\left\{\mathcal{G}^{(\ell_2)}_M\zeta^-\right\}\right]}{\pmb{\ell}^4_{2\perp}},\\
%\end{eqnarray}
%\begin{eqnarray}
\mathfrak{e}&=&\frac{4}{\left(1+\chi\right)^2}\left[\left[y-\frac{\left(1+y\right)^2+\left(1+y^2\right)\left(1+\kappa y^2 \chi\right)}{\left(1+\chi\right)}+\frac{2\left(1+y+y^2\right)\left(1+\kappa y^2 \chi\right)}{\left(1+\chi\right)^2}\right]\frac{\left[2-2\cos\left\{\mathcal{G}^{(\ell_2)}_M\zeta^-\right\}\right]}{\pmb{\ell}^4_{2\perp}}\right.\nonumber\\
 &&\quad\quad\quad\quad\quad\left. +2\beta\left[\left(2+y+\kappa y^2 \chi\right)-\frac{2\left(1+y\right)\left(1+\kappa y^2 \chi\right)}{\left(1+\chi\right)}\right]\frac{\sin\left\{\mathcal{G}^{(\ell_2)}_M\zeta^-\right\}}{\pmb{\ell}^2_{2\perp}}\right.\nonumber\\
 &&\quad\quad\quad\quad\quad\left. +2\beta^2\left(1+\kappa y^2 \chi\right)\cos\left\{\mathcal{G}^{(\ell_2)}_M\zeta^-\right\}\right],\\
\mathfrak{f}&=&\frac{4}{\left(1+\chi\right)^2}\left[2\left[\frac{2\left(1+\kappa y^2 \chi\right)}{\left(1+\chi\right)^2}-\frac{(2+\kappa y^2 \chi)}{\left(1+\chi\right)}\right]\frac{\left[2-2\cos\left\{\mathcal{G}^{(\ell_2)}_M\zeta^-\right\}\right]}{\pmb{\ell}^4_{2\perp}}\right.\nonumber\\
&&\quad\quad\quad\quad\quad\left. +2\beta\left[2+\kappa y^2 \chi-\frac{2\left(1+\kappa y^2 \chi\right)}{\left(1+\chi\right)}\right]\frac{\sin\left\{\mathcal{G}^{(\ell_2)}_M\zeta^-\right\}}{\pmb{\ell}^2_{2\perp}} +2\beta^2\left(1+\kappa y^2 \chi\right)\cos\left\{\mathcal{G}^{(\ell_2)}_M\zeta^-\right\} -2\beta^2\left(1+\kappa y^2 \chi\right)\right.\nonumber\\
&&\left.\quad\quad\quad\quad\quad +n\left[\frac{2\left(1+\kappa y^2 \chi\right)}{\left(1+\chi\right)^2}-\frac{(2+\kappa y^2 \chi)}{\left(1+\chi\right)}\right]\frac{\left[4\cos\left\{\mathcal{G}^{(\ell_2)}_M\zeta^-\right\}-2\right]}{\pmb{\ell}^4_{2\perp}}+2n\beta^2\left(1+\kappa y^2 \chi\right)\right],\\
\mathfrak{g}&=&0,
\end{eqnarray}
and
\begin{eqnarray}
\chi&=&\frac{y^2M^2}{\pmb{\ell}^2_{2\perp}},\label{eq:chi_def_main}\\
\beta&=&\frac{\zeta^-}{2y\left(1-y\right)q^-}. \label{eq:beta_def_main}
\end{eqnarray}
The expressions for $\mathfrak{a}$ through $\mathfrak{g}$ are directly comparable to the results in Ref.~\cite{Sirimanna:2021sqx}, after taking the massless limit $\chi\to0$. There are three main differences between our result and that of Ref.~\cite{Sirimanna:2021sqx}. The first discrepancy occurs at the term $\mathfrak{c}$, which originates from the diagram in Fig.~\ref{fig:kernel1_b_c}(b) and corresponds to $H_{10}$ in Ref.~\cite{Sirimanna:2021sqx}. Reference \cite{Sirimanna:2021sqx} found that if one sets the quark mass to zero before Taylor expanding in $k_\perp$, the process in Fig.~\ref{fig:kernel1_b_c}(b) does not contribute to the medium-modified kernel-1. Our result shows instead that if one performs the Taylor expansion in $k_\perp$ first and then takes the massless limit $\chi\to 0$ the opposite conclusion is reached: The process in Fig.~\ref{fig:kernel1_b_c}(b) does contribute to medium modifications to kernel-1. Furthermore, the QCD Casimir coefficient associated with $\mathfrak{d}$ is different in Ref.~\cite{Sirimanna:2021sqx}. Finally, $\mathfrak{f}$ differs from Ref.~\cite{Sirimanna:2021sqx} even in the massless limit: Eq. (67) in Ref.~\cite{Sirimanna:2021sqx} quotes a phase factor of $\left[2\cos\left\{\mathcal{G}^{(\ell_2)}_0\zeta^-\right\}-2\cos\left\{\Delta\mathcal{G}_0\zeta^-\right\}\right]$, while we obtain $\left[4\cos\left\{\mathcal{G}^{(\ell_2)}_0\zeta^-\right\}-2\cos\left\{\Delta\mathcal{G}_0\zeta^-\right\}\right]$ in the massless limit from the left and right cut interference term. Thus, the third line in $\mathfrak{f}$ differs from Eq.~(68) in Ref.~\cite{Sirimanna:2021sqx}. To summarize: The first two lines of $\mathfrak{f}$ are derived from central cut diagrams, while the left- and right-cut interference terms are in the last line.

The expression in Eq.~(\ref{eq:R_1T2}) can be simplified further by combining all $C_A$ terms together; and similarly for $C_F$ terms. Doing so yields the following:
\begin{equation}
\begin{split}
\mathcal{R}^{\left(Q;Q,g\right)}_{\left(1;T,2\right)}&=C_A \left[\frac{1 +\left(1-y\right)^2}{y}\right] \mathfrak{A}+C_F\left[\frac{1+\left(1-y\right)^2}{y}\right]\mathfrak{B},\\
\label{eq:R_1T2_final}
\end{split}
\end{equation}
where
\begin{eqnarray}
\mathfrak{A}&=&\frac{4}{\left(1+\chi\right)^3}\left[\frac{\left\{2-y-\left[2+4n-(5+2n)\kappa y^2+2\kappa y^3\right]\chi+\left[2-y-(1+2n)\kappa y^2\right]\chi^2\right\}}{2\left(1+\chi\right)}\frac{\left[2-2\cos\left\{\mathcal{G}^{(\ell_2)}_M\zeta^-\right\}\right]}{\pmb{\ell}^4_{2\perp}}\right.\\
&&\left.\quad\quad\quad\quad\quad-\beta\left(2-y\right)\left[1-\left(1-2\kappa y^2\right)\chi\right]\frac{\sin\left\{\mathcal{G}^{(\ell_2)}_M\zeta^-\right\}}{\pmb{\ell}^2_{2\perp}}+n\left[\frac{\chi\left(2-\kappa y^2 +\kappa y^2\chi\right)}{\left(1+\chi\right)}\right]\frac{1}{\pmb{\ell}^4_{2\perp}}+(1-n)\beta^2(1+\chi)(1+\kappa y^2 \chi)\right],\nonumber\\
\mathfrak{B}&=&\frac{4y^2}{\left(1+\chi\right)^4}\left[\left[\frac{1-4\left(1-\kappa y^2\right)\chi+\left(1-2\kappa y^2\right)\chi^2}{\pmb{\ell}^4_{2\perp}}\right]+\left[\chi\left(2-\kappa y^2+\kappa y^2 \chi\right)\right]\frac{\left[2-2\cos\left\{\mathcal{G}^{(\ell_2)}_M\zeta^-\right\}\right]}{\pmb{\ell}^4_{2\perp}}\right].
\end{eqnarray}
In terms of the longitudinal direction, 
\begin{equation}
\begin{split}
\mathcal{R}^{\left(Q;Q,g\right)}_{\left(1;L,1\right)}&=C_A \left[\frac{1 +\left(1-y\right)^2}{y}\right] \mathfrak{a}'+C_A \left[\frac{1 +\left(1-y\right)^2}{y}\right] \mathfrak{b}'+C_F\left[\frac{1+(1-y)^2}{y}\right]\mathfrak{c}'+\left[\frac{C_A}{2}-C_F\right]\left[\frac{1+(1-y)^2}{y}\right]\mathfrak{d}'\\
&-\frac{C_A}{2}\left[\frac{1+\left(1-y\right)^2}{y}\right]\mathfrak{e}'-\frac{C_A}{2}\left[\frac{1 +\left(1-y\right)^2}{y}\right]\mathfrak{f}'-C_F\left[\frac{1 +\left(1-y\right)^2}{y}\right]\mathfrak{g}'
\end{split}
\label{eq:R_1L1}
\end{equation}
with
\begin{eqnarray}
\mathfrak{a}'&=&\frac{1}{yq^-\left(1+\chi\right)^2}\left[\left[1+\kappa y^2 \chi+\frac{4\chi\left(1+\kappa y^2\chi\right)}{1+\chi}\right]\frac{\left[2-2\cos\left\{\mathcal{G}^{(\ell_2)}_M\zeta^-\right\}\right]}{\pmb{\ell}^2_{2\perp}}\right.\nonumber\\
&&\quad\quad\quad\quad\quad\quad\left.+2\beta\left(1+\kappa y^2\chi\right)\left[\left(1+\chi\right)\left(\frac{1-2y}{1-y}\right)-2\chi\right]\sin\left\{\mathcal{G}^{(\ell_2)}_M\zeta^-\right\}\right],\\
\mathfrak{b}'&=&0,\\ 
\mathfrak{c}'&=&\frac{1}{yq^-\left(1+\chi\right)^2}\left[(2-y)y\kappa\left(1+\kappa y^2\chi\right)+y\left[1-\frac{2\left(1+\kappa y^2 \chi\right)}{1+\chi}\right]\right]\frac{2}{\pmb{\ell}^2_{2\perp}},\\
\mathfrak{d}'&=&\frac{1}{yq^-\left(1+\chi\right)^2}\left[(2-y)y\kappa\left(1+\kappa y^2\chi\right)+y\left[1-\frac{2\left(1+\kappa y^2 \chi\right)}{1+\chi}\right]\right]\frac{\left[2-2\cos\left\{\mathcal{G}^{(\ell_2)}_M\zeta^-\right\}\right]}{\pmb{\ell}^2_{2\perp}},\\
\mathfrak{e}'&=&\frac{1}{yq^-\left(1+\chi\right)^2}\left[\left[\frac{1+\kappa y^2\chi}{2}+\frac{2\left(\chi-y\right)\left(1+\kappa y^2\chi\right)}{1+\chi}\right]\frac{\left[2-2\cos\left\{\mathcal{G}^{(\ell_2)}_M\zeta^-\right\}\right]}{\pmb{\ell}^2_{2\perp}}\right.\nonumber\\
&&\quad\quad\quad\quad\quad\quad+\left.2\beta\left(1+\kappa y^2 \chi\right)\left[\left(1+\chi\right)\left(\frac{1-2y}{1-y}\right)-2\chi\right]\sin\left\{\mathcal{G}^{(\ell_2)}_M\zeta^-\right\}\right],\\
\mathfrak{f}'&=&\frac{1}{yq^-\left(1+\chi\right)^2}\left[\left[1+\kappa y^2 \chi+\frac{4\chi\left(1+\kappa y^2\chi\right)}{\left(1+\chi\right)}\right]\frac{\left[2-2\cos\left\{\mathcal{G}^{(\ell_2)}_M\zeta^-\right\}\right]}{\pmb{\ell}^2_{2\perp}}\right.\nonumber\\
&&\left.\quad\quad\quad\quad\quad\quad+2\beta(1+\kappa y^2\chi)\left[\left(1+\chi\right)\left(\frac{1-2y}{1-y}\right)-2\chi\right]\sin\left\{\mathcal{G}^{(\ell_2)}_M\zeta^-\right\}\right.\nonumber\\
&&\left.\quad\quad\quad\quad\quad\quad-n\left[\frac{\left(1+\kappa y^2\chi\right)}{2}-\frac{2\chi\left(1+\kappa y^2\chi\right)}{(1+\chi)}\right]\frac{\left[4\cos\left\{\mathcal{G}^{(\ell_2)}_M\zeta^-\right\}-2\right]}{\pmb{\ell}^2_{2\perp}}\right],\\
\mathfrak{g}'&=&0,
\end{eqnarray}
where $\beta$ is defined in Eq.~(\ref{eq:beta_def_main}).
%and the contribution for each group of diagrams is explicit. 
Note that the first two lines $\mathfrak{f}'$ stem from central cut diagrams, while the last line is from the left- and right-cut interference term. Simplifying Eq.~(\ref{eq:R_1L1}) yields
\begin{equation}
\begin{split}
\mathcal{R}^{\left(Q;Q,g\right)}_{\left(1;L,1\right)}&=C_A \left[\frac{1 +\left(1-y\right)^2}{y}\right] \mathfrak{A}'+C_F \left[\frac{1 +\left(1-y\right)^2}{y}\right] \mathfrak{B}',
\end{split}
\end{equation}
where
\begin{eqnarray}
\mathfrak{A}'&=&\frac{1}{4yq^-\left(1+\chi\right)^3}\left[\left[ a+ b\chi + c\chi^2 \right]\frac{\left[2-2\cos\left\{\mathcal{G}^{(\ell_2)}_M\zeta^-\right\}\right]}{\pmb{\ell}^2_{2\perp}}+2n\left(1-3\chi\right)\frac{(1+\kappa y^2\chi)}{\pmb{\ell}^2_{2\perp}} \right],\nonumber\\
a&=&1 -2n + \left(2+4\kappa\right)y - 2\kappa y^2,\nonumber\\
b&=&5+6n+\left(2+4\kappa\right)y-\left(1+2n\right)\kappa y^2 +2\left(2-y\right)\kappa^2 y^3,\nonumber\\
c&=&\left(5+6n+4\kappa y - 2\kappa y^2\right)\kappa y^2,
\end{eqnarray}
and
\begin{equation}
\begin{split}
\mathfrak{B}'&=\frac{1}{yq^-\left(1+\chi\right)^2}\left[\left(2-y\right)y\kappa\left(1+\kappa y^2 \chi\right) +y\left[1-\frac{2\left(1+\kappa y^2 \chi\right)}{1+\chi}\right]\right]\frac{\left[2\cos\left\{\mathcal{G}^{(\ell_2)}_M\zeta^-\right\}\right]}{\pmb{\ell}^2_{2\perp}}.
\end{split}
\end{equation}
%
%%%%%%%%%%%%%%%%%%%%%%%%%%%%%%%%%%%%%%%%%%%%%%%%%%%%%%%%%%%%%%%%%%%%%%%%%%%%%%%%%%%%%%%%%%%%%%%%
%%%%%%%%%%%%%%%%%%%%%%%%%%%%%%%%%%%%%%%%%%%%%%%%%%%%%%%%%%%%%%%%%%%%%%%%%%%%%%%%%%%%%%%%%%%%%%%%
\subsection{Expansion of the two-gluon final state}
\label{subsec:two_gluon_expanded}
%%%%%%%%%%%%%%%%%%%%%%%%%%%%%%%%%%%%%%%%%%%%%%%%%%%%%%%%%%%%%%%%%%%%%%%%%%%%%%%%%%%%%%%%%%%%%%%%
%%%%%%%%%%%%%%%%%%%%%%%%%%%%%%%%%%%%%%%%%%%%%%%%%%%%%%%%%%%%%%%%%%%%%%%%%%%%%%%%%%%%%%%%%%%%%%%%
To extract the jet-medium transport coefficient associated with $q\to g+g$, one Taylor expands $\mathcal{S}_2$ around $k_\perp=0$ giving 
\begin{equation}
\begin{split}
\mathcal{R}^{\left(q;g,g\right)}_{\left(2;T,2\right)}&=C_AC^2_F\mathfrak{a}+2C^2_AC_F\mathfrak{b}-\frac{C^2_AC_F}{2}\mathfrak{c}-\frac{C^2_AC_F}{2}\mathfrak{d}-C_A C_F\left(\frac{C_A}{2}-C_F\right)\mathfrak{e},
\end{split}
\end{equation}
where
\begin{eqnarray}
\mathfrak{a}&=&\frac{4\left(1+y^2\right)}{yq^-\left(1-y\right)}\left[\frac{\left[2-2\cos\left\{\mathcal{G}^{(\ell_2)}_{0} \zeta^-\right\}\right]}{\pmb{\ell}^4_{2\perp}}-\frac{2\beta}{\pmb{\ell}^2_{2\perp}}\sin\left\{\mathcal{G}^{(\ell_2)}_{0} \zeta^-\right\}+2\beta^2\cos\left\{\mathcal{G}^{(\ell_2)}_{0} \zeta^-\right\}\right],\\
\mathfrak{b}&=&\left[\frac{y}{1-y}+\frac{1-y}{y}+y\left(1-y\right)\right]\frac{4y^2}{q^-\pmb{\ell}^4_{2\perp}},\\ 
\mathfrak{c}&=&0,\\
\mathfrak{d}&=&\frac{4\left(1+y^3\right)}{q^{-}\left(1-y\right)}\left[\frac{\left[2-2\cos\left\{\mathcal{G}^{(\ell_2)}_{0} \zeta^-\right\}\right]}{\pmb{\ell}^4_{2\perp}}-\frac{2\beta}{\pmb{\ell}^2_{2\perp}}\sin\left\{\mathcal{G}^{(\ell_2)}_{0} \zeta^-\right\}+\frac{2\beta^2}{y}\cos\left\{\mathcal{G}^{(\ell_2)}_{0} \zeta^-\right\}\right],\\
\mathfrak{e}&=&\frac{4\beta^2}{yq^-}\left[2-2\cos\left\{\mathcal{G}^{(\ell_2)}_{0} \zeta^-\right\}\right].
\end{eqnarray}
Performing a Taylor expansion of $\mathcal{S}_2$ around $k^-=0$ yields
\begin{equation}
\begin{split}
\mathcal{R}^{\left(q;g,g\right)}_{\left(2;L,1\right)}&=C_AC^2_F\mathfrak{a}'+2C^2_AC_F\mathfrak{b}'-\frac{C^2_AC_F}{2}\mathfrak{c}'-\frac{C^2_AC_F}{2}\mathfrak{d}'+C_A C_F\left(\frac{C_A}{2}-C_F\right)\mathfrak{e}',
\end{split}
\end{equation}
where
\begin{eqnarray}
\mathfrak{a}'&=&-\frac{3}{y\left[\left(1-y\right)q^-\right]^2\pmb{\ell}^2_{2\perp}}\left[2-2\cos\left\{\mathcal{G}^{(\ell_2)}_{0} \zeta^-\right\}\right]+\frac{2\beta\left(1+y^2\right)}{\left(yq^-\right)^2\left(1-y\right)}\left(\frac{1-2y}{1-y}\right)\sin\left\{\mathcal{G}^{(\ell_2)}_{0} \zeta^- \right\},\\
\mathfrak{b}'&=&\frac{1}{\left(q^-\right)^2\pmb{\ell}^2_{2\perp}}\left\{y+\frac{1}{y}-\frac{y}{(1-y)^2}-2\left[\frac{y}{1-y}+\frac{1-y}{y}+2y\left(1-y\right)\right]\right\},\\ 
\mathfrak{c}'&=&\left[\frac{2y\left[1+\left(1-y\right)^3\right]-3}{y\left(1-y\right)^2}\right]\frac{\left[2-2\cos\left\{\mathcal{G}^{(\ell_2)}_{0} \zeta^- \right\}\right]}{\left(q^-\right)^2\pmb{\ell}^2_{2\perp}},\\
\mathfrak{d}'&=&\frac{\left(1+y^3\right)}{y\left(1-y\right)\left(q^{-}\right)^2}\left[\frac{2\beta}{y}\left(\frac{1-2y}{1-y}\right)\sin\left\{\mathcal{G}^{(\ell_2)}_{0} \zeta^- \right\}-\left(\frac{3-2y}{1-y}\right)\frac{\left[2-2\cos\left\{\mathcal{G}^{(\ell_2)}_{0} \zeta^- \right\}\right]}{\pmb{\ell}^2_{2\perp}}\right],\\
\mathfrak{e}'&=&\frac{2}{y(1-y)\left(q^-\right)^2}\left[\frac{\left[2-2\cos\left\{\mathcal{G}^{(\ell_2)}_{0} \zeta^-\right\}\right]}{\pmb{\ell}^2_{2\perp}}+ \frac{\beta(1-2y)}{y}\sin\left\{\mathcal{G}^{(\ell_2)}_{0} \zeta^-\right\}\right].
\end{eqnarray}

%%%%%%%%%%%%%%%%%%%%%%%%%%%%%%%%%%%%%%%%%%%%%%%%%%%%%%%%%%%%%%%%%%%%%%%%%%%%%%%%%%%%%%%%%%%%%%%%
%%%%%%%%%%%%%%%%%%%%%%%%%%%%%%%%%%%%%%%%%%%%%%%%%%%%%%%%%%%%%%%%%%%%%%%%%%%%%%%%%%%%%%%%%%%%%%%%
\subsection{Expansion of the one quark and one antiquark final state}
\label{subsec:quark_antiquark_expanded}
%%%%%%%%%%%%%%%%%%%%%%%%%%%%%%%%%%%%%%%%%%%%%%%%%%%%%%%%%%%%%%%%%%%%%%%%%%%%%%%%%%%%%%%%%%%%%%%%
%%%%%%%%%%%%%%%%%%%%%%%%%%%%%%%%%%%%%%%%%%%%%%%%%%%%%%%%%%%%%%%%%%%%%%%%%%%%%%%%%%%%%%%%%%%%%%%%
For the case where a heavy quark $Q$ undergoes $Q\to Q+\bar q'$ illustrated in Fig.~\ref{fig:kernel-3_all}(a)
\begin{equation}
\begin{split}
\mathcal{R}^{\left(Q;Q,\bar q'\right)}_{\left(3;T,2\right)}&=\left[\frac{C_F C_A}{2}\right] \left[\frac{1 +\left(1-y\right)^2}{y}\right] \frac{1}{yq^-}\frac{4}{\left(1+\chi\right)^2}\\
&\times\left[\left[1-2\left(\frac{3+\kappa y^2\chi}{1+\chi}\right)+6\frac{\left(1+\kappa y^2\chi\right)}{\left(1+\chi\right)^2}\right]\frac{\left[2-2\cos\left\{\mathcal{G}^{(\ell_2)}_M\zeta^-\right\}\right]}{\pmb{\ell}^4_{2\perp}}\right.\\
&\left.\,\,\,\,\,\,\,\,\,+2\beta\left[3+\kappa y^2\chi-4\left(\frac{1+\kappa y^2 \chi}{1+\chi}\right)\right]\frac{\sin\left\{\mathcal{G}^{(\ell_2)}_M\zeta^-\right\}}{\pmb{\ell}^2_{2\perp}}\right.\\
&\left.\,\,\,\,\,\,\,\,\,+2\beta^2\left(1+\kappa y^2\chi\right)\cos\left\{\mathcal{G}^{(\ell_2)}_M\zeta^-\right\}\right],
\end{split}
\end{equation}
while
\begin{equation}
\begin{split}
\mathcal{R}^{\left(Q;Q,\bar q'\right)}_{\left(3;L,1\right)}&=\left[\frac{C_F C_A}{2}\right] \left[\frac{1 +\left(1-y\right)^2}{y}\right] \frac{1}{\left(yq^-\right)^2} \frac{2\left(1+\kappa y^2\chi\right)}{\left(1+\chi\right)^2}\times\\
&\times\left[\frac{2\chi}{\left(1+\chi\right)}\frac{\left[2-2\cos\left\{\mathcal{G}^{(\ell_2)}_M\zeta^-\right\}\right]}{\pmb{\ell}^2_{2\perp}}+\beta\left[\left(1+\chi\right)\left(\frac{1-2y}{1-y}\right)-2\chi\right]\sin\left\{\mathcal{G}^{(\ell_2)}_M\zeta^-\right\}\right].
\end{split}
\end{equation}
For the case a light quark $q$ undergoes $q\to q+\bar q'$ illustrated in Fig.~\ref{fig:kernel-3_all}(a), one simply sets $M=0$ the equations above.
%%
%\begin{equation}
%\begin{split}
%\mathcal{R}^{\left(q;q,\bar q'\right)}_{\left(3;T,2\right)}&=\left[\frac{C_F N_c}{2}\right] \left[\frac{1 +\left(1-y\right)^2}{y}\right] \frac{4}{yq^-}\left[\frac{\left[2-2\cos\left\{\mathcal{G}^{(\ell_2)}_0\zeta^-\right\}\right]}{\pmb{\ell}^4_{2\perp}} -2\beta\frac{\sin\left\{\mathcal{G}^{(\ell_2)}_0\zeta^-\right\}}{\pmb{\ell}^2_{2\perp}}+2\beta^2\cos\left\{\mathcal{G}^{(\ell_2)}_0\zeta^-\right\}\right],
%\end{split}
%\end{equation}
%%
%and
%%
%\begin{equation}
%\begin{split}
%\mathcal{R}^{\left(q;q,\bar q'\right)}_{\left(3;L,1\right)}&=\left[\frac{C_F N_c}{2}\right] \left[\frac{1 +\left(1-y\right)^2}{y}\right] \frac{1}{yq^-}\left[\left(\frac{1-2y}{1-y}\right)\frac{2\beta}{q^-}\sin\left\{\mathcal{G}^{(\ell_2)}_0\zeta^-\right\}\right].
%\end{split}
%\end{equation}
%%

In the situation where a light quark gets annihilated in the medium, as explored in $q\to q'+\bar q'$ illustrated in Fig.~\ref{fig:kernel-3_all}(b), one obtains
\begin{equation}
\begin{split}
\mathcal{R}^{\left(q;q',\bar q'\right)}_{\left(3;T,2\right)}&=\left[\frac{C_{F}C_A}{2} \right]\frac{1}{q^{-}}\frac{4y^2(1-2y+2y^2)}{\pmb{\ell}^4_{2\perp}}\frac{\left(1-\chi'\right)}{\left(1+\chi'\right)^3},
\label{eq:R_q_qprime_bar-qprime}
\end{split}
\end{equation}
where 
\begin{equation}
\chi'=\frac{M^2_{f'}}{\pmb{\ell}^2_{2\perp}}. 
\end{equation}
At sufficiently high $\pmb{\ell}^2_{2\perp}$, Eq.~(\ref{eq:R_q_qprime_bar-qprime}) implies that producing heavy flavor $Q'$ is equally likely as light flavor $q'$, that is
\begin{equation}
\lim_{\pmb{\ell}^2_{2\perp}\to\infty}\left[\frac{\mathcal{R}^{\left(q;Q',\bar Q'\right)}_{\left(3;T,2\right)}}{\mathcal{R}^{\left(q;q',\bar q'\right)}_{\left(3;T,2\right)}}\right]=1.
\end{equation}
The longitudinal drag gives
\begin{equation}
\begin{split}
\mathcal{R}^{\left(q;q',\bar q'\right)}_{\left(3;L,1\right)}&=-\left[\frac{C_{F}C_A}{2} \right]\frac{1}{\left(q^{-}\right)^2}\frac{2\left(1-3y+4y^2\right)}{\pmb{\ell}^2_{2\perp}\left(1+\chi'\right)}.
\end{split}
\end{equation}
Finally, in the situation where all diagrams in Fig.~\ref{fig:kernel-3_all} participate, i.e. a light quark undergoes $q\to q+\bar q$, one obtains the following:
\begin{equation}
\begin{split}
\mathcal{R}^{\left(q;q,\bar q\right)}_{\left(3;T,2\right)}&=\left[\frac{C_F C_A}{2}\right] \left[\frac{1 +\left(1-y\right)^2}{y}\right] \frac{4}{yq^-} \left[\frac{1}{\pmb{\ell}^4_{2\perp}}\left[2-2\cos\left\{\mathcal{G}^{(\ell_2)}_0\zeta^-\right\}\right] -\frac{2\beta}{\pmb{\ell}^2_{2\perp}}\sin\left\{\mathcal{G}^{(\ell_2)}_0\zeta^-\right\}+2\beta^2\cos\left\{\mathcal{G}^{(\ell_2)}_0\zeta^-\right\}\right]\\
&+\left[\frac{C_F C_A}{2}\right]\frac{4}{q^{-}}\frac{y^2\left(1-2y+2y^2\right)}{\pmb{\ell}^4_{2\perp}}\\
&+\left[ C_{F}C_A \left(C_F - \frac{C_A}{2}\right)\right] \frac{4\left(1-y\right)^2}{yq^{-}}\left[\frac{y}{\pmb\ell^4_{2\perp}}\left[2-2\cos\left\{\mathcal{G}^{(\ell_2)}_0\zeta^-\right\}\right] -\frac{2\beta y}{\pmb{\ell}^2_{2\perp}}\sin\left\{\mathcal{G}^{(\ell_2)}_0\zeta^-\right\}+ 2\beta^2\cos\left\{\mathcal{G}^{(\ell_2)}_0\zeta^-\right\}\right],
\end{split}
\end{equation}
and
\begin{equation}
\begin{split}
\mathcal{R}^{\left(q;q,\bar q\right)}_{\left(3;L,1\right)}&=\left[\frac{C_F C_A}{2}\right] \left[\frac{1 +\left(1-y\right)^2}{y}\right] \frac{1}{\left(yq^-\right)^2} \left[2\beta\left(\frac{1-2y}{1-y}\right)\sin\left\{\mathcal{G}^{(\ell_2)}_0\zeta^-\right\}\right]\\
&-\left[\frac{C_F C_A}{2}\right]\frac{2}{\left(q^{-}\right)^2}\left[\frac{1-3y+4y^2}{\pmb{\ell}^2_{2\perp}}\right]\\
&+\left[ C_{F}C_A \left(C_F - \frac{C_A}{2}\right)\right] \frac{\left(1-y\right)^2}{\left(yq^{-}\right)^2}\left[\frac{1-2y+2y^2}{1-y}\frac{\left[2-2\cos\left\{\mathcal{G}^{(\ell_2)}_0\zeta^-\right\}\right]}{\pmb\ell^2_{2\perp}}+2\beta\left(\frac{1-2y}{1-y}\right)\sin\left\{\mathcal{G}^{(\ell_2)}_0\zeta^-\right\}\right].
\end{split}
\end{equation}
%
%%%%%%%%%%%%%%%%%%%%%%%%%%%%%%%%%%%%%%%%%%%%%%%%%%%%%%%%%%%%%%%%%%%%%%%%%%%%%%%%%%%%%%%%%%%%%%%%
%%%%%%%%%%%%%%%%%%%%%%%%%%%%%%%%%%%%%%%%%%%%%%%%%%%%%%%%%%%%%%%%%%%%%%%%%%%%%%%%%%%%%%%%%%%%%%%%
\subsection{Expansion of the two-quark final state}
\label{subsec:two_quark_expanded}
%%%%%%%%%%%%%%%%%%%%%%%%%%%%%%%%%%%%%%%%%%%%%%%%%%%%%%%%%%%%%%%%%%%%%%%%%%%%%%%%%%%%%%%%%%%%%%%%
%%%%%%%%%%%%%%%%%%%%%%%%%%%%%%%%%%%%%%%%%%%%%%%%%%%%%%%%%%%%%%%%%%%%%%%%%%%%%%%%%%%%%%%%%%%%%%%%
For the case of a heavy quark $Q$ participating in the process $Q\to Q + q'$ depicted in Fig.~\ref{fig:kernel-4_all}(a)
\begin{equation}
\begin{split}
\mathcal{R}^{\left(Q;Q,q'\right)}_{\left(4;T,2\right)} & =  \left[ \frac{C_F C_A}{2} \right] \left[\frac{1+\left(1-y\right)^2 }{y}\right] \frac{1}{yq^-} \frac{4}{\left(1+\chi\right)^2} \\
&\times\left[\left[1-2\left(\frac{3+\kappa y^2\chi}{1+\chi}\right)+6\frac{\left(1+\kappa y^2\chi\right)}{\left(1+\chi\right)^2}\right]\frac{\left[2-2\cos\left\{\mathcal{G}^{(\ell_2)}_M\zeta^-\right\}\right]}{\pmb{\ell}^4_{2\perp}}\right.\\
&\left.\,\,\,\,\,\,\,\,\,+2\beta\left[3+\kappa y^2\chi-4\left(\frac{1+\kappa y^2 \chi}{1+\chi}\right)\right]\frac{\sin\left\{\mathcal{G}^{(\ell_2)}_M\zeta^-\right\}}{\pmb{\ell}^2_{2\perp}}\right.\\
&\left.\,\,\,\,\,\,\,\,\,+2\beta^2\left(1+\kappa y^2\chi\right)\cos\left\{\mathcal{G}^{(\ell_2)}_M\zeta^-\right\}\right],
\end{split}     \label{}
\end{equation}
and
\begin{equation}
\begin{split}
\mathcal{R}^{\left(Q;Q,q'\right)}_{\left(4;L,1\right)} & =  \left[ \frac{C_F C_A}{2} \right] \left[\frac{1+\left(1-y\right)^2 }{y}\right] \frac{1}{\left(yq^-\right)^2} \frac{2\left(1+\kappa y^2\chi\right)}{\left(1+\chi\right)^2}\\
& \times\left[\frac{2\chi}{\left(1+\chi\right)} \frac{\left[2-2\cos\left\{\mathcal{G}^{(\ell_2)}_M\zeta^-\right\}\right]}{\pmb{\ell}^2_{2\perp}}+\beta\left[\left(1+\chi\right)\left(\frac{1-2y}{1-y}\right)-2\chi\right]\sin\left\{\mathcal{G}^{(\ell_2)}_M\zeta^-\right\}\right].
\end{split}     \label{}
\end{equation}
For the case of a light quark $q$ participating in the process $q\to q+q'$ illustrated in Fig.~\ref{fig:kernel-4_all}(a), one simply sets $M=0$ in the two equations above. 

Finally, in the case of a light quark $q$ participating in the process $q\to q+q$, then both channels in Fig.~\ref{fig:kernel-4_all} contribute giving
\begin{equation}
\begin{split}
\mathcal{R}^{\left(q;q,q\right)}_{\left(4;T,2\right)}&= \left[\frac{C_F C_A}{2}\right] \left[\frac{1 +\left(1-y\right)^2 }{y}\right] \frac{4}{yq^-} \left[\frac{\left[2-2\cos\left\{\mathcal{G}^{(\ell_2)}_0\zeta^-\right\}\right]}{\pmb{\ell}^4_{2\perp}} -2\beta\frac{\sin\left\{\mathcal{G}^{(\ell_2)}_0\zeta^-\right\}}{\pmb{\ell}^2_{2\perp}}+2\beta^2\cos\left\{\mathcal{G}^{(\ell_2)}_0\zeta^-\right\}\right]\\
%&+ \frac{C_F N_c}{2} \left[ \frac{ 1 + y^2 }{1-y}\right] \frac{1}{q^-} \frac{1}{\left(1-y+\eta y\right)\pmb{\ell}^2_{2\perp}}  \left[ 2 -2 \cos\left\{\mathcal{G}^{(\ell_2)}_{0}\zeta^-\right\} \right]\nonumber\\
&+\frac{C_F C_A\left(C_F - \frac{C_A}{2} \right)}{(1-y)yq^-}\, 4\beta^2\left[2-2\cos\left\{\mathcal{G}^{(\ell_2)}_0\zeta^-\right\}\right],
\end{split}
\end{equation}
with
\begin{equation}
\begin{split}
\mathcal{R}^{\left(q;q,q\right)}_{\left(4;L,1\right)}&= \left[\frac{C_F C_A}{2}\right] \left[\frac{1 +\left(1-y\right)^2 }{y}\right] \frac{1}{\left(yq^-\right)^2} \left[2\beta\left(\frac{1-2y}{1-y}\right)\sin\left\{\mathcal{G}^{(\ell_2)}_0\zeta^-\right\}\right]\\
&-\left[\frac{C_F C_A}{2}\right] \left[ \frac{ 1 + y^2 }{1-y}\right] \frac{1}{\left(q^-\right)^2} \frac{1}{\left(1-y\right)^2}  \frac{\left[ 2 -2 \cos\left\{\mathcal{G}^{(\ell_2)}_{0}\zeta^-\right\} \right]}{\pmb{\ell}^2_{2\perp}}\\
&-\frac{C_F C_A\left(C_F - \frac{C_A}{2} \right)}{(1-y)}\, \frac{2}{\left(yq^-\right)^2}\left[\frac{\left[2-2\cos\left\{\mathcal{G}^{(\ell_2)}_{0}\zeta^-\right\}\right]}{\pmb{\ell}^2_{2\perp}}+\beta\left(\frac{1-2y}{1-y}\right)\sin\left\{\mathcal{G}^{(\ell_2)}_0\zeta^-\right\}\right],
\end{split}
\end{equation}
where $\beta$ is defined in Eq.~(\ref{eq:beta_def_main}).
%%%%%%%%%%%%%%%%%%%%%%%%%%%%%%%%%%%%%%%%%%%%%%%%%%%%%%%%%%%%%%%%%%%%%%%%%%%%%%%%%%%%%%%%%%%%%%%%
%%%%%%%%%%%%%%%%%%%%%%%%%%%%%%%%%%%%%%%%%%%%%%%%%%%%%%%%%%%%%%%%%%%%%%%%%%%%%%%%%%%%%%%%%%%%%%%%
\subsection{Length dependence of energy loss and its implications for jet phenomenology}
%%%%%%%%%%%%%%%%%%%%%%%%%%%%%%%%%%%%%%%%%%%%%%%%%%%%%%%%%%%%%%%%%%%%%%%%%%%%%%%%%%%%%%%%%%%%%%%%
%%%%%%%%%%%%%%%%%%%%%%%%%%%%%%%%%%%%%%%%%%%%%%%%%%%%%%%%%%%%%%%%%%%%%%%%%%%%%%%%%%%%%%%%%%%%%%%%
We perform a numerical evaluation of the perturbative coefficients $\mathcal{R}^{(a;b,c)}_{(i;T,2)}$ arising from the Taylor expansion of the full scattering kernel for kernel-1 and kernel-3. The collinear expansion [Eq.~(\ref{eq:kernel1-coll-expansion})] outlined in the preceding subsections enables us to decouple $k_{\perp}$ and $k^-$ dependencies, incorporating them into the definition of the jet transport coefficients. The length ($\zeta^-$) dependence only appears in the phase factors, such as $[2-2\cos\{\mathcal{G}^{(\ell_2)}_{M}\zeta^-\}]$, $\beta$, and the two-point correlators $\hat{ \mathcal{A}}_{(T,2)} $ or $\hat{ \mathcal{F}}_{(T,2)} $. Below, we discuss the length dependence based on the assumption that the two-point correlators $\hat{ \mathcal{A}}_{(T,2)} $ or $\hat{ \mathcal{F}}_{(T,2)} $ are invariant under translation around $\zeta^-$, thus focusing solely the $\zeta^-$ dependence in $\mathcal{R}$.

\begin{figure}[!h] 
\begin{subfigure}[t]{0.49\textwidth}
\includegraphics[width=\textwidth]{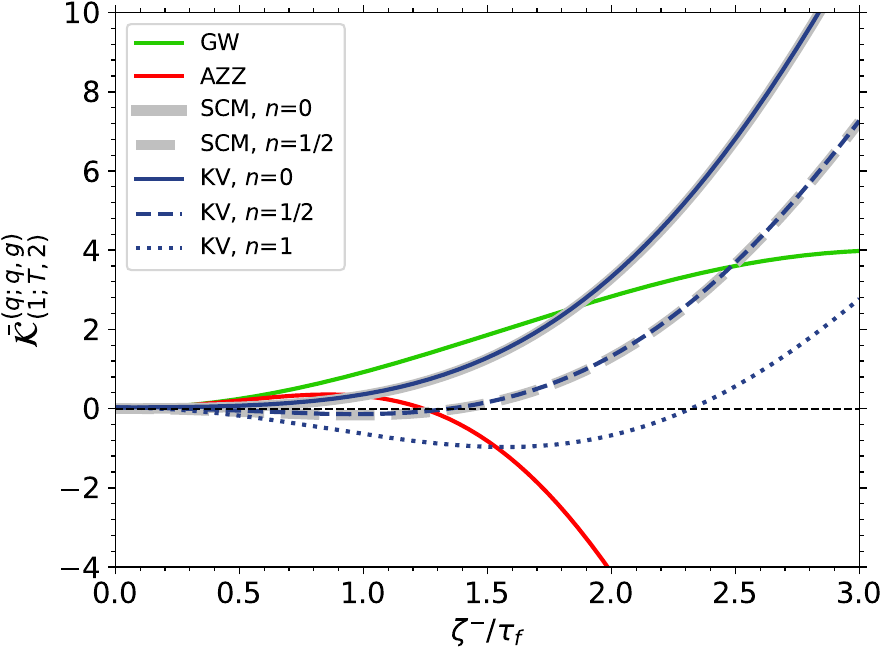}
\caption{$\bar{\mathcal{K}}^{(q;q,g)}_{(1;T,2)}$ at $y=0.25$}
\end{subfigure}
\begin{subfigure}[t]{0.49\textwidth}        
\includegraphics[width=\textwidth]{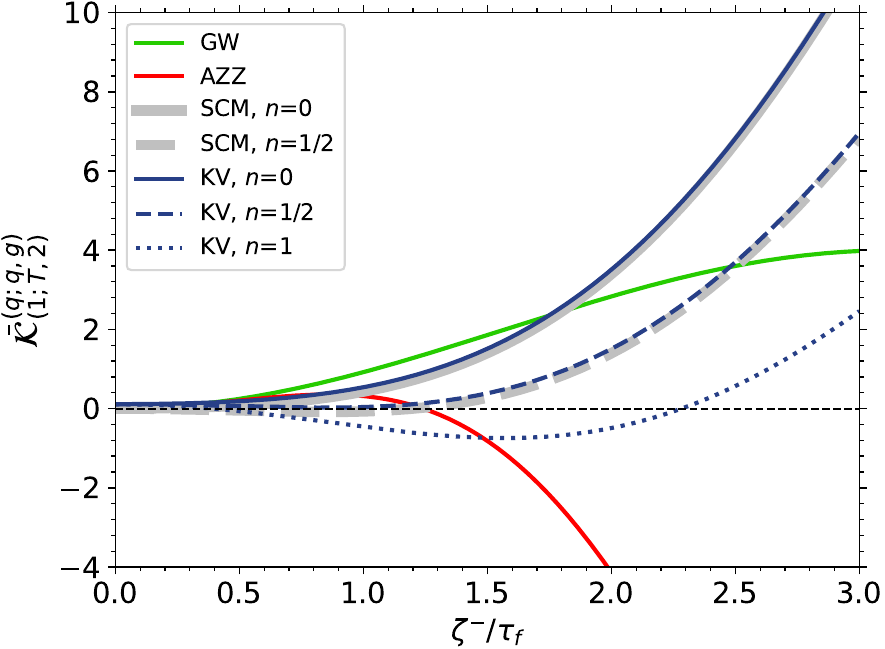}
\caption{$\bar{\mathcal{K}}^{(q;q,g)}_{(1;T,2)}$ at $y=0.5$}
\end{subfigure}
\begin{subfigure}[t]{0.49\textwidth}   
\includegraphics[width=\textwidth]{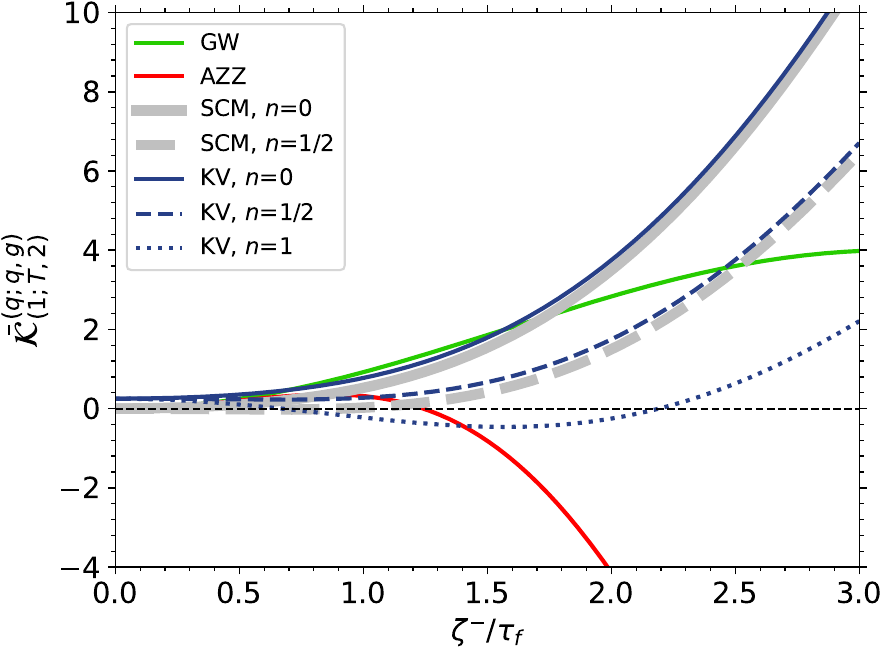}
\caption{$\bar{\mathcal{K}}^{(q;q,g)}_{(1;T,2)}$ at $y=0.75$}
\end{subfigure}
\caption{Comparison of  $\bar{\mathcal{K}}^{(q;q,g)}_{(1;T,2)}$ among various approaches for massless jet quark.} 
\label{fig:kernel1_comp_sirimanna}
\end{figure}

First, a comparison between the results in Fig.~14 of Ref.~\cite{Sirimanna:2021sqx}, and our results for kernel-1 is presented in Fig.~\ref{fig:kernel1_comp_sirimanna}. 
To match our kernel with $\bar{K}$ given in Eq.~(76) of Ref.~\cite{Sirimanna:2021sqx}, we define $x=\pmb{\ell}^2_{2\perp}(1+\chi)\beta=\mathcal{G}^{(\ell_2)}_M\zeta^-=\frac{\zeta^-}{\tau_f}$, factorize the overall coefficient $C_A\left[\frac{1+(1-y)^2}{y}\right]\frac{4}{(1+\chi)^3\pmb{\ell}^4_{2\perp}}$ from Eq.~(\ref{eq:R_1T2_final}), and take the massless $\chi\to0$ limit to give:
\begin{equation}
\bar{\mathcal{K}}^{(q;q,g)}_{(1;T,2)}=\lim_{\chi\to 0^+}\left[\frac{(1+\chi)^3\pmb{\ell}^4_{2\perp}\mathcal{R}^{(Q;Q,g)}_{(1;T,2)}}{4C_A\left[\frac{1+(1-y)^2}{y}\right]}\right]=\frac{2-y}{2}\left[2-2\cos\left\{x\right\}-2x\sin\{x\}\right]+(1-n)x^2+\frac{C_F}{C_A}y^2.
\label{eq:K-bar}
\end{equation}
Note that $\bar{\mathcal{K}}^{(q;q,g)}_{(1;T,2)}$ almost reduces to $\bar{K}$ in Ref.~\cite{Sirimanna:2021sqx}, except for the additional $\frac{C_F}{C_A}y^2$ term stemming from the non-vanishing process depicted in Fig.~\ref{fig:kernel-1_all}(c). For reference, the Aurenche, Zakharov and Zaraket (AZZ) \cite{Aurenche:2008hm,Aurenche:2008mq}, as well as the Guo and Wang (GW) \cite{Guo:2000nz,Wang:2001ifa} curves are shown, with $\bar{\mathcal{K}}_{\rm GW}=2-2\cos{x}$ and $\bar{\mathcal{K}}_{\rm AZZ}=2-2\cos{x}-2x\sin{x}+2x^2\cos{x}$. Our results are consistent with Sirimanna-Cao-Majumder (SCM) results for $y=0$. However, as $y$ increases, such as $y=0.75$ [Fig.~\ref{fig:kernel1_comp_sirimanna}(c)], slight variations can be observed. Note that our calculation (KV, $n=1$) disfavors $n > 1/2$, as the collision kernel $\bar{\mathcal{K}}^{(q;q,g)}_{(1;T,2)}$ becomes negative in region $0.5< \zeta^{-}/\tau_f < 2.4$. This is consistent with the phase-space constraint pointed out in Eq.~(\ref{eq:maintext_factor_n}), which puts the bound as $0\leq n< 1/2$.
\begin{figure}[!h] 
\begin{subfigure}[t]{0.49\textwidth}
\includegraphics[width=\textwidth]{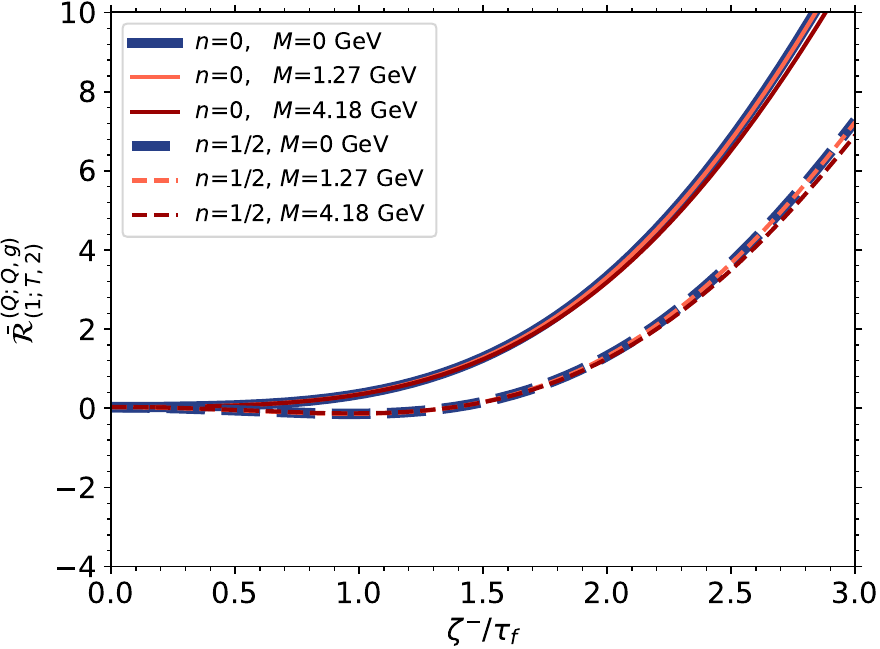}
\caption{$\bar{\mathcal{R}}^{(Q;Q,g)}_{(1;T,2)}$ at $y=0.25$}
\end{subfigure}
\begin{subfigure}[t]{0.49\textwidth}        
\includegraphics[width=\textwidth]{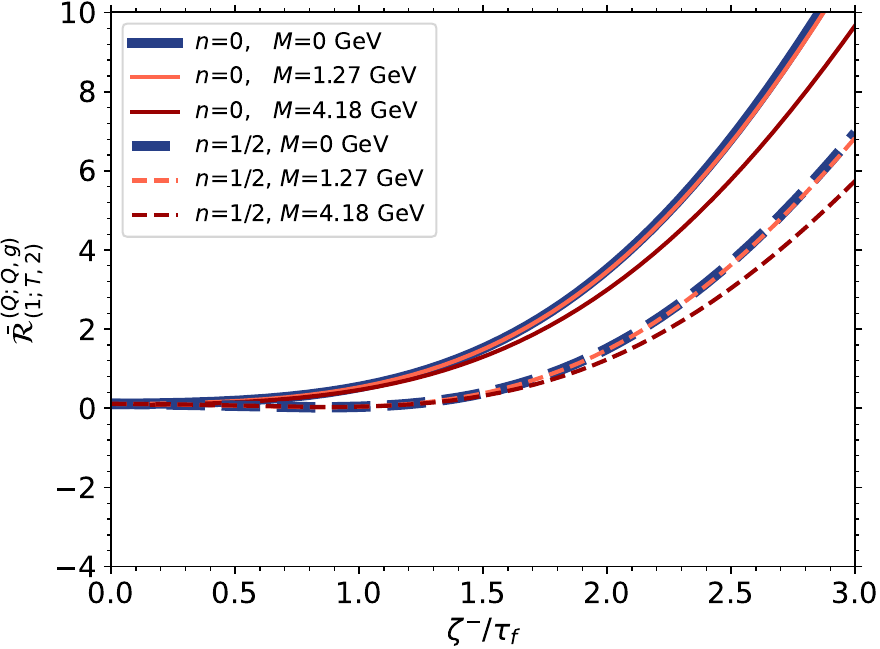}
\caption{$\bar{\mathcal{R}}^{(Q;Q,g)}_{(1;T,2)}$ at $y=0.5$}
\end{subfigure}
\begin{subfigure}[t]{0.49\textwidth}   
\includegraphics[width=\textwidth]{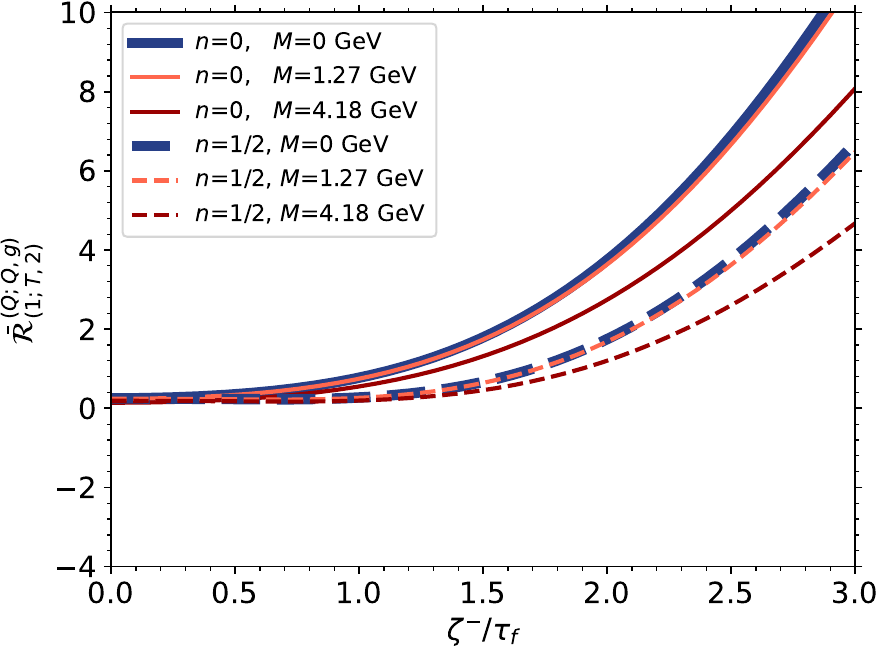}
\caption{$\bar{\mathcal{R}}^{(Q;Q,g)}_{(1;T,2)}$ at $y=0.75$}
\end{subfigure}
\caption{The mass dependence of $\bar{\mathcal{R}}^{(Q;Q,g)}_{(1;T,2)}$, a prefactor in front of $\hat{ \mathcal{A}}_{(T,2)} $, defined in Eq.~(\ref{eq:R-bar}) assuming $\ell_{2\perp}=10$ GeV, and $\tau^{-1}_f=\mathcal{G}^{(\ell_2)}_M$.} 
\label{fig:kernel-1_w_mass}
\end{figure}

To explore heavy-flavor energy loss in kernel-1, an interesting dimensionless quantity to look at is 
\begin{equation}
\bar{\mathcal{R}}^{(Q;Q,g)}_{(1;T,2)}=\frac{\pmb{\ell}^4_{2\perp}\mathcal{R}^{(Q;Q,g)}_{(1;T,2)}}{4C_A\left[\frac{1+(1-y)^2}{y}\right]}.
\label{eq:R-bar}
\end{equation}
with $\mathcal{R}^{(Q;Q,g)}_{(1;T,2)}$ given in Eq.~(\ref{eq:R_1T2_final}). Figure~\ref{fig:kernel-1_w_mass} displays the mass dependence of $\bar{\mathcal{R}}^{(Q;Q,g)}_{(1;T,2)}$, a perturbative coefficient of the second-order gradient term $\hat{ \mathcal{A}}_{(T,2)} $, characterizing Glauber-gluon~\cite{Idilbi:2008vm} induced transverse momentum diffusion of a (heavy) quark as it traverses the nuclear medium.
Within the ${\rm \overline{MS}}$~\cite{PDG_qmass} renormalization scheme, the heavy-quark masses are taken as $M=1.27$ GeV for the charm quark and $M=4.18$ GeV for the bottom quark. Our analysis shows that the charm quark exhibits no significant deviations in behavior when compared to lighter quarks, at a fixed $\ell_{2\perp}=10$ GeV. Of course, given Eq.~(\ref{eq:R_1T2_final}), this is not expected to hold as $\ell_{2\perp}\to 1.27$ GeV. The bottom quark displays a pronounced impact, at $\ell_{2\perp}=10$ GeV and for momentum fraction values exceeding $y>0.25$, highlighting the relevance of heavy-quark mass corrections in this kinematic regime. Similar behavior was observed in our earlier work for the photon Bremsstrahlung-based quark energy loss \cite{Kumar:2025egh}.
\begin{figure}[!h] 
\begin{subfigure}[t]{0.49\textwidth}
\includegraphics[width=\textwidth]{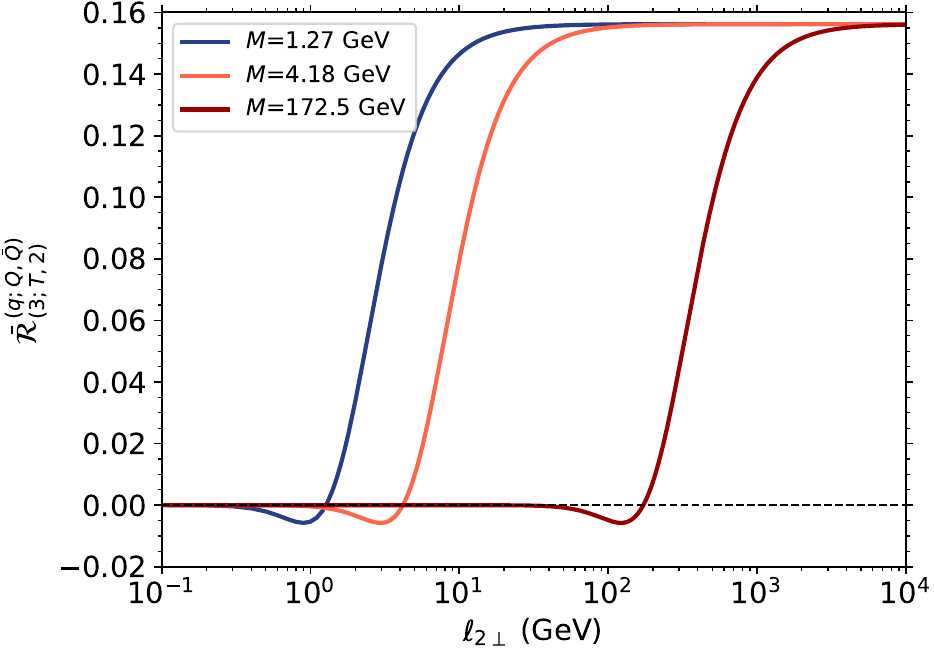}
\caption{$\bar{\mathcal{R}}^{(q;Q,\bar{Q})}_{(3;T,2)}$ at $y=0.25$}
\end{subfigure}
\begin{subfigure}[t]{0.49\textwidth}        
\includegraphics[width=\textwidth]{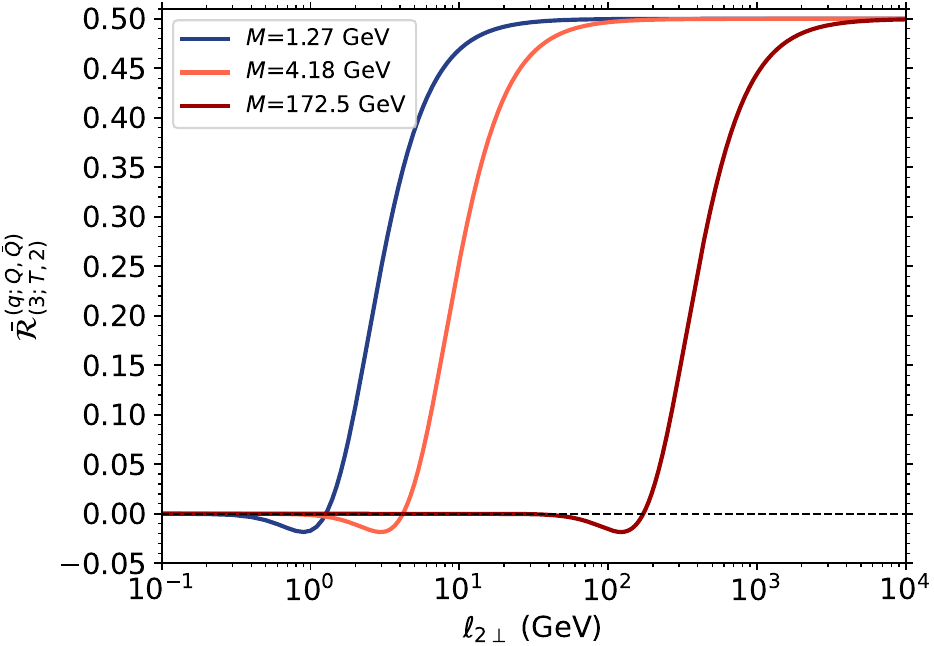}
\caption{$\bar{\mathcal{R}}^{(q;Q,\bar{Q})}_{(3;T,2)}$ at $y=0.5$}
\end{subfigure}
\begin{subfigure}[t]{0.49\textwidth}   
\includegraphics[width=\textwidth]{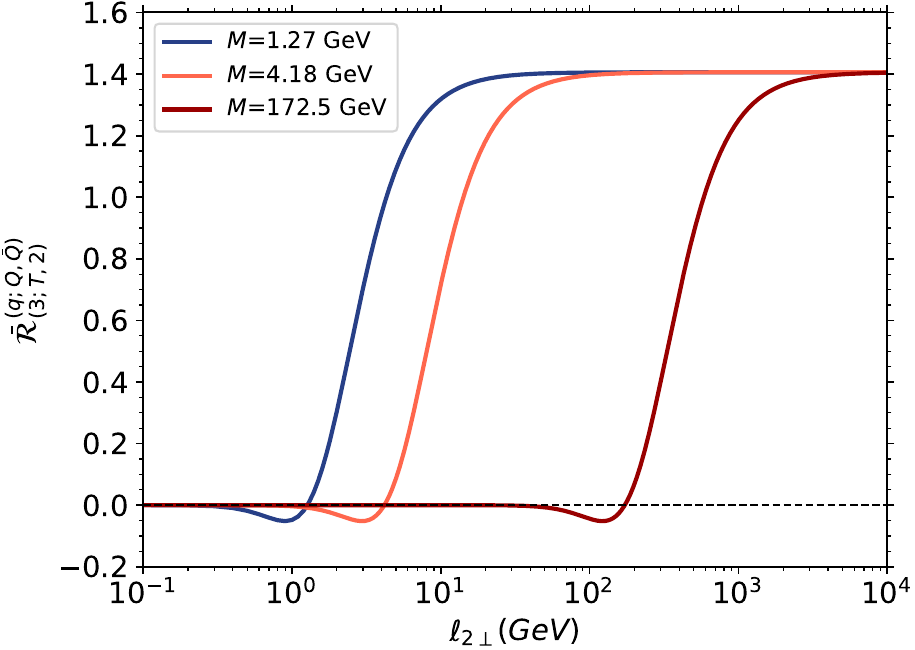}
\caption{$\bar{\mathcal{R}}^{(q;Q,\bar{Q})}_{(3;T,2)}$ at $y=0.75$}
\end{subfigure}
\caption{The mass dependence of $\bar{\mathcal{R}}^{(q;Q,\bar{Q})}_{(3;T,2)}$ defined in Eq.~(\ref{eq:R-bar-3}).} 
\label{fig:kernel-3_heavy_flavor}
\end{figure}

Interactions with light-flavored Glauber (anti)quarks in the nuclear medium admit a new channel for heavy-quark production depicted in Fig.~\ref{fig:kernel-3_all}(b), which is not available to jets showering in the vacuum, such as those created in proton-(anti)proton collision. Any quark-induced enhancement of heavy flavor in relativistic heavy-ion collisions is sensitive to the quark content within the nuclear medium. Of course, there is another source of heavy quark production, one where the gluon with momentum $\ell$ is produced by a gluon-gluon fusion, with momenta $p_1$ and $k$, respectively, in a process similar to Fig.~\ref{fig:kernel-3_amplt_square}(b). In the case of the QGP, the process of Fig.~\ref{fig:kernel-3_all}(b) and Fig.~\ref{fig:kernel-3_amplt_square}(b), is only allowed following the initial glasma dynamics, i.e. once light quarks have started to hydrodynamize heading towards the QGP. Defining a dimensionless quantity $\bar{\mathcal{R}}^{(q;Q,\bar{Q})}_{(3;T,2)}$ for a heavy-quark production channel in kernel-3 [Fig.~\ref{fig:kernel-3_all}(b)] as 
\begin{equation}
\bar{\mathcal{R}}^{(q;Q,\bar{Q})}_{(3;T,2)}=\frac{q^-\pmb{\ell}^4_{2\perp}\mathcal{R}^{(q;Q,\bar{Q})}_{(3;T,2)}}{\left[\frac{C_FC_A}{2}\right]},
\label{eq:R-bar-3}
\end{equation}
allows to investigate medium-induced heavy-flavor production, including top quarks, as depicted in Fig.~\ref{fig:kernel-3_heavy_flavor}.\footnote{Note that the fact that $\bar{\mathcal{R}}^{(q;Q,\bar{Q})}_{(3;T,2)}$ goes negative is not cause for concern, as there is a non-vanishing zeroth order term contributing, i.e. $\bar{\mathcal{R}}^{(q;Q,\bar{Q})}_{(3;0)}$, that should be combined with the result in Fig.~\ref{fig:kernel-3_heavy_flavor}.} In Eq.~(\ref{eq:R-bar-3}), $\bar{\mathcal{R}}^{(q;Q,\bar{Q})}_{(3;T,2)}$ is a prefactor in front of the transverse momentum diffusion correlator $\hat{\mathcal{F}}_{(T,2)}$ defined in Eq.~(\ref{eq:kernel1-coll-expansion}) and Eq.~(\ref{eq:F_hat_T2T2_kernel_2_3}). The $y$ dependence of $\bar{\mathcal{R}}^{(q;Q,\bar{Q})}_{(3;T,2)}$ is also explored in that figure. Recall that $\bar{\mathcal{R}}^{(q;Q,\bar{Q})}_{(3;T,2)}\propto \frac{1-\chi'}{(1+\chi')^3}$ [Eq.~(\ref{eq:R_q_qprime_bar-qprime})], where $\chi'=\frac{M^2}{\pmb{\ell}^2_{2\perp}}$, such at as $\ell_{2\perp}\gg M$,  $\chi'\to0$, thus $\bar{\mathcal{R}}^{(q;Q,\bar{Q})}_{(3;T,2)}$ becomes independent of $\ell_{2\perp}$ and $M$.

The results in Fig.~\ref{fig:kernel-3_heavy_flavor} complement the work in Refs.~\cite{Attems:2022ubu,Attems:2022otp}, by exploring heavy-flavor production for highly virtual (anti)quarks interacting with Glauber (anti)quarks.
The increase in available phase space due to the presence of highly virtual quarks, as evidenced by the possibility of producing top quark–antiquark pairs in this context, underscores the need for a more comprehensive analysis. To accurately quantify the extent to which the QGP enhances heavy-flavor production, it is essential to perform a Monte Carlo simulation that incorporates both our process and those outlined in Refs.~\cite{Attems:2022ubu,Attems:2022otp}, in conjunction with a realistic hydrodynamical description of the QGP. Such a combined framework would allow for a systematic evaluation of medium effects on heavy-quark yields and provide a more reliable connection to experimental observables. The main ingredient missing for such an assessment to be complete is the contribution to heavy-quark pair production from gluon-gluon fusion, as alluded to earlier. A full Monte Carlo simulation should also analyze its results using energy-energy correlators \cite{Barata:2025uxp} to track medium-induced heavy-quark pair production. Furthermore, no Monte Carlo simulation of jets in the QGP is complete without the associated photon production \cite{Kumar:2025egh}, to ensure that flavor hydrodynamization dynamics, which $\mathcal{K}_{i>1}$ are sensitive to both herein and in Ref.\cite{Kumar:2025egh}, is accounted for in a manner consistent with data.

Conversion processes proceeding through $2\to 2$ scattering have already been shown to affect the composition of jets in modern Monte Carlo event generators \cite{Sirimanna:2022zje}. The results in Ref.~\cite{Sirimanna:2022zje} however did not include how conversion in $1\to 2$ processes, specifically those explored in kernel-2 through kernel-4, affect the composition of jets. Thus, a follow-up study will examine how conversion processes in the QGP, both in $2\to2$ scattering as well as $1\to 2$ in-medium radiative processes, affect parton composition of the jet at high-virtuality, through event generators such as MATTER~\cite{Majumder:2013re,Cao:2017qpx}, as well as low-virtuality, using event generators such as LBT and MARTINI.\footnote{The radiative energy loss in LBT~\cite{Cao:2016gvr,JETSCAPE:2017eso} is based on the single-scattering induced emission, and accounts for heavy flavor energy loss. MARTINI uses on the Arnold-Moore-Yaffe formalism \cite{Arnold:2002ja,Jeon:2003gi}, including its next-to-leading order extensions and nonperturbative effects \cite{Yazdi:2022bru,Modarresi-Yazdi:2024vfh}. However, the Landau-Pomeranchuk-Migdal resummations for heavy quark energy loss has yet to be performed. A comparison of these two approches can be found in Ref.~\cite{JETSCAPE:2017eso}.} Furthermore, as mentioned in our previous study \cite{Kumar:2025egh}, conversion processes are sensitive to the composition of the nuclear medium, specifically the dynamical generation and evolution of quarks as the nuclear medium transitions from early glasma-like dynamics to hydrodynamics, where the quark composition and occupation number reach their near thermal equilibrium levels. Thus, photon production and conversion processes in jet showering, within the higher-twist formalism, should be used simultaneously to probe flavor hydrodynamizaion dynamics in relativistic heavy-ion collisions. 

As conversion processes in kernel-2 through kernel-4 were not included within the MATTER~\cite{Majumder:2013re,Cao:2017qpx} event generator used in a recent Bayesian constraint on $\hat{q}$ \cite{JETSCAPE:2024cqe}. A follow-up Bayesian study can use conversion processes herein as an estimate of theoretical systematic uncertainties. Unlike Bayesian constraints of QGP viscosities, where an important form of theoretical systematic uncertainties has been taken into account \cite{JETSCAPE:2020shq}, such uncertainties have yet to be included in constraints of jet-medium transport coefficients, e.g. $\hat q$. Future Bayesian efforts aimed at improving our understanding of $\hat q$ should include conversion processes as part of the high (and low) virtuality evolution of jet partons in the nuclear medium. 

%%%%%%%%%%%%%%%%%%%%%%%%%%%%%%%%%%%%%%%%%%%%%%%%%%%%%%%%%%%%%%%%%%%%%%%%%%%%%%%%%%%%%%%%%%%%%%%%
%%%%%%%%%%%%%%%%%%%%%%%%%%%%%%%%%%%%%%%%%%%%%%%%%%%%%%%%%%%%%%%%%%%%%%%%%%%%%%%%%%%%%%%%%%%%%%%%
%%%%%%%%%%%%%%%%%%%%%%%%%%%%%%%%%%%%%%%%%%%%%%%%%%%%%%%%%%%%%%%%%%%%%%%%%%%%%%%%%%%%%%%%%%%%%%%%
\section{Summary and Outlook}\label{sec:summary_outlook}
%%%%%%%%%%%%%%%%%%%%%%%%%%%%%%%%%%%%%%%%%%%%%%%%%%%%%%%%%%%%%%%%%%%%%%%%%%%%%%%%%%%%%%%%%%%%%%%%
%%%%%%%%%%%%%%%%%%%%%%%%%%%%%%%%%%%%%%%%%%%%%%%%%%%%%%%%%%%%%%%%%%%%%%%%%%%%%%%%%%%%%%%%%%%%%%%%
%%%%%%%%%%%%%%%%%%%%%%%%%%%%%%%%%%%%%%%%%%%%%%%%%%%%%%%%%%%%%%%%%%%%%%%%%%%%%%%%%%%%%%%%%%%%%%%%
In this paper, an improved calculation of the scattering kernel for quark jet energy loss at NLO in a nuclear medium is presented. The contributions from both in-medium Glauber quark and gluon~\cite{Idilbi:2008vm} have been accounted for the first time in a single uniform framework of the HT formalism that includes the effects of the heavy-quark masses, full phase factors, and fermion-to-boson conversion processes.

The virtual partons produced in the initial-state hard scattering are more likely to interact with gluons in the early stage, as the nuclear medium created in heavy-ion collisions is gluon dominant. As the system transitions from the initial glasma dynamics to the hydrodynamical QGP, the light flavors hydrodynamize and jet partons start interacting with quarks in the nascent QGP, through channels in $\mathcal{K}_2$ through $\mathcal{K}_4$. This changes the partonic composition of jets, relative to their vacuum composition or their composition in the glasma. The results presented herein, once implemented in a jet Monte Carlo, will allow to appreciate the change in the chemical composition of jets as the system hydrodynamizes.

The results presented herein thus account for interactions of highly virtual quarks with both quarks and gluons in the QGP. In so doing, all interactions with Glauber (anti)quarks and gluons have been accounted for within four distinct kernels. While kernel-1 deals with interactions between an incoming quark and Glauber gluons~\cite{Idilbi:2008vm}, kernel-2 and kernel-3 account for interactions with Glauber antiquarks, while kernel-4 accounts for interactions with Glauber quarks. The four kernels thus account for the following medium-induced radiative processes: (i) $q\to q+g$; (ii) $q\to g+g$; (iii) $q\to q+\bar{q}'$; as well as $q\to q'+\bar{q}'$, where $q'$ may be a different quark flavor from $q$; and (iv) $q\to q+q'$, where $q'$ may be a different quark flavor than $q$ once more. Furthermore, the quark mass scales were taken into account for all the relevant cases, namely assuming that the QGP does not contain any charm or heavier quarks. Should relativistic heavy-ion collisions ever reach temperatures that allow for charm-quark (or even bottom-quark) thermalization in the QGP, as has been explored in lattice QCD calculations \cite{Weber:2021hro,MILC:2013ops}, the results herein should be revised accordingly.  

As was the case in our previous calculation \cite{Kumar:2025egh}, the phase factors $\left[2-2\cos\left\{\mathcal{G}^{(\ell_2)}_M\zeta^-\right\}\right]$ and $e^{-i\Delta z^-\mathcal{H}^{(\ell_2,p_2)}_M}$ that contain explicit dependence on $\ell_{2\perp}$ have been kept when defining all scattering kernels. Furthermore, the basic assumption of the Glauber parton~\cite{Idilbi:2008vm} having transverse momentum greater than its light-cone momenta, i.e. $k_\perp\gg k^-, k^+$, is used throughout, and all processes involving a Glauber quark are found to be suppressed by $1/q^-$ or its higher powers. The scattering kernels $\mathcal{S}_i$ are presented before their Taylor expansion in $k$ is performed, specifically around $k_\perp=0$ and $k^-=0$. In addition to obtaining the complete NLO $O\left(\alpha^2_s\right)$ and next-to-leading-twist scattering kernels, the nonperturbative operators, i.e., $\hat{\mathcal{A}}_i$ and $\hat{\mathcal{F}}_i$ are linked to transverse-momentum-dependent parton distribution functions (TMD-PDFs).

Phenomenological studies by the JET and JETSCAPE Collaborations, together with estimates of $\hat{q}$ in cold nuclear matter,  indicate that $\hat{q}$ is an order of magnitude larger in hot QGP than in cold nuclear matter~\cite{JET:2013cls,JETSCAPE:2024cqe,Deng:2009ncl,Ru:2019qvz}. Assuming this trend remains the same for jet transport coefficients $\hat{\mathcal{F}}_i$, kernels 2–4 are expected to play a more significant role for jets traversing QGP than in cold nuclear environments. For these scattering kernels to be implemented within realistic Monte Carlo simulations of jet partons in the QGP, the transport coefficients $\hat{\mathcal{A}}_i$ and $\hat{\mathcal{F}}_i$ need to be computed, either in finite-temperature field theory or using lattice gauge theory. 
Having access to the scattering kernels herein, together with those in Ref.~\cite{Kumar:2025egh}, allows us to use photons and jets together when constraining flavor hydrodynamization dynamics. Still missing is a scattering kernel producing a virtual initial-state gluon, stemming from gluon-gluon fusion diagrams and quark-antiquark anihilation diagrams at the onset of the jet shower. Such diagrams are to be considered next, especially pertaining to the heavy-quark pair-production, which have become interesting \cite{Barata:2025uxp}.

Owing to the recent results establishing a correspondence between the HT and CGC formalisms~\cite{Fu:2023jqv,Fu:2024sba}, our findings on medium-induced modifications of hard parton evolution in e-A collisions, derived within the HT framework, are expected to be consistent with the CGC description.
Consequently, the medium modification of the hard parton presented in this work, can serve as a complementary probe of the same underlying physics that governs parton propagation through the glasma in the early stages of heavy-ion collisions, thereby providing a useful bridge between the HT formalism and saturation physics captured by CGC. Through the separation of the perturbative $\mathcal{R}^{(a;b,c)}_{(i;j)}$, and the non-perturbative two-point functions $\hat O_{(i;j)}$, our scattering kernels are able to describe both hot and cold nuclear media. Indeed, intricate details of the nuclear medium are solely encapsulated in $\hat O_{(i;j)}$. Thus, jet-medium studies using ongoing relativistic heavy-ion collisions, and those to be explored at the upcoming EIC, equally benefit from the findings herein.
At the Electron-Ion Collider, through model-to-data comparisons of leading hadron and reconstructed jet observables measured in coincidence with the scattered electron, one can directly access the underlying transport dynamics of the nuclear medium. Such observables are sensitive not only to jet quenching dynamics but also to the underlying partonic structure, including TMD parton distributions and phase-space correlations encoded in Wigner functions. Consequently, jet quenching dynamics studies establish a direct connection between HT-based energy-loss calculations and the multidimensional imaging of partons inside nuclei. 
To achieve these goals, a comprehensive Monte Carlo event generator is required to elucidate all aspects of EIC physics \cite{Accardi:2012qut, AbdulKhalek:2021gbh, Kim:2025clp}. Such a tool should consistently incorporate initial-state effects through TMD parton distributions together with in-medium Glauber quark and gluon distribution functions describing parton propagation through nuclear matter phase space. Such a Monte Carlo event generator will provide a benchmarking tool for interpreting future EIC measurements and for performing precise model-to-data comparisons of hadron and jet observables.

%%%%%%%%%%%%%%%%%%%%%%%%%%%%%%%%%%%%%%%%%%%%%%%%%%%%%%%%%%%%%%%%%%%%%%%%%%%%%%%%%%%%%%%%%%%%%%%%
%%%%%%%%%%%%%%%%%%%%%%%%%%%%%%%%%%%%%%%%%%%%%%%%%%%%%%%%%%%%%%%%%%%%%%%%%%%%%%%%%%%%%%%%%%%%%%%%
%%%%%%%%%%%%%%%%%%%%%%%%%%%%%%%%%%%%%%%%%%%%%%%%%%%%%%%%%%%%%%%%%%%%%%%%%%%%%%%%%%%%%%%%%%%%%%%%
\section*{ACKNOWLEDGMENTS}
%%%%%%%%%%%%%%%%%%%%%%%%%%%%%%%%%%%%%%%%%%%%%%%%%%%%%%%%%%%%%%%%%%%%%%%%%%%%%%%%%%%%%%%%%%%%%%%%
%%%%%%%%%%%%%%%%%%%%%%%%%%%%%%%%%%%%%%%%%%%%%%%%%%%%%%%%%%%%%%%%%%%%%%%%%%%%%%%%%%%%%%%%%%%%%%%%
%%%%%%%%%%%%%%%%%%%%%%%%%%%%%%%%%%%%%%%%%%%%%%%%%%%%%%%%%%%%%%%%%%%%%%%%%%%%%%%%%%%%%%%%%%%%%%%%
This work was supported by the Canada Research Chair under Grant No. CRC-2022-00146 and the Natural Sciences and Engineering Research Council (NSERC) of Canada under Grant No. SAPIN-2023-00029. This work was also supported in part by the National Science Foundation (NSF) within the framework of the JETSCAPE collaboration, under Grant No. OAC-2004571 (CSSI:X-SCAPE).

%%%%%%%%%%%%%%%%%%%%%%%%%%%%%%%%%%%%%%%%%%%%%%%%%%%%%%%%%%%%%%%%%%%%%%%%%%%%%%%%%%%%%%%%%%%%%%%%
%%%%%%%%%%%%%%%%%%%%%%%%%%%%%%%%%%%%%%%%%%%%%%%%%%%%%%%%%%%%%%%%%%%%%%%%%%%%%%%%%%%%%%%%%%%%%%%%
%%%%%%%%%%%%%%%%%%%%%%%%%%%%%%%%%%%%%%%%%%%%%%%%%%%%%%%%%%%%%%%%%%%%%%%%%%%%%%%%%%%%%%%%%%%%%%%%
\section*{Data Availability}
%%%%%%%%%%%%%%%%%%%%%%%%%%%%%%%%%%%%%%%%%%%%%%%%%%%%%%%%%%%%%%%%%%%%%%%%%%%%%%%%%%%%%%%%%%%%%%%%
%%%%%%%%%%%%%%%%%%%%%%%%%%%%%%%%%%%%%%%%%%%%%%%%%%%%%%%%%%%%%%%%%%%%%%%%%%%%%%%%%%%%%%%%%%%%%%%%
%%%%%%%%%%%%%%%%%%%%%%%%%%%%%%%%%%%%%%%%%%%%%%%%%%%%%%%%%%%%%%%%%%%%%%%%%%%%%%%%%%%%%%%%%%%%%%%%
No data were created or analyzed in this study. 

%%%%%%%%%%%%%%%%%%%%%%%%%%%%%%%%%%%%%%%%%%%%%%%%%%%%%%%%%%%%%%%%%%%%%%%%%%%%%%%%%%%%%%%%%%%%%%%%
%%%%%%%%%%%%%%%%%%%%%%%%%%%%%%%%%%%%%%%%%%%%%%%%%%%%%%%%%%%%%%%%%%%%%%%%%%%%%%%%%%%%%%%%%%%%%%%%
%%%%%%%%%%%%%%%%%%%%%%%%%%%%%%%%%%%%%%%%%%%%%%%%%%%%%%%%%%%%%%%%%%%%%%%%%%%%%%%%%%%%%%%%%%%%%%%%
%%  Add appendices %%%%%%
\include{Appendix}
%%%%%%%%%%%%%%%%%%%%%%%%%%%%%%%%%%%%%%%%%%%%%%%%%%%%%%%%%%%%%%%%%%%%%%%%%%%%%%%%%%%%%%%%%%%%%%%%
%%%%%%%%%%%%%%%%%%%%%%%%%%%%%%%%%%%%%%%%%%%%%%%%%%%%%%%%%%%%%%%%%%%%%%%%%%%%%%%%%%%%%%%%%%%%%%%%
%%%%%%%%%%%%%%%%%%%%%%%%%%%%%%%%%%%%%%%%%%%%%%%%%%%%%%%%%%%%%%%%%%%%%%%%%%%%%%%%%%%%%%%%%%%%%%%%
\bibliography{references}
\end{document}

%% file: Appendix.tex
\begin{appendices}
%%%%%%%%%%%%%%%%%%%%%%%%%%%%%%%%%%%%%%%%%%%%%%%%%%%%%%%%%%%%%%%%%%%%%%%%%%%%%%%%%%%%%%%%%%%%%%%%
%%%%%%%%%%%%%%%%%%%%%%%%%%%%%%%%%%%%%%%%%%%%%%%%%%%%%%%%%%%%%%%%%%%%%%%%%%%%%%%%%%%%%%%%%%%%%%%%
%%%%%%%%%%%%%%%%%%%%%%%%%%%%%%%%%%%%%%%%%%%%%%%%%%%%%%%%%%%%%%%%%%%%%%%%%%%%%%%%%%%%%%%%%%%%%%%%%
\section{$\mbox{}$\!\!\!\!\!\!: List of variables}
\label{append:list_of_function}
%%%%%%%%%%%%%%%%%%%%%%%%%%%%%%%%%%%%%%%%%%%%%%%%%%%%%%%%%%%%%%%%%%%%%%%%%%%%%%%%%%%%%%%%%%%%%%%%
%%%%%%%%%%%%%%%%%%%%%%%%%%%%%%%%%%%%%%%%%%%%%%%%%%%%%%%%%%%%%%%%%%%%%%%%%%%%%%%%%%%%%%%%%%%%%%%%%
%

\begin{equation}
    \kappa = \frac{1}{1+(1-y)^2} ,
    \label{eq:kappa_append}
\end{equation}

\begin{equation}
   \eta = \frac{k^-}{yq^-} ,
    \label{eq:eta_append}
\end{equation}
\begin{equation}
    J_1 = \left[ \left(1+\eta y\right)\pmb{\ell}_{2\perp} - y\pmb{k}_{\perp} \right]^2  + y^2 M^2 ,
\label{eq:J1_append}
\end{equation}
\begin{equation}
    J_2  = \pmb{\ell}^2_{2\perp}\left\{ -1 + y - \eta y \left(1 - y +\eta y\right)\right\} +y\pmb{k}^2_\perp \left\{-1 +y-\eta y\right\} + \left(\pmb{k}_\perp \cdot \pmb{\ell}_{2\perp}\right)\left\{1-y^2 + 2\eta y + \eta^2 y^2\right\},
\label{eq:J2_append}
\end{equation}
\begin{equation}
    \mathcal{G}^{(p_2)}_0 = p^{+}_{2} + \frac{\pmb{p}^{2}_{2\perp}}{2(q^{-}-p^{-}_{2})}  = \frac{\left(\pmb{\ell}_{2\perp}-\pmb{k}_{\perp}\right)^2}{2y\left(1-y+\eta y\right)\left(1-\eta\right) q^-},
    \label{eq:GP2M0_append}
\end{equation}
\begin{equation}
    \mathcal{G}^{(p_2)}_M = p^{+}_{2}+\frac{\pmb{p}^{2}_{2\perp}}{2\left(q^{-}-p^{-}_{2}\right)}-\frac{M^2}{2q^{-}} = \frac{\left(\pmb{\ell}_{2\perp}-\pmb{k}_{\perp}\right)^2 + y^2\left(1-\eta\right)^2M^2}{2y\left(1-y+\eta y\right)\left(1-\eta\right)q^-},
    \label{eq:GMP2_append}
\end{equation}
\begin{equation}
    \mathcal{G}^{(\ell_2)}_0 = \ell^{+}_{2} + \frac{\pmb{\ell}^{2}_{2\perp}}{2(q^{-}-\ell^{-}_{2})}  = \frac{\pmb{\ell}^2_{2\perp} }{2y(1-y) q^-},
    \label{eq:GL20_append}
\end{equation}
\begin{equation}
    \mathcal{G}^{(\ell_2)}_M =  \frac{\pmb{\ell}^2_{2\perp} +y^2M^2}{2y(1-y) q^-},
    \label{eq:GL2M_append}
\end{equation}
\begin{equation}
\mathcal{G}^{(\ell_2,p_2,k)}_{M} = \ell^+_2 + \frac{\pmb{p}^2_{2\perp}+M^2}{2p^-_2} - \frac{\pmb{k}^2_{\perp} + M^2}{2(1+\eta y)q^-} = \frac{J_{1}}{2(1+\eta y)y(1-y+\eta y)q^-},
\label{eq:GML2_p2k_J1_append}
\end{equation}
\begin{equation}
\mathcal{G}^{(\ell,k)}_{1} = \frac{\left[\pmb{\ell}_{\perp}(1+\eta y) + (1-y)\pmb{k}_{\perp}\right]^2 + y^2M^2(1+\eta)^2}{2y(1-y)q^{-}(1+\eta)(1+\eta y)},
\label{eq:G1ML_k_etay_append}
\end{equation}
\begin{equation}
\mathcal{G}^{(\ell,k)}_{2} = \frac{\left(\pmb{\ell}_{\perp} + \pmb{k}_{\perp}\right)^2 + y^2M^2(1+\eta)^2}{2y(1-y-\eta y)(1+\eta)q^{-}},
\label{eq:G2_ell_k_append}
\end{equation}
\begin{equation}
\mathcal{H}^{(\ell_2,p_2)}_{M} = \ell^{+}_{2} + p^{+}_{2} - \frac{M^2}{2q^-} =  \frac{ \pmb{\ell}^{2}_{2\perp} -yM^2 }{2yq^{-}}  + \frac{ (\pmb{\ell}_{2\perp} - \pmb{k}_{\perp})^2 + M^2  }{2q^{-}(1-y + \eta y)},
\label{eq:HL2P2M_append}
\end{equation}
\begin{equation}    \mathcal{H}^{(\ell_2,p_2)}_{0} = \ell^+_2 + p^+_2 = \frac{\pmb{\ell}^2_{2\perp}}{2yq^-} + \frac{(\pmb{\ell}_{2\perp} - \pmb{k}_\perp)^2 }{2(1-y+\eta y)q^-} ,
\label{eq:HL2P2M0_append}
\end{equation}
\begin{equation}    
\mathcal{H}^{(\ell_2,p_2)}_{1} = \ell^{+}_{2} + p^{+}_{2}  =  \frac{\pmb{\ell}^{2}_{2\perp}+M^2}{2yq^{-}}  + \frac{ (\pmb{\ell}_{2\perp} - \pmb{k}_{\perp})^2 +M^2}{2(1-y+\eta y)q^{-}} ,
\label{eq:H1L2P2_append}
\end{equation}
\begin{equation}    
\mathcal{H}^{(\ell_2,p_2)}_{2} =   \frac{\pmb{\ell}^{2}_{2\perp}+M^2}{2(1-y)q^{-}}  + \frac{ (\pmb{\ell}_{2\perp} + \pmb{k}_{\perp})^2 }{2(1+\eta)yq^{-}}  - \frac{M^2}{2q^-}  ,
\label{eq:H2L2P2_append}
\end{equation}

%%%%%%%%%%%%%%%%%%%%%%%%%%%%%%%%%%%%%%%%%%%%%%%%%%%%%%%%%%%%%%%%%%%%%%%%%%%%%%%%%%%%%%%%%%%%%%%%%
\section{$\mbox{}$\!\!\!\!\!\!: Single-emission single-scattering kernel: one gluon and one quark in the final state }
\label{append:kernel-1}
%%%%%%%%%%%%%%%%%%%%%%%%%%%%%%%%%%%%%%%%%%%%%%%%%%%%%%%%%%%%%%%%%%%%%%%%%%%%%%%%%%%%%%%%%%%%%%%%
%%%%%%%%%%%%%%%%%%%%%%%%%%%%%%%%%%%%%%%%%%%%%%%%%%%%%%%%%%%%%%%%%%%%%%%%%%%%%%%%%%%%%%%%%%%%%%%%%

In this section, we summarize the calculation of all possible diagrams at NLO contributing to  kernel-1 with a gluon and a quark in the final state. We discuss singularity structure, contour integrations, and involved traces in the final calculation of the hadronic tensor.

\begin{figure}[h!]
    \centering
    \includegraphics[width=0.45\textwidth]{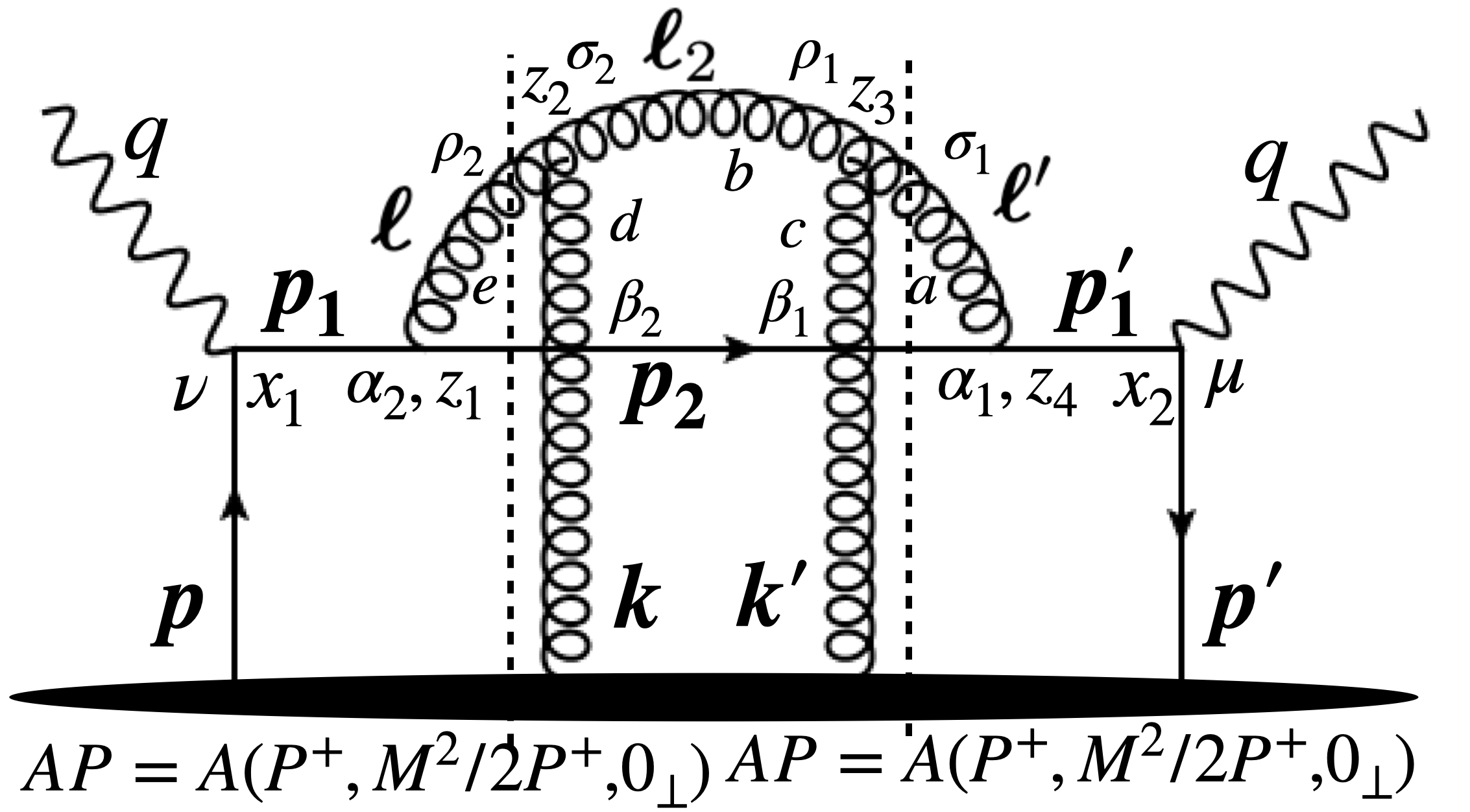}
    \caption{Interference diagrams in which the radiated gluon undergoes double-gluon scattering, contributing to  kernel-1. }
    \label{fig:kernel-1_dgg_scatt_left-rightcut}
\end{figure}
%%%%%%%%%%%%%%%%%%%%%%%%%%%%%%%%%%%%%%%%%%%%%%%%%%%%%%%%%%%%%%%%%
%%%%%%%%%%%%%%%%%%%%%%%%%%%%%%%%%%%%%%%%%%%%%%%%%%%%%%%%%%%%%%%%%
\subsection{Figure~\ref{fig:kernel-1_all}(a) left and right cuts}
%%%%%%%%%%%%%%%%%%%%%%%%%%%%%%%%%%%%%%%%%%%%%%%%%%%%%%%%%%%%%%%%%
%%%%%%%%%%%%%%%%%%%%%%%%%%%%%%%%%%%%%%%%%%%%%%%%%%%%%%%%%%%%%%%%%

Figure~\ref{fig:kernel-1_dgg_scatt_left-rightcut} represents a forward scattering diagram contributing to kernel-1. The left-cut give rise to an interference between the single-gluon emission with no scattering process and single gluon emission with double in-medium gluon scattering. The hadronic tensor for the left-cut diagram (Fig.~\ref{fig:kernel-1_dgg_scatt_left-rightcut}) is given by 
\begin{equation}
\begin{split}
W^{\mu\nu}_{1,\ell} & = \sum_{f} e^2_f g^4_{s} \int d^4 x_1 d^4 x_2 d^4 z_{2} d^4 z_{3} \int \frac{d^4 \ell}{(2\pi)^4} \frac{d^4 p'}{(2\pi)^4}  \frac{d^4 \ell_{2}}{(2\pi)^4} \frac{d^4 p_{2}}{(2\pi)^4} e^{-ip'x_{2}} e^{i\left(-q+\ell+p_2\right)x_{1}}\left\langle P \left| \bar{\psi}_{_f}(x_2) \frac{\gamma^{+}}{4} \psi_{_f}(x_{1})\right| P\right\rangle  \\
  & \times e^{i\left(q+p'-p_{2}-\ell_{2}\right) z_{3}} e^{i(\ell_{2} - \ell)z_{2}} \left\langle P_{A-1} \left| A^{+c}(z_3)A^{+d}(z_2)\right| P_{A-1}\right\rangle (i)^2 f^{bed} f^{abc} {\rm Tr}\left[t^a t^e\right] g^{\sigma_2 \rho_2} g^{\sigma_1 \rho_1} \left(\ell^{-}_{2}+\ell^{-}\right)^2\\
& \times \frac{  {\rm Tr} \left[\gamma^- \gamma^{\mu} \left(\slashed{q}+\slashed{p}'+M\right) \gamma^{\alpha_{1}}\left(  \slashed{p}_2 +  M\right) \gamma^{\alpha_2} \left( \slashed{\ell} + \slashed{p}_2+M\right)\gamma^{\nu} \right]
}{  \left[\left(q+p'\right)^{2}-M^2-i\epsilon\right] \left[\left(\ell+p_2\right)^2  - M^2 + i\epsilon\right]   \left[\left(q+p'-p_2\right)^2 -i\epsilon\right] \left[\ell_2^2-i\epsilon\right]  } \\
& \times d^{\left(q+p'-p_2\right)}_{\sigma_{1} \alpha_{1}} d^{(\ell_2)}_{\sigma_{2} \rho_{1}} d^{(\ell)}_{\alpha_{2} \rho_{2}} (2 \pi) \delta\left(\ell^{2}\right) (2 \pi)  \delta\left(p^{2}_{2}-M^{2}\right).
\end{split} \label{eq:kernel-1_dgg_scatt_leftcut_wi}
\end{equation}
The above expression has a singularity when the denominator of the propagator for $\ell_2$, $\ell'$ and $p_1$ becomes on-shell. There is one pole for the momentum variable $\ell^+_2$ and two poles for $p'^+$. The contour integration for $\ell_2^+$ gives
\begin{equation}
    C_{1} = \oint \frac{d\ell^+_2}{(2\pi)} \frac{e^{-i\ell^+_2(z^{-}_3-z^-_2)} }{[ \ell^2_2  - i\epsilon]} =  \oint \frac{d\ell^+_2}{(2\pi)} \frac{e^{-i\ell^+_2(z^{-}_3-z^{-}_{2})} }{ 2\ell^-_2 \left[\ell^{+}_2 - \frac{\pmb{\ell}^2_{2\perp}}{2\ell^-_2} - i\epsilon\right]} = \frac{(-2\pi i)}{2\pi}\frac{\theta( z^{-}_3 - z^{-}_{2} )}{2\ell^-_2}  e^{-i\left( \frac{\pmb{\ell}^2_{2\perp}}{2\ell^-_2}\right)(z^{-}_3-z^{-}_{2})}.
\end{equation}
Similarly, the contour integration for $p'^+$ gives
\begin{equation}
\begin{split}
C_{2} & = \oint \frac{dp'^{+}}{(2\pi)} \frac{e^{-ip'^{+}\left(x^{-}_{2}-z^{-}_{3}\right)}}{\left[\left(q+p'-p_2\right)^{2} - i\epsilon\right]\left[\left(q+p'\right)^{2}-M^2 -i\epsilon\right]} \\
      & = \frac{(-2\pi i)}{2\pi} \frac{\theta\left(x^{-}_{2} - z^{-}_{3}\right)}{4q^{-}\ell^-}  e^{i\left(q^{+}-\frac{M^2}{2q^-}\right)\left(x^{-}_{2}-z^{-}_{3}\right)}\left[ \frac{ -1 + e^{-i \mathcal{G}^{(\ell)}_M\left(x^{-}_{2}-z^{-}_{3}\right)} }{ \mathcal{G}^{(\ell)}_M}  \right],
\end{split}
\end{equation}
where $\mathcal{G}^{(\ell)}_M$ is defined in Eq.~(\ref{eq:GL2M_append}).

The trace in the numerator of the third line of Eq.~(\ref{eq:kernel-1_dgg_scatt_leftcut_wi}) gives
\begin{equation}
    \begin{split}
       & {\rm Tr} \left[\gamma^- \gamma^{\mu} \left(\slashed{q}+\slashed{p}'+M\right) \gamma^{\alpha_{1}}\left(  \slashed{p}_2 +  M\right) \gamma^{\alpha_2} \left( \slashed{\ell} + \slashed{p}_2+M\right)\gamma^{\nu} \right] d^{(q+p'-p_2)}_{\sigma_{1} \alpha_{1}} d^{(\ell_2)}_{\sigma_{2} \rho_{1}} d^{(\ell)}_{\alpha_{2} \rho_{2}} g^{\sigma_2 \rho_2} g^{\sigma_1 \rho_1} \\
       & = \frac{8q^{-}[-g^{\mu\nu}_{\perp\perp}]}{y(1-y)} \left[\frac{1+\left(1-y\right)^2}{y}\right] \left[\pmb{\ell}^2_{\perp} + \kappa y^4 M^2 \right],
    \end{split}
\end{equation}
where $\kappa$ is defined in Eq.~(\ref{eq:kappa_append}). Finally, the hadronic tensor for the left-cut diagram (Fig.~\ref{fig:kernel-1_dgg_scatt_left-rightcut}) reduces to the following form:
\begin{eqnarray}
W^{\mu\nu}_{1,\ell} & = & C_{A} \sum_f 2 \left[-g^{\mu\nu}_{\perp\perp}\right]  e^2_f g^4_{s} \int d (\Delta x^{-}) e^{i \left(q^{+} - \frac{M^2}{2q^-} \right) \Delta x^{-}} \left\langle P \left| \bar{\psi}_{_f}(\Delta x^-) \frac{\gamma^{+}}{4} \psi_{_f}(0)\right| P\right\rangle \nonumber\\
  & \times & n \int d (\Delta z^{-})d^2 \Delta z_{\perp} \frac{dy}{2\pi}\frac{d^2 \ell_{\perp}}{(2\pi)^2} \frac{d^2 k_{\perp}}{(2\pi)^2} e^{-i\Delta z^{-}\mathcal{H}^{(\ell,p_2)}_2} e^{i\pmb{k}_{\perp}\cdot \Delta\pmb{z}_{\perp} } \nonumber\\ 
  & \times & \int d \zeta^- \theta(z^{-}_{3}-z^{-}_{2}) \theta(x^{-}_{2}-z^{-}_{3}) \left[\frac{1+ \left(1-y\right)^2}{y}\right] \left[ \frac{\left(1+\frac{\eta}{2}\right)^2}{\left(1+\eta\right)}\right]\frac{\left[\pmb{\ell}^2_{\perp} + \kappa y^4 M^2\right]}{\left[\pmb{\ell}^2_{\perp} + y^2 M^2\right]^2} \nonumber \\
  & \times & \left[1 - e^{-i\mathcal{G}^{(\ell)}_M\left(x^{-}_{2}-z^{-}_{3}\right)} \right] e^{i\mathcal{G}^{(\ell)}_{M}\left(x^{-}_{1}-z^{-}_{2}\right)}\left\langle P_{A-1} \left\vert {\rm Tr}\left[A^{+}(\zeta^-, \Delta z^-, \Delta z_{\perp})A^+(\zeta^-,0)\right]\right\vert P_{A-1}\right\rangle,
\label{eq:K1_W_final_leftcut}
\end{eqnarray}
where $\mathcal{H}^{(\ell,p_2)}_{2}$ is defined in Eq.~(\ref{eq:H2L2P2_append}) and $\mathcal{G}^{(\ell)}_{M}$ is defined in Eq.~(\ref{eq:GL2M_append}).
In above Eq.~(\ref{eq:K1_W_final_leftcut}), the factor $n$ originates from the time ordering of the two gluons originating from the plasma. Since the left-cut diagram (Fig.~\ref{fig:kernel-1_dgg_scatt_left-rightcut}) gives rise to two gluons on the same side of the cut-line, it imposes a phase space constraint  $\theta(\Delta z^{+})$ in the calculation of the hadonic tensor. The effective integral over $d(\Delta z^+)$  becomes
\begin{eqnarray}
    \int d(\Delta z^+) \delta(\Delta z^+)  \theta(\Delta z^{+}) = \theta\left(\Delta z^+=0\right) = n.
    \label{eq:factor_n}
\end{eqnarray} 
This integral arises when we perform phase-space integration over $d(\Delta z^+)$, such as shown in Eq.~(\ref{eq:k1_y_l2_kperp_2}). In this paper, we define the integral shown in Eq.~(\ref{eq:factor_n}) as $n$, which ranges between 0 and 1, depending on the definition of $\theta(0)$. A detailed discussion can be found in Appendix A of Ref.~\cite{Sirimanna:2021sqx}.

The right-cut diagram shown in Fig.~\ref{fig:kernel-1_dgg_scatt_left-rightcut} is a complex conjugate of the left-cut diagram. The final expression of the hadronic tensor for the right-cut diagram reduces to the following form:
\begin{eqnarray}
W^{\mu\nu}_{1,r} & = &C_{A} \sum_f 2 \left[-g^{\mu\nu}_{\perp\perp}\right]  e^2_f g^4_{s} \int d (\Delta x^{-}) e^{i \left(q^{+} -\frac{M^2}{2q^-}\right)\Delta x^{-}} \left\langle P \left| \bar{\psi}_{_f}(\Delta x^-) \frac{\gamma^{+}}{4} \psi_{_f}(0)\right| P\right\rangle \nonumber\\
  & \times & \int d (\Delta z^{-})d^2 \Delta z_{\perp} \frac{dy}{2\pi}\frac{d^2 \ell'_{\perp}}{(2\pi)^2} \frac{d^2 k_{\perp}}{(2\pi)^2} e^{-i\Delta z^{-}\mathcal{H}^{(\ell',p_2)}_2} e^{i\pmb{k}_{\perp}\cdot \Delta\pmb{z}_{\perp} } \nonumber\\ 
  & \times & n \int d \zeta^- \theta(z^{-}_{2}-z^{-}_{3}) \theta(x^{-}_{1}-z^{-}_{2}) \left[\frac{1+ \left(1-y\right)^2}{y}\right] \left[ \frac{\left(1+\frac{\eta}{2}\right)^2}{\left(1+\eta\right)}\right]\frac{\left[\pmb{\ell}'^2_{\perp} + \kappa y^4 M^2  \right]}{\left[\pmb{\ell}'^2_{\perp}   + y^2  M^2 \right]^2}\nonumber \\
  & \times & \left[1 - e^{i \mathcal{G}^{(\ell')}_M\left(x^{-}_{1}-z^{-}_{2}\right)} \right] e^{-i\mathcal{G}^{(\ell')}_{M}\left(x^{-}_{2}-z^{-}_{3}\right)}\left\langle P_{A-1}\left\vert {\rm Tr}\left[A^{+}\left(\zeta^-, \Delta z^-, \Delta z_{\perp}\right)A^+\left(\zeta^-,0\right)\right]\right\vert P_{A-1}\right\rangle,
\label{eq:K1_W_final_rightcut}
\end{eqnarray}
where $\mathcal{H}^{(\ell',p_2)}_{2}$ is defined in Eq.~(\ref{eq:H2L2P2_append}) and $\mathcal{G}^{(\ell')}_{M}$ is defined in Eq.~(\ref{eq:GL2M_append}). The factor $n$ in Eq.~(\ref{eq:K1_W_final_rightcut}) is described in Eq.~(\ref{eq:factor_n}). Adding the left-cut and right-cut diagrams, i.e., Eq.~(\ref{eq:K1_W_final_leftcut}) and Eq.~(\ref{eq:K1_W_final_rightcut}), respectively, gives the following expression of the hadronic tensor:
\begin{eqnarray}
W^{\mu\nu}_{1,\ell+r} & = &C_{A} \sum_f 2 \left[-g^{\mu\nu}_{\perp\perp}\right]  e^2_f g^4_{s} \int d (\Delta x^{-}) e^{i \left(q^{+} -\frac{M^2}{2q^-}\right)\Delta x^{-}} \left\langle P \left| \bar{\psi}_{_f}(\Delta x^-) \frac{\gamma^{+}}{4} \psi_{_f}(0)\right| P\right\rangle \nonumber\\
  & \times & n \int d (\Delta z^{-})d^2 \Delta z_{\perp} \frac{dy}{2\pi}\frac{d^2 \ell_{\perp}}{(2\pi)^2} \frac{d^2 k_{\perp}}{(2\pi)^2} e^{-i\Delta z^{-}\mathcal{H}^{(\ell,p_2)}_2} e^{i\pmb{k}_{\perp}\cdot \Delta\pmb{z}_{\perp} }   \nonumber\\ 
  & \times & \int d \zeta^-  \theta(\zeta^-) \left[\frac{1+ \left(1-y\right)^2}{y}\right] \left[ \frac{\left(1+\frac{\eta}{2}\right)^2}{\left(1+\eta\right)}\right]\frac{\left[\pmb{\ell}^2_{\perp} + \kappa y^4 M^2  \right]}{\left[\pmb{\ell}^2_{\perp} + y^2 M^2 \right]^2}\left[ \cos\left\{\mathcal{G}^{(\ell)}_{M}\zeta^{-}\right\} -1 \right]\nonumber\\
  & \times & \left\langle P_{A-1} \left\vert {\rm Tr}\left[A^{+}\left(\zeta^-, \Delta z^-, \Delta z_{\perp}\right)A^+\left(\zeta^-,0\right)\right]\right\vert P_{A-1}\right\rangle.
\label{eq:K1_W_final_add_left-rightcut}
\end{eqnarray}
\begin{figure}[!h]
    \centering 
    \begin{subfigure}[t]{0.48\textwidth}
        \includegraphics[height=1.7in]{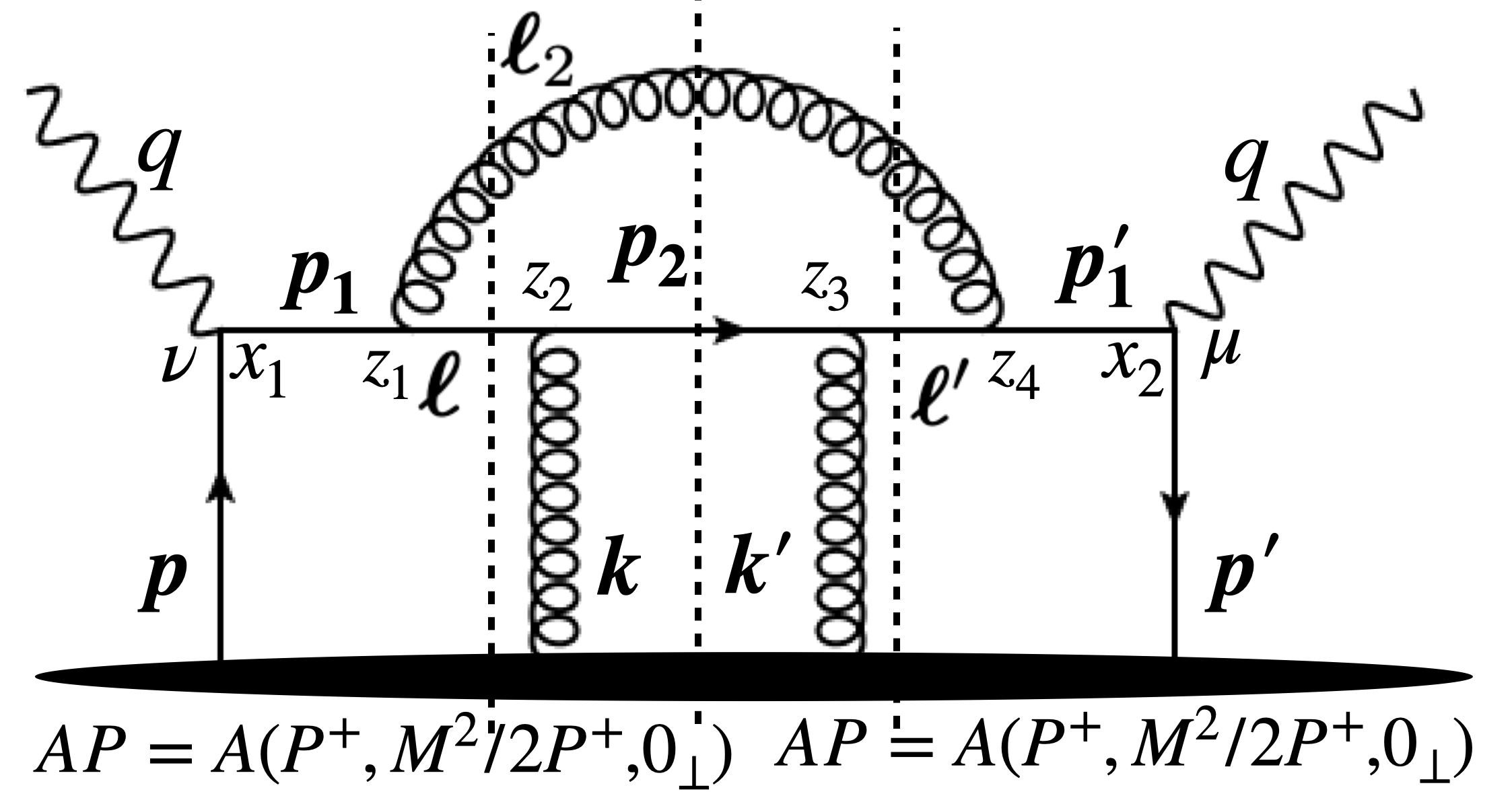}
        \caption{  }
    \end{subfigure}%
    \begin{subfigure}[t]{0.46\textwidth}
        \includegraphics[height=1.6in]{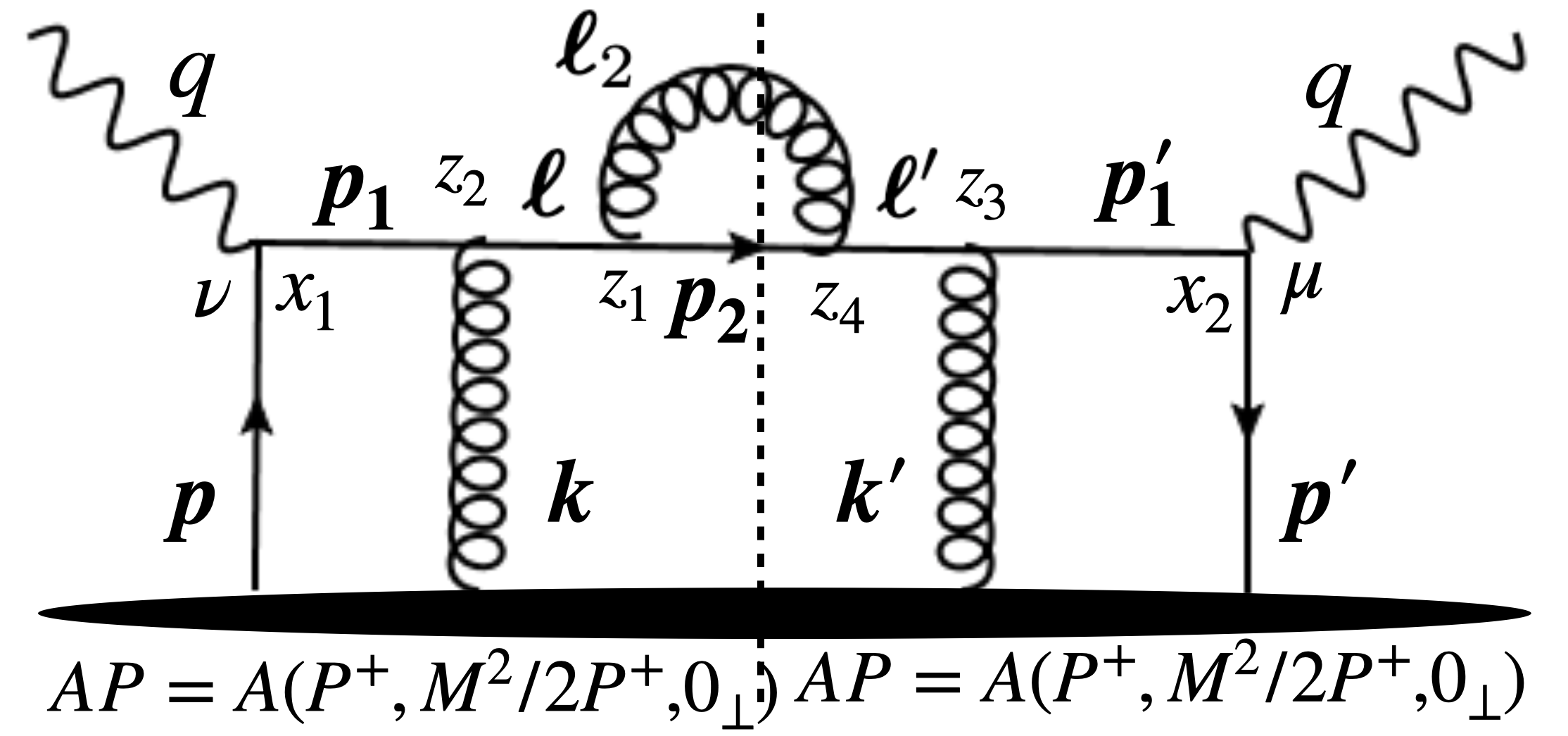}
        \caption{}
    \end{subfigure}
\caption{Scattering diagrams with gluon in-medium scattering, contributing to kernel-1. (a) Gluon scattering with the medium occurs after gluon radiation. (b) Gluon scattering with the medium occurs before gluon radiation.}
\label{fig:kernel1_b_c}
\end{figure}
%%%%%%%%%%%%%%%%%%%%%%%%%%%%%%%%%%%%%%%%%%%%%%%%%%%%%%%%%%%%%%%%%%%%%%%%%%%%%%%%%%%%%
%%%%%%%%%%%%%%%%%%%%%%%%%%%%%%%%%%%%%%%%%%%%%%%%%%%%%%%%%%%%%%%%%%%%%%%%%%%%%%%%%%%%%
\subsection{Figures~\ref{fig:kernel-1_all}(b) and \ref{fig:kernel-1_all}(c)}
%%%%%%%%%%%%%%%%%%%%%%%%%%%%%%%%%%%%%%%%%%%%%%%%%%%%%%%%%%%%%%%%%%%%%%%%%%%%%%%%%%%%%
%%%%%%%%%%%%%%%%%%%%%%%%%%%%%%%%%%%%%%%%%%%%%%%%%%%%%%%%%%%%%%%%%%%%%%%%%%%%%%%%%%%%%
Figure~\ref{fig:kernel1_b_c} identifies all position and momenta variables needed to compute the Feynman diagram in Figure~\ref{fig:kernel-1_all}(b) and Figure~\ref{fig:kernel-1_all}(c). The hadronic tensor for the central-cut diagram [Fig.~\ref{fig:kernel1_b_c}(a)] is 
\begin{eqnarray}
W^{\mu\nu}_{1,c} & = & \sum_f  e^2_f g^4_{s} \int d^4 x_{1} d^4 x_{2} d^4 z_{2} d^4 z_{3} \int \frac{d^4 p}{(2\pi)^4} \frac{d^4 p'}{(2\pi)^4}  \frac{d^4 \ell_{2}}{(2\pi)^4} \frac{d^4 p_{2}}{(2\pi)^4} e^{-ip'x_2} e^{ipx_1}\left\langle P \left| \bar{\psi}_{_f}(x_2) \frac{\gamma^{+}}{4} \psi_{_f}(x_1)\right| P\right\rangle \nonumber \\
& \times & e^{iz_3\left(q+p' - p_{2}-\ell_{2}\right) } e^{iz_2(\ell_2 + p_2 - q-p)} \left\langle P_{A-1} \left| {\rm Tr}\left[t^{d}A^{+c}t^{c}(z_3)A^{+b}(z_2)t^{b}t^{a}\right]\right| P_{A-1} \right\rangle \delta^{ad} (2 \pi) \delta\left(\ell_{2}^{2}\right) (2 \pi) \delta\left(p_{2}^{2}-M^{2}\right) d^{(\ell_2)}_{\sigma_{1} \sigma_{4}} \nonumber \\
& \times & \frac{{\rm Tr} \left[\gamma^{-} \gamma^{\mu} \left(\slashed{q}+\slashed{p}'+M\right) \gamma^{\sigma_{4}}\left(\slashed{q} +\slashed{p}' - \slashed{\ell}_2 + M\right)\gamma^-  \left(\slashed{p}_2+M\right) \gamma^-\left(\slashed{q}+\slashed{p}-\slashed{\ell}_2+M\right)\gamma^{\sigma_1} \left(\slashed{q}+\slashed{p}+M\right)\gamma^{\nu} \right]}{\left[\left(q+p'\right)^{2}-M^2-i\epsilon\right] \left[\left(q+p'-\ell_2\right)^2 - M^2 - i\epsilon\right] \left[\left(q+p-\ell_2\right)^2 -M^2+i\epsilon\right]\left[\left(q+p\right)^2-M^2+i\epsilon\right]}.
\label{eq:K1_bc_a_Wmunu_ccut_i}
\end{eqnarray}
The above expression [Eq.~(\ref{eq:K1_bc_a_Wmunu_ccut_i})] becomes singular when the denominator of the quark propagator for $p_1$, $\ell$, $\ell'$, and $p'_1$ vanishes. Computing this integral is easiest in the complex plane of $p^{+}$ and $p'^{+}$, where both $p^{+}$ and $p'^{+}$ have two simple poles.
The contour integration for $p^{+}$ can be carried out as
\begin{equation}
\begin{split}
C_{1} & = \oint \frac{dp^{+}}{(2\pi)} \frac{e^{ip^{+}\left(x^{-}_{1}-z^{-}_{2}\right)}}{\left[\left(q+p\right)^{2}-M^{2}+i\epsilon\right]\left[\left(q+p-\ell_{2}\right)^{2} -M^2 + i\epsilon\right]} \\
      & = \frac{(2\pi i)}{2\pi} \frac{\theta(x^{-}_{1} - z^{-}_{2})}{4q^{-}(q^{-}-\ell^{-}_{2})}  e^{i\left(-q^{+}+\frac{M^2}{2q^-}\right)\left(x^{-}_{1}-z^-_2\right)}\left[ \frac{ -1 + e^{i \mathcal{G}^{(\ell_2)}_M\left(x^{-}_{1}-z^{-}_2\right)} }{ \mathcal{G}^{(\ell_2)}_M}  \right],
\end{split}
\end{equation}
where $\mathcal{G}^{(\ell_2)}_M$ is defined in Eq.~(\ref{eq:GL2M_append}). 
The contour integration for $p'^+$ proceeds analogously, giving
\begin{equation}
\begin{split}
C_{2} & = \oint \frac{dp'^{+}}{(2\pi)} \frac{e^{-ip'^{+}\left(x^{-}_{2}-z^{-}_{3}\right)}}{\left[\left(q+p'\right)^{2} - M^{2} - i \epsilon\right]\left[\left(q+p'-\ell_{2}\right)^{2} -M^2 - i\epsilon \right]} \\
      & = \frac{(-2\pi i)}{2\pi} \frac{\theta\left(x^{-}_{2} - z^{-}_{3}\right)}{4q^{-}(q^{-}-\ell^{-}_{2})} e^{-i\left(-q^{+} + \frac{M^2}{2q^-}\right)\left(x^{-}_{2}-z^-_3\right)}\left[\frac{ -1 + e^{-i\mathcal{G}^{(\ell_2)}_M\left(x^{-}_{2}-z^-_3\right)}}{\mathcal{G}^{(\ell_2)}_M}\right].
\end{split}
\end{equation}
The trace in Eq.~(\ref{eq:K1_bc_a_Wmunu_ccut_i}) can be simplified to get
\begin{equation}
\begin{split}
    & {\rm Tr} \left[\gamma^{-} \gamma^{\mu} \left(\slashed{q}+\slashed{p}'+M\right)\gamma^{\sigma_{4}}\left(\slashed{q} +\slashed{p}' - \slashed{\ell}_2 + M\right)\gamma^- \left(\slashed{p}_2+M\right) \gamma^{-}\left(\slashed{q}+\slashed{p}-\slashed{\ell}_2+M\right)\gamma^{\sigma_{1}} \left(\slashed{q}+\slashed{p}+M\right)\gamma^{\nu} \right] d^{(\ell_2)}_{\sigma_{1} \sigma_{4}} \\ 
  & = 32 (q^-)^3 \left[ -g^{\mu\nu}_{\perp\perp} \right] \left[\frac{1-y+\eta y}{y} \right] \left[ \frac{1+ \left(1-y\right)^2}{y}\right] \left[ \pmb{\ell}^2_{2\perp} + \kappa y^4 M^2 \right].
\end{split}
\end{equation}
The final expression of the hadronic tensor for the central-cut diagram [Fig.~\ref{fig:kernel1_b_c}(a)] becomes
\begin{eqnarray}
\begin{split}
W^{\mu\nu}_{1,c} & = C_{F} \sum_f 2 \left[-g^{\mu\nu}_{\perp\perp}\right]  e^2_f g^4_{s} \int d (\Delta x^{-}) e^{i\Delta x^{-}\left(q^{+}-\frac{M^2}{2q^-}\right)} \left\langle P \left| \bar{\psi}_{_f}(\Delta x^-) \frac{\gamma^{+}}{4} \psi_{_f}(0)\right| P\right\rangle\\
  & \times  \int d (\Delta z^{-})d^2 \Delta z_{\perp} \frac{dy}{2\pi}\frac{d^2 \ell_{2\perp}}{(2\pi)^2} \frac{d^2 k_{\perp}}{(2\pi)^2} e^{-i\Delta z^{-}\mathcal{H}^{(\ell_2,p_2)}_M} e^{i\pmb{k}_{\perp}\cdot \Delta\pmb{z}_{\perp} } \\ 
  & \times \int d \zeta^- \theta(\zeta^-)  \left[\frac{1+ \left(1-y\right)^2}{y}\right] \frac{[\pmb{\ell}^2_{2\perp} + \kappa y^4 M^2]}{ \left[\pmb{\ell}^2_{2\perp} + y^2 M^2\right]^2} \left[2 - 2\cos\left\{\mathcal{G}^{(\ell_2)}_{M}\zeta^-\right\}\right] \\
  & \times \left\langle P_{A-1} \left| {\rm Tr}\left[A^{+}\left(\zeta^-, \Delta z^-, \Delta z_{\perp}\right)A^+\left(\zeta^-,0\right)\right] \right| P_{A-1} \right\rangle,
\end{split}
\label{eq:K1_bc_a_central_fi}
\end{eqnarray}
where $C_F=(N^2_c -1)/(2N_c)$ and $\mathcal{H}^{(\ell_2,p_2)}_M$ is defined in Eq.~(\ref{eq:HL2P2M_append}).

The left-cut [Fig.~\ref{fig:kernel1_b_c}(a)] gives rise to an interference between the single-gluon emission with no scattering process and single gluon preemission with double in-medium gluon scattering. The hadronic tensor for the left-cut diagram [Fig.~\ref{fig:kernel1_b_c}(a)] is given as 
\begin{eqnarray}
W^{\mu\nu}_{1,\ell} & =& \sum_{f}  e^2_f g^4_{s} \int d^4 x_{1} d^4 x_{2} d^4 z_{2} d^4 z_{3} \int \frac{d^4 \ell}{(2\pi)^4} \frac{d^4 p'}{(2\pi)^4}  \frac{d^4 \ell_{2}}{(2\pi)^4} \frac{d^4 p_{2}}{(2\pi)^4} e^{-ip'x_2} e^{i(-q+\ell+\ell_2)x_1}\left\langle P \left| \bar{\psi}_{_f}(x_2) \frac{\gamma^{+}}{4} \psi_{_f}(x_1)\right| P\right\rangle  \nonumber \\
  & & \times e^{i\left(q+p'-p_{2}-\ell_{2}\right) z_{3}} e^{i(p_{2} - \ell)z_{2}} \left\langle P_{A-1} \left| {\rm Tr}\left[t^{d}A^{+c}(z_3)t^{c}A^{+b}(z_2)t^{b}t^a\right]\right| P_{A-1} \right\rangle \delta^{ad} d^{(\ell_2)}_{\sigma_{1} \sigma_{4}} (2 \pi) \delta\left(\ell_{2}^{2}\right) (2 \pi)  \delta\left(\ell^{2}-M^{2}\right) \nonumber \\
& & \times \frac{  {\rm Tr} \left[\gamma^- \gamma^{\mu} \left(\slashed{q}+\slashed{p}'+M\right) \gamma^{\sigma_{4}}\left( \slashed{q} +\slashed{p}' - \slashed{\ell}_2 +  M\right)\gamma^{-}\left(\slashed{p}_2+M\right) \gamma^{-}\left(\slashed{\ell} +M\right)\gamma^{\sigma_1} \left(\slashed{\ell}_2 +\slashed{\ell}+M\right)\gamma^{\nu} \right]
}{  \left[\left(q+p'\right)^{2}-M^2-i\epsilon\right] \left[\left(q+p'-\ell_2\right)^2 - M^2 - i\epsilon\right]   \left[p_2^2 -M^2-i\epsilon\right] \left[\left(\ell_2+\ell\right)^2-M^2+i\epsilon\right]  }. \label{eq:kernel1_bc_a_wmunui_leftcut}
\end{eqnarray}
Equation~(\ref{eq:kernel1_bc_a_wmunui_leftcut}) admits singularities owing to the presence of two simple poles for momentum variable $p'^+$ and one simple pole for $p^+_2$. The contour integration for momentum $p^+_2$ gives
\begin{equation}
    C_{1} = \oint \frac{dp^+_2}{(2\pi)} \frac{e^{-ip^+_2(z^{-}_3-z^-_2)} }{[ p^2_2 -M^2 - i\epsilon]} =  \oint \frac{dp^+_2}{(2\pi)} \frac{e^{-ip^+_2(z^{-}_3-z^{-}_{2})} }{ 2p^-_2 \left[p^{+}_2 - \frac{\pmb{p}^2_{2\perp}+M^2}{2p^-_2} - i\epsilon\right]} = \frac{(-2\pi i)}{2\pi}\frac{\theta( z^{-}_3 - z^{-}_{2} )}{2p^-_2}  e^{-i\left( \frac{\pmb{p}^2_{2\perp}+M^2}{2p^-_2}\right)(z^{-}_3-z^{-}_{2})}.
\end{equation}
Similarly, the contour integration for $p'^+$ can be done
\begin{equation}
\begin{split}
C_{2} & = \oint \frac{dp'^{+}}{(2\pi)} \frac{e^{-ip'^{+}(x^{-}_{2}-z^{-}_{3})}}{\left[\left(q+p'\right)^{2} - M^{2} - i \epsilon\right]\left[\left(q+p'-\ell_{2}\right)^{2} -M^2 - i\epsilon\right]} \\
      & = \frac{(-2\pi i)}{2\pi} \frac{\theta(x^{-}_{2} - z^{-}_{3})}{4q^{-}(q^{-}-\ell^{-}_{2})}  e^{i\left(q^{+} - \frac{M^2}{2q^{-}})(x^{-}_{2}-z^{-}_{3}\right)}\left[ \frac{ -1 + e^{-i \mathcal{G}^{(\ell_2)}_M(x^{-}_{2}-z^{-}_{3})} }{ \mathcal{G}^{(\ell_2)}_M}  \right],
\end{split}
%\label{eq:}
\end{equation}
where $\mathcal{G}^{(\ell_2)}_M$ is defined in Eq.~(\ref{eq:GL2M_append}). The trace in the numerator of the third line of Eq.~(\ref{eq:kernel1_bc_a_wmunui_leftcut}) yields 
\begin{equation}
 \begin{split}
 &   {\rm Tr} \left[\gamma^- \gamma^{\mu} \left(\slashed{q}+\slashed{p}'+M\right) \gamma^{\sigma_{4}}\left( \slashed{q} +\slashed{p}' - \slashed{\ell}_2 +  M\right)\gamma^{-}\left(\slashed{p}_2+M\right) \gamma^{-}\left(\slashed{\ell} +M\right)\gamma^{\sigma_1} \left(\slashed{\ell}_2 +\slashed{\ell}+M\right)\gamma^{\nu} \right] d^{(\ell_2)}_{\sigma_1 \sigma_4} \\
 & = 32 (q^-)^3 [-g^{\mu\nu}_{\perp\perp}] \left[ \frac{1-y+\eta y}{y}\right] \left[ \frac{1 + \left(1-y\right)^2}{y}\right]  \left[ \pmb{\ell}^2_{2\perp} + \kappa y^4 M^2 \right].
 \end{split}
\end{equation}
The final expression of the hadronic tensor for the left-cut diagram [Fig.~\ref{fig:kernel1_b_c}(a)] is given by
\begin{equation}
\begin{split}
    W^{\mu\nu}_{1,\ell} =& C_{F} \sum_f 2 [-g^{\mu\nu}_{\perp \perp }] e^{2}_{f}g^{4}_{s}\int d (\Delta x^{-})   e^{i\left(q^{+}-\frac{M^2}{2q^-}\right)(\Delta x^{-} )} 
  \left\langle P \left| \bar{\psi}_{_f}( \Delta x^{-}) \frac{\gamma^{+}}{4} \psi_{_f}(0)\right| P\right\rangle  \\
  & \times n  \int  d \zeta^{-} d (\Delta z^{-}) d^2 \Delta z_{\perp} \frac{dy}{2\pi} \frac{d^2 \ell_{2\perp}}{(2\pi)^2} \frac{d^2 k_{\perp}}{(2\pi)^2}   e^{-i (\Delta z^{-})\mathcal{H}^{(\ell_2,p_2)}_{M}}     e^{i \pmb{k}_{\perp}\cdot\Delta\pmb{z}_{\perp}} \theta( z^{-}_3-z^{-}_2) \theta( x^{-}_{2}-z^{-}_3) \\
  & \times \left[ \frac{ 1 + \left(1-y\right)^2 }{y}\right]\frac{\left[  \pmb{\ell}^2_{2\perp} + \kappa y^4 M^2 \right]}{\left[  \pmb{\ell}^2_{2\perp} + y^2 M^2\right]^2} \left[ 1 - e^{-i\mathcal{G}^{(\ell_2)}_{M}(x^{-}_{2}-z^-_3)} \right] e^{i\mathcal{G}^{(\ell_2)}_{M}(x^{-}_{1}-z^{-}_{2})} \\
  &\times  \langle P_{A-1} |{\rm Tr}[ A^+(\zeta^-, \Delta z^-, \Delta z_{\perp}) A^+(\zeta^-,0)] | P_{A-1} \rangle   ,
\end{split} 
\label{eq:K1_Wmunu_bc_a_leftcut_final}
\end{equation}
where $\mathcal{G}^{(\ell_2)}_M$ is defined in Eq.~(\ref{eq:GL2M_append}) and $\mathcal{H}^{(\ell_2,p_2)}_{M}$ is defined in Eq.~(\ref{eq:HL2P2M_append}). The factor $n$ in Eq.~(\ref{eq:K1_Wmunu_bc_a_leftcut_final}) is described in Eq.~(\ref{eq:factor_n}).

The right-cut diagram shown in Fig.~\ref{fig:kernel1_b_c}(a) is a complex-conjugate of the left-cut diagram. The final expression of the hadronic tensor for the right-cut diagram [Fig.~\ref{fig:kernel1_b_c}(a)] reduces to the following form
\begin{equation}
\begin{split}
W^{\mu\nu}_{1,r} = & C_F \sum_f 2 [-g^{\mu\nu}_{\perp \perp }] e^2_f g^4_s \int d (\Delta x^{-})   e^{i\left(q^{+}-\frac{M^2}{2q^-}\right)(\Delta x^{-} )} 
  \left\langle P \left| \bar{\psi}_{_f}( \Delta x^{-}) \frac{\gamma^{+}}{4} \psi_{_f}(0)\right| P\right\rangle  \\
  & \times n \int  d \zeta^{-} d (\Delta z^{-}) d^2 \Delta z_{\perp} \frac{dy}{2\pi} \frac{d^2 \ell_{2\perp}}{(2\pi)^2} \frac{d^2 k_{\perp}}{(2\pi)^2}   e^{-i  (\Delta z^{-})\mathcal{H}^{(\ell_2,p_2)}_{M}}     e^{i \pmb{k}_{\perp}\cdot\Delta\pmb{z}_{\perp}}  \theta( -z^{-}_3+z^{-}_2) \theta( x^{-}_{1}-z^{-}_2) \\
  & \times \left[ \frac{ 1 + \left(1-y\right)^2 }{y}\right]  \frac{\left[\pmb{\ell}^2_{2\perp} + \kappa y^4 M^2 \right]}{\left[  \pmb{\ell}^2_{2\perp} + y^2 M^2\right]^2} \left[ 1 - e^{i\mathcal{G}^{(\ell_2)}_{M}(x^{-}_{1}-z^-_2)} \right] e^{-i\mathcal{G}^{(\ell_2)}_{M} (x^{-}_{2}-z^{-}_{3})} \\
  & \times \left\langle P_{A-1} \left|{\rm Tr}\left[ A^+\left(\zeta^-, \Delta z^-, \Delta z_{\perp}\right) A^+\left(\zeta^-,0\right)\right] \right| P_{A-1} \right\rangle,
\label{eq:K1_Wmunu_bc_a_rightcut_final}
\end{split} 
\end{equation}
where $\mathcal{G}^{(\ell_2)}_M$ is defined in Eq.~(\ref{eq:GL2M_append}) and $\mathcal{H}^{(\ell_2,p_2)}_{M}$ is defined in Eq.~(\ref{eq:HL2P2M_append}). The factor $n$ in Eq.~(\ref{eq:K1_Wmunu_bc_a_rightcut_final}) is described in Eq.~(\ref{eq:factor_n}). Adding the left-cut and right-cut diagrams, i.e., Eq.~(\ref{eq:K1_Wmunu_bc_a_leftcut_final}) and Eq.~(\ref{eq:K1_Wmunu_bc_a_rightcut_final}), respectively, gives the following expression of the hadronic tensor:
\begin{eqnarray}
    W^{\mu\nu}_{1,\ell+r} & =&C_F \sum_f 2 [-g^{\mu\nu}_{\perp \perp }] e^2_f g^4_s \int d (\Delta x^{-})   e^{i\left(q^{+}-\frac{M^2}{2q^-}\right)(\Delta x^{-} )}  
  \left\langle P \left| \bar{\psi}_{_f}( \Delta x^{-}) \frac{\gamma^{+}}{4} \psi_{_f}(0)\right| P\right\rangle \nonumber \\
  & &\times n \int  d \zeta^{-} d (\Delta z^{-}) d^2 \Delta z_{\perp} \frac{dy}{2\pi} \frac{d^2 \ell_{2\perp}}{(2\pi)^2} \frac{d^2 k_{\perp}}{(2\pi)^2} e^{-i  (\Delta z^{-})\mathcal{H}^{(\ell_2,p_2)}_{M}}     e^{i \pmb{k}_{\perp}\cdot\Delta\pmb{z}_{\perp}} \theta( \zeta^-) \label{eq:K1_Wmunu_bc_a_add_left_rightcut_final} \\
  & &\times  \left[ \frac{ 1 + \left(1-y\right)^2 }{y}\right] \frac{\left[\pmb{\ell}^2_{2\perp} + \kappa y^4 M^2 \right]}{\left[\pmb{\ell}^2_{2\perp} + y^2 M^2\right]^2} \left[\cos\left\{\mathcal{G}^{(\ell_2)}_{M}\zeta^-\right\} -1 \right]\left\langle P_{A-1} \left|{\rm Tr}\left[ A^+\left(\zeta^-, \Delta z^-, \Delta z_{\perp}\right) A^+\left(\zeta^-,0\right)\right] \right| P_{A-1} \right\rangle \nonumber.
\end{eqnarray}

The hadronic tensor for the central-cut diagram shown in Fig.~\ref{fig:kernel1_b_c}(b) can be written as
\begin{eqnarray}
W^{\mu\nu}_{1,c} & =&  \sum_f  e^2_f g^4_{s} \int d^4 x_{1} d^4 x_{2} d^4 z_{2} d^4 z_{3} \int \frac{d^4 p}{(2\pi)^4} \frac{d^4 p'}{(2\pi)^4}  \frac{d^4 \ell_{2}}{(2\pi)^4} \frac{d^4 p_{2}}{(2\pi)^4} e^{-ip'x_2} e^{ipx_1}\left\langle P \left| \bar{\psi}_{_f}(x_2) \frac{\gamma^{+}}{4} \psi_{_f}(x_1)\right| P\right\rangle \nonumber \\
  & & \times e^{i\left(q+p'-p_{2}-\ell_{2}\right) z_{3}} e^{i(\ell_{2}+p_{2} - q - p)z_{2}} \left\langle P_{A-1} \left| {\rm Tr}\left[t^{c}A^{+c}(z_3)t^{d}t^{a}t^{b}A^{+b}(z_2)\right] \right| P_{A-1} \right\rangle \delta^{ad} d^{(\ell_2)}_{\sigma_{1} \sigma_{4}} (2 \pi) \delta\left(\ell_{2}^{2}\right) (2 \pi)  \delta\left(p_{2}^{2}-M^{2}\right) \nonumber \\
& & \times \frac{  {\rm Tr} \left[\gamma^- \gamma^{\mu} \left(\slashed{q}+\slashed{p}'+M\right) \gamma^-\left( \slashed{\ell}_2 + \slashed{p}_2 + M\right)\gamma^{\sigma_{4}}\left(\slashed{p}_2+M\right) \gamma^{\sigma_1}\left(\slashed{\ell}_2 + \slashed{p}_2+M\right)\gamma^- \left(\slashed{q}+\slashed{p}+M\right)\gamma^{\nu} \right]
}{  \left[\left(q+p'\right)^{2}-M^2-i\epsilon\right] \left[\left(\ell_2+p_2\right)^2 - M^2 - i\epsilon\right]   \left[\left(\ell_2+p_2\right)^2 -M^2+i\epsilon\right] \left[\left(q+p\right)^2-M^2+i\epsilon\right]  }.
\label{eq:kernel1_bc_b_wi_central}
\end{eqnarray}
Equation~(\ref{eq:kernel1_bc_b_wi_central}) has singularity arising from the denominator of the quark propagator with momentum $p_{1}$ and $p'_{1}$. We identify one pole for each momentum variable $p^+$ and $p'^+$. 
 The contour integration for $p^+$ in the complex plane is given by
\begin{equation}
    C_{1} = \oint \frac{dp^+}{(2\pi)} \frac{e^{ip^+(x^{-}_{1}-z^-_2)} }{\left[\left(q+p\right)^2 -M^2 + i\epsilon\right]} =  \oint \frac{dp^+}{(2\pi)} \frac{e^{ip^+(x^{-}_{1}-z^{-}_{2})} }{ 2q^{-}[ q^{+}+p^{+} - [M^2/(2q^-)]+ i\epsilon]} = \frac{(2\pi i)}{2\pi}\frac{\theta\left( x^{-}_{1} - z^{-}_{2}\right)}{2q^{-}}  e^{i\left(-q^{+} + \frac{M^2}{2q^-}\right)(x^{-}_{1}-z^{-}_{2})}. 
\end{equation}
Similarly, the contour integration for momentum $p'^+$ is carried out 
\begin{equation}
    C_{2} = \oint \frac{dp'^+}{(2\pi)} \frac{e^{-ip'^+(x^{-}_{2}-z^{-}_3)} }{\left[\left(q+p'\right)^2 -M^2 -i\epsilon\right]} =  \oint \frac{dp'^+}{(2\pi)} \frac{e^{-ip'^+(x^{-}_{2}-z^{-}_3)} }{2q^- [ q^{+}+p'^{+} - [M^2/(2q^-)]-i\epsilon]} = \frac{(-2\pi i)}{2\pi} \frac{\theta\left( x^{-}_{2} - z^{-}_{3}\right)}{2q^-}  e^{i\left(q^+ - \frac{M^2}{2q^-}\right)(x^{-}_{2}-z^{-}_3)}.
\end{equation}
Including mass corrections, the trace yields
\begin{equation}
\begin{split}
& {\rm Tr} \left[\gamma^- \gamma^{\mu} \left(\slashed{q}+\slashed{p}'+M\right) \gamma^-\left( \slashed{\ell}_2 + \slashed{p}_2 + M\right) \gamma^{\sigma_{4}}  \left(\slashed{p}_2+M\right) \gamma^{\sigma_1} \left(\slashed{\ell}_2 + \slashed{p}_2+M\right) \gamma^-\left(\slashed{q}+\slashed{p}+M\right)\gamma^{\nu} \right] d^{(\ell_2)}_{\sigma_1 \sigma_4} \\
& = \frac{32 [-g^{\mu\nu}_{\perp\perp}](q^-)^3}{y\left(1-y+\eta y\right)} \left[ \frac{\left(1+\eta y\right)^2 + \left(1-y + \eta y\right)^2}{y}\right] \left[ \left\{\left(1+\eta y\right)\pmb{\ell}_{2\perp} - y\pmb{k}_{\perp}\right\}^2 + \kappa y^4 M^2 \right].
\end{split}
\end{equation}
The final expression of the hadronic tensor for the central-cut [Fig.~\ref{fig:kernel1_b_c}(b)] is
\begin{equation}
\begin{split}
    W^{\mu\nu}_{1,c} & = C_F \sum_f 2 [-g^{\mu\nu}_{\perp \perp }] e^2_f g^4_s \int d (\Delta x^{-})   e^{i\left(q^{+}-\frac{M^2}{2q^{-}}\right)(\Delta x^{-} )} 
  \left\langle P \left| \bar{\psi}_{_f}( \Delta x^{-}) \frac{\gamma^{+}}{4} \psi_{_f}(0)\right| P\right\rangle  \\
  & \times   \int  d \zeta^{-} d (\Delta z^{-}) d^2 \Delta z_{\perp} \frac{dy}{2\pi} \frac{d^2 \ell_{2\perp}}{(2\pi)^2} \frac{d^2 k_{\perp}}{(2\pi)^2}  e^{-i \mathcal{H}^{(\ell_2,p_2)}_{M} (\Delta z^{-})}     e^{i \pmb{k}_{\perp}\cdot\Delta\pmb{z}_{\perp}} \theta(\zeta^-)\\%\theta( x^{-}_{1}-z^{-}_2) \theta( x^{-}_{2}-z^{-}_3)  \\
%  & \times \frac{\theta( x^{-}_{1}-z^{-}_2) \theta( x^{-}_{2}-z^{-}_3)\left[\left\{\left(1+\eta y\right)\pmb{\ell}_{2\perp} - y\pmb{k}_{\perp}\right\}^{2} + \kappa y^4 M^2 \right]}{ \left[\left(\pmb{\ell}_{2\perp} - y\pmb{k}_{\perp}\right)^2 +2y\eta  (\pmb{\ell}^2_{2\perp}- y\pmb{\ell}_{2\perp}\cdot \pmb{k}_{\perp})+ \eta^2 y^2 \pmb{\ell}^2_{2\perp} + y^2 M^2\right]^{2}}   
%  \left\langle P_{A-1} \left| {\rm Tr}\left[A^+\left(\zeta^-, \Delta z^-, \Delta z_{\perp}\right) A^+\left(\zeta^-,0\right)\right] \right| P_{A-1} \right\rangle  \\
%   & \times \frac{\theta( x^{-}_{1}-z^{-}_2) \theta( x^{-}_{2}-z^{-}_3)\left[\left\{\left(1+\eta y\right)\pmb{\ell}_{2\perp} - y\pmb{k}_{\perp}\right\}^{2} + \kappa y^4 M^2 \right]}{ \left[\left\{\left(1+\eta y\right)\pmb{\ell}_{2\perp} - y\pmb{k}_{\perp}\right\}^{2} + y^2 M^2\right]^{2}}   
%  \left\langle P_{A-1} \left| {\rm Tr}\left[A^+\left(\zeta^-, \Delta z^-, \Delta z_{\perp}\right) A^+\left(\zeta^-,0\right)\right] \right| P_{A-1} \right\rangle  \\
   & \times\left[ \frac{ \left(1+\eta y\right)^2 + \left(1-y+\eta y\right)^2}{y}\right] \frac{\left[\left\{\left(1+\eta y\right)\pmb{\ell}_{2\perp} - y\pmb{k}_{\perp}\right\}^{2} + \kappa y^4 M^2 \right]}{ J_1^{2}} \\
  & \times \left\langle P_{A-1} \left| {\rm Tr}\left[A^+\left(\zeta^-, \Delta z^-, \Delta z_{\perp}\right) A^+\left(\zeta^-,0\right)\right] \right| P_{A-1} \right\rangle  ,
\end{split} 
\label{eq:K1_bc_b_fi_wmunu}
\end{equation}
where $\mathcal{H}^{(\ell_2,p_2)}_{M}$ is defined in Eq.~(\ref{eq:HL2P2M_append}).
%%%%%%%%%%%%%%%%%%%%%%%%%%%%%%%%%%%%%%%%%%%%%%%%%%%%%%%%%%%%%%%%%%%%%%%%%%%%%%%%%%%%%
%%%%%%%%%%%%%%%%%%%%%%%%%%%%%%%%%%%%%%%%%%%%%%%%%%%%%%%%%%%%%%%%%%%%%%%%%%%%%%%%%%%%%
\subsection{Figures~\ref{fig:kernel-1_all}(d) and \ref{fig:kernel-1_all}(e)}
%%%%%%%%%%%%%%%%%%%%%%%%%%%%%%%%%%%%%%%%%%%%%%%%%%%%%%%%%%%%%%%%%%%%%%%%%%%%%%%%%%%%%
%%%%%%%%%%%%%%%%%%%%%%%%%%%%%%%%%%%%%%%%%%%%%%%%%%%%%%%%%%%%%%%%%%%%%%%%%%%%%%%%%%%%%
%Write kernel-1 h and i diagram
\begin{figure}[!h]
    \centering 
    \begin{subfigure}[t]{0.48\textwidth}
        \includegraphics[height=1.7in]{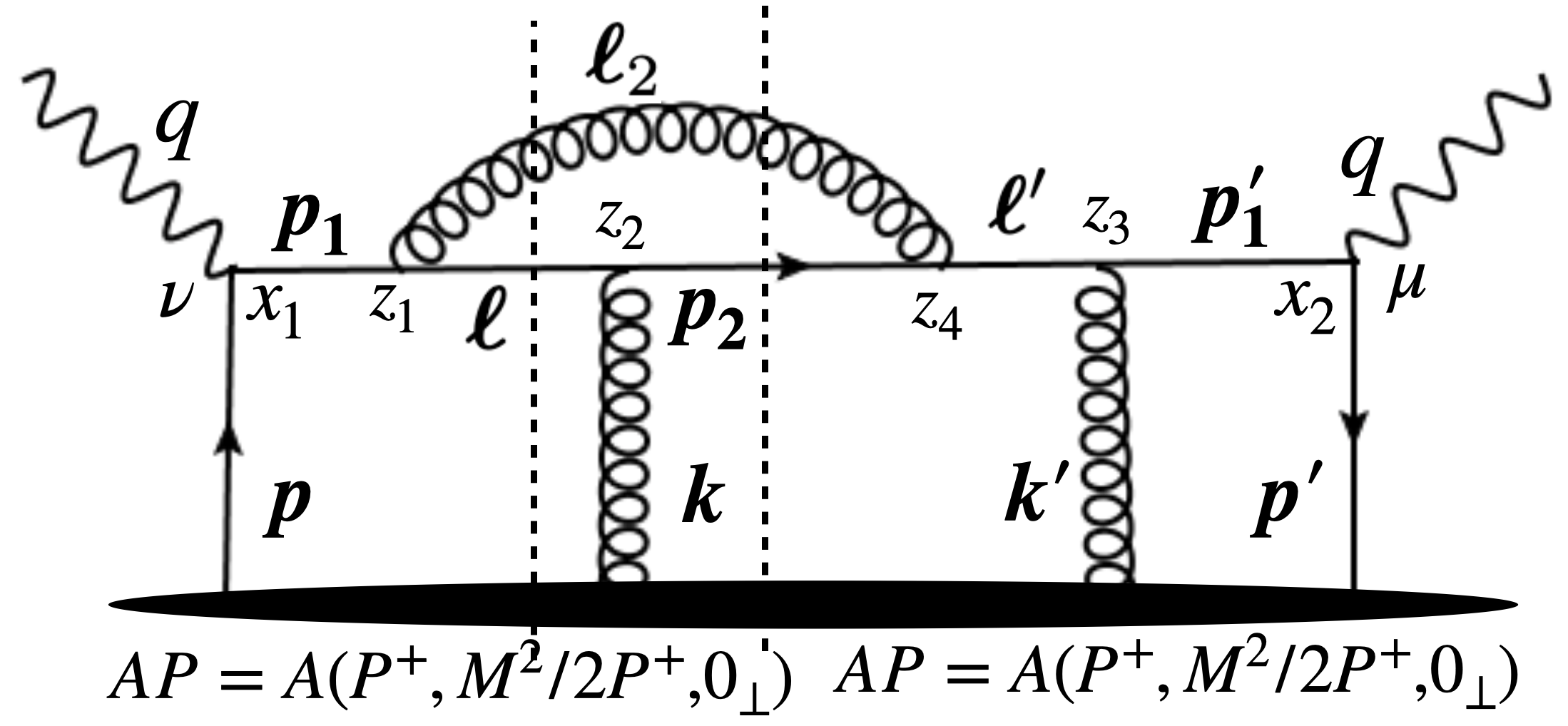}
        \caption{  }
    \end{subfigure}%
    \begin{subfigure}[t]{0.46\textwidth}
        \includegraphics[height=1.6in]{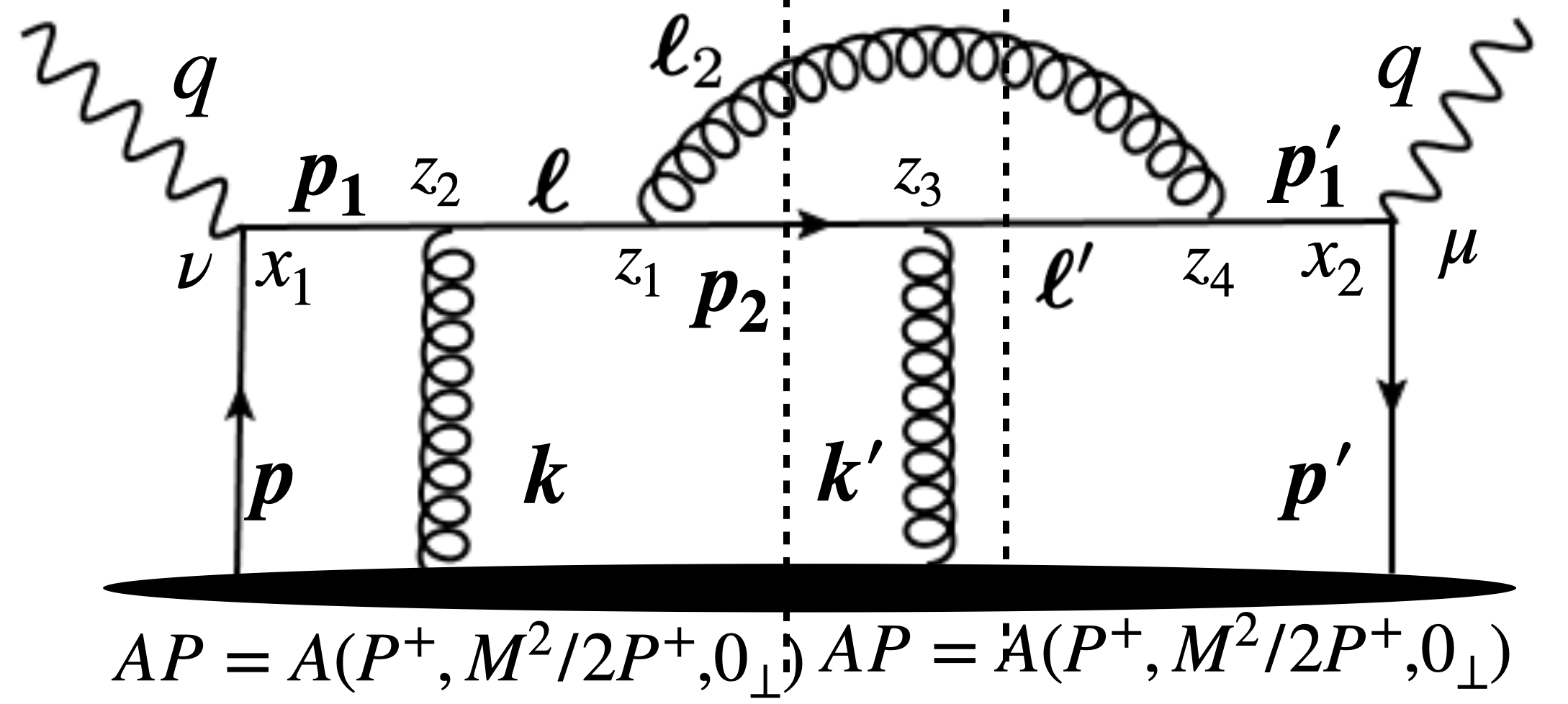}
        \caption{}
    \end{subfigure}
\caption{Scattering diagrams contributing to kernel-1. (a) The right-cut gives an interference between pre-emission scattering and post-emission scattering processes. The left-cut gives an interference between the gluon emission process in vacuum and the gluon emission with two in-medium gluon scattering. (b) Complex-conjugate of the processes on the left panel.}
\label{fig:kernel1_h_i}
\end{figure}

The hadronic tensor for the right-cut diagram shown in Fig.~\ref{fig:kernel1_h_i}(a) is
\begin{eqnarray}
W^{\mu\nu}_{1,r_1} & =& \sum_f  e^2_f g^4_{s} \int d^4 x_{1} d^4 x_{2} d^4 z_{2} d^4 z_{3} \int \frac{d^4 p}{(2\pi)^4} \frac{d^4 p'}{(2\pi)^4}  \frac{d^4 \ell_{2}}{(2\pi)^4} \frac{d^4 p_{2}}{(2\pi)^4} e^{-ip'x_{2}} e^{ipx_{1}}\left\langle P \left| \bar{\psi}_{_f}(x_2) \frac{\gamma^{+}}{4} \psi_{_f}(x_1)\right| P\right\rangle \nonumber \\
  & & \times e^{i\left(q+p'-p_{2}-\ell_{2}\right) z_{3}} e^{i(\ell_{2}+p_{2} - q - p)z_{2}} \left\langle P_{A-1} \left| {\rm Tr}\left[t^{c}A^{+c}(z_3)t^{d}t^{b}A^{+b}(z_2)t^{a}\right] \right| P_{A-1} \right\rangle \delta^{ad} d^{(\ell_2)}_{\sigma_{1} \sigma_{4}} (2 \pi) \delta\left(\ell_{2}^{2}\right) (2 \pi)  \delta\left(p_{2}^{2}-M^{2}\right) \nonumber \\
&  &\times\frac{ {\rm Tr} \left[\gamma^- \gamma^{\mu} \left(\slashed{q}+\slashed{p}'+M\right) \gamma^-\left( \slashed{\ell}_2 + \slashed{p}_2 + M\right)\gamma^{\sigma_{4}}\left(\slashed{p}_2+M\right) \gamma^-\left(\slashed{q} + \slashed{p} - \slashed{\ell}_2+M\right)\gamma^{\sigma_1}  \left(\slashed{q}+\slashed{p}+M\right)\gamma^{\nu} \right] 
}{  \left[\left(q+p'\right)^{2}-M^2-i\epsilon\right] \left[\left(\ell_2+p_2\right)^2 - M^2 - i\epsilon\right]  \left[\left(q+p-\ell_2\right)^2 -M^2+i\epsilon\right] \left[\left(q+p\right)^2-M^2+i\epsilon\right]  }. \label{eq:kernel-1_hi_a_wmunui_right}
\end{eqnarray}
Equation~(\ref{eq:kernel-1_hi_a_wmunui_right}) has singularity arising from the denominator of the quark propagator with momentum $p_{1}$, $\ell$ and $p'_{1}$. It has two simple poles for the momentum variable $p^+$ and one simple pole for $p'^+$. %We compute the integral in the complex plane of $p^+$ and $p'^{+}$.
The contour integration for momentum $p^+$ in the complex plane gives
\begin{equation}
\begin{split}
C_{1} & = \oint \frac{dp^{+}}{(2\pi)} \frac{e^{ip^{+}\left(x^{-}_{1}-z^{-}_{2}\right)}}{\left[\left(q+p\right)^{2} - M^{2} + i \epsilon\right]\left[\left(q+p-\ell_{2}\right)^{2} -M^2 + i\epsilon\right]} \\
      & = \frac{(2\pi i)}{2\pi} \frac{\theta\left(x^{-}_{1} - z^{-}_{2}\right)}{4q^{-}(q^{-}-\ell^{-}_{2})}  e^{i\left(-q^{+} + \frac{M^2}{2q^{-}}\right)(x^{-}_{1}-z^{-}_{2})}\left[ \frac{ -1 + e^{i \mathcal{G}^{(\ell_2)}_M(x^{-}_{1}-z^{-}_{2})} }{ \mathcal{G}^{(\ell_2)}_M}  \right],
\end{split}
\end{equation}
where $ \mathcal{G}^{(\ell_2)}_M $ is defined in Eq.~(\ref{eq:GL2M_append}).

Similarly, the contour integration for momentum $p'^+$ is carried out as 
\begin{equation}
    C_{2} = \oint \frac{dp'^+}{(2\pi)} \frac{e^{-ip'^+\left(x^{-}_{2}-z^{-}_3\right)} }{\left[\left(q+p'\right)^2 -M^2 -i\epsilon\right]} = \frac{(-2\pi i)}{2\pi} \frac{\theta( x^{-}_{2} - z^{-}_{3} )}{2q^-}  e^{i\left(q^+ - \frac{M^2}{2q^-}\right)(x^{-}_{2}-z^{-}_3)}.  
\end{equation}
Including mass corrections, the trace yields
\begin{equation}
\begin{split}
& {\rm Tr} \left[\gamma^- \gamma^{\mu} \left(\slashed{q}+\slashed{p}'+M\right) \gamma^-\left( \slashed{\ell}_2 + \slashed{p}_2 + M\right) \gamma^{\sigma_{4}}  \left(\slashed{p}_2+M\right) \gamma^- \left(\slashed{q}+ \slashed{p} - \slashed{\ell}_2+M\right) \gamma^{\sigma_1} \left(\slashed{q}+\slashed{p}+M\right)\gamma^{\nu} \right] d^{(\ell_2)}_{\sigma_1 \sigma_4} \\
& = \frac{32 [-g^{\mu\nu}_{\perp\perp}]\left(q^-\right)^3}{y} \left[ \frac{ 1  + \left(1-y\right)^2 + \eta y (2-y)}{y}\right] \left[ \left(1+\eta y\right)\pmb{\ell}^2_{2\perp} - y\pmb{k}_{\perp}\cdot \pmb{\ell}_{2\perp}  + \kappa y^4 M^2 \right].
\end{split}\label{eq:trace_right-cut_ph_qqgm_ph_qqgm}
\end{equation}
The final expression of the hadronic tensor for the right-cut [Fig.~\ref{fig:kernel1_h_i}(a)] is given as
\begin{equation}
\begin{split}
    W^{\mu\nu}_{1,r_1} =& \left(C_F - \frac{C_A}{2}\right) \sum_f 2 [-g^{\mu\nu}_{\perp \perp }] e^2_f g^4_s \int d (\Delta x^{-})   e^{iq^{+}(\Delta x^{-} )} 
    e^{-i[M^{2}/(2q^{-})](\Delta x^{-} )}
  \left\langle P \left| \bar{\psi}_{_f}( \Delta x^{-}) \frac{\gamma^{+}}{4} \psi_{_f}(0)\right| P\right\rangle  \\
  & \times  \int  d \zeta^{-} d (\Delta z^{-}) d^2 \Delta z_{\perp} \frac{dy}{2\pi} \frac{d^2 \ell_{2\perp}}{(2\pi)^2} \frac{d^2 k_{\perp}}{(2\pi)^2} e^{-i \mathcal{H}^{(\ell_2,p_2)}_{M} (\Delta z^{-})} e^{i \pmb{k}_{\perp}\cdot\Delta\pmb{z}_{\perp}} \theta( x^{-}_{1}-z^{-}_2) \theta( x^{-}_{2}-z^{-}_3)  \\
  & \times \left[ \frac{ 1 + \left(1-y\right)^2 +\eta y(2-y)}{y}\right] \frac{\left[  (1+\eta y)\pmb{\ell}^2_{2\perp} - y\pmb{k}_{\perp}\cdot\pmb{\ell}_{2\perp}   + \kappa y^4 M^2 \right]}{\left[  \pmb{\ell}^2_{2\perp} + y^2 M^2\right] J_1}\left[ -1 + e^{i\mathcal{G}^{(\ell_2)}_{M}(x^{-}_{1}-z^-_2)} \right] \\
  & \times \left\langle P_{A-1} \left| {\rm Tr}\left[A^+(\zeta^-, \Delta z^-, \Delta z_{\perp}) A^+(\zeta^-,0)\right] \right| P_{A-1} \right\rangle,
\end{split} 
\end{equation}
where $\mathcal{G}^{(\ell_2)}_M$ is defined in Eq.~(\ref{eq:GL2M_append}), $\mathcal{H}^{(\ell_2,p_2)}_{M}$ defined in Eq.~(\ref{eq:HL2P2M_append}), and $J_1$ is defined in Eq.~(\ref{eq:J1_append}). 

Now, we consider the left-cut diagram shown in Fig.~\ref{fig:kernel1_h_i}(a). Its hadronic tensor can be written as
\begin{equation}
\begin{split}
W^{\mu\nu}_{1,\ell_1} & =\sum_f  e^2_f g^4_{s} \int d^4 x_{1} d^4 x_{2} d^4 z_{2} d^4 z_{3} \int \frac{d^4 \ell}{(2\pi)^4} \frac{d^4 p'}{(2\pi)^4}  \frac{d^4 \ell_{2}}{(2\pi)^4} \frac{d^4 p_{2}}{(2\pi)^4} e^{-ip'x_2} e^{i(\ell_2 + \ell -q)x_1}\left\langle P \left| \bar{\psi}_{_f}(x_2) \frac{\gamma^{+}}{4} \psi_{_f}(x_1)\right| P\right\rangle  \\
  & \times e^{i\left(q+p'-p_{2}-\ell_{2}\right) z_{3}} e^{i( p_{2} - \ell )z_{2}} \left\langle P_{A-1} \left| {\rm Tr}\left[t^{c}A^{+c}(z_3)t^{d}t^{b}A^{+b}(z_2)t^{a}\right] \right| P_{A-1} \right\rangle \delta^{ad} d^{(\ell_2)}_{\sigma_{1} \sigma_{4}} (2 \pi) \delta\left(\ell_{2}^{2}\right) (2 \pi)  \delta\left(\ell^{2}-M^{2}\right) \\
& \times \frac{ {\rm Tr} \left[ \gamma^- \gamma^{\mu} \left(\slashed{q}+\slashed{p}'+M\right) \gamma^-\left( \slashed{\ell}_2 + \slashed{p}_2 + M\right)\gamma^{\sigma_{4}} \left(\slashed{p}_2+M\right)\gamma^-\left(\slashed{\ell} +M\right) \gamma^{\sigma_1}\left(\slashed{\ell}_2+\slashed{\ell}+M\right)\gamma^{\nu} \right] }{ \left[\left(q+p'\right)^{2}-M^2-i\epsilon\right] \left[\left(\ell_2+p_2\right)^2 - M^2 - i\epsilon\right] \left[p^2_2 -M^2-i\epsilon\right] \left[\left(\ell_2+\ell\right)^2-M^2+i\epsilon\right] } .
\end{split} \label{eq:kernel-1_ph_qqgm_ph_qqgm_both_wi_a_left}
\end{equation}
Equation~(\ref{eq:kernel-1_ph_qqgm_ph_qqgm_both_wi_a_left}) has singularity arising from the denominator of the quark propagator with momentum $p'_{1}$, $\ell'$ and $p_{2}$. We identify two simple poles for the momentum variable $p^+_2$ and one simple pole for $p'^+$. We compute the integral in the complex plane of $p^+_2$ and $p'^{+}$.
The contour integration for momentum $p'^+$ is carried out as 
\begin{equation}
    C_{1} = \oint \frac{dp'^+}{(2\pi)} \frac{e^{-ip'^+(x^{-}_{2}-z^{-}_3)} }{\left[\left(q+p'\right)^2 -M^2 -i\epsilon\right]}  = \frac{(-2\pi i)}{2\pi} \frac{\theta\left( x^{-}_{2} - z^{-}_{3}\right)}{2q^-}  e^{i\left(q^+ - \frac{M^2}{2q^-}\right)(x^{-}_{2}-z^{-}_3)}  .
\end{equation}
Similarly, the contour integration for momentum $p^+_2$ is carried out as
\begin{equation}
\begin{split}
    C_{2} & = \oint \frac{dp^+_2}{(2\pi)} \frac{e^{-ip^+_2(z^{-}_3-z^{-}_2)} }{\left[ p^2_2 -M^2 -i\epsilon\right] \left[\left(\ell_2+p_2\right)^2 -M^2 - i\epsilon \right] } \\
    & = \oint \frac{dp^+_2}{(2\pi)} \frac{e^{-ip^+_2(z^{-}_3-z^{-}_2)}}{\left[2p^-_2 p^+_2 - \left(\pmb{p}^2_{2\perp} + M^2\right) - i \epsilon\right] \left[ 2\left(1+\eta y\right)q^-\left(\ell^+_2+p^+_2\right) - \left(\pmb{k}^2_\perp +M^2\right) -i\epsilon\right]}  \\ 
    & = \frac{(-2\pi i)}{2\pi} \frac{\theta\left( z^{-}_3 - z^{-}_{2} \right)}{4p^-_2 \left(1+\eta y\right)q^-} e^{-i\left[ \frac{\left(\pmb{\ell}_{2\perp}-\pmb{k}_{\perp}\right)^2+M^2}{2\left(1-y+\eta y\right)q^-}\right](z^{-}_3-z^{-}_2)} \left[ \frac{1 - e^{i\mathcal{G}^{(\ell_2,p_2,k)}_{M}(z^-_3 - z^-_2)}}{\mathcal{G}^{(\ell_2,p_2,k)}_{M}}\right],
    \end{split}
\end{equation}
where $\mathcal{G}^{(\ell_2,p_2,k)}_{M}$ is defined in Eq.~(\ref{eq:GML2_p2k_J1_append}). The trace in the numerator of the third line of Eq.~(\ref{eq:kernel-1_ph_qqgm_ph_qqgm_both_wi_a_left}) is the same as the trace for the right-cut diagram given in Eq.~(\ref{eq:trace_right-cut_ph_qqgm_ph_qqgm}). The final expression of the hadronic tensor for the left-cut diagram [Fig.~\ref{fig:kernel1_h_i}(a)] is given by
\begin{equation}
\begin{split}
W^{\mu\nu}_{1,\ell_1} & =\left(C_F - \frac{C_A}{2}\right)\sum_f 2 [-g^{\mu\nu}_{\perp \perp }] e^2_f g^4_s \int d (\Delta x^{-})   e^{i\left( q^{+}-\frac{M^2}{2q^-}\right)(\Delta x^{-} )} 
  \left\langle P \left| \bar{\psi}_{_f}( \Delta x^{-}) \frac{\gamma^{+}}{4} \psi_{_f}(0)\right| P\right\rangle  \\
  & \times  n \int  d \zeta^{-} d (\Delta z^{-}) d^2 \Delta z_{\perp} \frac{dy}{2\pi} \frac{d^2 \ell_{2\perp}}{(2\pi)^2} \frac{d^2 k_{\perp}}{(2\pi)^2} e^{-i \Delta z^{-}\mathcal{H}^{(\ell_2,p_2)}_{M}} e^{i \pmb{k}_{\perp}\cdot \Delta\pmb{z}_{\perp}} \theta\left(z^{-}_3-z^{-}_2\right) \theta\left(x^{-}_{2}-z^{-}_3\right)  \\
  & \times \left[ \frac{ 1 + \left(1-y\right)^2 +\eta y(2-y)}{y}\right]   \frac{\left[\left(1+\eta y\right)\pmb{\ell}^2_{2\perp} - y\pmb{k}_{\perp}\cdot\pmb{\ell}_{2\perp} + \kappa y^4 M^2 \right]}{\left[\pmb{\ell}^2_{2\perp} + y^2 M^2\right] J_1} \left[ -1 + e^{i\mathcal{G}^{(\ell_2,p_2,k)}_{M}(z^{-}_3-z^-_2)} \right] e^{i\mathcal{G}^{(\ell_2)}_{M}(x^{-}_{1}-z^-_{2})}\\
  &\times \left\langle P_{A-1} \left| {\rm Tr}\left[A^+(\zeta^-, \Delta z^-, \Delta z_{\perp}) A^+(\zeta^-, 0)\right] \right| P_{A-1} \right\rangle,
\end{split}
\label{eq:K1_wmnunu_hi_a_leftcut_final}
\end{equation}
where $\mathcal{G}^{(\ell_2,p_2,k)}_{M}$  is defined in Eq.~(\ref{eq:GML2_p2k_J1_append}). The factor $n$ in Eq.~(\ref{eq:K1_wmnunu_hi_a_leftcut_final}) is described in Eq.~(\ref{eq:factor_n}).

Subsequently, the diagram shown in the right panel of Fig.~\ref{fig:kernel1_h_i} is considered next. The topology of the diagram is the same as the diagram on the left panel. Moreover, they are complex conjugate of each other. The hadronic tensor for the left-cut diagram shown in Fig.~\ref{fig:kernel1_h_i}(b) can be written as
\begin{equation}
\begin{split}
    W^{\mu\nu}_{1,\ell_2} & = \left( C_F - \frac{C_A}{2}\right) \sum_f 2 [-g^{\mu\nu}_{\perp \perp }] e^2_f g^4_s \int d (\Delta x^{-})   e^{iq^{+}(\Delta x^{-} )} 
    e^{-i[M^{2}/(2q^{-})](\Delta x^{-} )}
  \left\langle P \left| \bar{\psi}_{_f}( \Delta x^{-}) \frac{\gamma^{+}}{4} \psi_{_f}(0)\right| P\right\rangle  \\
  & \times   \int  d \zeta^{-} d (\Delta z^{-}) d^2 \Delta z_{\perp} \frac{dy}{2\pi} \frac{d^2 \ell_{2\perp}}{(2\pi)^2} \frac{d^2 k_{\perp}}{(2\pi)^2}  e^{-i \mathcal{H}^{(\ell_2,p_2)}_{M} (\Delta z^{-})}     e^{i \pmb{k}_{\perp}\cdot\Delta\pmb{z}_{\perp}} \theta\left( x^{-}_{1}-z^{-}_2\right) \theta\left( x^{-}_{2}-z^{-}_3\right) \\
  & \times \left[ \frac{ 1 + \left(1-y\right)^2 +\eta y\left(2-y\right)}{y}\right]\frac{ \left[\left(1+\eta y\right)\pmb{\ell}^2_{2\perp} - y\pmb{k}_{\perp}\cdot\pmb{\ell}_{2\perp} + \kappa y^4 M^2 \right]}{\left[  \pmb{\ell}^2_{2\perp} + y^2 M^2\right] J_1}\left[ -1 + e^{-i\mathcal{G}^{(\ell_2)}_{M}(x^{-}_{2}-z^-_3)} \right] \\
  & \times \left\langle P_{A-1} \left| {\rm Tr}\left[A^+\left(\zeta^-, \Delta z^-, \Delta z_{\perp}\right) A^+\left(\zeta^-,0\right)\right] \right| P_{A-1} \right\rangle, 
\end{split} 
\end{equation}
where $\mathcal{G}^{(\ell_2)}_M$ is defined in Eq.~(\ref{eq:GL2M_append}), $\mathcal{H}^{(\ell_2,p_2)}_{M}$ defined in Eq.~(\ref{eq:HL2P2M_append}), and $J_1$ is defined in Eq.~(\ref{eq:J1_append}). 
Similarly, the final expression of the hadronic tensor for the right-cut diagram [Fig.~\ref{fig:kernel1_h_i}(b)] is given as
\begin{equation}
\begin{split}
    W^{\mu\nu}_{1,r_2} & = \left( C_F - \frac{C_A}{2}\right) \sum_f 2 [-g^{\mu\nu}_{\perp \perp }] e^2_f g^4_s \int d (\Delta x^{-})   e^{i\left(q^{+}-\frac{M^2}{2q^-}\right)(\Delta x^{-} )} 
  \left\langle P \left| \bar{\psi}_{_f}( \Delta x^{-}) \frac{\gamma^{+}}{4} \psi_{_f}(0)\right| P\right\rangle  \\
  & \times n  \int  d \zeta^{-} d (\Delta z^{-}) d^2 \Delta z_{\perp} \frac{dy}{2\pi} \frac{d^2 \ell_{2\perp}}{(2\pi)^2} \frac{d^2 k_{\perp}}{(2\pi)^2} e^{-i (\Delta z^{-})\mathcal{H}^{(\ell_2,p_2)}_{M}}     e^{i \pmb{k}_{\perp}\cdot \Delta\pmb{z}_{\perp}} \theta\left(-z^{-}_3 + z^{-}_2\right) \theta\left(x^{-}_{1}-z^{-}_2\right) \\
  & \times \left[ \frac{ 1 + \left(1-y\right)^2 +\eta y\left(2-y\right)}{y}\right] \frac{\left[\left(1+\eta y\right)\pmb{\ell}^2_{2\perp} - y\pmb{k}_{\perp}\cdot\pmb{\ell}_{2\perp} + \kappa y^4 M^2 \right]}{\left[\pmb{\ell}^2_{2\perp} + y^2 M^2 \right] J_1} \left[ -1 + e^{i\mathcal{G}^{(\ell_2,p_2,k)}_{M}(z^{-}_3-z^-_2)} \right] e^{-i\mathcal{G}^{(\ell_2)}_{M}(x^{-}_{2}-z^{-}_{3})} \\
  & \times  \left\langle P_{A-1} \left|{\rm Tr}\left[A^+\left(\zeta^-, \Delta z^-, \Delta z_{\perp}\right) A^+\left(\zeta^-, 0\right)\right] \right| P_{A-1} \right\rangle,
\label{eq:K1_wmnunu_hi_b_rightcut_final}
\end{split} 
\end{equation}
where $\mathcal{G}^{(\ell_2,p_2,k)}_{M}$ is defined in Eq.~(\ref{eq:GML2_p2k_J1_append}). The factor $n$ in Eq.~(\ref{eq:K1_wmnunu_hi_b_rightcut_final}) is described in Eq.~(\ref{eq:factor_n}).

Note that our definition of the length-integration variable is $\zeta^{-}=x^{-}_{1}-z^{-}_{2}=x^{-}_{2}-z^{-}_{3}$ which becomes $\zeta^{-}=-z^{-}_{2}=-z^{-}_{3}$ when $x^{-}_{1}$ and $x^{-}_{2}$ are initialized as the origin. This leads the term $\left[ -1 + e^{i\mathcal{G}^{(\ell_2,p_2,k)}_{M}(z^{-}_3-z^-_2)} \right]$ in Eq.~(\ref{eq:K1_wmnunu_hi_a_leftcut_final}) and Eq.~(\ref{eq:K1_wmnunu_hi_b_rightcut_final}) to vanish. Therefore, adding the left-cut diagram [Fig.~\ref{fig:kernel1_h_i}(a)] and right-cut diagram [Fig.~\ref{fig:kernel1_h_i}(b)] one obtains zero, and hence no contribution to the energy-loss kernel. This implies the hadronic tensor for the left-cut diagram [Fig.~\ref{fig:kernel1_h_i}(a)] and right-cut diagram [Fig.~\ref{fig:kernel1_h_i}(b)] is 
\begin{eqnarray}
    W^{\mu\nu}_{1,\ell_1} = W^{\mu\nu}_{1,r_2} =0.
    \label{eq:K1_de_vanishing}
\end{eqnarray}
However, the right-cut diagram [Fig.~\ref{fig:kernel1_h_i}(a)] and left-cut diagram [Fig.~\ref{fig:kernel1_h_i}(b)] have non-vanishing contributions, and their total hadronic tensor is given by
\begin{equation}
\begin{split}
    W^{\mu\nu}_{1,r_1+\ell_2} & = -\left( C_F - \frac{C_A}{2}\right) \sum_f 2 [-g^{\mu\nu}_{\perp \perp }] e^2_f g^4_s \int d (\Delta x^{-})   e^{iq^{+}(\Delta x^{-} )} 
    e^{-i[M^{2}/(2q^{-})](\Delta x^{-} )}
  \left\langle P \left| \bar{\psi}_{_f}( \Delta x^{-}) \frac{\gamma^{+}}{4} \psi_{_f}(0)\right| P\right\rangle  \\
  & \times   \int  d \zeta^{-} d (\Delta z^{-}) d^2 \Delta z_{\perp} \frac{dy}{2\pi} \frac{d^2 \ell_{2\perp}}{(2\pi)^2} \frac{d^2 k_{\perp}}{(2\pi)^2}  e^{-i \mathcal{H}^{(\ell_2,p_2)}_{M} (\Delta z^{-})}     e^{i \pmb{k}_{\perp}\cdot\Delta\pmb{z}_{\perp}} \theta\left( \zeta^{-}\right) \\
  & \times \left[ \frac{ 1 + \left(1-y\right)^2 +\eta y\left(2-y\right)}{y}\right]\frac{ \left[\left(1+\eta y\right)\pmb{\ell}^2_{2\perp} - y\pmb{k}_{\perp}\cdot\pmb{\ell}_{2\perp} + \kappa y^4 M^2 \right]}{\left[  \pmb{\ell}^2_{2\perp} + y^2 M^2\right] J_1}\left[ 2 - 2\cos\left\{\mathcal{G}^{(\ell_2)}_{M}\zeta^{-}\right\} \right] \\
  & \times \left\langle P_{A-1} \left| {\rm Tr}\left[A^+\left(\zeta^-, \Delta z^-, \Delta z_{\perp}\right) A^+\left(\zeta^-,0\right)\right] \right| P_{A-1} \right\rangle.
\end{split} 
\label{eq:K1_de_nonvan}
\end{equation}
%
%%%%%%%%%%%%%%%%%%%%%%%%%%%%%%%%%%%%%%%%%%%%%%%%%%%%%%%%%%%%%%%%%%%%%%%%%%%%%%%%%%%%%
%%%%%%%%%%%%%%%%%%%%%%%%%%%%%%%%%%%%%%%%%%%%%%%%%%%%%%%%%%%%%%%%%%%%%%%%%%%%%%%%%%%%%
\subsection{Figures~\ref{fig:kernel-1_all}(f) and \ref{fig:kernel-1_all}(g)}
%%%%%%%%%%%%%%%%%%%%%%%%%%%%%%%%%%%%%%%%%%%%%%%%%%%%%%%%%%%%%%%%%%%%%%%%%%%%%%%%%%%%%
%%%%%%%%%%%%%%%%%%%%%%%%%%%%%%%%%%%%%%%%%%%%%%%%%%%%%%%%%%%%%%%%%%%%%%%%%%%%%%%%%%%%%
%Triple gluon vertex diagram
%
\begin{figure}[!h]
    \centering 
    \begin{subfigure}[t]{0.495\textwidth}
        \centering        \includegraphics[height=1.6in]{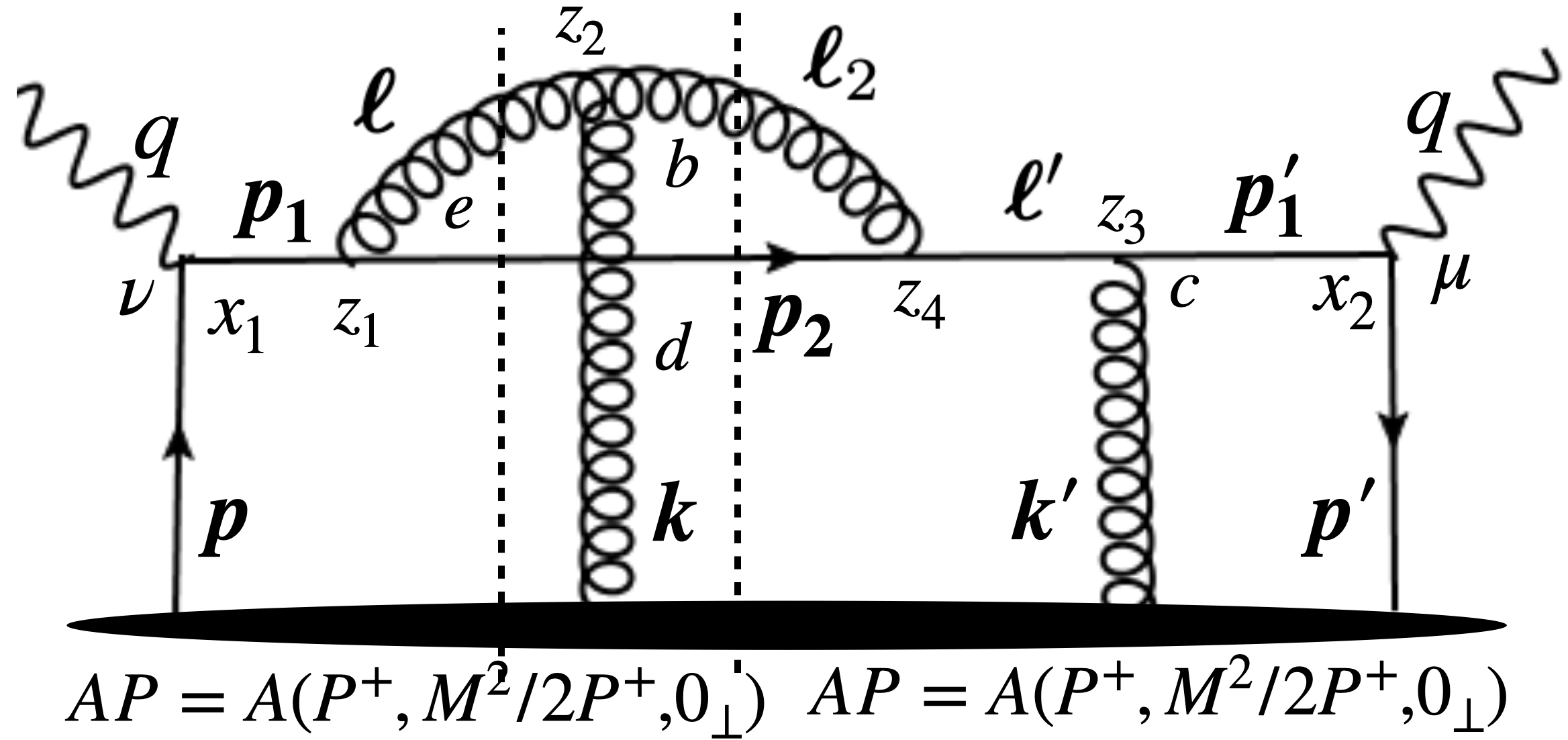}
        \caption{ }
    \end{subfigure}%
    \begin{subfigure}[t]{0.495\textwidth}
        \centering        \includegraphics[height=1.6in]{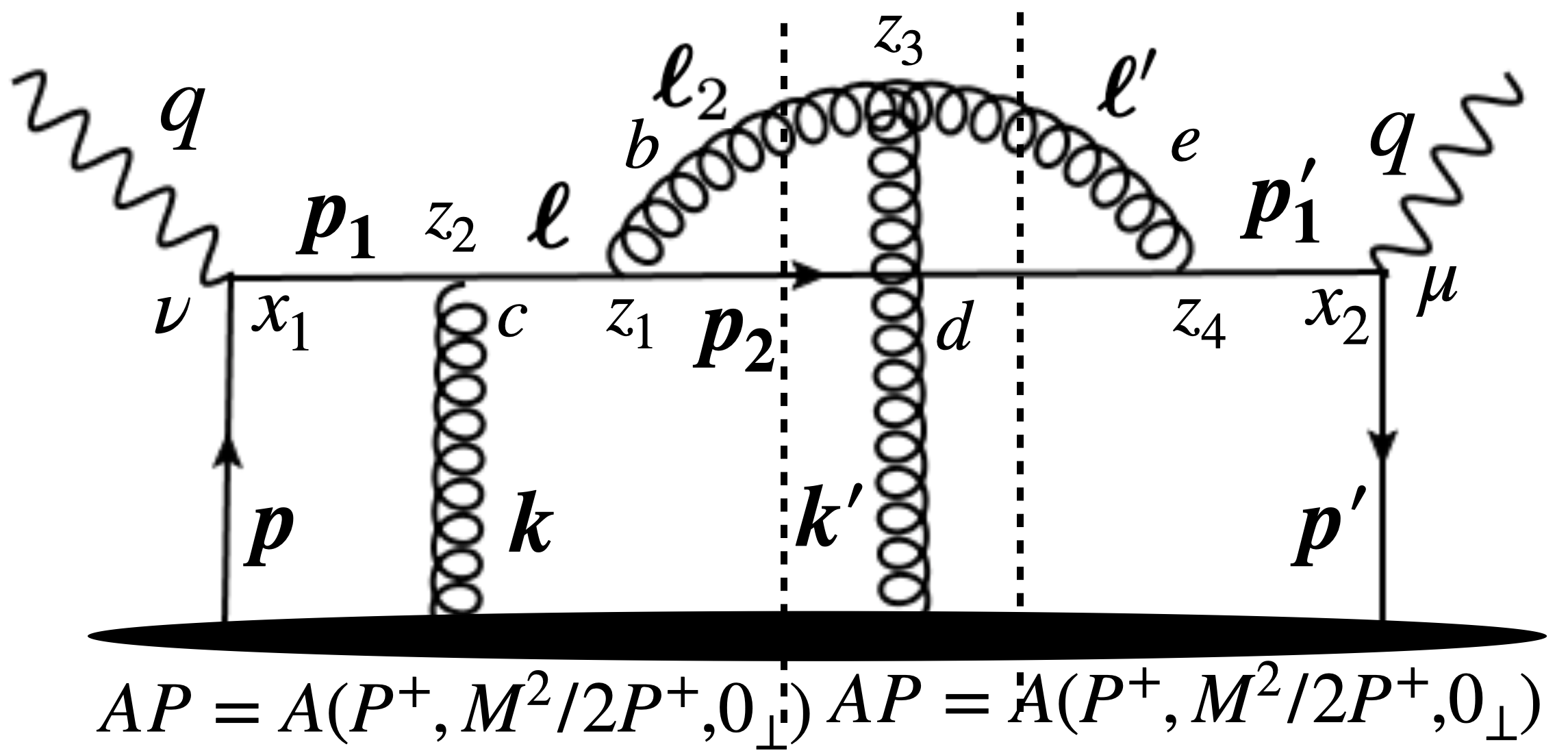}
        \caption{}
    \end{subfigure}
\caption{A forward scattering diagram contributing to kernel-1.}
\label{fig:kernel1_f_g}
\end{figure}

The interference diagrams with a triple-gluon vertex are now considered, where variables needed to perform the calcultion, such as position and momentum, are labeled in Fig.~\ref{fig:kernel1_f_g}. The right-cut in the forward scattering diagram [Fig.~\ref{fig:kernel1_f_g}(a)] represents an interference between single-gluon emission with gluon-gluon in-medium scattering and single-gluon emission post quark-gluon scattering process. The hadronic tensor for the right-cut diagram [Fig.~\ref{fig:kernel1_f_g}(a)] is given as 
\begin{equation}
\begin{split}
W^{\mu\nu}_{1,r_1} & = \sum_{f} e^2_f g^4_{s} \int d^4 x_1 d^4 x_2 d^4 z_{2} d^4 z_{3} \int \frac{d^4 p}{(2\pi)^4} \frac{d^4 p'}{(2\pi)^4}  \frac{d^4 \ell_{2}}{(2\pi)^4} \frac{d^4 p_{2}}{(2\pi)^4} e^{-ip'x_{2}} e^{ipx_{1}}\left\langle P \left| \bar{\psi}_{_f}(x_2) \frac{\gamma^{+}}{4} \psi_{_f}(x_{1})\right| P\right\rangle  \\
  & \times e^{i\left(q+p'-p_{2}-\ell_{2}\right) z_{3}} e^{i(\ell_{2}+p_2 -q -p)z_{2}} \left\langle P_{A-1} \left| A^{+c}(z_3)A^{+d}(z_2)\right| P_{A-1}\right\rangle (i)f^{bed}  {\rm Tr}\left[t^c t^b t^e\right] g^{\sigma_2 \rho_2} \left(-\ell^{-}_{2}-\ell^{-}\right)\\
& \times \frac{  {\rm Tr} \left[\gamma^- \gamma^{\mu} \left(\slashed{q}+\slashed{p}'+M\right) \gamma^{-}\left( \slashed{\ell}_{2} + \slashed{p}_2 +  M\right) \gamma^{\alpha_1} \left(\slashed{p}_2 +  M\right) \gamma^{\alpha_2} \left( \slashed{q} + \slashed{p}+M\right)\gamma^{\nu} \right]
}{  \left[\left(q+p'\right)^{2}-M^2-i\epsilon\right] \left[\left(\ell_2+p_2\right)^2  - M^2 - i\epsilon\right]   \left[\left(q+p\right)^2 -M^2 + i\epsilon\right] \left[\left(q+p-p_2\right)^2 + i\epsilon\right]  } \\
& \times d^{\left(\ell_2\right)}_{\sigma_{2} \alpha_{1}}  d^{(q+p-p_2)}_{\alpha_{2} \rho_{2}} (2 \pi) \delta\left(\ell^{2}_2\right) (2 \pi)  \delta\left(p^{2}_{2}-M^{2}\right).
\end{split}
\label{eq:kernel1_f_rcut_wi}
\end{equation}
Equation~(\ref{eq:kernel1_f_rcut_wi}) has singularity arising from the denominator of the quark propagator with momentum $p_{1}$, $\ell$ and $p'_{1}$. We identify one pole for momentum variable $p'^+$ and two poles for $p^+$.
The contour integration for momentum $p^+$ is given as
\begin{equation}
\begin{split}
C_{1} & = \oint \frac{dp^{+}}{(2\pi)} \frac{e^{ip^{+}(x^{-}_{1}-z^{-}_{2})}}{\left[\left(q+p\right)^{2} - M^{2} + i \epsilon\right]\left[\left(q+p-p_{2}\right)^{2}+ i\epsilon\right]} \\
      & = \frac{(2\pi i)}{2\pi} \frac{\theta\left(x^{-}_{1} - z^{-}_{2}\right)}{4q^{-}\left(q^{-}-p^{-}_{2}\right)}  e^{i\left[-q^{+} + \frac{M^2}{2q^{-}}\right]\left(x^{-}_{1}-z^{-}_{2}\right)}\left[ \frac{ -1 + e^{i \mathcal{G}^{(p_2)}_M\left(x^{-}_{1} - z^{-}_{2}\right)} }{ \mathcal{G}^{(p_2)}_M}  \right]. 
      \end{split}
\end{equation}
Similarly, the contour integration for $p'^+$ can be done
\begin{equation}
\begin{split}
C_{2} = \oint \frac{dp'^+}{(2\pi)} \frac{e^{-ip'^+(x^{-}_{2}-z^-_3)} }{\left[\left(q+p'\right)^2 -M^2 - i\epsilon\right]} = \frac{(-2\pi i)}{2\pi}\frac{\theta\left(x^{-}_{2} - z^{-}_{3}\right)}{2q^{-}}  e^{i\left(q^{+}-\frac{M^2}{2q^-}\right)\left(x^{-}_{2}-z^{-}_{3}\right)} .
\end{split}
%\label{eq:}
\end{equation}
The trace in the numerator of the third line of Eq.~(\ref{eq:kernel1_f_rcut_wi}) gives
\begin{equation}
    \begin{split}
       & {\rm Tr} \left[\gamma^- \gamma^{\mu} \left(\slashed{q}+\slashed{p}'+M\right) \gamma^{-}\left( \slashed{\ell}_{2} + \slashed{p}_2 +  M\right) \gamma^{\alpha_1} \left(\slashed{p}_2 +  M\right) \gamma^{\alpha_2} \left( \slashed{q} + \slashed{p}+M\right)\gamma^{\nu} \right] d^{\left(\ell_2\right)}_{\sigma_{2} \alpha_{1}}  d^{(q+p-p_2)}_{\alpha_{2} \rho_{2}} g^{\sigma_{2}\rho_{2}} \\
       & = \frac{16[-g^{\mu\nu}_{\perp\perp}]\left(q^-\right)^2 \left[1+\left(1-y\right)^2\right] (1-y) }{(-1+\eta)y^2[1-y+\eta y]^2}  \left[\left(\pmb{\ell}_{2\perp}-\pmb{k}_{\perp}\right)\cdot\left(\pmb{\ell}_{2\perp}-y\pmb{k}_{\perp}\right) + \kappa y^4 M^2 \right].
    \end{split}
\end{equation}
The final expression of the hadronic tensor for the right-cut diagram [Fig.~\ref{fig:kernel1_f_g}(a)] is given as
\begin{equation}
\begin{split}
W^{\mu\nu}_{1,r_1} & =  \frac{C_A}{2} \sum_f 2 [-g^{\mu\nu}_{\perp \perp }] e^2_f g^4_s \int d (\Delta x^{-})   e^{i\left( q^{+}-\frac{M^2}{2q^-}\right)(\Delta x^{-} )} 
  \left\langle P \left| \bar{\psi}_{_f}( \Delta x^{-}) \frac{\gamma^{+}}{4} \psi_{_f}(0)\right| P\right\rangle  \\
  & \times   \int  d \zeta^{-} d (\Delta z^{-}) d^2 \Delta z_{\perp} \frac{dy}{2\pi} \frac{d^2 \ell_{2\perp}}{(2\pi)^2} \frac{d^2 k_{\perp}}{(2\pi)^2}  e^{-i (\Delta z^{-})\mathcal{H}^{(\ell_2,p_2)}_{M} }     e^{i \pmb{k}_{\perp}\cdot \Delta\pmb{z}_{\perp}} \theta( x^{-}_{1}-z^{-}_2) \theta( x^{-}_2 - z^{-}_3)  \\
  & \times  \left[ \frac{ 1 + (1-y)^2 }{y}\right] \left[\frac{1-\frac{\eta}{2}}{1-\eta}\right]\frac{\left[  \left(\pmb{\ell}_{2\perp} - \pmb{k}_{\perp}\right)\cdot\left(\pmb{\ell}_{2\perp} - y\pmb{k}_{\perp}\right)   + \kappa y^4 M^2 \right]}{\left[  \left(\pmb{\ell}_{2\perp} - \pmb{k}_{\perp}\right)^2 + y^2(1-\eta)^2 M^2\right] J_1}  \left[ -1 + e^{i\mathcal{G}^{(p_2)}_{M}(x^{-}_1-z^-_2)} \right]  \\
  & \times \left\langle P_{A-1} \left|{\rm Tr}\left[A^+\left(\zeta^-, \Delta z^-, \Delta z_{\perp}\right) A^+\left(\zeta^-, 0\right)\right] \right| P_{A-1} \right\rangle,
\end{split} 
\end{equation}
where $\mathcal{G}^{(p_2)}_M$ is defined in Eq.~(\ref{eq:GMP2_append}), $\mathcal{H}^{(\ell_2,p_2)}_{M}$ defined in Eq.~(\ref{eq:HL2P2M_append}), and $J_1$ is defined in Eq.~(\ref{eq:J1_append}).

Furthermore, the left-cut diagram is shown in Fig.~\ref{fig:kernel1_f_g}(a). Its hadronic tensor can be written as
\begin{equation}
\begin{split}
W^{\mu\nu}_{1,\ell_1} & = \sum_{f} e^2_f g^4_{s} \int d^4 x_1 d^4 x_2 d^4 z_{2} d^4 z_{3} \int \frac{d^4 \ell}{(2\pi)^4} \frac{d^4 p'}{(2\pi)^4}  \frac{d^4 \ell_{2}}{(2\pi)^4} \frac{d^4 p_{2}}{(2\pi)^4} e^{-ip'x_{2}} e^{i(\ell+p_2-q)x_{1}}\left\langle P \left| \bar{\psi}_{_f}(x_2) \frac{\gamma^{+}}{4} \psi_{_f}(x_{1})\right| P\right\rangle  \\
  & \times e^{i\left(q+p'-p_{2}-\ell_{2}\right) z_{3}} e^{i(\ell_{2}-\ell)z_{2}} \left\langle P_{A-1} \left| A^{+c}(z_3)A^{+d}(z_2)\right| P_{A-1}\right\rangle (i)f^{bed}  {\rm Tr}\left[t^c t^b t^e\right] g^{\sigma_2 \rho_2} \left(-\ell^{-}_{2}-\ell^{-}\right)\\
& \times \frac{  {\rm Tr} \left[\gamma^- \gamma^{\mu} \left(\slashed{q}+\slashed{p}'+M\right) \gamma^{-}\left( \slashed{\ell}_{2} + \slashed{p}_2 +  M\right) \gamma^{\alpha_1} \left(\slashed{p}_2 +  M\right) \gamma^{\alpha_2} \left( \slashed{\ell} + \slashed{p}_{2}+M\right)\gamma^{\nu} \right]
}{  \left[\left(q+p'\right)^{2}-M^2-i\epsilon\right] \left[\left(\ell_2+p_2\right)^2  - M^2 - i\epsilon\right]   \left[\ell^2_{2}  - i\epsilon\right] \left[(\ell+p_2)^2 -M^2+ i\epsilon\right]  } \\
& \times d^{\left(\ell_2\right)}_{\sigma_{2} \alpha_{1}}  d^{(\ell)}_{\alpha_{2} \rho_{2}} (2 \pi) \delta\left(\ell^{2}\right) (2 \pi)  \delta\left(p^{2}_{2}-M^{2}\right).
\end{split}
\label{eq:kernel1_f_leftcut_wi}
\end{equation}
Equation~(\ref{eq:kernel1_f_leftcut_wi}) has singularity arising from the denominator of the quark propagator with momentum $\ell'$, $\ell_2$ and $p'_{1}$. We identify one pole for momentum variable $p'^+$ and two poles for $\ell^{+}_2$.
The contour integration for momentum $\ell^{+}_2$ is given as
\begin{equation}
\begin{split}
C_{1} & = \oint \frac{d\ell^{+}_2}{(2\pi)} \frac{e^{-i\ell^{+}_2(z^{-}_{3}-z^{-}_{2})}}{\left[\left(\ell_2+p_2\right)^{2} - M^{2} - i \epsilon\right]\left[ \ell^{2}_{2} - i\epsilon\right]} \\
      & = \frac{(-2\pi i)}{2\pi} \frac{\theta\left(z^{-}_{3} - z^{-}_{2}\right)}{4\ell^{-}_2\left(\ell^{-}_{2}+p^{-}_{2}\right)}  e^{-i\left[ \frac{\pmb{\ell}^2_{2\perp}}{2\ell^{-}_2}\right]\left(z^{-}_{3}-z^{-}_{2}\right)}\left[ \frac{ 1 - e^{i \mathcal{G}^{(\ell,k)}_1\left(z^{-}_{3} - z^{-}_{2}\right)} }{ \mathcal{G}^{(\ell,k)}_1}  \right],
      \end{split}
\end{equation}
where $\mathcal{G}^{(\ell,k)}_1$ is defined in Eq.~(\ref{eq:G1ML_k_etay_append}). Similarly, the contour integration for $p'^+$ can be done
\begin{equation}
\begin{split}
C_{2} = \oint \frac{dp'^+}{(2\pi)} \frac{e^{-ip'^+(x^{-}_{2}-z^-_3)} }{\left[\left(q+p'\right)^2 -M^2 - i\epsilon\right]} = \frac{(-2\pi i)}{2\pi}\frac{\theta\left(x^{-}_{2} - z^{-}_{3}\right)}{2q^{-}}  e^{i\left(q^{+}-\frac{M^2}{2q^-}\right)\left(x^{-}_{2}-z^{-}_{3}\right)} .
\end{split}
%\label{eq:}
\end{equation}
The trace in the numerator of the third line of Eq.~(\ref{eq:kernel1_f_leftcut_wi}) gives
\begin{equation}
    \begin{split}
       & {\rm Tr} \left[\gamma^- \gamma^{\mu} \left(\slashed{q}+\slashed{p}'+M\right) \gamma^{-}\left( \slashed{\ell}_{2} + \slashed{p}_2 +  M\right) \gamma^{\alpha_1} \left(\slashed{p}_2 +  M\right) \gamma^{\alpha_2} \left( \slashed{\ell} + \slashed{p}_{2}+M\right)\gamma^{\nu} \right] d^{\left(\ell_2\right)}_{\sigma_{2} \alpha_{1}}  d^{(\ell)}_{\alpha_{2} \rho_{2}} g^{\sigma_{2}\rho_{2}} \\
       & = \frac{16[-g^{\mu\nu}_{\perp\perp}]\left(q^-\right)^2\left[1+\left(1-y\right)^2\right] (1-y) }{(-1+\eta)y^2[1-y+\eta y]^2} \left[\pmb{\ell}_{\perp}\cdot\left(\pmb{\ell}_{\perp}+ (1-y)\pmb{k}_{\perp}\right) + \kappa y^4 M^2 \right].
    \end{split}
\end{equation}
The final expression of the hadronic tensor for the left-cut diagram [Fig.~\ref{fig:kernel1_f_g}(a)] is given as
\begin{equation}
\begin{split}
    W^{\mu\nu}_{1,\ell_1} & = - \frac{C_A}{2} \sum_f 2 [-g^{\mu\nu}_{\perp \perp }] e^2_f g^4_s \int d (\Delta x^{-})   e^{i\left( q^{+}-\frac{M^2}{2q^-}\right)(\Delta x^{-} )} 
  \left\langle P \left| \bar{\psi}_{_f}( \Delta x^{-}) \frac{\gamma^{+}}{4} \psi_{_f}(0)\right| P\right\rangle  \\
  & \times n  \int  d \zeta^{-} d (\Delta z^{-}) d^2 \Delta z_{\perp} \frac{dy}{2\pi} \frac{d^2 \ell_{\perp}}{(2\pi)^2} \frac{d^2 k_{\perp}}{(2\pi)^2}  e^{-i (\Delta z^{-})\mathcal{H}^{(\ell,p_2)}_{2} }     e^{i \pmb{k}_{\perp}\cdot \Delta\pmb{z}_{\perp}} \theta\left(z^{-}_{3}-z^{-}_2\right) \theta\left(x^{-}_2 - z^{-}_3\right)  \\
  & \times \left[ \frac{ 1 + \left(1-y\right)^2 }{y}\right]\left[1+\frac{\eta}{2}\right] \frac{\left[  \pmb{\ell}_{\perp}\cdot\left(\pmb{\ell}_{\perp} +\left(1-y\right)\pmb{k}_{\perp}\right) + \kappa y^4 M^2 \right]}{\left[  \pmb{\ell}^2_{\perp}  + M^2y^2 \right] \left[ \left\{\pmb{\ell}_{\perp}\left(1+\eta y\right)+\left(1-y\right)\pmb{k}_{\perp}\right\}^2 + y^2M^2 \right]}  \\
  &\times \left[ 1 - e^{i\mathcal{G}^{(\ell,k)}_{1}(z^{-}_{3}-z^-_{2})} \right]  e^{i\mathcal{G}^{(\ell)}_{M}(x^{-}_{1}-z^{-}_{2})} \left\langle P_{A-1} \left|{\rm Tr}\left[A^+(\zeta^-, \Delta z^-, \Delta z_{\perp}) A^+(\zeta^-, 0)\right] \right| P_{A-1} \right\rangle,
\end{split} 
\label{eq:K1_wmnunu_fg_a_leftcut_final}
\end{equation}
where $\mathcal{G}^{(\ell,k)}_1$ is defined in Eq.~(\ref{eq:G1ML_k_etay_append}), $\mathcal{H}^{(\ell,p_2)}_{2}$ defined in Eq.~(\ref{eq:H2L2P2_append}). The factor $n$ in Eq.~(\ref{eq:K1_wmnunu_fg_a_leftcut_final}) is described in Eq.~(\ref{eq:factor_n}).

Following this, we consider the diagram shown in the right panel of Fig.~\ref{fig:kernel1_f_g}. The topology of the diagram is the same as the diagram on the left panel. Moreover, they are complex conjugate of each other. The final expression of the hadronic tensor for the left-cut diagram shown in Fig.~\ref{fig:kernel1_f_g}(b) can be written as
\begin{equation}
\begin{split}
    W^{\mu\nu}_{1,\ell_2} & =  \frac{C_A}{2} \sum_f 2 [-g^{\mu\nu}_{\perp \perp }] e^2_f g^4_s \int d (\Delta x^{-})   e^{i\left( q^{+}-\frac{M^2}{2q^-}\right)(\Delta x^{-} )} 
  \left\langle P \left| \bar{\psi}_{_f}( \Delta x^{-}) \frac{\gamma^{+}}{4} \psi_{_f}(0)\right| P\right\rangle  \\
  & \times   \int  d \zeta^{-} d (\Delta z^{-}) d^2 \Delta z_{\perp} \frac{dy}{2\pi} \frac{d^2 \ell_{2\perp}}{(2\pi)^2} \frac{d^2 k_{\perp}}{(2\pi)^2} e^{-i (\Delta z^{-})\mathcal{H}^{(\ell_2,p_2)}_{M} }     e^{i \pmb{k}_{\perp}\cdot \Delta\pmb{z}_{\perp}}  \theta( x^{-}_{1}-z^{-}_2) \theta( x^{-}_2 - z^{-}_3)  \\
  & \times \left[ \frac{ 1 + (1-y)^2 }{y}\right] \left[\frac{1-\frac{\eta}{2}}{1-\eta}\right]\frac{ \left[  \left(\pmb{\ell}_{2\perp} - \pmb{k}_{\perp}\right)\cdot\left(\pmb{\ell}_{2\perp} - y\pmb{k}_{\perp}\right) + \kappa y^4 M^2 \right]}{\left[\left(\pmb{\ell}_{2\perp} - \pmb{k}_{\perp}\right)^2 + M^2y^2(1-\eta)^2 \right] J_1} \left[ -1 + e^{-i\mathcal{G}^{(p_2)}_{M}(x^{-}_2-z^-_3)} \right]\\
  & \times \left\langle P_{A-1} |{\rm Tr}[A^+(\zeta^-, \Delta z^-, \Delta z_{\perp}) A^+(\zeta^-, 0)] | P_{A-1} \right\rangle,
\end{split} 
\end{equation}
where $\mathcal{G}^{(p_2)}_M$ is defined in Eq.~(\ref{eq:GMP2_append}), $\mathcal{H}^{(\ell_2,p_2)}_{M}$ defined in Eq.~(\ref{eq:HL2P2M_append}), and $J_1$ is defined in Eq.~(\ref{eq:J1_append}).

Similarly, the hadronic tensor for the right-cut diagram in Fig.~\ref{fig:kernel1_f_g}(b) is given as
\begin{equation}
\begin{split}
    W^{\mu\nu}_{1,r_2} & = - \frac{C_A}{2} \sum_f 2 [-g^{\mu\nu}_{\perp \perp }] e^2_f g^4_s \int d (\Delta x^{-})   e^{i\left( q^{+}-\frac{M^2}{2q^{-}}\right)(\Delta x^{-} )} 
  \left\langle P \left| \bar{\psi}_{_f}( \Delta x^{-}) \frac{\gamma^{+}}{4} \psi_{_f}(0)\right| P\right\rangle  \\
  & \times n  \int  d \zeta^{-} d (\Delta z^{-}) d^2 \Delta z_{\perp} \frac{dy}{2\pi} \frac{d^2 \ell'_{\perp}}{(2\pi)^2} \frac{d^2 k_{\perp}}{(2\pi)^2} e^{-i (\Delta z^{-})\mathcal{H}^{(\ell',p_2)}_{2} } e^{i \pmb{k}_{\perp}\cdot \Delta\pmb{z}_{\perp}} \theta( -z^{-}_{3}+z^{-}_2) \theta( x^{-}_1 - z^{-}_2)\\
  & \times \left[ \frac{ 1 + (1-y)^2 }{y}\right]\left[1+\frac{\eta}{2}\right] \frac{\left[  \pmb{\ell}'_{\perp}\cdot\left(\pmb{\ell}'_{\perp} +(1-y)\pmb{k}_{\perp}\right)   + \kappa y^4 M^2 \right]}{\left[  \pmb{\ell}'^2_{\perp}  + y^2 M^2 \right] \left[ \left\{\pmb{\ell}'_{\perp}(1+\eta y)+(1-y)\pmb{k}_{\perp}\right\}^2 + y^2M^2 \right]} \\
  &\times  \left[ 1 - e^{i\mathcal{G}^{(\ell',k)}_{1}(z^{-}_{3}-z^-_{2})} \right] e^{-i\mathcal{G}^{(\ell')}_{M}(x^{-}_{2}-z^{-}_{3})}  \left\langle P_{A-1} \left|{\rm Tr}\left[A^+(\zeta^-, \Delta z^-, \Delta z_{\perp}) A^+(\zeta^-, 0)\right] \right| P_{A-1} \right\rangle,
\end{split} 
\label{eq:K1_wmnunu_fg_b_rightcut_final}
\end{equation}
where $\mathcal{G}^{(\ell',k)}_1$ is defined in Eq.~(\ref{eq:G1ML_k_etay_append}), $\mathcal{H}^{(\ell',p_2)}_{2}$ defined in Eq.~(\ref{eq:H2L2P2_append}). The factor $n$ in Eq.~(\ref{eq:K1_wmnunu_fg_b_rightcut_final}) is described in Eq.~(\ref{eq:factor_n}).

Note that our definition of the length integration variable is $\zeta^{-}=x^{-}_{1}-z^{-}_{2}=x^{-}_{2}-z^{-}_{3}$ which becomes $\zeta^{-}=-z^{-}_{2}=-z^{-}_{3}$ when $x^{-}_{1}$ and $x^{-}_{2}$ are initialized as the origin. This leads the term $\left[ 1 - e^{i\mathcal{G}^{(\ell,k)}_{1}(z^{-}_3-z^-_2)} \right]$ in Eq.~(\ref{eq:K1_wmnunu_fg_a_leftcut_final}) and Eq.~(\ref{eq:K1_wmnunu_fg_b_rightcut_final}) to vanish.
 Therefore, the left-cut diagram [Fig.~\ref{fig:kernel1_f_g}(a)] and right-cut diagram [Fig.~\ref{fig:kernel1_f_g}(b)] do not contribute to the energy loss kernel. This implies the hadronic tensor for the left-cut diagram [Fig.~\ref{fig:kernel1_f_g}(a)] and right-cut diagram [Fig.~\ref{fig:kernel1_f_g}(b)] is 
\begin{eqnarray}
    W^{\mu\nu}_{1,\ell_1} = W^{\mu\nu}_{1,r_2} =0.
    \label{eq:K1_fg_vanishing}
\end{eqnarray}
However, the right-cut diagram [Fig.~\ref{fig:kernel1_f_g}(a)] and left-cut diagram [Fig.~\ref{fig:kernel1_f_g}(b)] have nonvanishing contributions, and their total hadronic tensor is given by
\begin{equation}
\begin{split}
    W^{\mu\nu}_{1,r_1+\ell_2}  =&  -\frac{C_A}{2} \sum_f 2 [-g^{\mu\nu}_{\perp \perp }] e^2_f g^4_s \int d (\Delta x^{-})   e^{i\left( q^{+}-\frac{M^2}{2q^-}\right)(\Delta x^{-} )} 
  \left\langle P \left| \bar{\psi}_{_f}( \Delta x^{-}) \frac{\gamma^{+}}{4} \psi_{_f}(0)\right| P\right\rangle  \\
  & \times   \int  d \zeta^{-} d (\Delta z^{-}) d^2 \Delta z_{\perp} \frac{dy}{2\pi} \frac{d^2 \ell_{2\perp}}{(2\pi)^2} \frac{d^2 k_{\perp}}{(2\pi)^2} e^{-i (\Delta z^{-})\mathcal{H}^{(\ell_2,p_2)}_{M} }     e^{i \pmb{k}_{\perp}\cdot \Delta\pmb{z}_{\perp}}  \theta(\zeta^-)  \\
  & \times \left[ \frac{ 1 + (1-y)^2 }{y}\right] \left[\frac{1-\frac{\eta}{2}}{1-\eta}\right]\frac{ \left[  \left(\pmb{\ell}_{2\perp} - \pmb{k}_{\perp}\right)\cdot\left(\pmb{\ell}_{2\perp} - y\pmb{k}_{\perp}\right) + \kappa y^4 M^2 \right]}{\left[\left(\pmb{\ell}_{2\perp} - \pmb{k}_{\perp}\right)^2 + y^2(1-\eta)^2M^2 \right] J_1} \left[ 2 -2\cos\left\{\mathcal{G}^{(p_2)}_{M}\zeta^{-}\right\} \right]\\
  & \times \left\langle P_{A-1} |{\rm Tr}[A^+(\zeta^-, \Delta z^-, \Delta z_{\perp}) A^+(\zeta^-, 0)] | P_{A-1} \right\rangle.
\end{split} 
\label{eq:K1_fg_non-vanishing}
\end{equation}
%
%%%%%%%%%%%%%%%%%%%%%%%%%%%%%%%%%%%%%%%%%%%%%%%%%%%%%%%%%%%%%%%%%%%%%%%%%%%%%%%%%%%%%
%%%%%%%%%%%%%%%%%%%%%%%%%%%%%%%%%%%%%%%%%%%%%%%%%%%%%%%%%%%%%%%%%%%%%%%%%%%%%%%%%%%%%
\subsection{Figures~\ref{fig:kernel-1_all}(h) and \ref{fig:kernel-1_all}(i)}
%%%%%%%%%%%%%%%%%%%%%%%%%%%%%%%%%%%%%%%%%%%%%%%%%%%%%%%%%%%%%%%%%%%%%%%%%%%%%%%%%%%%%
%%%%%%%%%%%%%%%%%%%%%%%%%%%%%%%%%%%%%%%%%%%%%%%%%%%%%%%%%%%%%%%%%%%%%%%%%%%%%%%%%%%%%
\begin{figure}[!h]
    \centering 
    \begin{subfigure}[t]{0.495\textwidth}
        \centering        \includegraphics[height=1.6in]{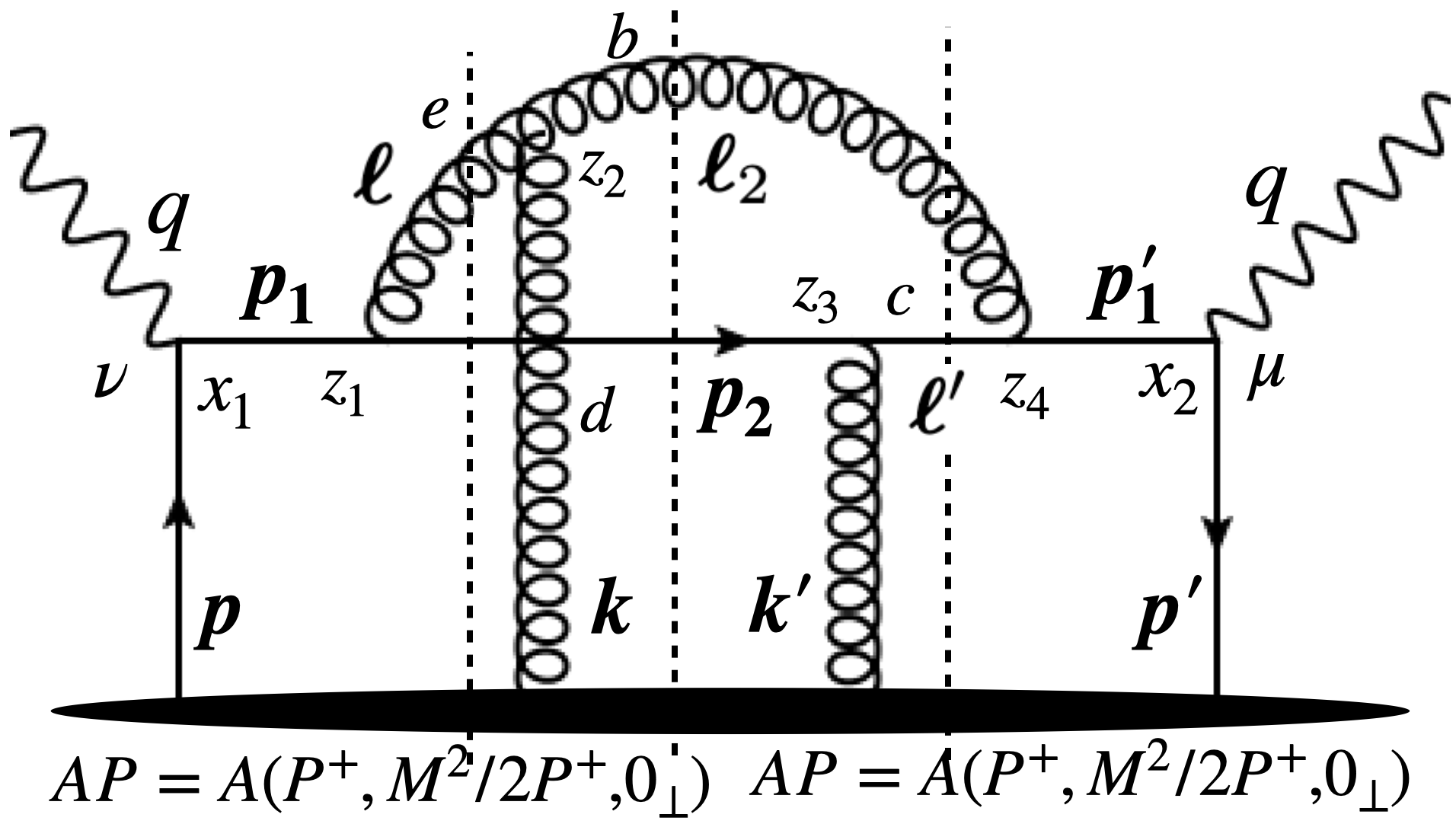}
        \caption{ }
    \end{subfigure}%
    \begin{subfigure}[t]{0.495\textwidth}
        \centering        \includegraphics[height=1.6in]{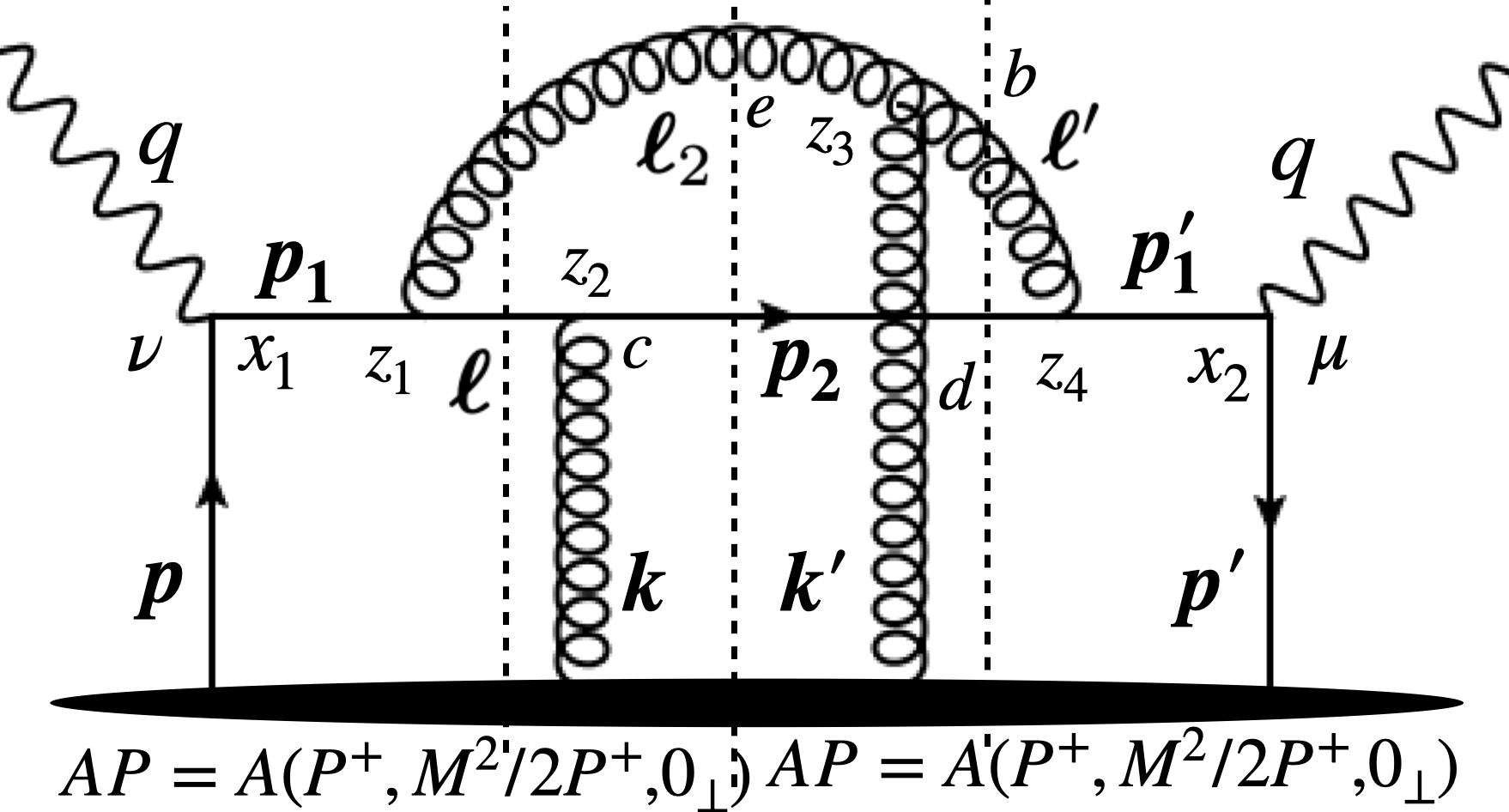}
        \caption{}
    \end{subfigure}
\caption{A forward scattering diagram contributing to kernel-1.}
\label{fig:kernel-1_de_both}
\end{figure}
The forward scattering diagrams shown in Fig.~\ref{fig:kernel-1_de_both} depicts the variables needed to perform the calcualtion. The center-cut [Fig.~\ref{fig:kernel-1_de_both}(a)] leads to a diagram with gluon-gluon scattering on the amplitude side and postemission quark scattering with the medium on the complex conjugate side. The hadronic tensor for the center-cut diagram [Fig.~\ref{fig:kernel-1_de_both}(a)] is given as
\begin{equation}
\begin{split}
W^{\mu\nu}_{1,c_1} & = \sum_{f} e^2_f g^4_{s} \int d^4 x_1 d^4 x_2 d^4 z_{2} d^4 z_{3} \int \frac{d^4 p}{(2\pi)^4} \frac{d^4 p'}{(2\pi)^4}  \frac{d^4 \ell_{2}}{(2\pi)^4} \frac{d^4 p_{2}}{(2\pi)^4} e^{-ip'x_{2}} e^{ipx_{1}}\left\langle P \left| \bar{\psi}_{_f}(x_2) \frac{\gamma^{+}}{4} \psi_{_f}(x_{1})\right| P\right\rangle  \\
  & \times e^{i\left(q+p'-p_{2}-\ell_{2}\right) z_{3}} e^{i(\ell_{2}+p_2 -q -p)z_{2}} \left\langle P_{A-1} \left| A^{+c}(z_3)A^{+d}(z_2)\right| P_{A-1}\right\rangle (-i)f^{bed}  {\rm Tr}\left[t^b t^c t^e\right] g^{\sigma_2 \rho_2} \left(-\ell^{-}_{2}-\ell^{-}\right)\\
& \times \frac{  {\rm Tr} \left[\gamma^- \gamma^{\mu} \left(\slashed{q}+\slashed{p}'+M\right) \gamma^{\alpha_1}\left( \slashed{q} + \slashed{p}' -\slashed{\ell}_2 +  M\right) \gamma^{-} \left(\slashed{p}_2 +  M\right) \gamma^{\alpha_2} \left( \slashed{q} + \slashed{p}+M\right)\gamma^{\nu} \right]
}{  \left[\left(q+p'\right)^{2}-M^2-i\epsilon\right] \left[\left(q+p'-\ell_2\right)^2  - M^2 - i\epsilon\right]   \left[\left(q+p\right)^2 -M^2 + i\epsilon\right] \left[\left(q+p-p_2\right)^2 + i\epsilon\right]  } \\
& \times d^{\left(\ell_2\right)}_{\sigma_{2} \alpha_{1}}  d^{(q+p-p_2)}_{\alpha_{2} \rho_{2}} (2 \pi) \delta\left(\ell^{2}_2\right) (2 \pi)  \delta\left(p^{2}_{2}-M^{2}\right).
\end{split}
\label{eq:kernel1_d_centercut_wi}
\end{equation}
The above expression of the hadonic tensor has singularity when the denominator of the propagator for $p_1$, $\ell$, $\ell'$ and $p'_1$ becomes on-shell. It contains two simple poles for $p^+$ and $p'^+$. The contour integration for $p^+$ gives
\begin{equation}
\begin{split}
C_{1} & = \oint \frac{dp^{+}}{(2\pi)} \frac{e^{ip^{+}(x^{-}_{1}-z^{-}_{2})}}{\left[\left(q+p\right)^{2} - M^{2} + i\epsilon\right]\left[\left(q+p-p_{2}\right)^{2} + i\epsilon\right]} \\
      & = \frac{(2\pi i)}{2\pi} \frac{\theta\left(x^{-}_{1} - z^{-}_{2}\right)}{4q^{-}(q^{-}-p^{-}_{2})}  e^{i\left[-q^{+} + \frac{M^2}{2q^{-}}\right]\left(x^{-}_{1}-z^{-}_{2}\right)}\left[ \frac{ -1 + e^{i \mathcal{G}^{(p_2)}_{M}\left(x^{-}_{1}-z^{-}_{2}\right)} }{ \mathcal{G}^{(p_2)}_{M}}  \right],
\end{split}
%\label{eq:}
\end{equation}
where $\mathcal{G}^{(p_2)}_{M}$ is defined in Eq.~(\ref{eq:GMP2_append}). Similarly, the contour integration for $p'^+$ can be done
\begin{equation}
\begin{split}
C_{2} & = \oint \frac{dp'^{+}}{(2\pi)} \frac{e^{-ip'^{+}\left(x^{-}_{2}-z^{-}_{3}\right)}}{\left[\left(q+p'\right)^{2} - M^{2} - i \epsilon\right]\left[\left(q+p'-\ell_{2}\right)^{2} -M^2 - i\epsilon \right]} \\
      & = \frac{(-2\pi i)}{2\pi} \frac{\theta\left(x^{-}_{2} - z^{-}_{3}\right)}{4q^{-}(q^{-}-\ell^{-}_{2})} e^{-i\left(-q^{+} + \frac{M^2}{2q^-}\right)\left(x^{-}_{2}-z^-_3\right)}\left[\frac{ -1 + e^{-i\mathcal{G}^{(\ell_2)}_M\left(x^{-}_{2}-z^-_3\right)}}{\mathcal{G}^{(\ell_2)}_M}\right],
\end{split}
\end{equation}
where $\mathcal{G}^{(\ell_2)}_{M}$ is defined in Eq.~(\ref{eq:GL2M_append}). The trace in the numerator of the third line of Eq.~(\ref{eq:kernel1_d_centercut_wi}) gives
\begin{equation}
    \begin{split}
       & {\rm Tr} \left[\gamma^- \gamma^{\mu} \left(\slashed{q}+\slashed{p}'+M\right) \gamma^{\alpha_1}\left( \slashed{q} + \slashed{p}' -\slashed{\ell}_2 +  M\right) \gamma^{-} \left(\slashed{p}_2 +  M\right) \gamma^{\alpha_2} \left( \slashed{q} + \slashed{p}+M\right)\gamma^{\nu} \right] d^{\left(\ell_2\right)}_{\sigma_{2} \alpha_{1}}  d^{(q+p-p_2)}_{\alpha_{2} \rho_{2}} g^{\sigma_2 \rho_2} \\
       & = -16[-g^{\mu\nu}_{\perp\perp}](1-y)^2\left(q^-\right)^2\left[\frac{1+\left(1-y\right)^2}{(1-\eta)y^2\left(1-y+\eta y\right)^2}\right] \left[\left(\pmb{\ell}_{2\perp}-\pmb{k}_{\perp}\right)\cdot\pmb{\ell}_{2\perp} + \kappa y^4 M^2 \right].
    \end{split}
\end{equation}
The final expression of the hadronic tensor for center-cut diagram [Fig.~\ref{fig:kernel-1_de_both}(a)] becomes
\begin{eqnarray}
\begin{split}
W^{\mu\nu}_{1,c_1} & = -\frac{C_{A}}{2} \sum_f 2 \left[-g^{\mu\nu}_{\perp\perp}\right]  e^2_f g^4_{s} \int d (\Delta x^{-}) e^{i\Delta x^{-}\left(q^{+}-\frac{M^2}{2q^-}\right)} \left\langle P \left| \bar{\psi}_{_f}(\Delta x^-) \frac{\gamma^{+}}{4} \psi_{_f}(0)\right| P\right\rangle\\
  & \times \int d \zeta^-  d (\Delta z^{-})d^2 \Delta z_{\perp} \frac{dy}{2\pi}\frac{d^2 \ell_{2\perp}}{(2\pi)^2} \frac{d^2 k_{\perp}}{(2\pi)^2} 
  e^{-i\Delta z^{-}\mathcal{H}^{(\ell_2,p_2)}_M} e^{i\pmb{k}_{\perp}\cdot \Delta\pmb{z}_{\perp} } \theta\left(x^{-}_{1}-z^{-}_{2}\right) \theta\left(x^{-}_{2}-z^{-}_{3}\right)\\ 
  & \times \left[\frac{1+ \left(1-y\right)^2}{y}\right] \left[ \frac{\left(1-\frac{\eta}{2}\right)}{\left(1-\eta\right)}\right]  \frac{\left[\left(\pmb{\ell}_{2\perp} - \pmb{k}_{\perp}\right)\cdot\pmb{\ell}_{2\perp} + \kappa y^4 M^2\right]}{\left[\pmb{\ell}^2_{2\perp}  + y^2 M^2 \right] \left[\left(\pmb{\ell}_{2\perp} - \pmb{k}_{\perp}\right)^2  + y^2\left(1-\eta\right)^2  M^2 \right]}  \\
  & \times \left[ -1 + e^{i \mathcal{G}^{(p_2)}_{M}\left(x^{-}_{1}-z^{-}_{2}\right)}\right]\left[ -1 + e^{-i \mathcal{G}^{(\ell_2)}_{M}\left(x^{-}_{2}-z^{-}_{3}\right)}\right] \left\langle P_{A-1} \left\vert {\rm Tr}\left[A^{+}\left(\zeta^-, \Delta z^-, \Delta z_{\perp}\right)A^+\left(\zeta^-,0\right)\right]\right\vert P_{A-1}\right\rangle.
\end{split}
\label{eq:K1_d_centercut_final}
\end{eqnarray}
Note that the topology of the center-cut diagram shown in Fig.~\ref{fig:kernel-1_de_both}(b) is the same as the center-cut diagram on the left panel Fig.~\ref{fig:kernel-1_de_both}(a). Moreover, they are complex conjugate of each other. The final expression of the hadronic tensor for the center-cut diagram shown in Fig.~\ref{fig:kernel-1_de_both}(b) can be written as
\begin{eqnarray}
\begin{split}
W^{\mu\nu}_{1,c_2} & = -\frac{C_{A}}{2} \sum_f 2 \left[-g^{\mu\nu}_{\perp\perp}\right]  e^2_f g^4_{s} \int d (\Delta x^{-}) e^{i\Delta x^{-}\left(q^{+}-\frac{M^2}{2q^-}\right)} \left\langle P \left| \bar{\psi}_{_f}(\Delta x^-) \frac{\gamma^{+}}{4} \psi_{_f}(0)\right| P\right\rangle\\
  & \times \int d \zeta^-  d (\Delta z^{-})d^2 \Delta z_{\perp} \frac{dy}{2\pi}\frac{d^2 \ell_{2\perp}}{(2\pi)^2} \frac{d^2 k_{\perp}}{(2\pi)^2} 
  e^{-i\Delta z^{-}\mathcal{H}^{(\ell_2,p_2)}_M} e^{i\pmb{k}_{\perp}\cdot \Delta\pmb{z}_{\perp} } \theta\left(x^{-}_{1}-z^{-}_{2}\right) \theta\left(x^{-}_{2}-z^{-}_{3}\right) \\ 
  & \times \left[\frac{1+ \left(1-y\right)^2}{y}\right] \left[ \frac{\left(1-\frac{\eta}{2}\right)}{\left(1-\eta\right)}\right]  \frac{ \left[\left(\pmb{\ell}_{2\perp} - \pmb{k}_{\perp}\right)\cdot\pmb{\ell}_{2\perp} + \kappa y^4 M^2\right]}{\left[\pmb{\ell}^2_{2\perp}  + y^2 M^2 \right] \left[\left(\pmb{\ell}_{2\perp} - \pmb{k}_{\perp}\right)^2  + y^2\left(1-\eta\right)^2  M^2 \right]} \\
  & \times \left[ -1 + e^{i \mathcal{G}^{(\ell_2)}_{M}\left(x^{-}_{1}-z^{-}_{2}\right)}\right]\left[ -1 + e^{-i \mathcal{G}^{(p_2)}_{M}\left(x^{-}_{2}-z^{-}_{3}\right)}\right] \left\langle P_{A-1} \left\vert {\rm Tr}\left[A^{+}\left(\zeta^-, \Delta z^-, \Delta z_{\perp}\right)A^+\left(\zeta^-,0\right)\right]\right\vert P_{A-1}\right\rangle.
\end{split}
\label{eq:K1_e_centercut_final}
\end{eqnarray}

Adding the hadronic tensors for the center-cut diagram in Fig.~\ref{fig:kernel-1_de_both}(a) and Fig.~\ref{fig:kernel-1_de_both}(b), the final hadronic tensor given as
\begin{eqnarray}
\begin{split}
W^{\mu\nu}_{1,c_1+c_2} & = -\frac{C_{A}}{2} \sum_f 2 \left[-g^{\mu\nu}_{\perp\perp}\right]  e^2_f g^4_{s} \int d (\Delta x^{-}) e^{i\Delta x^{-}\left(q^{+}-\frac{M^2}{2q^-}\right)} \left\langle P \left| \bar{\psi}_{_f}(\Delta x^-) \frac{\gamma^{+}}{4} \psi_{_f}(0)\right| P\right\rangle\\
  & \times \int d \zeta^-  d (\Delta z^{-})d^2 \Delta z_{\perp} \frac{dy}{2\pi}\frac{d^2 \ell_{2\perp}}{(2\pi)^2} \frac{d^2 k_{\perp}}{(2\pi)^2} 
  e^{-i\Delta z^{-}\mathcal{H}^{(\ell_2,p_2)}_M} e^{i\pmb{k}_{\perp}\cdot \Delta\pmb{z}_{\perp} } \theta\left(\zeta^-\right) \\ 
  & \times \left[\frac{1+ \left(1-y\right)^2}{y}\right] \left[ \frac{\left(1-\frac{\eta}{2}\right)}{\left(1-\eta\right)}\right]  \frac{ \left[\left(\pmb{\ell}_{2\perp} - \pmb{k}_{\perp}\right)\cdot\pmb{\ell}_{2\perp} + \kappa y^4 M^2\right]}{\left[\pmb{\ell}^2_{2\perp}  + y^2 M^2 \right] \left[\left(\pmb{\ell}_{2\perp} - \pmb{k}_{\perp}\right)^2  + y^2\left(1-\eta\right)^2  M^2 \right]} \\
  & \times \left[2 - 2\cos\left\{\mathcal{G}^{(\ell_2)}_{0}\zeta^-\right\}  -2 \cos\left\{\mathcal{G}^{(p_2)}_{0}\zeta^-\right\} + 2\cos\left\{\left(\mathcal{G}^{(p_2)}_{0}-\mathcal{G}^{(\ell_2)}_{0}\right)\zeta^-\right\}\right] \\
  & \times \left\langle P_{A-1} \left\vert {\rm Tr}\left[A^{+}\left(\zeta^-, \Delta z^-, \Delta z_{\perp}\right)A^+\left(\zeta^-,0\right)\right]\right\vert P_{A-1}\right\rangle.
\end{split}
\label{eq:K1_hi_centercut_final}
\end{eqnarray}
The left-cut diagram of Fig.~\ref{fig:kernel-1_de_both}(a) contains singularity arising from the denominator of the propagators of $\ell_2$, $\ell'$, and $p'_1$. It involves nested integrals over variables $p'^+$ and $\ell'^+$. The integrals are carried out using the method of contour integration as follows
\begin{eqnarray}
C & =& \oint \oint \frac{d\ell'^{+}}{(2\pi)} \frac{dp'^{+}}{(2\pi)} \frac{e^{i\ell'^{+}(z^{-}_{3}-z^{-}_{2})} e^{-ip'^{+}(x^{-}_{2}-z^{-}_{2})} }{\left[\left(q+p'\right)^{2} - M^{2} - i\epsilon\right]\left[\left(q+p'-\ell'\right)^{2}  -i\epsilon\right] \left[\ell'^2 -M^2 -i\epsilon \right]} \nonumber \\
&  =& \oint \oint \frac{d\ell'^{+}}{(2\pi)} \frac{dp'^{+}}{(2\pi)} \frac{e^{i\ell'^{+}(z^{-}_{3}-z^{-}_{2})} e^{-ip'^{+}(x^{-}_{2}-z^{-}_{2})} }{ 8 q^{-}(q^{-}-\ell'^{-})\ell'^{-}\left[ q^{+} +p'^{+} -\frac{M^2}{2q^-} -i\epsilon\right]\left[ q^{+} +p'^{+} -\ell'^{+} - \frac{\pmb{\ell}'^2_{\perp}}{2(q^{-}-\ell'^{-})} -i\epsilon\right] \left[ \ell'^{+}  - \frac{\pmb{\ell}'^2_{\perp}+M^2}{2\ell'^{-}} -i\epsilon\right]} \nonumber \\
      & =& \frac{(-2\pi i)}{2\pi} \frac{(-2\pi i)}{2\pi}  \frac{\theta\left(x^{-}_{2} - z^{-}_{2}\right)}{8q^{-}(q^{-}-\ell'^{-}) \ell'^{-}}  \frac{1}{\mathcal{G}^{(\ell,-k)}_{2}} e^{i\left[q^{+} - \frac{M^2}{2q^{-}}\right]\left(x^{-}_{2}-z^{-}_{2}\right)} e^{ i \frac{\pmb{\ell}'^2_{\perp}+M^2}{2\ell'^{-}}(z^{-}_{3}-z^{-}_{2})} \nonumber \\
    & \times & \left[ -\theta(-z^{-}_{3}+z^{-}_{2}) - \theta(z^{-}_{3}-z^{-}_{2}) e^{-i\mathcal{G}^{(\ell,-k)}_{2}(z^{-}_{3}-z^{-}_{2}) } +   \theta(x^{-}_{2}-z^{-}_{3}) e^{-i\mathcal{G}^{(\ell,-k)}_{2}(x^{-}_{2}-z^{-}_{2}) } \{ 1 -  e^{i\mathcal{G}^{(\ell,-k)}_{2}(x^{-}_{2}-z^{-}_{3}) } \}\right].  
\end{eqnarray}
The final hadronic tensor for left-cut diagram [Fig.~\ref{fig:kernel-1_de_both}(a)] is given as
\begin{eqnarray}
\begin{split}
W^{\mu\nu}_{1,\ell} & =- \frac{C_{A}}{2} \sum_f 2 \left[-g^{\mu\nu}_{\perp\perp}\right]  e^2_f g^4_{s} \int d (\Delta x^{-}) e^{i\Delta x^{-}\left(q^{+}-\frac{M^2}{2q^-}\right)} \left\langle P \left| \bar{\psi}_{_f}(\Delta x^-) \frac{\gamma^{+}}{4} \psi_{_f}(0)\right| P\right\rangle\\
  & \times n \int d \zeta^-  d (\Delta z^{-})d^2 \Delta z_{\perp} \frac{dy}{2\pi}\frac{d^2 \ell_{\perp}}{(2\pi)^2} \frac{d^2 k_{\perp}}{(2\pi)^2} 
  e^{-i\Delta z^{-}\mathcal{H}^{(\ell,p_2)}_M} e^{i\pmb{k}_{\perp}\cdot \Delta\pmb{z}_{\perp} } e^{i\mathcal{G}^{(\ell)}_{M}\left(x^{-}_{1}-z^{-}_{3}\right)} \theta(x^{-}_{2}-z^{-}_{2}) \\ 
  & \times  \left[\frac{1+ \left(1-y\right)^2}{y}\right] \left[ 1-\frac{\eta}{2}\right] \frac{\left[\left(\pmb{\ell}_{\perp} - \pmb{k}_{\perp}\right)\cdot\pmb{\ell}_{\perp} +\kappa y^4 M^2 \right]}{\left[\pmb{\ell}^2_{\perp}  + y^2 M^2 \right] \left[\left(\pmb{\ell}_{\perp} - \pmb{k}_{\perp}\right)^2  + y^2\left(1-\eta\right)^2  M^2 \right]}\\
  & \times \left[ \theta(-z^{-}_{3}+z^{-}_{2})  + \theta(z^{-}_{3}-z^{-}_{2}) e^{-i\mathcal{G}^{(\ell,-k)}_{2}(z^{-}_{3}-z^{-}_{2}) } +   \theta(x^{-}_{2}-z^{-}_{3}) e^{-i\mathcal{G}^{(\ell,-k)}_{2}(x^{-}_{2}-z^{-}_{2}) } \left\{ -1 +  e^{i\mathcal{G}^{(\ell,-k)}_{2}(x^{-}_{2}-z^{-}_{3}) } \right\} \right]\\
  & \times \left\langle P_{A-1} \left\vert {\rm Tr}\left[A^{+}\left(\zeta^-, \Delta z^-, \Delta z_{\perp}\right)A^+\left(\zeta^-,0\right)\right]\right\vert P_{A-1}\right\rangle,
\end{split}
\label{eq:K1_d_leftcut_final}
\end{eqnarray}
where $\mathcal{G}^{(\ell,-k)}_{2}$ is defined in Eq.~(\ref{eq:G2_ell_k_append}) and $\mathcal{H}^{(\ell,p_2)}_M$ is defined in Eq.~(\ref{eq:HL2P2M_append}). The factor $n$ in Eq.~(\ref{eq:K1_d_leftcut_final}) is described in Eq.~(\ref{eq:factor_n}).

The right-cut diagram of Fig.~\ref{fig:kernel-1_de_both}(a) contains singularity arising from the denominator of the propagators of $p_2$, $\ell$, and $p_1$. It involves nested integrals over variables $p^+$ and $p^+_2$.
The integrals are carried out using the method of contour integration as follows:
\begin{eqnarray}
C &=& \oint \oint \frac{dp_2^{+}}{(2\pi)} \frac{dp^{+}}{(2\pi)} \frac{e^{-ip^{+}_2(z^{-}_{3}-z^{-}_{2})} e^{ip^{+}(x^{-}_{1}-z^{-}_{2})} }{\left[\left(q+p\right)^{2} - M^{2} + i\epsilon\right]\left[\left(q+p-p_{2}\right)^{2}  +i\epsilon\right] \left[p^2_2-M^2 + i\epsilon \right]} \\
 &=& \oint \oint \frac{dp_2^{+}}{(2\pi)} \frac{dp^{+}}{(2\pi)} \frac{e^{-ip^{+}_2(z^{-}_{3}-z^{-}_{2})} e^{ip^{+}(x^{-}_{1}-z^{-}_{2})} }{ 8 q^{-}(q^{-}-p^{-}_{2})p^{-}_{2}\left[ q^{+} +p^{+} -\frac{M^2}{2q^-} +i\epsilon\right]\left[ q^{+} + p^{+} -p^{+}_{2} - \frac{\pmb{p}^2_{2\perp}}{2(q^{-}-p^{-}_2)} +i\epsilon\right] \left[ p^{+}_{2} - \frac{\pmb{p}^2_{2\perp}+M^2}{2p^{-}_2} + i\epsilon\right]} \nonumber \\
&=& \frac{(2\pi i)}{2\pi} \frac{(2\pi i)}{2\pi}  \frac{\theta\left(x^{-}_{1} - z^{-}_{2}\right)}{8q^{-}(q^{-}-p^{-}_{2}) p^{-}_{2}}  \frac{1}{\mathcal{G}^{(p_2)}_{M}} e^{i\left[-q^{+} + \frac{M^2}{2q^{-}}\right]\left(x^{-}_{1}-z^{-}_{2}\right)} e^{ -i \frac{\pmb{p}^2_{2\perp}+M^2}{2p^{-}_{2}}(z^{-}_{3}-z^{-}_{2})} \nonumber \\
&& \times \left[ -\theta(-z^{-}_{3}+z^{-}_{2}) - \theta(z^{-}_{3}-z^{-}_{2}) e^{i\mathcal{G}^{(p_2)}_{M}(z^{-}_{3}-z^{-}_{2}) } + \theta(x^{-}_{1}-z^{-}_{3}) e^{i\mathcal{G}^{(p_2)}_{M}(x^{-}_{1}-z^{-}_{2})}\left\{1 - e^{-i\mathcal{G}^{(p_2)}_{M}(x^{-}_{1}-z^{-}_{3}) }\right\}\right]. \nonumber
\end{eqnarray}
The final hadronic tensor for right-cut diagram [Fig.~\ref{fig:kernel-1_de_both}(a)] is given as
\begin{eqnarray}
\begin{split}
W^{\mu\nu}_{1,r} & =- \frac{C_{A}}{2} \sum_f 2 \left[-g^{\mu\nu}_{\perp\perp}\right]  e^2_f g^4_{s} \int d (\Delta x^{-}) e^{i\Delta x^{-}\left(q^{+}-\frac{M^2}{2q^-}\right)} \left\langle P \left| \bar{\psi}_{_f}(\Delta x^-) \frac{\gamma^{+}}{4} \psi_{_f}(0)\right| P\right\rangle\\
  & \times n \int d \zeta^-  d (\Delta z^{-})d^2 \Delta z_{\perp} \frac{dy}{2\pi}\frac{d^2 \ell_{2\perp}}{(2\pi)^2} \frac{d^2 k_{\perp}}{(2\pi)^2} e^{-i\Delta z^{-}\mathcal{H}^{(\ell_2,p_2)}_M} e^{i\pmb{k}_{\perp}\cdot \Delta\pmb{z}_{\perp} } e^{-i\mathcal{G}^{(\ell_2)}_{M}\left(x^{-}_{2}-z^{-}_{3}\right)} \theta(x^{-}_{1}-z^{-}_{2})\\ 
  & \times    \left[\frac{1+ \left(1-y\right)^2}{y}\right] \left[1-\frac{\eta}{2}\right] \frac{\left[\left(\pmb{\ell}_{2\perp} -  \pmb{k}_{\perp}\right)\cdot\pmb{\ell}_{2\perp} + \kappa y^4 M^2\right]}{\left[\pmb{\ell}^2_{2\perp}  + y^2 M^2 \right] \left[\left(\pmb{\ell}_{2\perp} - \pmb{k}_{\perp}\right)^2  + y^2\left(1-\eta\right)^2  M^2 \right]}  \\
  & \times\left[\theta(-z^{-}_{3}+z^{-}_{2}) + \theta(z^{-}_{3}-z^{-}_{2}) e^{i\mathcal{G}^{(p_2)}_{M}(z^{-}_{3}-z^{-}_{2})}+\theta(x^{-}_{1}-z^{-}_{3}) e^{i\mathcal{G}^{(p_2)}_{M}(x^{-}_{1}-z^{-}_{2})} \left\{-1 + e^{-i\mathcal{G}^{(p_2)}_{M}(x^{-}_{1}-z^{-}_{3})} \right\}\right],\\
  & \times \left\langle P_{A-1} \left\vert {\rm Tr}\left[A^{+}\left(\zeta^-, \Delta z^-, \Delta z_{\perp}\right)A^+\left(\zeta^-,0\right)\right]\right\vert P_{A-1}\right\rangle,
\end{split}
\label{eq:K1_d_rightcut_final}
\end{eqnarray}
where $\mathcal{G}^{(p_2)}_{M}$ is defined in Eq.~(\ref{eq:GMP2_append}), $\mathcal{G}^{(\ell_2)}_{M}$ is defined in Eq.~(\ref{eq:GL2M_append}) and $\mathcal{H}^{(\ell_2,p_2)}_M$ is defined in Eq.~(\ref{eq:HL2P2M_append}). The factor $n$ in Eq.~(\ref{eq:K1_d_rightcut_final}) is described in Eq.~(\ref{eq:factor_n}). The left-cut and right-cut expressions given in Eq.~(\ref{eq:K1_d_leftcut_final}) and Eq.~(\ref{eq:K1_d_rightcut_final}) can be added together using $\zeta^{-}=x^{-}_{1}-z^{-}_{2}=x^{-}_{2}-z^{-}_{3}$ and further initializing $x^{-}_{1}$ and $x^{-}_{2}$ to the origin, yielding
\begin{eqnarray}
\begin{split}
W^{\mu\nu}_{1,\ell+r} & = -\frac{C_{A}}{2} \sum_f 2 \left[-g^{\mu\nu}_{\perp\perp}\right]  e^2_f g^4_{s} \int d (\Delta x^{-}) e^{i\Delta x^{-}\left(q^{+}-\frac{M^2}{2q^-}\right)} \left\langle P \left| \bar{\psi}_{_f}(\Delta x^-) \frac{\gamma^{+}}{4} \psi_{_f}(0)\right| P\right\rangle\\
  & \times n \int d \zeta^-  d (\Delta z^{-})d^2 \Delta z_{\perp} \frac{dy}{2\pi}\frac{d^2 \ell_{2\perp}}{(2\pi)^2} \frac{d^2 k_{\perp}}{(2\pi)^2} e^{-i\Delta z^{-}\mathcal{H}^{(\ell_2,p_2)}_M} e^{i\pmb{k}_{\perp}\cdot \Delta\pmb{z}_{\perp} } \theta(\zeta^-)  \\ 
  & \times \left[\frac{1+ \left(1-y\right)^2}{y}\right] \left[1-\frac{\eta}{2}\right]  \frac{  \left[\left(\pmb{\ell}_{2\perp} - \pmb{k}_{\perp}\right)\cdot\pmb{\ell}_{2\perp} + \kappa y^4 M^2\right]}{\left[\pmb{\ell}^2_{2\perp}  + y^2 M^2 \right] \left[\left(\pmb{\ell}_{2\perp} - \pmb{k}_{\perp}\right)^2  + y^2\left(1-\eta\right)^2  M^2 \right]}\\
  & \times \left[ 4 \cos\left\{\mathcal{G}^{(\ell_2)}_{M}\zeta^-\right\}  - 2\cos\left\{\left(\mathcal{G}^{(p_2)}_{M}-\mathcal{G}^{(\ell_2)}_{M}\right)\zeta^-\right\} \right]\left\langle P_{A-1} \left\vert {\rm Tr}\left[A^{+}\left(\zeta^-, \Delta z^-, \Delta z_{\perp}\right)A^+\left(\zeta^-,0\right)\right]\right\vert P_{A-1}\right\rangle.
\end{split}
\label{eq:K1_d_rightcut_final_x1_x2_origin}
\end{eqnarray}
Note that the left-cut diagram in Fig.~\ref{fig:kernel-1_de_both}(a) is identical to the left-cut diagram in Fig.~\ref{fig:kernel-1_de_both}(b). This is mainly because  $\theta(-z^{-}_{3}+z^{-}_{2})$ and $\theta(z^{-}_{3}-z^{-}_{2})$ in Eq.~(\ref{eq:K1_d_leftcut_final}) account for both the possibility, i.e. $z^{-}_{3}-z^{-}_{2} > 0 $ and $z^{-}_{3}-z^{-}_{2} < 0 $. Hence, it is not added as it would lead to double-counting. Similarly, the right-cut diagram in Fig.~\ref{fig:kernel-1_de_both}(a) is identical to the right-cut diagram in Fig.~\ref{fig:kernel-1_de_both}(b) and hence it would not be added.
%%%%%%%%%%%%%%%%%%%%%%%%%%%%%%%%%%%%%%%%%%%%%%%%%%%%%%%%%%%%%%%%%%%%%%%%%%%%%%%%%%%%%
%%%%%%%%%%%%%%%%%%%%%%%%%%%%%%%%%%%%%%%%%%%%%%%%%%%%%%%%%%%%%%%%%%%%%%%%%%%%%%%%%%%%%
\subsection{Figures~\ref{fig:kernel-1_all}(j) and \ref{fig:kernel-1_all}(k)}
%%%%%%%%%%%%%%%%%%%%%%%%%%%%%%%%%%%%%%%%%%%%%%%%%%%%%%%%%%%%%%%%%%%%%%%%%%%%%%%%%%%%%
%%%%%%%%%%%%%%%%%%%%%%%%%%%%%%%%%%%%%%%%%%%%%%%%%%%%%%%%%%%%%%%%%%%%%%%%%%%%%%%%%%%%%
%
\begin{figure}[!h]
    \centering 
    \begin{subfigure}[t]{0.495\textwidth}
        \centering        \includegraphics[height=1.6in]{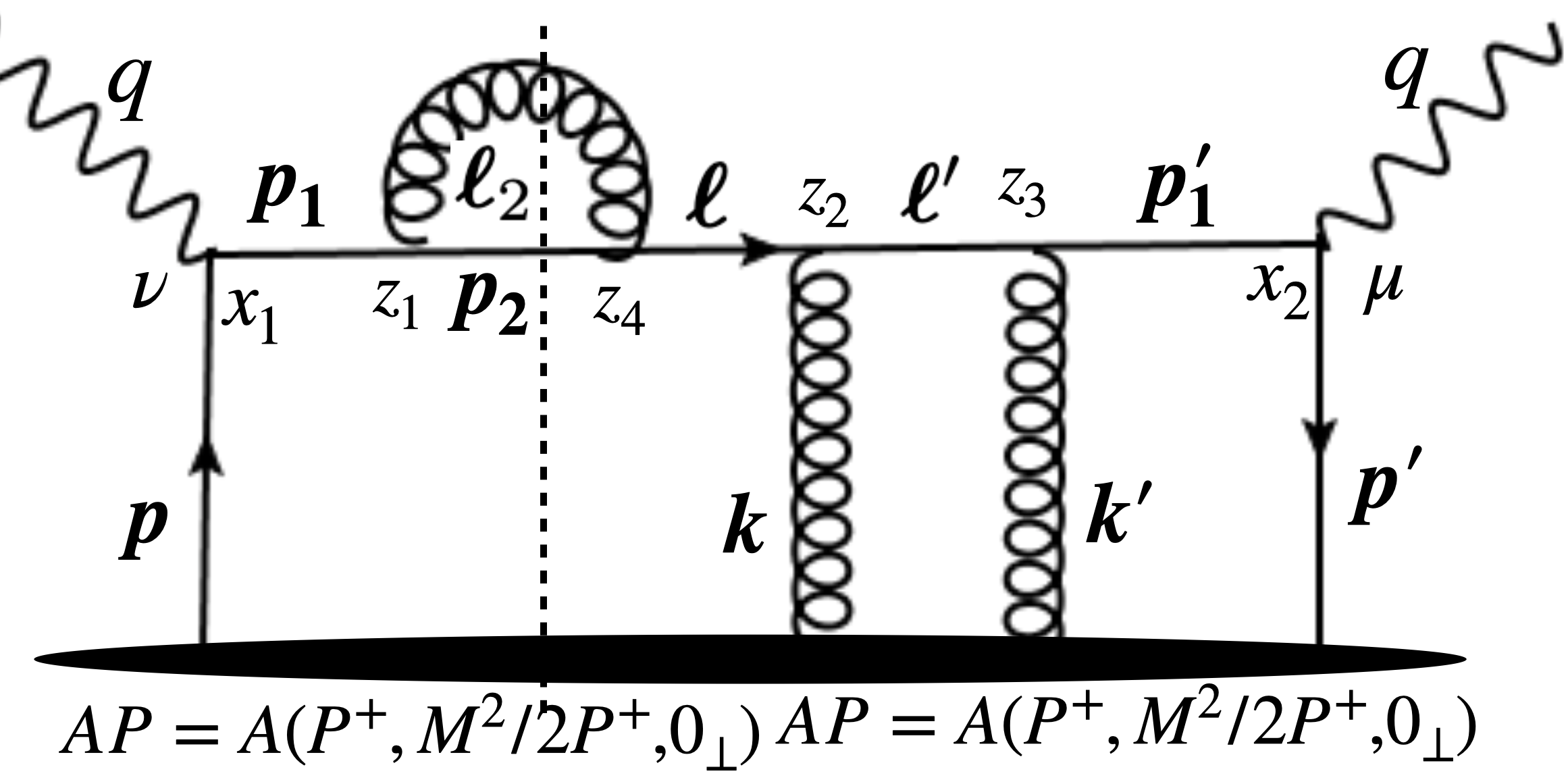}
        \caption{ }
    \end{subfigure}%
    \begin{subfigure}[t]{0.495\textwidth}
        \centering        \includegraphics[height=1.6in]{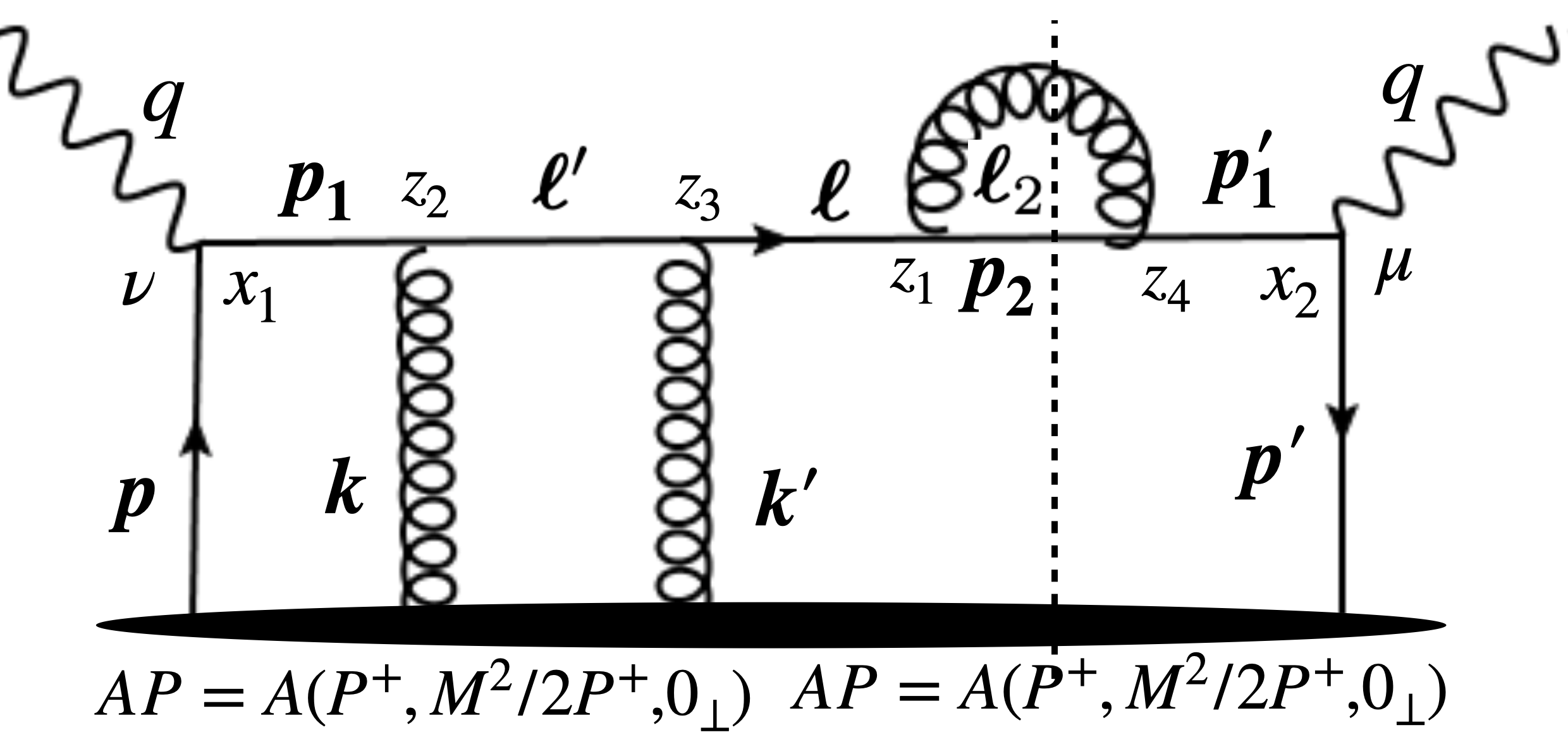}
        \caption{}
    \end{subfigure}
\caption{A forward scattering diagram contributing to kernel-1 with a gluon emission post two successive gluon scattering.}
\label{fig:kernel-1_last_2_successive_scatt}
\end{figure}

The last diagram to consider in kernel-1 is presented in Fig.~\ref{fig:kernel-1_last_2_successive_scatt}, along with the variables used in the calculation. The cut diagram in Fig.~\ref{fig:kernel-1_last_2_successive_scatt}(a) contains final-state gluon radiation on the amplitude side and two successive gluon scatterings followed by the gluon radiation on the complex-conjugate side. The hadronic tensor for the diagram in Fig.~\ref{fig:kernel-1_last_2_successive_scatt}(a) is given by
\begin{equation}
\begin{split}
W^{\mu\nu}_{1,\ell} & = \sum_{f} e^2_f g^4_{s} \int d^4 x_1 d^4 x_2 d^4 z_{2} d^4 z_{3} \int \frac{d^4 \ell'}{(2\pi)^4} \frac{d^4 p'}{(2\pi)^4}  \frac{d^4 \ell_{2}}{(2\pi)^4} \frac{d^4 p_{2}}{(2\pi)^4} e^{-ip'x_{2}} e^{i(\ell_2+p_2-q)x_{1}}\left\langle P \left| \bar{\psi}_{_f}(x_2) \frac{\gamma^{+}}{4} \psi_{_f}(x_{1})\right| P\right\rangle  \\
  & \times e^{i\left(q+p'-\ell'\right) z_{3}} e^{i(\ell'-\ell_2-p_2)z_{2}} \left\langle P_{A-1} \left| A^{+c}(z_3)A^{+b}(z_2)\right| P_{A-1}\right\rangle  {\rm Tr}\left[t^c t^b t^d t^a\right] \delta^{ad}  \\
& \times \frac{  {\rm Tr} \left[\gamma^- \gamma^{\mu} \left(\slashed{q}+\slashed{p}'+M\right) \gamma^{-}\left( \slashed{\ell}' +  M\right) \gamma^{-} \left(\slashed{\ell}_2 + \slashed{p}_2 +  M\right) \gamma^{\sigma_4} \left( \slashed{p}_2 + M\right)\gamma^{\sigma_1}  \left(\slashed{\ell}_2 + \slashed{p}_2 +  M\right) \gamma^{\nu} \right]
}{  \left[\left(q+p'\right)^{2}-M^2-i\epsilon\right] \left[\left(\ell'\right)^2  - M^2 - i\epsilon\right]   \left[\left(\ell_2 + p_2\right)^2 - M^2 - i\epsilon\right] \left[\left(\ell_2+p_2\right)^2 -M^2 + i\epsilon\right]  } \\
& \times d^{\left(\ell_2\right)}_{\sigma_{1} \sigma_{4}}  (2 \pi) \delta\left(\ell^{2}_2\right) (2 \pi)  \delta\left(p^{2}_{2}-M^{2}\right).
\end{split}
\label{eq:kernel1_lastdiagram_a_wmunu_initial}
\end{equation}
Equation~(\ref{eq:kernel1_lastdiagram_a_wmunu_initial}) has singularity arising from the denominator of the quark propagator with momentum $\ell'$ and $p'_{1}$. We identify one pole for each momentum variable $p'^+$ and $\ell'^{+}$.
The contour integration for momentum $p'^{+}$ is given as
\begin{equation}
\begin{split}
C_{1} = \oint \frac{dp'^+}{(2\pi)} \frac{e^{-ip'^+(x^{-}_{2}-z^-_3)} }{\left[\left(q+p'\right)^2 -M^2 - i\epsilon\right]} = \frac{(-2\pi i)}{2\pi}\frac{\theta\left(x^{-}_{2} - z^{-}_{3}\right)}{2q^{-}}  e^{i\left(q^{+}-\frac{M^2}{2q^-}\right)\left(x^{-}_{2}-z^{-}_{3}\right)} .
      \end{split}
\end{equation}
Similarly, the contour integration for $\ell'^+$ can be done giving:
\begin{equation}
\begin{split}
C_{2} & = \oint \frac{d\ell'^{+}}{(2\pi)} \frac{e^{-i\ell'^{+}(z^{-}_{3}-z^{-}_{2})}}{\left[ \ell'^{2} -M^2 -i\epsilon\right]}  = \frac{(-2\pi i)}{2\pi} \frac{\theta\left(z^{-}_{3} - z^{-}_{2}\right)}{2\ell'^{-}}  e^{-i\left[ \frac{\pmb{k}^2_{\perp}+M^2}{2\ell'^{-}}\right]\left(z^{-}_{3}-z^{-}_{2}\right)}. 
\end{split}
%\label{eq:}
\end{equation}
The trace in the numerator of the third line of Eq.~(\ref{eq:kernel1_lastdiagram_a_wmunu_initial}) gives
\begin{equation}
    \begin{split}
       &  {\rm Tr} \left[\gamma^- \gamma^{\mu} \left(\slashed{q}+\slashed{p}'+M\right) \gamma^{-}\left( \slashed{\ell}' +  M\right) \gamma^{-} \left(\slashed{\ell}_2 + \slashed{p}_2 +  M\right) \gamma^{\sigma_4} \left( \slashed{p}_2 + M\right)\gamma^{\sigma_1}  \left(\slashed{\ell}_2 + \slashed{p}_2 +  M\right) \gamma^{\nu} \right]  d^{\left(\ell_2\right)}_{\sigma_{1} \sigma_{4}} \\
       & = \frac{32[-g^{\mu\nu}_{\perp\perp}](1+\eta y)\left(q^-\right)^3 }{(1-y)y}\left[ \frac{1+\left(1-y\right)^2}{y} \right] \left[\pmb{\ell}^2_{2\perp}  + \kappa y^4 M^2 \right].
    \end{split}
\end{equation}
The final hadronic tensor for the diagram in Fig.~\ref{fig:kernel-1_last_2_successive_scatt}(a) is given by
\begin{eqnarray}
\begin{split}
W^{\mu\nu}_{1,\ell} & =- C_{F} \sum_f 2 \left[-g^{\mu\nu}_{\perp\perp}\right]  e^2_f g^4_{s} \int d (\Delta x^{-}) e^{i\Delta x^{-}\left(q^{+}-\frac{M^2}{2q^-}\right)} \left\langle P \left| \bar{\psi}_{_f}(\Delta x^-) \frac{\gamma^{+}}{4} \psi_{_f}(0)\right| P\right\rangle\\
  & \times n \int d \zeta^-  d (\Delta z^{-})d^2 \Delta z_{\perp} \frac{dy}{2\pi}\frac{d^2 \ell_{2\perp}}{(2\pi)^2} \frac{d^2 k_{\perp}}{(2\pi)^2} e^{-i\Delta z^{-}\left\{\frac{\pmb{k}^2_{\perp}+M^2}{2q^-(1+\eta y)} - \frac{M^2}{2q^-}\right\} } e^{i\pmb{k}_{\perp}\cdot \Delta\pmb{z}_{\perp} } \theta(z^{-}_{3}-z^{-}_{2}) \theta(x^{-}_{2}-z^{-}_{3}) \\ 
  & \times \left[\frac{1+ \left(1-y\right)^2}{y}\right] \frac{\left[ \pmb{\ell}^2_{2\perp}  + \kappa y^4 M^2\right]}{\left[\pmb{\ell}^2_{2\perp}  + y^2 M^2 \right]^2 } e^{i\mathcal{G}^{(\ell_2)}_{M}\left(x^{-}_{1}-z^{-}_{2}\right)} \\
  & \times \left\langle P_{A-1} \left\vert {\rm Tr}\left[A^{+}\left(\zeta^-, \Delta z^-, \Delta z_{\perp}\right)A^+\left(\zeta^-,0\right)\right]\right\vert P_{A-1}\right\rangle.
\end{split}
\label{eq:kernel1_lastdiagram_a_wmunu_final}
\end{eqnarray}
The factor $n$ in Eq.~(\ref{eq:kernel1_lastdiagram_a_wmunu_final}) is described in Eq.~(\ref{eq:factor_n}). Note, the diagram shown in Fig.~\ref{fig:kernel-1_last_2_successive_scatt}(b) is a complex-conjugate of the diagram in Fig.~\ref{fig:kernel-1_last_2_successive_scatt}(a). The final expression of the hadronic tensor for Fig.~\ref{fig:kernel-1_last_2_successive_scatt}(b) is given as
\begin{eqnarray}
\begin{split}
W^{\mu\nu}_{1,r} & =- C_{F} \sum_f 2 \left[-g^{\mu\nu}_{\perp\perp}\right]  e^2_f g^4_{s} \int d (\Delta x^{-}) e^{i\Delta x^{-}\left(q^{+}-\frac{M^2}{2q^-}\right)} \left\langle P \left| \bar{\psi}_{_f}(\Delta x^-) \frac{\gamma^{+}}{4} \psi_{_f}(0)\right| P\right\rangle\\
  & \times n \int d \zeta^-  d (\Delta z^{-})d^2 \Delta z_{\perp} \frac{dy}{2\pi}\frac{d^2 \ell_{2\perp}}{(2\pi)^2} \frac{d^2 k_{\perp}}{(2\pi)^2} 
  e^{-i\Delta z^{-}\left\{\frac{\pmb{k}^2_{\perp}+M^2}{2q^-(1+\eta y)} - \frac{M^2}{2q^-}\right\} } e^{i\pmb{k}_{\perp}\cdot \Delta\pmb{z}_{\perp} }  \theta(-z^{-}_{3}+z^{-}_{2}) \theta(x^{-}_{1}-z^{-}_{2}) \\ 
  & \times \left[\frac{1+ \left(1-y\right)^2}{y}\right]  \frac{ \left[ \pmb{\ell}^2_{2\perp}  + \kappa y^4 M^2\right]}{\left[\pmb{\ell}^2_{2\perp}  + y^2 M^2 \right]^2 } e^{-i\mathcal{G}^{(\ell_2)}_{M}\left(x^{-}_{2}-z^{-}_{3}\right)} \left\langle P_{A-1} \left\vert {\rm Tr}\left[A^{+}\left(\zeta^-, \Delta z^-, \Delta z_{\perp}\right)A^+\left(\zeta^-,0\right)\right]\right\vert P_{A-1}\right\rangle.
\end{split}
\label{eq:kernel1_lastdiagram_b_wmunu_final}
\end{eqnarray}
Adding the hadronic tensors of the forward scattering diagrams in Fig.~\ref{fig:kernel-1_last_2_successive_scatt}(a) and Fig.~\ref{fig:kernel-1_last_2_successive_scatt}(b) gives
\begin{eqnarray}
\begin{split}
W^{\mu\nu}_{1,\ell+r} & =- C_{F} \sum_f 2 \left[-g^{\mu\nu}_{\perp\perp}\right]  e^2_f g^4_{s} \int d (\Delta x^{-}) e^{i\Delta x^{-}\left(q^{+}-\frac{M^2}{2q^-}\right)} \left\langle P \left| \bar{\psi}_{_f}(\Delta x^-) \frac{\gamma^{+}}{4} \psi_{_f}(0)\right| P\right\rangle\\
  & \times n \int d \zeta^-  d (\Delta z^{-})d^2 \Delta z_{\perp} \frac{dy}{2\pi}\frac{d^2 \ell_{2\perp}}{(2\pi)^2} \frac{d^2 k_{\perp}}{(2\pi)^2}  
  e^{-i\Delta z^{-}\left\{\frac{\pmb{k}^2_{\perp}+M^2}{2q^-(1+\eta y)} - \frac{M^2}{2q^-}\right\} } e^{i\pmb{k}_{\perp}\cdot \Delta\pmb{z}_{\perp} } \theta(\zeta^-) \\ 
  & \times \left[\frac{1+ \left(1-y\right)^2}{y}\right] \frac{\left[\pmb{\ell}^2_{2\perp} + \kappa y^4 M^2\right]}{\left[\pmb{\ell}^2_{2\perp} + y^2 M^2\right]^2} \cos\left\{\mathcal{G}^{(\ell_2)}_{M} \zeta^- \right\} \left\langle P_{A-1} \left\vert {\rm Tr}\left[A^{+}\left(\zeta^-, \Delta z^-, \Delta z_{\perp}\right)A^+\left(\zeta^-,0\right)\right]\right\vert P_{A-1}\right\rangle,
%  & \times ,
\end{split}
\label{eq:kernel1_lastdiagram_ab_wmunu_sum}
\end{eqnarray}
where the factor $n$ is defined in Eq.~(\ref{eq:factor_n}). Notice that the third line  in Eq.~(\ref{eq:kernel1_lastdiagram_ab_wmunu_sum}) does not depend on $\pmb{k}_{\perp}$ and $k^-$, therefore, the diagrams in Fig.~\ref{fig:kernel-1_last_2_successive_scatt}(a) and Fig.~\ref{fig:kernel-1_last_2_successive_scatt}(b) do not contribute to the energy loss.
%%%%%%%%%%%%%%%%%%%%%%%%%%%%%%%%%%%%%%%%%%%%%%%%%%%%%%%%%%%%%%%%%%%%%%%%%%%%%%%%%%%%%%%%%%%%%%%%%%%%%%%%%%
%%%%%%%%%%%%%%%%%%%%%%%%%%%%%%%%%%%%%%%%%%%%%%%%%%%%%%%%%%%%%%%%%%%%%%%%%%%%%%%%%%%%%%%%%%%%%%%%%%%%%%%%%%
%%%%%%%%%%%%%%%%%%%%%%%%%%%%%%%%%%%%%%%%%%%%%%%%%%%%%%%%%%%%%%%%%%%%%%%%%%%%%%%%%%%%%%%%%%%%%%%%%%%%%%%%%%
\section{$\mbox{}$\!\!\!\!\!\!: Single-emission single-scattering kernel: Two gluons in the final state }
\label{append:kernel-2}
%%%%%%%%%%%%%%%%%%%%%%%%%%%%%%%%%%%%%%%%%%%%%%%%%%%%%%%%%%%%%%%%%%%%%%%%%%%%%%%%%%%%%%%%%%%%%%%%%%%%%%%%%%
%%%%%%%%%%%%%%%%%%%%%%%%%%%%%%%%%%%%%%%%%%%%%%%%%%%%%%%%%%%%%%%%%%%%%%%%%%%%%%%%%%%%%%%%%%%%%%%%%%%%%%%%%%
%%%%%%%%%%%%%%%%%%%%%%%%%%%%%%%%%%%%%%%%%%%%%%%%%%%%%%%%%%%%%%%%%%%%%%%%%%%%%%%%%%%%%%%%%%%%%%%%%%%%%%%%%%
In this section, we summarize the calculation of all possible diagrams at NLO  contributing to  kernel-2 with two gluons in the final state. We discuss singularity structure, contour integrations, and involved traces in the final calculation of the hadronic tensor.

\begin{figure}[!h]
    \centering 
    \begin{subfigure}[t]{0.495\textwidth}
        \centering        \includegraphics[width=\textwidth]{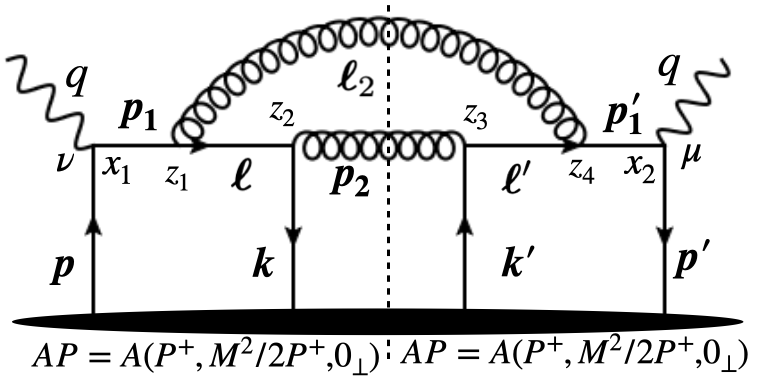}
        \caption{Quark anti-quark scattering channel. }
    \end{subfigure}%
    \begin{subfigure}[t]{0.495\textwidth}
        \centering        \includegraphics[width=\textwidth]{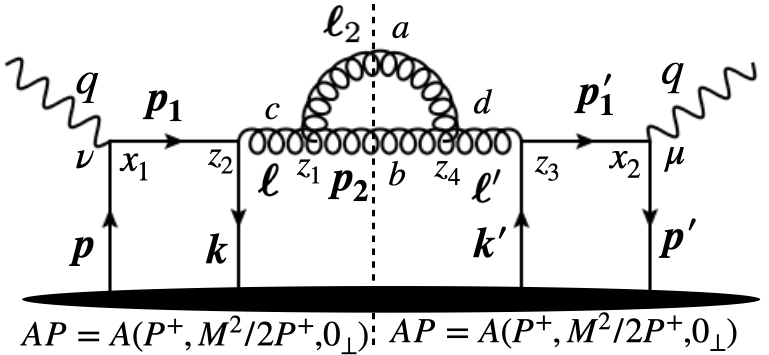}
        \caption{Quark anti-quark annihilation diagram.}
    \end{subfigure}
\caption{A forward scattering diagram contributing to kernel-2.}
\label{fig:kernel-2_ab_amplt_square}
\end{figure}
%
%%%%%%%%%%%%%%%%%%%%%%%%%%%%%%%%%%%%%%%%%%%%%%%%%%%%%%%%%%%%%%%%%%%%%%%%%%%%%%%%%%%%%
%%%%%%%%%%%%%%%%%%%%%%%%%%%%%%%%%%%%%%%%%%%%%%%%%%%%%%%%%%%%%%%%%%%%%%%%%%%%%%%%%%%%%
\subsection{Figures~\ref{fig:kernel-2_all}(a) and \ref{fig:kernel-2_all}(b)}
%%%%%%%%%%%%%%%%%%%%%%%%%%%%%%%%%%%%%%%%%%%%%%%%%%%%%%%%%%%%%%%%%%%%%%%%%%%%%%%%%%%%%
%%%%%%%%%%%%%%%%%%%%%%%%%%%%%%%%%%%%%%%%%%%%%%%%%%%%%%%%%%%%%%%%%%%%%%%%%%%%%%%%%%%%%
The hadronic tensor for Fig.~\ref{fig:kernel-2_ab_amplt_square}(a) has the following form
\begin{equation}
\begin{split}
W^{\mu\nu}_{2,c} & = \sum_f e^2_f g^4_{s} \int d^4 x_{1} d^4 x_{2} d^4 z_{2} d^4 z_{3} \int \frac{d^4 p}{(2\pi)^4} \frac{d^4 p'}{(2\pi)^4}  \frac{d^4 \ell_{2}}{(2\pi)^4} \frac{d^4 p_{2}}{(2\pi)^4} e^{-ip'x_2} e^{ipx_1}\left\langle P \left| \bar{\psi}_{_f}(x_2) \frac{\gamma^{+}}{4} \psi_{_f}(x_1)\right| P\right\rangle \delta^{ad} \delta^{bc} \\
  & \times e^{i\left(q+p'-p_{2}-\ell_{2}\right) z_{3}} e^{i(\ell_{2}+p_{2} - q - p)z_{2}}\left\langle P_{A-1} \left| \bar{\psi}_{_f}(z_2)\frac{\gamma^+}{4}\psi_{_f}(z_3)\right| P_{A-1} \right\rangle (2 \pi) \delta\left(\ell_{2}^{2}\right) (2 \pi)  \delta\left(p_{2}^{2}\right) {\rm Tr}[t^{d}t^{c}t^{b}t^{a}]   \\
& \times \frac{ {\rm Tr} \left[\gamma^- \gamma^{\mu} \left(\slashed{q}+\slashed{p}'\right) \gamma^{\sigma_4}\left(\slashed{q}+\slashed{p}' - \slashed{\ell}_2\right)\gamma^{\sigma_{3}} \gamma^{-} \gamma^{\sigma_2} \left( \slashed{q} + \slashed{p} -\slashed{\ell}_2 \right) \gamma^{\sigma_1} \left( \slashed{q} + \slashed{p} \right) \gamma^{\nu} \right] 
}{  \left[\left(q+p'\right)^{2}-i\epsilon\right] \left[\left(q+p\right)^2  + i\epsilon\right]  \left[\left(q+p'-\ell_2\right)^2 - i\epsilon\right] \left[\left(q+p -\ell_2\right)^2+i\epsilon\right]  } d^{(\ell_2)}_{\sigma_{1} \sigma_{4}} d^{(p_2)}_{\sigma_{3} \sigma_{2}} .
\end{split} \label{eq:kernel2_wi_a}
\end{equation}
The above expression of the hadonic tensor has singularity when the denominator of the propagator for $p_1$, $\ell$, $\ell'$ and $p'_1$ becomes on-shell. It contains two simple poles for $p^+$ and $p'^+$. The contour integration for $p^+$ gives
\begin{equation}
\begin{split}
C_{1} & = \oint \frac{dp^{+}}{(2\pi)} \frac{e^{ip^{+}(x^{-}_{1}-z^{-}_{2})}}{\left[\left(q+p\right)^{2}  + i \epsilon\right]\left[\left(q+p-\ell_{2}\right)^{2}  + i\epsilon\right]} \\
      & = \oint \frac{dp^{+}}{(2\pi)} \frac{e^{ip^{+}(x^{-}_{1} - z^{-}_{2}) }}{2q^{-}[ q^{+} + p^{+}  + i \epsilon] 2(q^{-} -\ell^{-}_{2} )\left[ q^{+} + p^{+} -\ell^{+}_{2} - \frac{\pmb{\ell}^{2}_{2\perp}}{2(q^{-}-\ell^{-}_{2})} + i \epsilon\right]}   \\
      & = \frac{(2\pi i)}{2\pi} \frac{\theta(x^{-}_{1} - z^{-}_{2})}{4q^{-}(q^{-}-\ell^{-}_{2})}  e^{-iq^{+} (x^{-}_{1}-z^{-}_{2})}\left[ \frac{ -1 + e^{i \mathcal{G}^{(\ell_2)}_0(x^{-}_{1}-z^{-}_{2})} }{ \mathcal{G}^{(\ell_2)}_0}  \right],
\end{split}
\end{equation}
where $\mathcal{G}^{(\ell_2)}_0$ is defined in Eq.~(\ref{eq:GL20_append}). Similarly, the contour integration for $p'^+$ gives
\begin{equation}
\begin{split}
C_{2} & = \oint \frac{dp'^{+}}{(2\pi)} \frac{e^{-ip'^{+}(x^{-}_{2}-z^{-}_{3})}}{\left[\left(q+p'\right)^{2}  - i \epsilon\right]\left[\left(q+p'-\ell_{2}\right)^{2}  - i\epsilon\right]} \\
      & = \oint \frac{dp'^{+}}{(2\pi)} \frac{e^{-ip'^{+}(x^{-}_{2} - z^{-}_{3}) }}{2q^{-}[ q^{+} + p'^{+} 
      - i \epsilon] 2(q^{-} -\ell^{-}_{2} )\left[ q^{+} + p'^{+} -\ell^{+}_{2} - \frac{\pmb{\ell}^{2}_{2\perp}}{2(q^{-}-\ell^{-}_{2})} - i \epsilon\right]}   \\
      & = \frac{(-2\pi i)}{2\pi} \frac{\theta(x^{-}_{2} - z^{-}_{3})}{4q^{-}(q^{-}-\ell^{-}_{2})}  e^{iq^{+}(x^{-}_{2}-z^{-}_{3})}\left[ \frac{ -1 + e^{-i \mathcal{G}^{(\ell_2)}_0(x^{-}_{2}-z^{-}_{3})} }{ \mathcal{G}^{(\ell_2)}_0}  \right] .
\end{split}
\end{equation}
The trace in the third line of Eq.~(\ref{eq:kernel2_wi_a}) simplifies to
\begin{equation}
\begin{split}    
&  {\rm Tr} \left[\gamma^- \gamma^{\mu} \left(\slashed{q}+\slashed{p}'\right) \gamma^{\sigma_4}\left(\slashed{q}+\slashed{p}' - \slashed{\ell}_2\right)\gamma^{\sigma_{3}} \gamma^{-} \gamma^{\sigma_2} \left( \slashed{q} + \slashed{p} -\slashed{\ell}_2 \right) \gamma^{\sigma_1} \left( \slashed{q} + \slashed{p} \right) \gamma^{\nu} \right] d^{(\ell_2)}_{\sigma_{1} \sigma_{4}} d^{(p_2)}_{\sigma_{3} \sigma_{2}} \\
& = 32 [-g^{\mu\nu}_{\perp\perp}](q^-)^2 \left[ \frac{1 + \left(1-y\right)^2}{y^2} \right]  \pmb{\ell}^2_{2\perp} .
\end{split}
\end{equation}
Finally, the hadronic tensor [Fig.~\ref{fig:kernel-2_ab_amplt_square}(a)] reduces to the following form
\begin{equation}
\begin{split}
    W^{\mu\nu}_{2,c} & = \left[C^2_F C_A \right] \sum_f  2[-g^{\mu\nu}_{\perp \perp }] e^2_f g^4_s \int d (\Delta X^{-})   e^{iq^{+}(\Delta X^{-} )}     
  \left\langle P \left| \bar{\psi}_{_f}( \Delta X^{-}) \frac{\gamma^{+}}{4} \psi_{_f}(0)\right| P\right\rangle  \\
  & \times   \int  d \zeta^{-} d (\Delta z^{-}) d^2 \Delta z_{\perp} \frac{dy}{2\pi} \frac{d^2 \ell_{2\perp}}{(2\pi)^2} \frac{d^2 k_{\perp}}{(2\pi)^2}  e^{-i (\Delta z^{-})\mathcal{H}^{(\ell_2 , p_2)}_{0} }     e^{i \pmb{k}_{\perp} \cdot \Delta\pmb{z}_{\perp}}  \theta( x^{-}_{1}-z^{-}_2) \theta( x^{-}_{2}-z^{-}_3) \\
  & \times \left[ \frac{ 1 + \left(1-y\right)^2 }{y}\right] \frac{ 1 }{  \pmb{\ell}^2_{2\perp}  } \frac{\left[ 2 -2 \cos\left\{\mathcal{G}^{(\ell_2)}_{0}\zeta^-\right\} \right]}{(1-y+\eta y)q^-}\left\langle P_{A-1} \left| \bar{\psi}_{_f}(\zeta^-, 0) \frac{\gamma^+}{4} \psi_{_f}(\zeta^-, \Delta z^-, \Delta z_{\perp})\right| P_{A-1} \right\rangle,
\end{split}     
\label{eq:kernel2-wmunu-final_a}
\end{equation}
where $ \mathcal{G}^{(\ell_2)}_0$ is defined in Eq.~(\ref{eq:GL20_append}) and $\mathcal{H}^{(\ell_2,p_2)}_{0}$ is given in Eq.~(\ref{eq:HL2P2M0_append}). 

Since the final state [Fig.~\ref{fig:kernel-2_ab_amplt_square}(a)] consists of two identical gluons, the momentum of the two gluons could be interchanged. After performing $p_2\leftrightarrow\ell_2$, the resulting hadronic tensor is given as
\begin{eqnarray}
\begin{split}
    W^{\mu\nu}_{2,c} & = \left[C^2_F C_A \right] \sum_f  2[-g^{\mu\nu}_{\perp \perp }] e^2_f g^4_s \int d (\Delta X^{-})   e^{iq^{+}(\Delta X^{-} )}     
  \left\langle P \left| \bar{\psi}_{_f}( \Delta X^{-}) \frac{\gamma^{+}}{4} \psi_{_f}(0)\right| P\right\rangle  \\
  & \times   \int  d \zeta^{-} d (\Delta z^{-}) d^2 \Delta z_{\perp} \frac{dy}{2\pi} \frac{d^2 \ell_{2\perp}}{(2\pi)^2} \frac{d^2 k_{\perp}}{(2\pi)^2}  e^{-i (\Delta z^{-})\mathcal{H}^{(\ell_2 , p_2)}_{0} }     e^{i \pmb{k}_{\perp} \cdot \Delta\pmb{z}_{\perp}}  \theta( x^{-}_{1}-z^{-}_2) \theta( x^{-}_{2}-z^{-}_3) \\
  & \times  \left[ \frac{ 1 + y^2 +\eta y^2 (\eta -2) }{(1-y+\eta y)}\right] \frac{1}{  \left(\pmb{\ell}_{2\perp} - \pmb{k}_{\perp} \right)^2 } \frac{\left[ 2 -2 \cos\left\{\mathcal{G}^{(p_2)}_{0}\zeta^-\right\} \right]}{yq^-}\left\langle P_{A-1} \left| \bar{\psi}_{_f}(\zeta^-, 0) \frac{\gamma^+}{4} \psi_{_f}(\zeta^-, \Delta z^-, \Delta z_{\perp})\right| P_{A-1} \right\rangle,   
\end{split}
\label{eq:kernel2-wmunu-final_a_l2p2Interchanged}
\end{eqnarray}
where $ \mathcal{G}^{(p_2)}_0$ is defined in Eq.~(\ref{eq:GP2M0_append}).

Next, we consider the diagram in Fig.~\ref{fig:kernel-2_ab_amplt_square}(b). It consists of a triple-gluon vertex on both the amplitude and complex-conjugate sides.
The hadronic tensor for the diagram in Fig.~\ref{fig:kernel-2_ab_amplt_square}(b) is given as
\begin{eqnarray}
W^{\mu\nu}_{2,c} & =& \sum_f e^2_f g^4_{s} \int d^4 x_{1} d^4 x_{2} d^4 z_{2} d^4 z_{3} \int \frac{d^4 p}{(2\pi)^4} \frac{d^4 p'}{(2\pi)^4}  \frac{d^4 \ell_{2}}{(2\pi)^4} \frac{d^4 p_{2}}{(2\pi)^4} e^{-ip'x_2} e^{ipx_1}\left\langle P \left| \bar{\psi}_{_f}(x_2) \frac{\gamma^{+}}{4} \psi_{_f}(x_1)\right| P\right\rangle {\rm Tr} [t^d t^c] \nonumber\\
  & \times & e^{i\left(q+p'-p_{2}-\ell_{2}\right) z_{3}} e^{i(\ell_{2}+p_{2} - q - p)z_{2}}\left\langle P_{A-1} \left| \bar{\psi}_{_f}(z_2)\frac{\gamma^+}{4}\psi_{_f}(z_3)\right| P_{A-1} \right\rangle (2 \pi) \delta\left(\ell_{2}^{2}\right) (2 \pi)  \delta\left(p_{2}^{2}\right)  \nonumber \\
& \times & \frac{ {\rm Tr} \left[\gamma^- \gamma^{\mu} \left(\slashed{q}+\slashed{p}'\right) \gamma^{\alpha_3} \gamma^{-} \gamma^{\alpha_2} \left( \slashed{q} + \slashed{p}  \right)  \gamma^{\nu} \right]  f^{acb} \left[ g^{\rho_{2}\sigma_{2}} \left(-2\ell_2 -p_2\right)^{\beta_2} + g^{\sigma_{2}\beta_{2}} \left(\ell_2+2p_2\right)^{\rho_2} + g^{\beta_{2}\rho_{2}} \left(-p_2 +\ell_2\right)^{\sigma_2} \right]
}{  \left[\left(q+p'\right)^{2}-i\epsilon\right] \left[\left(q+p\right)^2  + i\epsilon\right]  \left[\left(\ell_2+p_2\right)^2 - i\epsilon\right] \left[\left(\ell_2+p_2\right)^2+i\epsilon\right]  }  \nonumber \\
& \times & (-f^{abd}) \left[ g^{\rho_{1}\beta_{1}} \left(\ell_2 -p_2\right)^{\sigma_1} + g^{\beta_{1}\sigma_{1}} \left(2p_2+\ell_2\right)^{\rho_1} + g^{\sigma_{1}\rho_{1}} \left( -2\ell_2 - p_2\right)^{\beta_1} \right] d^{(\ell_2)}_{\rho_{1}\rho_{2}} d^{(p_2)}_{\beta_{1} \beta_{2}} d^{(\ell_2+p_2)}_{\sigma_{1} \alpha_{3}} d^{(\ell_2+p_2)}_{\sigma_{2} \alpha_{2}}.
\label{eq:kernel2_wi_b}
\end{eqnarray}
Equation~(\ref{eq:kernel2_wi_b}) has singularity arising from the denominator of the quark propagator with momentum $p_{1}$ and $p'_{1}$. We identify one pole for each momentum variable $p^+$ and $p'^+$. 
 The contour integration for $p^+$ in the complex plane is given by
\begin{equation}
    C_{1} = \oint \frac{dp^+}{(2\pi)} \frac{e^{ip^+(x^{-}_{1}-z^-_2)} }{\left[\left(q+p\right)^2  + i\epsilon\right]} =  \oint \frac{dp^+}{(2\pi)} \frac{e^{ip^+(x^{-}_{1}-z^{-}_{2})} }{ 2q^{-}[ q^{+}+p^{+} + i\epsilon]} = \frac{(2\pi i)}{2\pi}\frac{\theta\left( x^{-}_{1} - z^{-}_{2}\right)}{2q^{-}}  e^{-iq^{+} (x^{-}_{1}-z^{-}_{2})}. 
\end{equation}
Similarly, the contour integration for momentum $p'^+$ is carried out 
\begin{equation}
    C_{2} = \oint \frac{dp'^+}{(2\pi)} \frac{e^{-ip'^+(x^{-}_{2}-z^{-}_3)} }{\left[\left(q+p'\right)^2  -i\epsilon\right]} =  \oint \frac{dp'^+}{(2\pi)} \frac{e^{-ip'^+(x^{-}_{2}-z^{-}_3)} }{2q^- [ q^{+}+p'^{+} -i\epsilon]} = \frac{(-2\pi i)}{2\pi} \frac{\theta\left( x^{-}_{2} - z^{-}_{3}\right)}{2q^-}  e^{iq^+(x^{-}_{2}-z^{-}_3)}.
\end{equation}
Simplifying the trace in Eq.~(\ref{eq:kernel2_wi_b}) yields
\begin{eqnarray}
&  {\rm Tr} & \left[\gamma^- \gamma^{\mu} \left(\slashed{q}+\slashed{p}'\right) \gamma^{\alpha_3} \gamma^{-} \gamma^{\alpha_2} \left( \slashed{q} + \slashed{p}  \right)  \gamma^{\nu} \right]  \left[ g^{\rho_{2}\sigma_{2}} \left(-2\ell_2 -p_2\right)^{\beta_2} + g^{\sigma_{2}\beta_{2}} \left(\ell_2+2p_2\right)^{\rho_2} + g^{\beta_{2}\rho_{2}} \left(-p_2 +\ell_2\right)^{\sigma_2} \right]
   \nonumber\\
& \times & \left[ g^{\rho_{1}\beta_{1}} \left(\ell_2 -p_2\right)^{\sigma_1} + g^{\beta_{1}\sigma_{1}} \left(2p_2+\ell_2\right)^{\rho_1} + g^{\sigma_{1}\rho_{1}} \left( -2\ell_2 - p_2\right)^{\beta_1} \right] d^{(\ell_2)}_{\rho_{1}\rho_{2}} d^{(p_2)}_{\beta_{1} \beta_{2}} d^{(\ell_2+p_2)}_{\sigma_{1} \alpha_{3}} d^{(\ell_2+p_2)}_{\sigma_{2} \alpha_{2}} \\
& = & \frac{64 [-g^{\mu\nu}_{\perp\perp}](q^-)^2}{y\left(1-y+\eta y\right)(1+\eta y)^2} \left[ \frac{ \left(1+\eta y\right)^2 y}{\left(1-y + \eta y\right)} + \frac{(1+2\eta y)(1-y+\eta y)}{y} + (1+\eta^2)y(1-y+\eta y) \right] \left[ \left\{\left(1+\eta y\right)\pmb{\ell}_{2\perp} - y\pmb{k}_{\perp}\right\}^2\right]. \nonumber
\end{eqnarray}
The final expression of the hadronic tensor for the central-cut diagram [Fig.~\ref{fig:kernel-2_ab_amplt_square}(b)] is
\begin{eqnarray}
    W^{\mu\nu}_{2,c} & =& \left[2C_F C^2_A \right] \sum_f  2[-g^{\mu\nu}_{\perp \perp }] e^2_f g^4_s \int d (\Delta X^{-})   e^{iq^{+}(\Delta X^{-} )}     
  \left\langle P \left| \bar{\psi}_{_f}( \Delta X^{-}) \frac{\gamma^{+}}{4} \psi_{_f}(0)\right| P\right\rangle \nonumber \\
  & \times &  \int  d \zeta^{-} d (\Delta z^{-}) d^2 \Delta z_{\perp} \frac{dy}{2\pi} \frac{d^2 \ell_{2\perp}}{(2\pi)^2} \frac{d^2 k_{\perp}}{(2\pi)^2}  e^{-i (\Delta z^{-})\mathcal{H}^{(\ell_2 , p_2)}_{0} }     e^{i \pmb{k}_{\perp} \cdot \Delta\pmb{z}_{\perp}} \theta(\zeta^-)\nonumber\\% \theta( x^{-}_{1}-z^{-}_2) \theta( x^{-}_{2}-z^{-}_3) \nonumber \\
  & \times & \left[ \frac{ \left(1+\eta y\right)^2 y}{\left(1-y + \eta y\right)} + \frac{(1+2\eta y)(1-y+\eta y)}{y} + (1+\eta^2)y(1-y+\eta y) \right] \nonumber \\
  & \times &  \frac{ 1 }{(1+\eta y)^2 q^-} \frac{1}{  \left[(1+\eta y)\pmb{\ell}_{2\perp} - y\pmb{k}_{\perp} \right]^2 } \left\langle P_{A-1} \left| \bar{\psi}_{_f}(\zeta^-, 0) \frac{\gamma^+}{4} \psi_{_f}(\zeta^-, \Delta z^-, \Delta z_{\perp})\right| P_{A-1} \right\rangle,   
\label{eq:kernel2_wfinal_b}
\end{eqnarray}
where $\mathcal{H}^{(\ell_2,p_2)}_{0}$ is given in Eq.~(\ref{eq:HL2P2M0_append}). Note that the quantity in the square bracket in the third line of Eq.~(\ref{eq:kernel2_wfinal_b}) is the medium-modified gluon-gluon splitting function.
%%%%%%%%%%%%%%%%%%%%%%%%%%%%%%%%%%%%%%%%%%%%%%%%%%%%%%%%%%%%%%%%%%%%%%%%%%%%%%%%%%%%%
%%%%%%%%%%%%%%%%%%%%%%%%%%%%%%%%%%%%%%%%%%%%%%%%%%%%%%%%%%%%%%%%%%%%%%%%%%%%%%%%%%%%%
\subsection{Figures~\ref{fig:kernel-2_all}(c) and \ref{fig:kernel-2_all}(d)}
%%%%%%%%%%%%%%%%%%%%%%%%%%%%%%%%%%%%%%%%%%%%%%%%%%%%%%%%%%%%%%%%%%%%%%%%%%%%%%%%%%%%%
%%%%%%%%%%%%%%%%%%%%%%%%%%%%%%%%%%%%%%%%%%%%%%%%%%%%%%%%%%%%%%%%%%%%%%%%%%%%%%%%%%%%%
%
\begin{figure}[!h]
    \centering 
    \begin{subfigure}[t]{0.495\textwidth}
        \centering        \includegraphics[width=\textwidth]{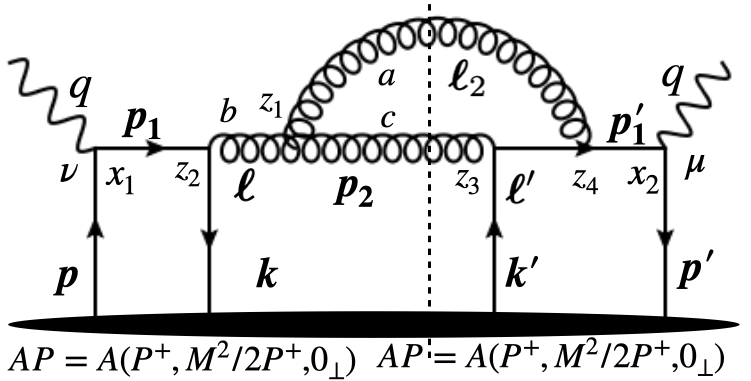}
        \caption{Interference diagram. }
    \end{subfigure}%
    \begin{subfigure}[t]{0.495\textwidth}
        \centering        \includegraphics[width=\textwidth]{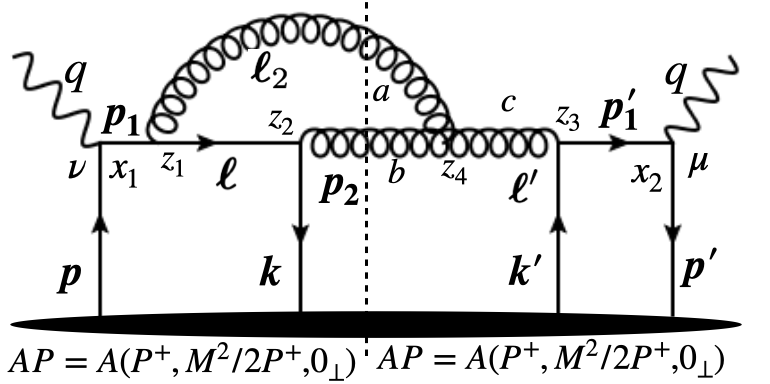}
        \caption{Complex conjugate of the diagram on the left panel.}
    \end{subfigure}
\caption{A forward scattering diagram contributing to kernel-2.}
\label{fig:kernel-2_cd_interference}
\end{figure}

Hereafter, we discuss interference diagrams contributing to kernel-2. 
The hadronic tensor for the interference diagram in Fig.~\ref{fig:kernel-2_cd_interference}(a) is given as
\begin{eqnarray}
W^{\mu\nu}_{2,c_1} & =& \sum_f e^2_f g^4_{s} \int d^4 x_{1} d^4 x_{2} d^4 z_{2} d^4 z_{3} \int \frac{d^4 p}{(2\pi)^4} \frac{d^4 p'}{(2\pi)^4}  \frac{d^4 \ell_{2}}{(2\pi)^4} \frac{d^4 p_{2}}{(2\pi)^4} e^{-ip'x_2} e^{ipx_1}\left\langle P \left| \bar{\psi}_{_f}(x_2) \frac{\gamma^{+}}{4} \psi_{_f}(x_1)\right| P\right\rangle {\rm Tr} [t^a t^c t^b] \nonumber\\
  & \times & e^{i\left(q+p'-p_{2}-\ell_{2}\right) z_{3}} e^{i(\ell_{2}+p_{2} - q - p)z_{2}}\left\langle P_{A-1} \left| \bar{\psi}_{_f}(z_2)\frac{\gamma^+}{4}\psi_{_f}(z_3)\right| P_{A-1} \right\rangle (2 \pi) \delta\left(\ell_{2}^{2}\right) (2 \pi)  \delta\left(p_{2}^{2}\right) d^{(\ell_2)}_{\rho_{1}\alpha_{4}} d^{(p_2)}_{\beta_{1} \alpha_{3}} d^{(\ell_2+p_2)}_{\sigma_{1} \alpha_{2}} \nonumber \\
& \times & \frac{ {\rm Tr} \left[\gamma^- \gamma^{\mu} \left(\slashed{q}+\slashed{p}'\right) \gamma^{\alpha_4} (\slashed{q}+\slashed{p}'-\slashed{\ell}_2) \gamma^{\alpha_3}\gamma^{-} \gamma^{\alpha_2} \left( \slashed{q} + \slashed{p}  \right)  \gamma^{\nu} \right]  
}{  \left[\left(q+p'\right)^{2}-i\epsilon\right] \left[\left(q+p\right)^2  + i\epsilon\right]  \left[\left(q+p'-\ell_2\right)^2 - i\epsilon\right] \left[\left(\ell_2+p_2\right)^2+i\epsilon\right]  }  \nonumber \\
& \times & (-if^{abc}) \left[ g^{\rho_{1}\sigma_{1}} \left(-2\ell_2 -p_2\right)^{\beta_1} + g^{\sigma_{1}\beta_{1}} \left(\ell_2+2p_2\right)^{\rho_1} + g^{\beta_{1}\rho_{1}} \left(-p_2 +\ell_2\right)^{\sigma_1} \right]  .
\label{eq:kernel2_wiinitial_c}
\end{eqnarray}
Equation (\ref{eq:kernel2_wiinitial_c}) has singularity arising from the denominator of the quark propagator with momentum $p_{1}$, $p'_{1}$, and $\ell'$. We identify one pole for the momentum variable $p^+$ and two poles for $p'^+$. 
 The contour integration for $p^+$ in the complex plane is given by
\begin{equation}
    C_{1} = \oint \frac{dp^+}{(2\pi)} \frac{e^{ip^+(x^{-}_{1}-z^-_2)} }{\left[\left(q+p\right)^2  + i\epsilon\right]} =  \oint \frac{dp^+}{(2\pi)} \frac{e^{ip^+(x^{-}_{1}-z^{-}_{2})} }{ 2q^{-}[ q^{+}+p^{+} + i\epsilon]} = \frac{(2\pi i)}{2\pi}\frac{\theta\left( x^{-}_{1} - z^{-}_{2}\right)}{2q^{-}}  e^{-iq^{+} (x^{-}_{1}-z^{-}_{2})}. 
\end{equation}
Similarly, the contour integration for momentum $p'^+$ is carried out 
\begin{equation}
\begin{split}
C_{2} & = \oint \frac{dp'^{+}}{(2\pi)} \frac{e^{-ip'^{+}(x^{-}_{2}-z^{-}_{3})}}{\left[\left(q+p'\right)^{2} - i \epsilon\right]\left[\left(q+p'-\ell_{2}\right)^{2}  - i\epsilon\right]} \\
      & = \frac{(-2\pi i)}{2\pi} \frac{\theta(x^{-}_{2} - z^{-}_{3})}{4q^{-}(q^{-}-\ell^{-}_{2})}  e^{iq^{+}\left(x^{-}_{2}-z^{-}_{3}\right)}\left[ \frac{ -1 + e^{-i \mathcal{G}^{(\ell_2)}_0(x^{-}_{2}-z^{-}_{3})} }{ \mathcal{G}^{(\ell_2)}_0}  \right],
\end{split}
\end{equation}
where $ \mathcal{G}^{(\ell_2)}_0$ is defined in Eq.~(\ref{eq:GL20_append}). Simplifying the trace in the third line of Eq.~(\ref{eq:kernel2_wiinitial_c}) yields
\begin{eqnarray}
&  {\rm Tr} & \left[\gamma^- \gamma^{\mu} \left(\slashed{q}+\slashed{p}'\right) \gamma^{\alpha_4} (\slashed{q}+\slashed{p}'-\slashed{\ell}_2) \gamma^{\alpha_3}\gamma^{-} \gamma^{\alpha_2} \left( \slashed{q} + \slashed{p}  \right)  \gamma^{\nu} \right] d^{(\ell_2)}_{\rho_{1}\alpha_{4}} d^{(p_2)}_{\beta_{1} \alpha_{3}} d^{(\ell_2+p_2)}_{\sigma_{1} \alpha_{2}} 
   \nonumber\\
& \times & \left[ g^{\rho_{1}\sigma_{1}} \left(-2\ell_2 -p_2\right)^{\beta_1} + g^{\sigma_{1}\beta_{1}} \left(\ell_2+2p_2\right)^{\rho_1} + g^{\beta_{1}\rho_{1}} \left(-p_2 +\ell_2\right)^{\sigma_1} \right] \\
& = & \frac{-32 [-g^{\mu\nu}_{\perp\perp}](q^-)^2}{(1+\eta y)} \left[ \frac{1 + \left(1-y + \eta y\right)^3}{y^2(1-y + \eta y)} \right] \left[ \left\{\left(1+\eta y\right)\pmb{\ell}^2_{2\perp} - y\pmb{\ell}_{2\perp}\cdot\pmb{k}_{\perp}\right\}  \right]. \nonumber
\end{eqnarray}

The final expression of the hadronic tensor for the central-cut diagram [Fig.~\ref{fig:kernel-2_cd_interference}(a)] is
\begin{eqnarray}
    W^{\mu\nu}_{2,c_1} & =& \left[\frac{C_F C^2_A}{2} \right] \sum_f  2[-g^{\mu\nu}_{\perp \perp }] e^2_f g^4_s \int d (\Delta X^{-})   e^{iq^{+}(\Delta X^{-} )}     
  \left\langle P \left| \bar{\psi}_{_f}( \Delta X^{-}) \frac{\gamma^{+}}{4} \psi_{_f}(0)\right| P\right\rangle \nonumber \\
  & \times &  \int  d \zeta^{-} d (\Delta z^{-}) d^2 \Delta z_{\perp} \frac{dy}{2\pi} \frac{d^2 \ell_{2\perp}}{(2\pi)^2} \frac{d^2 k_{\perp}}{(2\pi)^2}  e^{-i (\Delta z^{-})\mathcal{H}^{(\ell_2 , p_2)}_{0} }     e^{i \pmb{k}_{\perp} \cdot \Delta\pmb{z}_{\perp}}  \theta( x^{-}_{1}-z^{-}_2) \theta( x^{-}_{2}-z^{-}_3)  \nonumber \\
  & \times & \left[ \frac{1 + \left(1-y + \eta y\right)^3}{y(1-y + \eta y)}  \right] \frac{ \left[ -1 + e^{-i\mathcal{G}^{(\ell_2)}_{0}(x^{-}_{2}-z^{-}_{3})}\right] }{(1+\eta y) q^-} \frac{\left[(1+\eta y)\pmb{\ell}^2_{2\perp} - y\pmb{\ell}_{2\perp}\cdot\pmb{k}_{\perp} \right]}{ \pmb{\ell}^2_{2\perp}  \left[(1+\eta y)\pmb{\ell}_{2\perp} - y\pmb{k}_{\perp} \right]^2 } \nonumber\\
  & \times &   \left\langle P_{A-1} \left| \bar{\psi}_{_f}(\zeta^-, 0) \frac{\gamma^+}{4} \psi_{_f}(\zeta^-, \Delta z^-, \Delta z_{\perp})\right| P_{A-1} \right\rangle,
  \label{eqn:kernel2_wfinal_c_this}
\end{eqnarray}
where $\mathcal{H}^{(\ell_2,p_2)}_{0}$ is given in Eq.~(\ref{eq:HL2P2M0_append}).

Since the final state [Fig.~\ref{fig:kernel-2_cd_interference}(a)] consists of two identical gluons, the momentum of the two gluons could be interchanged. This gives rise to additional contributions to the hadronic tensor. After performing $p_2\leftrightarrow\ell_2$, the resulting hadronic tensor is given as
\begin{eqnarray}
    W^{\mu\nu}_{2,c_2} & =& \left[\frac{C_F C^2_A}{2} \right] \sum_f  2[-g^{\mu\nu}_{\perp \perp }] e^2_f g^4_s \int d (\Delta X^{-})   e^{iq^{+}(\Delta X^{-} )}     
  \left\langle P \left| \bar{\psi}_{_f}( \Delta X^{-}) \frac{\gamma^{+}}{4} \psi_{_f}(0)\right| P\right\rangle \nonumber \\
  & \times &  \int  d \zeta^{-} d (\Delta z^{-}) d^2 \Delta z_{\perp} \frac{dy}{2\pi} \frac{d^2 \ell_{2\perp}}{(2\pi)^2} \frac{d^2 k_{\perp}}{(2\pi)^2}  e^{-i (\Delta z^{-})\mathcal{H}^{(\ell_2 , p_2)}_{0} }     e^{i \pmb{k}_{\perp} \cdot \Delta\pmb{z}_{\perp}}  \theta( x^{-}_{1}-z^{-}_2) \theta( x^{-}_{2}-z^{-}_3)  \nonumber \\
  & \times & \left[ \frac{1 + y^3}{y(1-y + \eta y)}  \right] \frac{ \left[ -1 + e^{-i\mathcal{G}^{(p_2)}_{0}(x^{-}_{2}-z^{-}_{3})}\right] }{(1+\eta y) q^-} \frac{(\pmb{\ell}_{2\perp}-\pmb{k}_{\perp}) \cdot \left\{(1+\eta y)\pmb{\ell}_{2\perp} - y\pmb{k}_{\perp} \right\}}{ (\pmb{\ell}_{2\perp}-\pmb{k}_{\perp})^2  \left[(1+\eta y)\pmb{\ell}_{2\perp} - y\pmb{k}_{\perp} \right]^2 } \nonumber\\
  & \times &   \left\langle P_{A-1} \left| \bar{\psi}_{_f}(\zeta^-, 0) \frac{\gamma^+}{4} \psi_{_f}(\zeta^-, \Delta z^-, \Delta z_{\perp})\right| P_{A-1} \right\rangle,   
\label{eq:kernel2_wfinal_c_l2p2_interchange}
\end{eqnarray}
where $\mathcal{G}^{(p_2)}_{0}$ is defined in Eq.~(\ref{eq:GP2M0_append}).

Note that the diagrams in Fig.~\ref{fig:kernel-2_cd_interference}(a) and Fig.~\ref{fig:kernel-2_cd_interference}(b) are complex-conjugate of each other. They differ only in contour integration over variable $p^{+}$ and $p'^{+}$.
 The calculation of the hadronic tensor for Fig.~\ref{fig:kernel-2_cd_interference}(b) involves the contour integration for $p^{+}$ and is given as
\begin{equation}
\begin{split}
C_{1} & = \oint \frac{dp^{+}}{(2\pi)} \frac{e^{ip^{+}\left(x^{-}_{1}-z^{-}_{2}\right)}}{\left[\left(q+p\right)^{2}  + i \epsilon\right]\left[\left(q+p-\ell_{2}\right)^{2}  + i\epsilon\right]} \\
      & = \frac{(2\pi i)}{2\pi} \frac{\theta\left(x^{-}_{1} - z^{-}_{2}\right)}{4q^{-}(q^{-}-\ell^{-}_{2})}  e^{-iq^{+}(x^{-}_{1}-z^{-}_{2})}\left[ \frac{ -1 + e^{i \mathcal{G}^{(\ell_2)}_0(x^{-}_{1}-z^{-}_{2})} }{ \mathcal{G}^{(\ell_2)}_0}  \right],
\end{split}
\end{equation}
where $ \mathcal{G}^{(\ell_2)}_0 $ is defined in Eq.~(\ref{eq:GL20_append}). Similarly, the contour integration for momentum $p'^+$ is carried out as 
\begin{equation}
    C_{2} = \oint \frac{dp'^+}{(2\pi)} \frac{e^{-ip'^+\left(x^{-}_{2}-z^{-}_3\right)} }{\left[\left(q+p'\right)^2  -i\epsilon\right]} = \frac{(-2\pi i)}{2\pi} \frac{\theta( x^{-}_{2} - z^{-}_{3} )}{2q^-}  e^{iq^{+}(x^{-}_{2}-z^{-}_3)}.  
\end{equation}
The final expression of the hadronic tensor for the central-cut diagram [Fig.~\ref{fig:kernel-2_cd_interference}(b)] is
\begin{eqnarray}
    W^{\mu\nu}_{2,c_3} & =& \left[\frac{C_F C^2_A}{2} \right] \sum_f  2[-g^{\mu\nu}_{\perp \perp }] e^2_f g^4_s \int d (\Delta X^{-})   e^{iq^{+}(\Delta X^{-} )}     
  \left\langle P \left| \bar{\psi}_{_f}( \Delta X^{-}) \frac{\gamma^{+}}{4} \psi_{_f}(0)\right| P\right\rangle \nonumber \\
  & \times &  \int  d \zeta^{-} d (\Delta z^{-}) d^2 \Delta z_{\perp} \frac{dy}{2\pi} \frac{d^2 \ell_{2\perp}}{(2\pi)^2} \frac{d^2 k_{\perp}}{(2\pi)^2}  e^{-i (\Delta z^{-})\mathcal{H}^{(\ell_2 , p_2)}_{0} }     e^{i \pmb{k}_{\perp} \cdot \Delta\pmb{z}_{\perp}}  \theta( x^{-}_{1}-z^{-}_2) \theta( x^{-}_{2}-z^{-}_3)  \nonumber \\
  & \times & \left[ \frac{1 + \left(1-y + \eta y\right)^3}{y(1-y + \eta y)}  \right] \frac{ \left[ -1 + e^{i\mathcal{G}^{(\ell_2)}_{0}(x^{-}_{1}-z^{-}_{2})}\right] }{(1+\eta y) q^-} \frac{\left[(1+\eta y)\pmb{\ell}^2_{2\perp} - y\pmb{\ell}_{2\perp}\cdot\pmb{k}_{\perp} \right]}{ \pmb{\ell}^2_{2\perp}  \left[(1+\eta y)\pmb{\ell}_{2\perp} - y\pmb{k}_{\perp} \right]^2 } \nonumber \\
  & \times &   \left\langle P_{A-1} \left| \bar{\psi}_{_f}(\zeta^-, 0) \frac{\gamma^+}{4} \psi_{_f}(\zeta^-, \Delta z^-, \Delta z_{\perp})\right| P_{A-1} \right\rangle,  
\label{eq:kernel2_wfinal_d}
\end{eqnarray}
where $\mathcal{H}^{(\ell_2,p_2)}_{0}$ is given in Eq.~(\ref{eq:HL2P2M0_append}).
Since the final state [Fig.~\ref{fig:kernel-2_cd_interference}(b)] consists of two identical gluons, the momentum of the two gluons could be interchanged. This gives rise to additional contributions to the hadronic tensor. After performing $p_2\leftrightarrow\ell_2$, the resulting hadronic tensor is given as
\begin{eqnarray}
    W^{\mu\nu}_{2,c_4} & =& \left[\frac{C_F C^2_A}{2} \right] \sum_f  2[-g^{\mu\nu}_{\perp \perp }] e^2_f g^4_s \int d (\Delta X^{-})   e^{iq^{+}(\Delta X^{-} )}     
  \left\langle P \left| \bar{\psi}_{_f}( \Delta X^{-}) \frac{\gamma^{+}}{4} \psi_{_f}(0)\right| P\right\rangle \nonumber \\
  &  & \times   \int  d \zeta^{-} d (\Delta z^{-}) d^2 \Delta z_{\perp} \frac{dy}{2\pi} \frac{d^2 \ell_{2\perp}}{(2\pi)^2} \frac{d^2 k_{\perp}}{(2\pi)^2}  e^{-i (\Delta z^{-})\mathcal{H}^{(\ell_2 , p_2)}_{0} }     e^{i \pmb{k}_{\perp} \cdot \Delta\pmb{z}_{\perp}}  \theta( x^{-}_{1}-z^{-}_2) \theta( x^{-}_{2}-z^{-}_3)  \nonumber \\
  & & \times  \left[ \frac{1 + y^3}{y(1-y + \eta y)}  \right] \frac{ \left[ -1 + e^{i\mathcal{G}^{(p_2)}_{0}(x^{-}_{1}-z^{-}_{2})}\right] }{(1+\eta y) q^-} \frac{(\pmb{\ell}_{2\perp}-\pmb{k}_{\perp}) \cdot \left\{(1+\eta y)\pmb{\ell}_{2\perp} - y\pmb{k}_{\perp} \right\}}{ (\pmb{\ell}_{2\perp}-\pmb{k}_{\perp})^2  \left[(1+\eta y)\pmb{\ell}_{2\perp} - y\pmb{k}_{\perp} \right]^2 }   \nonumber \\
  & & \times    \left\langle P_{A-1} \left| \bar{\psi}_{_f}(\zeta^-, 0) \frac{\gamma^+}{4} \psi_{_f}(\zeta^-, \Delta z^-, \Delta z_{\perp})\right| P_{A-1} \right\rangle,
\label{eq:kernel2_wfinal_d_l2p2_interchange}
\end{eqnarray}
where $\mathcal{G}^{(p_2)}_{0}$ is defined in Eq.~(\ref{eq:GP2M0_append}). The contributions to the hadronic tensor from Eq.~(\ref{eqn:kernel2_wfinal_c_this}) and Eq.~(\ref{eq:kernel2_wfinal_d}) can be added together
\begin{eqnarray}
    W^{\mu\nu}_{2,c_1+c_3}  &= &   -\left[\frac{C_F C^2_A}{2} \right] \sum_f  2[-g^{\mu\nu}_{\perp \perp }] e^2_f g^4_s \int d (\Delta X^{-})   e^{iq^{+}(\Delta X^{-} )}     
  \left\langle P \left| \bar{\psi}_{_f}( \Delta X^{-}) \frac{\gamma^{+}}{4} \psi_{_f}(0)\right| P\right\rangle \nonumber \\
  & & \times   \int  d \zeta^{-} d (\Delta z^{-}) d^2 \Delta z_{\perp} \frac{dy}{2\pi} \frac{d^2 \ell_{2\perp}}{(2\pi)^2} \frac{d^2 k_{\perp}}{(2\pi)^2}  e^{-i (\Delta z^{-})\mathcal{H}^{(\ell_2 , p_2)}_{0} }     e^{i \pmb{k}_{\perp} \cdot \Delta\pmb{z}_{\perp}}  \theta(\zeta^-)  \nonumber \\
  & & \times  \left[ \frac{1 + \left(1-y + \eta y\right)^3}{y(1-y + \eta y)}  \right] \frac{ \left[2 - 2\cos\left\{\mathcal{G}^{(\ell_2)}_{0}\zeta^{-}\right\}\right] }{(1+\eta y) q^-} \frac{\left[(1+\eta y)\pmb{\ell}^2_{2\perp} - y\pmb{\ell}_{2\perp}\cdot\pmb{k}_{\perp} \right]}{ \pmb{\ell}^2_{2\perp}  \left[(1+\eta y)\pmb{\ell}_{2\perp} - y\pmb{k}_{\perp} \right]^2 } \nonumber\\
  & & \times    \left\langle P_{A-1} \left| \bar{\psi}_{_f}(\zeta^-, 0) \frac{\gamma^+}{4} \psi_{_f}(\zeta^-, \Delta z^-, \Delta z_{\perp})\right| P_{A-1} \right\rangle. 
\label{eq:kernel2_wfinal_c_d}
\end{eqnarray}
Similarly the contributions to the hadronic tensor from Eq.~(\ref{eq:kernel2_wfinal_c_l2p2_interchange}) and Eq.~(\ref{eq:kernel2_wfinal_d_l2p2_interchange}) can be added together
\begin{eqnarray}
    W^{\mu\nu}_{2,c_2+c_4} & =& -\left[\frac{C_F C^2_A}{2} \right] \sum_f  2[-g^{\mu\nu}_{\perp \perp }] e^2_f g^4_s \int d (\Delta X^{-})   e^{iq^{+}(\Delta X^{-} )}     
  \left\langle P \left| \bar{\psi}_{_f}( \Delta X^{-}) \frac{\gamma^{+}}{4} \psi_{_f}(0)\right| P\right\rangle \nonumber \\
  & \times &  \int  d \zeta^{-} d (\Delta z^{-}) d^2 \Delta z_{\perp} \frac{dy}{2\pi} \frac{d^2 \ell_{2\perp}}{(2\pi)^2} \frac{d^2 k_{\perp}}{(2\pi)^2}  e^{-i (\Delta z^{-})\mathcal{H}^{(\ell_2 , p_2)}_{0} }     e^{i \pmb{k}_{\perp} \cdot \Delta\pmb{z}_{\perp}}  \theta(\zeta^{-})  \nonumber \\
  & \times & \left[ \frac{1 + y^3}{y(1-y + \eta y)}  \right] \frac{ \left[ 2 - 2\cos\left\{\mathcal{G}^{(p_2)}_{0}\zeta^{-}\right\}\right] }{(1+\eta y) q^-} \frac{\left(\pmb{\ell}_{2\perp}-\pmb{k}_{\perp}\right)\cdot\left\{\left(1+\eta y\right)\pmb{\ell}_{2\perp} - y\pmb{k}_{\perp} \right\}}{ \left(\pmb{\ell}_{2\perp}-\pmb{k}_{\perp}\right)^2  \left[(1+\eta y)\pmb{\ell}_{2\perp} - y\pmb{k}_{\perp} \right]^2 } \nonumber\\
  & \times &   \left\langle P_{A-1} \left| \bar{\psi}_{_f}(\zeta^-, 0) \frac{\gamma^+}{4} \psi_{_f}(\zeta^-, \Delta z^-, \Delta z_{\perp})\right| P_{A-1} \right\rangle.
\label{eq:kernel2_wfinal_cd_l2p2_interchange}
\end{eqnarray}
\begin{figure}[!h]
    \centering
    \includegraphics[width=0.54\textwidth]{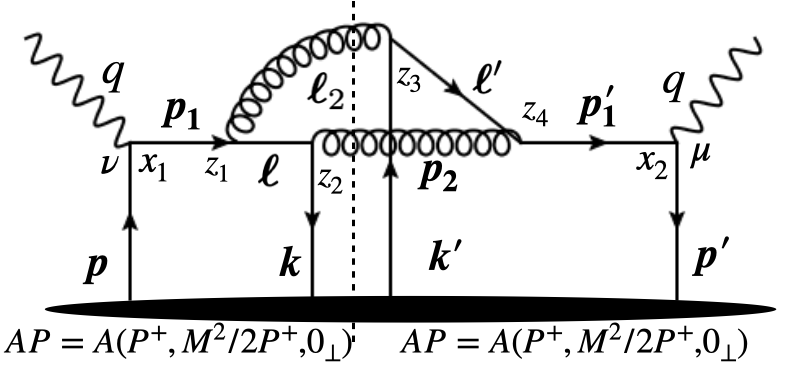}
    \caption{An interference diagram contributing to kernel-2. The cut-line (i.e., dashed line) represents the final state.}
    \label{fig:kernel-2_e_interference}
\end{figure}
%
%%%%%%%%%%%%%%%%%%%%%%%%%%%%%%%%%%%%%%%%%%%%%%%%%%%%%%%%%%%%%%%%%%%%%%%%%%%%%%%%%%%%%
%%%%%%%%%%%%%%%%%%%%%%%%%%%%%%%%%%%%%%%%%%%%%%%%%%%%%%%%%%%%%%%%%%%%%%%%%%%%%%%%%%%%%
\subsection{Figure~\ref{fig:kernel-2_all}(e)}
%%%%%%%%%%%%%%%%%%%%%%%%%%%%%%%%%%%%%%%%%%%%%%%%%%%%%%%%%%%%%%%%%%%%%%%%%%%%%%%%%%%%%
%%%%%%%%%%%%%%%%%%%%%%%%%%%%%%%%%%%%%%%%%%%%%%%%%%%%%%%%%%%%%%%%%%%%%%%%%%%%%%%%%%%%%
The final interference diagram contributing to kernel-2 is shown in Fig.~\ref{fig:kernel-2_e_interference}. The hadronic tensor for Fig.~\ref{fig:kernel-2_e_interference} is
\begin{equation}
\begin{split}
W^{\mu\nu}_{2,c_1} & = \sum_f  e^2_f g^4_{s} \int d^4 x_{1} d^4 x_{2} d^4 z_{2} d^4 z_{3} \int \frac{d^4 p}{(2\pi)^4} \frac{d^4 p'}{(2\pi)^4}  \frac{d^4 \ell_{2}}{(2\pi)^4} \frac{d^4 p_{2}}{(2\pi)^4} e^{-ip'x_2} e^{ipx_1}\left\langle P \left| \bar{\psi}_{_f}(x_2) \frac{\gamma^{+}}{4} \psi_{_f}(x_1)\right| P\right\rangle \delta^{ac}  \delta^{bd} {\rm Tr}[t^{d}t^{c}t^{b}t^{a}] \\
  & \times e^{i\left(q+p'-p_{2}-\ell_{2}\right) z_{3}} e^{i(\ell_{2}+p_{2} - q - p)z_{2}} \left\langle P_{A-1} \left| \bar{\psi}_{_f}(z_2)\frac{\gamma^+}{4}\psi_{_f}(z_3)\right| P_{A-1} \right\rangle d^{(\ell_2)}_{\sigma_{1} \sigma_{3}} d^{(p_2)}_{\sigma_{2} \sigma_{4}} (2 \pi) \delta\left(\ell_{2}^{2}\right) (2 \pi)  \delta\left(p_{2}^{2}\right) \\
& \times \frac{ {\rm Tr} \left[\gamma^- \gamma^{\mu} \left(\slashed{q}+\slashed{p}'\right) \gamma^{\sigma_4}\left(\slashed{q}+\slashed{p}' - \slashed{p}_2\right)\gamma^{\sigma_{3}} \gamma^{-} \gamma^{\sigma_2} \left( \slashed{q} + \slashed{p} -\slashed{\ell}_2 \right) \gamma^{\sigma_1} \left( \slashed{q} + \slashed{p} \right) \gamma^{\nu} \right] 
}{  \left[\left(q+p'\right)^{2}-i\epsilon\right] \left[\left(q+p\right)^2  + i\epsilon\right]  \left[\left(q+p'-p_2\right)^2 - i\epsilon\right] \left[\left(q+p -\ell_2\right)^2+i\epsilon\right]  }.
\end{split} \label{eq:kernel2_wiinitial_e}
\end{equation}
The above expression has singularity when the denominator of the propagator for $p_1$, $\ell$, $\ell'$ and $p'_1$ becomes on-shell. It contains two simple poles for $p^+$ and $p'^+$. The contour integration for $p^+$ gives
\begin{equation}
\begin{split}
C_{1} & = \oint \frac{dp^{+}}{(2\pi)} \frac{e^{ip^{+}(x^{-}_{1}-z^{-}_{2})}}{\left[\left(q+p\right)^{2}  + i \epsilon\right]\left[\left(q+p-\ell_{2}\right)^{2}  + i\epsilon\right]} \\
 %     & = \oint \frac{dp^{+}}{(2\pi)} \frac{e^{ip^{+}(x^{-}_{1} - z^{-}_{2}) }}{2q^{-}\left[q^{+} + p^{+}  + i \epsilon\right] 2(q^{-} -\ell^{-}_{2} )\left[ q^{+} + p^{+} -\ell^{+}_{2} - \frac{\pmb{\ell}^{2}_{2\perp}}{2(q^{-}-p^{-}_{2})} + i \epsilon\right]}   \\
      & = \frac{(2\pi i)}{2\pi} \frac{\theta(x^{-}_{1} - z^{-}_{2})}{4q^{-}(q^{-}-\ell^{-}_{2})}  e^{-iq^{+} (x^{-}_{1}-z^{-}_{2})}\left[ \frac{ -1 + e^{i \mathcal{G}^{(\ell_2)}_0(x^{-}_{1}-z^{-}_{2})} }{ \mathcal{G}^{(\ell_2)}_0}  \right],
\end{split}
\end{equation}

where $\mathcal{G}^{(\ell_2)}_0$ is defined in Eq.~(\ref{eq:GL20_append}). Similarly, the contour integration for $p'^+$ gives
\begin{equation}
\begin{split}
C_{2} & = \oint \frac{dp'^{+}}{(2\pi)} \frac{e^{-ip'^{+}(x^{-}_{2}-z^{-}_{3})}}{\left[\left(q+p'\right)^{2} - i\epsilon\right]\left[\left(q+p'-p_{2}\right)^{2}  - i\epsilon\right]} \\
%& = \oint \frac{dp'^{+}}{(2\pi)} \frac{e^{-ip'^{+}(x^{-}_{2} - z^{-}_{3}) }}{2q^{-}[ q^{+} + p'^{+}   - i \epsilon] 2(q^{-} -p^{-}_{2} )\left[ q^{+} + p'^{+} -p^{+}_{2} - \frac{\pmb{p}^{2}_{2\perp}}{2(q^{-}-p^{-}_{2})} - i \epsilon\right]}   \\
& = \frac{(-2\pi i)}{2\pi} \frac{\theta(x^{-}_{2} - z^{-}_{3})}{4q^{-}(q^{-}-p^{-}_{2})}  e^{iq^{+}(x^{-}_{2}-z^{-}_{3})}\left[ \frac{ -1 + e^{-i \mathcal{G}^{(p_2)}_0(x^{-}_{2}-z^{-}_{3})} }{ \mathcal{G}^{(p_2)}_0}  \right],
\end{split}
\end{equation}
where $\mathcal{G}^{(p_2)}_0 $ is defined in Eq.~(\ref{eq:GP2M0_append}). The trace in the third line of Eq.~(\ref{eq:kernel2_wiinitial_e}) simplifies to
\begin{equation}
    \begin{split}
         & {\rm Tr} \left[\gamma^- \gamma^{\mu} \left(\slashed{q}+\slashed{p}'\right) \gamma^{\sigma_4}\left(\slashed{q}+\slashed{p}' - \slashed{p}_2\right)\gamma^{\sigma_{3}} \gamma^{-} \gamma^{\sigma_2} \left( \slashed{q} + \slashed{p} -\slashed{\ell}_2 \right) \gamma^{\sigma_1} \left( \slashed{q} + \slashed{p} \right) \gamma^{\nu} \right] d^{(p_2)}_{\sigma_{2} \sigma_{4}} d^{(\ell_2)}_{\sigma_{3} \sigma_{1}} \\
         & = \frac{32 \left(q^-\right)^2 \left[ -g^{\mu\nu}_{\perp\perp}\right] (1-y+2\eta y) }{(1-y+\eta y)y} \left[ -\pmb{\ell}^{2}_{2\perp} +  \pmb{\ell}_{2\perp}\cdot \pmb{k}_{\perp} \right].
    \end{split}
\end{equation}
The final expression of the hadronic tensor for Fig.~\ref{fig:kernel-2_e_interference} is given by
\begin{equation}
\begin{split}
    W^{\mu\nu}_{2,c_1} & =  \left[\left(C_F - \frac{C_A}{2}\right)C_F C_A\right] \sum_f  2[-g^{\mu\nu}_{\perp \perp }] e^2_f g^4_s \int d (\Delta X^{-})   e^{iq^{+}(\Delta X^{-} )}     
  \left\langle P \left| \bar{\psi}_{_f}( \Delta X^{-}) \frac{\gamma^{+}}{4} \psi_{_f}(0)\right| P\right\rangle  \\
  & \times   \int  d \zeta^{-} d (\Delta z^{-}) d^2 \Delta z_{\perp} \frac{dy}{2\pi} \frac{d^2 \ell_{2\perp}}{(2\pi)^2} \frac{d^2 k_{\perp}}{(\pi)^2} \left[ -1 + e^{i\mathcal{G}^{(\ell_2)}_{0}(x^{-}_{1} - z^{-}_{2})} \right] \left[ -1 + e^{-i\mathcal{G}^{(p_2)}_{0}(x^{-}_{2} - z^{-}_{3})} \right] e^{-i (\Delta z^{-})\mathcal{H}^{(\ell_2 , p_2)}_{0} }    \\
  & \times \frac{\theta( x^{-}_{1}-z^{-}_2) \theta( x^{-}_{2}-z^{-}_3)  }{  (1-y+\eta y)q^-  } \frac{\left[ -\pmb{\ell}^{2}_{2\perp} +  \pmb{\ell}_{2\perp}\cdot \pmb{k}_{\perp} \right]}{\left(\pmb{\ell}_{2\perp}-\pmb{k}_{\perp} \right)^2\pmb{\ell}^2_{2\perp} }    e^{i \pmb{k}_{\perp} \cdot \Delta\pmb{z}_{\perp}} \left\langle P_{A-1} \left| \bar{\psi}_{_f}(\zeta^-, 0) \frac{\gamma^+}{4} \psi_{_f}(\zeta^-, \Delta z^-, \Delta z_{\perp})\right| P_{A-1} \right\rangle  \\
& \times \left[ \frac{(1-y+2\eta y)}{y} \right], 
\end{split}     
\label{eq:kernel2_wfinal_e}
\end{equation}
where $\mathcal{G}^{(\ell_2)}_0$, $\mathcal{G}^{(p_2)}_0$, and $\mathcal{H}^{(\ell_2,p_2)}_{0}$ are given in Eq.~(\ref{eq:GL20_append}), Eq.~(\ref{eq:GP2M0_append}), and Eq.~(\ref{eq:HL2P2M0_append}), respectively.
Since the final state [Fig.~\ref{fig:kernel-2_e_interference}] consists of two identical gluons, the momentum of the two gluons could be interchanged. This gives rise to additional contributions to the hadronic tensor. After performing $p_2\leftrightarrow\ell_2$, the resulting hadronic tensor is given as
\begin{equation}
\begin{split}
    W^{\mu\nu}_{2,c_2} & =  \left[\left(C_F - \frac{C_A}{2}\right)C_F C_A\right] \sum_f  2[-g^{\mu\nu}_{\perp \perp }] e^2_f g^4_s \int d (\Delta X^{-})   e^{iq^{+}(\Delta X^{-} )}     
  \left\langle P \left| \bar{\psi}_{_f}( \Delta X^{-}) \frac{\gamma^{+}}{4} \psi_{_f}(0)\right| P\right\rangle  \\
  & \times   \int  d \zeta^{-} d (\Delta z^{-}) d^2 \Delta z_{\perp} \frac{dy}{2\pi} \frac{d^2 \ell_{2\perp}}{(2\pi)^2} \frac{d^2 k_{\perp}}{(\pi)^2} \left[ -1 + e^{i\mathcal{G}^{(p_2)}_{0}(x^{-}_{1} - z^{-}_{2})} \right] \left[ -1 + e^{-i\mathcal{G}^{(\ell_2)}_{0}(x^{-}_{2} - z^{-}_{3})} \right] e^{-i (\Delta z^{-})\mathcal{H}^{(\ell_2 , p_2)}_{0} }    \\
  & \times \frac{\theta( x^{-}_{1}-z^{-}_2) \theta( x^{-}_{2}-z^{-}_3)  }{  (1-y+\eta y)q^-  } \frac{\left[ -\pmb{\ell}^{2}_{2\perp} +  \pmb{\ell}_{2\perp}\cdot \pmb{k}_{\perp} \right]}{\left(\pmb{\ell}_{2\perp}-\pmb{k}_{\perp} \right)^2\pmb{\ell}^2_{2\perp} }    e^{i \pmb{k}_{\perp} \cdot \Delta\pmb{z}_{\perp}} \left\langle P_{A-1} \left| \bar{\psi}_{_f}(\zeta^-, 0) \frac{\gamma^+}{4} \psi_{_f}(\zeta^-, \Delta z^-, \Delta z_{\perp})\right| P_{A-1} \right\rangle  \\
& \times \left[ \frac{(1-y+2\eta y)}{y} \right].
\end{split}     
\label{eq:kernel2_wfinal_e_complexconjugate}
\end{equation}

Adding Eq.~(\ref{eq:kernel2_wfinal_e}) and Eq.~(\ref{eq:kernel2_wfinal_e_complexconjugate}) 
together gives the final hadronic tensor as
\begin{equation}
\begin{split}
    W^{\mu\nu}_{2,c_1+c_2} & =  \left[\left(C_F - \frac{C_A}{2}\right)C_F C_A\right] \sum_f  2[-g^{\mu\nu}_{\perp \perp }] e^2_f g^4_s\int d (\Delta X^{-})   e^{iq^{+}(\Delta X^{-} )}     
  \left\langle P \left| \bar{\psi}_{_f}( \Delta X^{-}) \frac{\gamma^{+}}{4} \psi_{_f}(0)\right| P\right\rangle  \\
  & \times \int d \zeta^{-} d (\Delta z^{-}) d^2 \Delta z_{\perp} \frac{dy}{2\pi} \frac{d^2 \ell_{2\perp}}{(2\pi)^2} \frac{d^2 k_{\perp}}{(\pi)^2}  e^{-i (\Delta z^{-})\mathcal{H}^{(\ell_2 , p_2)}_{0} }  e^{i \pmb{k}_{\perp} \cdot \Delta\pmb{z}_{\perp}}\\
  & \times \frac{\theta( \zeta^-)}{(1-y+\eta y)q^-} \left[2 - 2\cos\left\{\mathcal{G}^{(\ell_2)}_{0}\zeta^-\right\}  -2 \cos\left\{\mathcal{G}^{(p_2)}_{0}\zeta^-\right\} + 2\cos\left\{\left(\mathcal{G}^{(p_2)}_{0}-\mathcal{G}^{(\ell_2)}_{0}\right)\zeta^-\right\}\right]\\ 
%  & \times \frac{\theta( x^{-}_{1}-z^{-}_2) \theta( x^{-}_{2}-z^{-}_3)  }{  (1-y+\eta y)q^-  }  \left[2 - 2\cos\left\{\mathcal{G}^{(\ell_2)}_{0}\zeta^-\right\}  -2 \cos\left\{\mathcal{G}^{(p_2)}_{0}\zeta^-\right\} + 2\cos\left\{\left(\mathcal{G}^{(p_2)}_{0}-\mathcal{G}^{(\ell_2)}_{0}\right)\zeta^-\right\}\right]\\
& \times \left[ \frac{(1-y+2\eta y)}{y} \right]\frac{\left[ -\pmb{\ell}^{2}_{2\perp} +  \pmb{\ell}_{2\perp}\cdot \pmb{k}_{\perp} \right]}{\left(\pmb{\ell}_{2\perp}-\pmb{k}_{\perp} \right)^2\pmb{\ell}^2_{2\perp} } \left\langle P_{A-1} \left| \bar{\psi}_{_f}(\zeta^-, 0) \frac{\gamma^+}{4} \psi_{_f}(\zeta^-, \Delta z^-, \Delta z_{\perp})\right| P_{A-1} \right\rangle .
\end{split}     
\label{eq:kernel2_wfinal_e_both_added}
\end{equation}
%

%%%%%%%%%%%%%%%%%%%%%%%%%%%%%%%%%%%%%%%%%%%%%%%%%%%%%%%%%%%%%%%%%%%%%%%%%%%%%%%%%%%%%%%%%%%%%%%%%
\section{$\mbox{}$\!\!\!\!\!\!: Single-emission single-scattering kernel: One quark and one antiquark in the final state }
\label{append:kernel-3}
%%%%%%%%%%%%%%%%%%%%%%%%%%%%%%%%%%%%%%%%%%%%%%%%%%%%%%%%%%%%%%%%%%%%%%%%%%%%%%%%%%%%%%%%%%%%%%%%
%%%%%%%%%%%%%%%%%%%%%%%%%%%%%%%%%%%%%%%%%%%%%%%%%%%%%%%%%%%%%%%%%%%%%%%%%%%%%%%%%%%%%%%%%%%%%%%%%
In this section, we summarize the calculation of all possible diagrams at NLO  contributing to  kernel-3 with a quark and antiquark in the final state. We discuss singularity structure, contour integrations, and involved traces in the final calculation of the hadronic tensor. As there are only four diagrams in this kernel, this appendix will not have subsections.
\begin{figure}[!h]
    \centering 
    \begin{subfigure}[t]{0.495\textwidth}
        \centering        \includegraphics[height=1.6in]{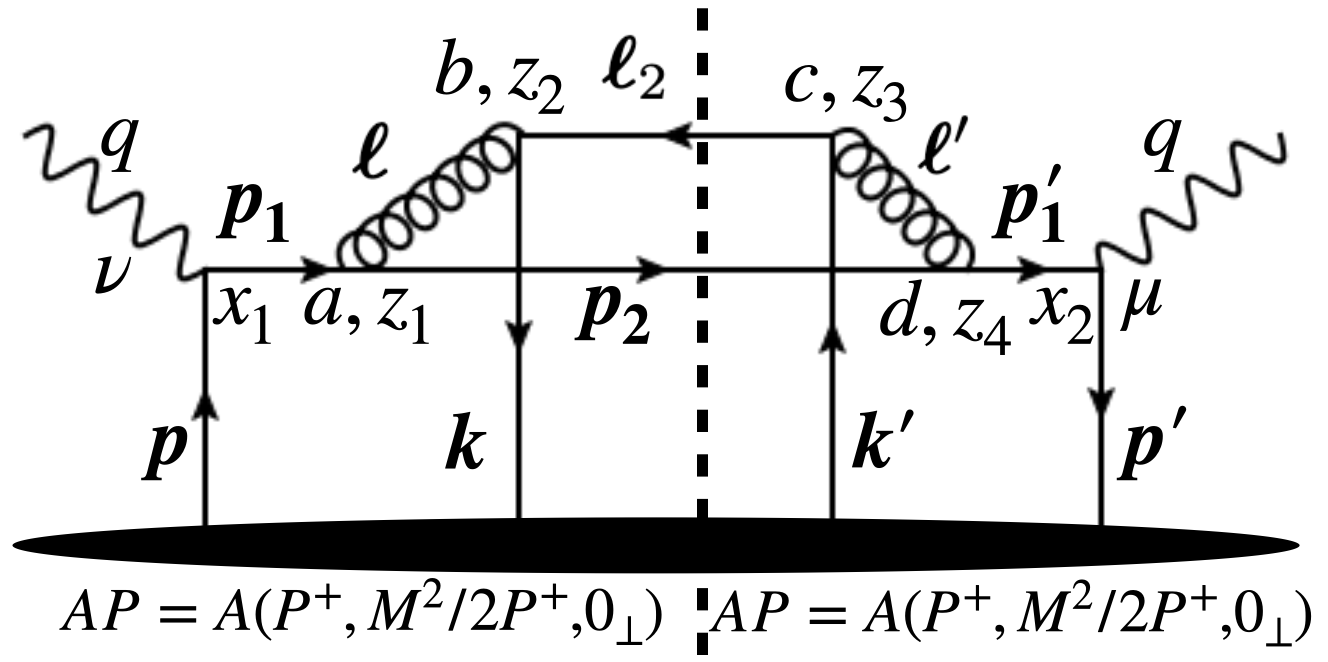}
        \caption{Quark anti-quark scattering channel. }
    \end{subfigure}%
    \begin{subfigure}[t]{0.495\textwidth}
        \centering        \includegraphics[height=1.6in]{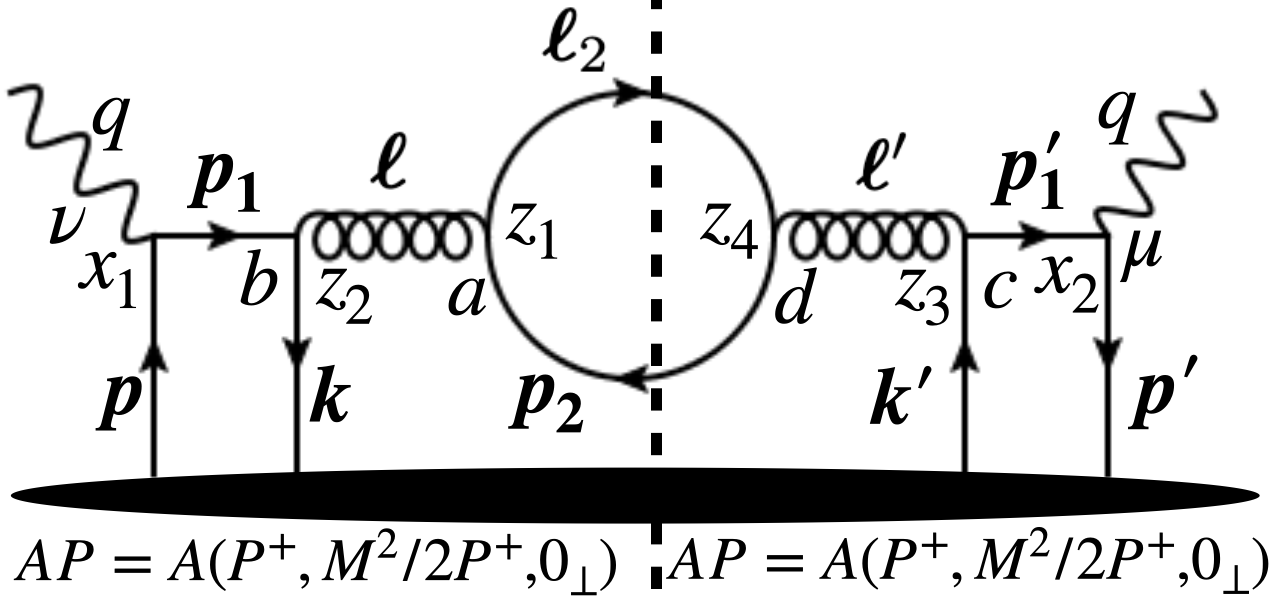}
        \caption{Quark anti-quark annihilation diagram.}
    \end{subfigure}
\caption{A forward scattering diagram contributing to kernel-3.}
\label{fig:kernel-3_amplt_square}
\end{figure}

The hadronic tensor for Fig.~\ref{fig:kernel-3_amplt_square}(a) has the following form
\begin{equation}
\begin{split}
W^{\mu\nu}_{3,c} & =\sum_f \sum_{f'}  e^2_f  g^4_{s} \int d^4 x_1 d^4 x_2 d^4 z_{2} d^4 z_{3} \int \frac{d^4 p}{(2\pi)^4} \frac{d^4 p'}{(2\pi)^4}  \frac{d^4 \ell_{2}}{(2\pi)^4} \frac{d^4 p_{2}}{(2\pi)^4} e^{-ip'x_2} e^{ipx_1}\left\langle P \left| \bar{\psi}_{_f}(x_2) \frac{\gamma^{+}}{4} \psi_{_f}(x_1)\right| P\right\rangle  \\
  & \times e^{i\left(q+p'-p_{2}-\ell_{2}\right) z_{3}} e^{i(\ell_{2}+p_{2} - q - p)z_{2}} \left\langle P_{A-1} \left| \bar{\psi}_{_{f'}}(z_2)\frac{\gamma^+}{4}\psi_{_{f'}}(z_3)\right| P_{A-1} \right\rangle d^{(q+p-p_2)}_{\sigma_{1} \sigma_{2}} d^{(q+p'-p_2)}_{\sigma_{3} \sigma_{4}} (2 \pi) \delta\left(\ell_{2}^{2}\right) (2 \pi)  \delta\left(p_{2}^{2}-M^{2}\right) \\
& \times\frac{ {\rm Tr} \left[\gamma^- \gamma^{\mu} \left(\slashed{q}+\slashed{p}'+M\right) \gamma^{\sigma_4}\left(\slashed{p}_2+M\right)\gamma^{\sigma_{1}} \left( \slashed{q} + \slashed{p} + M\right) \gamma^{\nu}  \right] {\rm Tr} \left[ \gamma^- \gamma^{\sigma_2} \slashed{\ell}_2 \gamma^{\sigma_3} \right] 
}{  \left[\left(q+p'\right)^{2}-M^2 -i\epsilon\right] \left[\left(q+p\right)^2 - M^2 + i\epsilon\right]  \left[\left(q+p'-p_2\right)^2 - i\epsilon\right] \left[\left(q+p -p_2\right)^2+i\epsilon\right]  } \delta^{ab}\delta^{cd} {\rm Tr}\left[t^a t^d\right]{\rm Tr}\left[t^b t^c\right],
\end{split} \label{eq:app_k3_qqbar_scat_wmnunu_1}
\end{equation}
where $M$ is the mass of the quark of flavor $f$. The above expression of the hadonic tensor has singularity when the denominator of the propagator for $p_1$, $\ell$, $\ell'$ and $p'_1$ becomes on-shell. It contains two simple poles for $p^+$ and $p'^+$. The contour integration for $p^+$ gives
\begin{equation}
\begin{split}
C_{1} & = \oint \frac{dp^{+}}{(2\pi)} \frac{e^{ip^{+}(x^{-}_{1}-z^{-}_{2})}}{\left[\left(q+p\right)^{2} - M^{2} + i\epsilon\right]\left[\left(q+p-p_{2}\right)^{2} + i\epsilon\right]} \\
      & = \frac{(2\pi i)}{2\pi} \frac{\theta\left(x^{-}_{1} - z^{-}_{2}\right)}{4q^{-}(q^{-}-p^{-}_{2})}  e^{i\left[-q^{+} + \frac{M^2}{2q^{-}}\right]\left(x^{-}_{1}-z^{-}_{2}\right)}\left[ \frac{ -1 + e^{i \mathcal{G}^{(p_2)}_{M}\left(x^{-}_{1}-z^{-}_{2}\right)} }{ \mathcal{G}^{(p_2)}_{M}}  \right],
\end{split}
%\label{eq:}
\end{equation}
where $\mathcal{G}^{(p_2)}_{M}$ is defined in Eq.~(\ref{eq:GMP2_append}). Similarly, the contour integration for $p'^+$ can be done
\begin{equation}
\begin{split}
C_{2} & = \oint \frac{dp'^{+}}{(2\pi)} \frac{e^{-ip'^{+}(x^{-}_{2}-z^{-}_{3})}}{\left[\left(q+p'\right)^{2} - M^{2} - i \epsilon\right]\left[\left(q+p'-p_{2}\right)^{2}- i\epsilon\right]} \\
      & = \frac{(-2\pi i)}{2\pi} \frac{\theta\left(x^{-}_{2} - z^{-}_{3}\right)}{4q^{-}\left(q^{-}-p^{-}_{2}\right)}  e^{i\left[q^{+} - \frac{M^2}{2q^{-}}\right]\left(x^{-}_{2}-z^{-}_{3}\right)}\left[ \frac{ -1 + e^{-i \mathcal{G}^{(p_2)}_M\left(x^{-}_{2}-z^{-}_{3}\right)} }{ \mathcal{G}^{(p_2)}_M}  \right].
\end{split}
%\label{eq:}
\end{equation}
The trace in the numerator of the third line of Eq.~(\ref{eq:app_k3_qqbar_scat_wmnunu_1}) gives
\begin{equation}
    \begin{split}
       & {\rm Tr} \left[\gamma^- \gamma^{\mu} \left(\slashed{q}+\slashed{p}'+M\right) \gamma^{\sigma_4}\left(\slashed{p}_2+M\right)\gamma^{\sigma_{1}} \left( \slashed{q} + \slashed{p} + M\right) \gamma^{\nu}  \right] {\rm Tr} \left[ \gamma^- \gamma^{\sigma_2} \slashed{\ell}_2 \gamma^{\sigma_3} \right] d^{\left(q+p-p_2\right)}_{\sigma_{1} \sigma_{2}} d^{\left(q+p'-p_2\right)}_{\sigma_{3} \sigma_{4}} \\
       & = 32[-g^{\mu\nu}_{\perp\perp}]\left(q^-\right)^2\left[\frac{1+\left(1-y\right)^2}{y\left(1-y+\eta y\right)}\right] \left[\left(\pmb{\ell}_{2\perp}-\pmb{k}_{\perp}\right)^2 + \kappa y^4 M^2 \right],
    \end{split}
\end{equation}
where $\kappa$ is defined in Eq.~(\ref{eq:kappa_append}). Finally, the hadronic tensor [Fig.~\ref{fig:kernel-3_amplt_square}(a)] reduces to the following form
\begin{equation}
\begin{split}
    W^{\mu\nu}_{3,c} & = \left[ \frac{C_F C_A}{2} \right] \sum_f \sum_{f'} 2 [-g^{\mu\nu}_{\perp \perp }] e^2_f g^4_s \int d (\Delta x^{-})   e^{iq^{+}(\Delta x^{-} )} 
    e^{-i[M^{2}/(2q^{-})](\Delta x^{-} )}
  \left\langle P \left| \bar{\psi}_{_f}( \Delta x^{-}) \frac{\gamma^{+}}{4} \psi_{_f}(0)\right| P\right\rangle  \\
  & \times    \int  d \zeta^{-} d (\Delta z^{-}) d^2 \Delta z_{\perp} \frac{dy}{2\pi} \frac{d^2 \ell_{2\perp}}{(2\pi)^2} \frac{d^2 k_{\perp}}{(2\pi)^2}  e^{-i (\Delta z^{-})\mathcal{H}^{(\ell_2 , p_2)}_{M} }     e^{i \pmb{k}_{\perp} \cdot \Delta\pmb{z}_{\perp}} \theta(\zeta^-)\\% \theta\left( x^{-}_{1}-z^{-}_2\right) \theta\left(x^{-}_{2}-z^{-}_3\right) \\
  & \times \left[ \frac{ 1 + \left(1-y\right)^2 }{y}\right] \frac{1}{yq^-} \frac{\left[\left(\pmb{\ell}_{2\perp} - \pmb{k}_{\perp}\right)^2  + \kappa y^4 M^2 \right]}{\left[  \left(\pmb{\ell}_{2\perp} -\pmb{k}_{\perp}\right)^2 + y^2\left(1-\eta\right)^2 M^2 \right]^2}\left[ 2 -2 \cos\left\{\mathcal{G}^{(p_2)}_{M}\zeta^-\right\} \right]\\
  &\times    \left\langle P_{A-1} \left| \bar{\psi}_{_{f'}}(\zeta^-,0) \frac{\gamma^+}{4} \psi_{_{f'}}(\zeta^-, \Delta z^-, \Delta z_{\perp})\right| P_{A-1} \right\rangle, 
\end{split}\label{eq:app_k3_qqbar_scat_wmnunu_final}
\end{equation}
where $\mathcal{G}^{(p_2)}_M$ is given in Eq.~(\ref{eq:GMP2_append}), $\mathcal{H}^{(\ell_2,p_2)}_{M}$ is defined in Eq.~(\ref{eq:HL2P2M_append}), and $M$ is the mass of the quark flavor $f$.

Now, we consider a central-cut diagram shown in Fig.~\ref{fig:kernel-3_amplt_square}(b). The hadronic tensor has the following form
\begin{equation}
\begin{split}
W^{\mu\nu}_{3,c} & =\sum_{f} \sum_{f'}  e^2_f g^4_{s} \int d^4 x_1 d^4 x_2 d^4 z_{2} d^4 z_{3} \int \frac{d^4 p}{(2\pi)^4} \frac{d^4 p'}{(2\pi)^4}  \frac{d^4 \ell_{2}}{(2\pi)^4} \frac{d^4 p_{2}}{(2\pi)^4} e^{-ip'x_2} e^{ipx_1}\left\langle P \left| \bar{\psi}_{_{f}}(x_2) \frac{\gamma^{+}}{4} \psi_{_{f}}(x_1)\right| P\right\rangle  \\
  & \times e^{i\left(q+p'-p_{2}-\ell_{2}\right) z_{3}} e^{i(\ell_{2}+p_{2} - q - p)z_{2}} \left \langle P_{A-1} \left| \bar{\psi}_{_{f}}(z_2) \frac{\gamma^+}{4}\psi_{_{f}}(z_3)\right| P_{A-1} \right\rangle  (2 \pi) \delta\left(\ell_{2}^{2}-M^2_{_{f'}}\right) (2 \pi)  \delta\left(p_{2}^{2}-M^{2}_{_{f'}}\right) \delta^{ab} \delta^{cd} \\
& \times \frac{  {\rm Tr} \left[\gamma^- \gamma^{\mu} \left(\slashed{q}+\slashed{p}'\right) \gamma^{\sigma_3}\gamma^-\gamma^{\sigma_2}\left( \slashed{q} + \slashed{p}\right) \gamma^{\nu} \right] {\rm Tr} \left[ \left(\slashed{\ell}_2+M_{_{f'}}\right) \gamma^{\sigma_{4}}\left(\slashed{p}_2+M_{_{f'}}\right) \gamma^{\sigma_1} \right]
}{  \left[\left(q+p'\right)^{2}-i\epsilon\right] \left[\left(\ell_2+p_2\right)^2  - i\epsilon\right]   \left[\left(\ell_2+p_2\right)^2 +i\epsilon\right] \left[\left(q+p\right)^2+i\epsilon\right]  } 
d^{(\ell_2+p_2)}_{\sigma_{1} \sigma_{2}} d^{(\ell_2+p_2)}_{\sigma_{3} \sigma_{4}}
 {\rm Tr}\left[t^b t^c\right] {\rm Tr}\left[t^a t^d\right],
\end{split} \label{eq:append_kernel3_qqbar_ann_wmunu_1}
\end{equation}
where $M_{_{f'}}$ is the mass of the quark flavor $f'$. Equation~(\ref{eq:append_kernel3_qqbar_ann_wmunu_1}) has singularity arising from the denominator of the quark propagator with momentum $p_{1}$ and $p'_{1}$. We identify one pole for each momentum variable $p^+$ and $p'^+$.
The contour integration for momentum $p^+$ is given as
\begin{equation}
    C_{1} = \oint \frac{dp^+}{(2\pi)} \frac{e^{ip^+(x^{-}_{1}-z^-_2)} }{\left[\left(q+p\right)^2 + i\epsilon\right]} = \frac{(2\pi i)}{2\pi}\frac{\theta\left(x^{-}_{1} - z^{-}_{2}\right)}{2q^{-}}  e^{-iq^{+}\left(x^{-}_{1}-z^{-}_{2}\right)} .
\end{equation}
Similarly, the contour integration for momentum $p'^+$ is carried out as 
\begin{equation}
    C_{2} = \oint \frac{dp'^+}{(2\pi)} \frac{e^{-ip'^+\left(x^{-}_{2}-z^{-}_3\right)} }{\left[\left(q+p'\right)^2-i\epsilon\right]} = \frac{(-2\pi i)}{2\pi} \frac{\theta\left(x^{-}_{2} - z^{-}_{3}\right)}{2q^-}  e^{iq^+ \left(x^{-}_{2}-z^{-}_3\right)}  .
\end{equation}
Including mass correction up to $\mathcal{O}(M^2)$, the trace yields
\begin{equation}
    \begin{split}
      &  \mathrm{Tr} \left[  \gamma^{-} \gamma^{\mu}  (\slashed{q} +\slashed{p}')   \gamma^{\sigma_{3}} \gamma^{-} \gamma^{\sigma_{2}} (\slashed{q} +\slashed{p})  \gamma^{\nu} \right]
\mathrm{Tr} \left[ (\slashed{\ell}_{2}+M_{_{f'}}) \gamma^{\sigma_{4}} (\slashed{p}_{2} +M_{_{f'}}) \gamma^{\sigma_{1}}  \right] d^{(\ell_{2}+p_{2})}_{\sigma_{4} \sigma_{3}}  d^{(\ell_{2}+p_{2})}_{\sigma_{1} \sigma_{2}}  \\
   & = 32 \left(q^{-}\right)^{2} [-g^{\mu\nu}_{\perp \perp}]  \left[ \frac{ \left\{\left(1+\eta y\right)\pmb{\ell}_{2\perp} - y\pmb{k}_{\perp}\right\}^2 + M^2_{_{f'}}\left(1+\eta y\right)^2    }{y\left(1-y+\eta y\right)\left(1+\eta y\right)^2}\right] \left[ y^2 + \left(1-y+\eta y\right)^2\right] . \\    
    \end{split}
\end{equation}
The final expression of the hadronic tensor for the central-cut [Fig.~\ref{fig:kernel-3_amplt_square}(b)] is given as
\begin{equation}
       \begin{split}
    W^{\mu\nu}_{3,c} & =\left[ \frac{C_{F}C_{A}}{2} \right]   \sum_f \sum_{f'} 2 \left[ -g^{\mu\nu}_{\perp \perp} \right] e^2_{f} g^4_s \int d (\Delta x^{-})   e^{iq^{+}(\Delta x^{-} )}    
  \left\langle P \left| \bar{\psi}_{_f}( \Delta x^{-}) \frac{\gamma^{+}}{4} \psi_{f}(0)\right| P\right\rangle  \\
  & \times   \int  d \zeta^{-} d (\Delta z^{-})d^2 \Delta z_{\perp} \frac{dy}{2\pi} \frac{d^2 \ell_{2\perp}}{(2\pi)^2} \frac{d^2 k_{\perp}}{(2\pi)^2}  e^{-i \mathcal{H}^{(\ell_2,p_2)}_{1} (\Delta z^{-})}     e^{-i \pmb{k}_{\perp}\cdot \Delta\pmb{z}_{\perp}}   \\
  & \times \frac{\theta(\zeta^-)}{ \left(1+\eta y\right)^2q^{-}}\frac{\left[ y^2 + \left(1-y+\eta y\right)^2\right]}{\left[ \left\{\left(1+\eta y\right)\pmb{\ell}_{2\perp} - y \pmb{k}_{\perp}\right\}^2 + M^2_{f'} \left(1+\eta y\right)^2\right]}  \\  
%  & \times \frac{\theta( x^{-}_{1} -z^{-}_2) \theta( x^{-}_{2} -z^{-}_{3})}{ \left(1+\eta y\right)^2q^{-}}\frac{\left[ y^2 + \left(1-y+\eta y\right)^2\right]}{\left[ \left\{\left(1+\eta y\right)\pmb{\ell}_{2\perp} - y \pmb{k}_{\perp}\right\}^2 + M^2_{f'} \left(1+\eta y\right)^2\right]}  \\
  &\times  \left\langle P_{A-1} \left|   \bar{\psi}_{_{f}}(\zeta^{-},0) \frac{\gamma^+}{4}  \psi_{_{f}}(\zeta^{-},\Delta z^{-}, \Delta \pmb{z}_{\perp}) \right| P_{A-1} \right\rangle, 
\end{split} 
\label{eq:K3_final_g_qqbar_wmunu}
\end{equation}
where $\mathcal{H}^{(\ell_2,p_2)}_{1}$ is defined in Eq.~(\ref{eq:H1L2P2_append}).

%comment

%
\begin{figure}[!h]
    \centering 
    \begin{subfigure}[t]{0.47\textwidth}
        \centering        \includegraphics[height=1.45in]{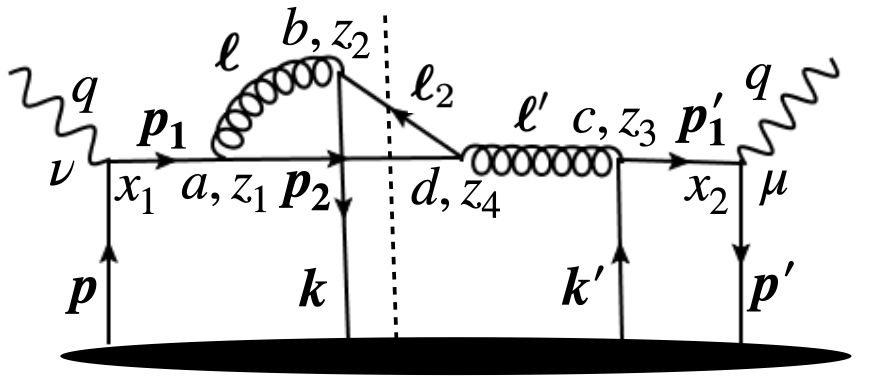}
        \caption{An interference diagram. }
    \end{subfigure}%
    \begin{subfigure}[t]{0.47\textwidth}
        \centering        \includegraphics[height=1.45in]{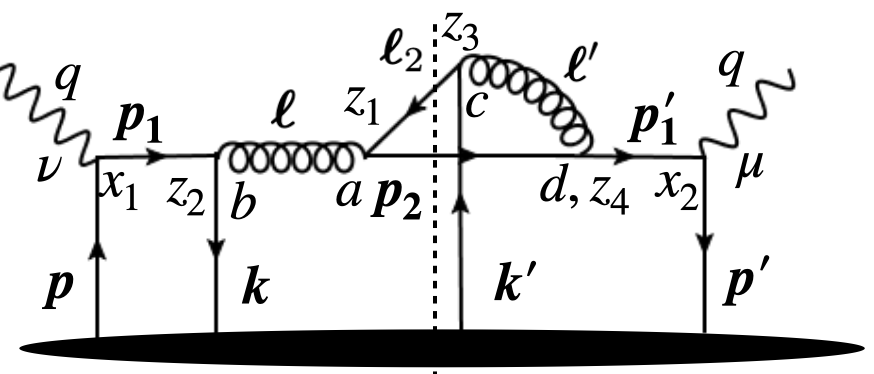}
        \caption{Complex-conjugate of the diagram on the left panel.}
    \end{subfigure}
\caption{A forward scattering diagram contributing to kernel-3.}
\label{fig:kernel-3_interference}
\end{figure}

Furthermore, we consider an interference diagram shown in Fig~\ref{fig:kernel-3_interference}(a). The hadronic tensor is given as
\begin{equation}
    \begin{split}
         W^{\mu \nu}_{3,c_1} & =\sum_{f}  e^2_f g^{4}_s  \int d^{4} x_1  d^{4} x_2  
 \int  d^{4} z_{2}  d^{4} z_{3}    \int \frac{d^{4} p}{(2 \pi)^{4}}  \frac{d^{4} p'}{(2 \pi)^{4}}    \frac{d^{4} \ell_{2}}{(2 \pi)^{4}}  \frac{d^{4} p_{2}}{(2 \pi)^{4}} 
e^{-ip' x_2}  e^{ip x_1}  \left\langle P\left\vert \bar{\psi}_{_f}\left(y\right) \frac{\gamma^+}{4} \psi_{_f}\left(x\right) \right\vert P \right\rangle\\
&  \times   e^{iz_{2}(\ell_{2} +p_{2} - q -p )}  e^{iz_{3}( -p_{2} -\ell_{2}  +q + p' )}  \left\langle P_{A-1} \left\vert \bar{\psi}_{_f}\left(z_{2}\right) 
\frac{\gamma^{+}}{4} 
\psi_{_f} \left( z_{3} \right) 
   \right\vert P_{A-1} \right\rangle  {\rm Tr}\left[t^c t^b t^d t^a\right] \delta^{ab} \delta^{cd}  \\
& \times  \frac{\mathrm{Tr} \left[  \gamma^{-} \gamma^{\mu} \left(\slashed{q} + \slashed{p}'\right)   \gamma^{\sigma_{3}} 
\gamma^{-}
\gamma^{\sigma_{2}}  \slashed{\ell}_{2}  
\gamma^{\sigma_{4}} \slashed{p}_{2}  
\gamma^{\sigma_{1}}  \left(\slashed{q} + \slashed{p}\right)  \gamma^{\nu} \right] d^{(\ell_2+p_2)} _{\sigma_{4}\sigma_{3}} d^{(q+p-p_2)}_{\sigma_{1} \sigma_{2}} }{ \left[\left(q+p'\right)^{2}-i\epsilon\right] \left[\left(q+p\right)^{2}+ i\epsilon\right] \left[\left(\ell_{2}+p_{2}\right)^{2}-i\epsilon\right] \left[\left(q+ p - p_{2}\right)^{2}+i \epsilon\right] } (2\pi) \delta(\ell^{ 2}_{2}) (2\pi) \delta(p^{ 2}_{2} ) .
    \end{split}
\end{equation}
The above expression has singularity when the denominator of the parton propagator for $p_{1}$, $\ell$ and $p'_{1}$ becomes zero. We identify two poles for the momentum variable $p$ and one pole for $p'$. We compute the integral in the complex plane of $p^{+}$ and $p'^{+}$.

In this central-cut diagram, the momenta for the final state partons are $\ell^{-}_{2}=yq^{-}$ and $p^{-}_{2}=(1-y + \eta y)q^{-}$.
The contour integration for $p^{+}$ is given as
\begin{equation}
    \begin{split}
    C_{1} & = \oint \frac{dp^{+}}{2\pi} \frac{e^{ip^{+}( x^{-}_{1} - z^{-}_{2})}}{\left[\left(q+ p\right)^2  +i\epsilon\right]\left[\left(q+ p - p_{2}\right)^2 + i\epsilon\right] } \\
    & = \left( \frac{2\pi i}{2\pi}\right) \frac{\theta\left(x^{-}_{1} - z^{-}_{2}\right)}{4q^{-} (1-\eta)yq^{-}} e^{-i q^{+} ( x^{-}_{1} - z^{-}_{2})}  \left[ \frac{-1 + e^{i\mathcal{G}^{(p_2)}_{0}(x^{-}_{1} - z^{-}_{2})}}{\mathcal{G}^{(p_2)}_{0}}  \right],
    \end{split}
\end{equation}
where  $\mathcal{G}^{(p_2)}_{0} $ is given in Eq.~(\ref{eq:GP2M0_append}). Similarly, the contour integration for $p'^{+}$ is 
\begin{equation}
    \begin{split}
    C_{2} & = \oint \frac{dp'^{+}}{2\pi} \frac{e^{ip'^{+}(-x^{-}_{2} + z^{-}_{3})}}{\left[\left(q+p'\right)^{2} - i\epsilon\right]} 
     = \left( \frac{-2\pi i}{2\pi}\right) \frac{\theta\left( x^{-}_{2} -z^{-}_{3}\right)}{2q^{-}}  e^{-iq^{+}(-x^{-}_{2} + z^{-}_{3})}.  
    \end{split}
\end{equation}
Simplifying the trace yields the following expression
\begin{equation}
    \begin{split}
       & \mathrm{Tr} \left[  \gamma^{-} \gamma^{\mu} \left(\slashed{q} + \slashed{p}'\right)   \gamma^{\sigma_{3}} 
\gamma^{-}
\gamma^{\sigma_{2}}  \slashed{\ell}_{2}  
\gamma^{\sigma_{4}} \slashed{p}_{2}  
\gamma^{\sigma_{1}}  \left(\slashed{q} + \slashed{p}\right)  \gamma^{\nu} \right] 
         \times   d^{(\ell_{2}+p_{2})}_{\sigma_{4} \sigma_{3}}  d^{(q+p-p_{2})}_{\sigma_{1} \sigma_{2}}  \\
       & = 32 \left(q^{-}\right)^{2} [-g^{\mu\nu}_{\perp\perp}]  \left[ \frac{J_2}{y(1-\eta)(1+\eta y)}  \right],
    \end{split}
\end{equation}
where $J_2$ is defined in Eq.~(\ref{eq:J2_append}). The final expression for the hadronic tensor [Fig~\ref{fig:kernel-3_interference}(a)] reduces to the following form:
\begin{equation}
       \begin{split}
    W^{\mu\nu}_{3,c_1} & = \left[ C_{F}C_{A} \left(C_F - 
 \frac{C_A}{2}\right)\right] \sum_{f} 2 \left[ -g^{\mu\nu}_{\perp \perp} \right] e^2_f g^4_s \int d (\Delta x^{-})   e^{iq^{+}(\Delta x^{-} )}    
  \left\langle P \left| \bar{\psi}_{_f}( \Delta x^{-}) \frac{\gamma^{+}}{4} \psi_{_f}(0)\right| P\right\rangle  \\
  & \times   \int  d \zeta^{-} d (\Delta z^{-})d^2 \Delta z_{\perp} \frac{dy}{2\pi} \frac{d^2 \ell_{2\perp}}{(2\pi)^2} \frac{d^2 k_{\perp}}{(2\pi)^2}  e^{-i \mathcal{H}^{(\ell_2,p_2)}_{0} (\Delta z^{-})}     e^{i \pmb{k}_{\perp}\cdot \Delta\pmb{z}_{\perp}}  \theta\left( x^{-}_{1} -z^{-}_2\right) \theta\left( x^{-}_{2} -z^{-}_{3}\right) \\
  & \times \frac{1}{ yq^{-} } \left[ \frac{1-y+\eta y}{\left(1+\eta y\right)\left(1-\eta\right)}\right] \frac{J_2}{ \left[ \pmb{\ell}_{2\perp} - \pmb{k}_\perp\right]^2 \left[ \left(1+\eta y\right)\pmb{\ell}_{2\perp} - y\pmb{k}_{\perp} \right]^2 } \left[ -1 + e^{i\mathcal{G}^{(p_2)}_{0}(x^{-}_{1}-z^-_2)} \right]   \\
  &
\times   \left\langle P_{A-1} \left| \bar{\psi}_{_f}(\zeta^{-},0) \frac{\gamma^+}{4}  \psi_{_f}(\zeta^{-},\Delta z^{-}, \Delta \pmb{z}_{\perp}) \right| P_{A-1} \right\rangle,
    \end{split}
\label{eq:k3_a_intf_q_gq_qqbar_g_q_wmunu} 
\end{equation}
where $\mathcal{G}^{(p_2)}_0$ is given in Eq.~(\ref{eq:GP2M0_append}) and $\mathcal{H}^{(\ell_2,p_2)}_{0}$ is given in Eq.~(\ref{eq:HL2P2M0_append}).

Note that the diagrams in Fig.~\ref{fig:kernel-3_interference}(b) and Fig.~\ref{fig:kernel-3_interference}(a) are complex-conjugate of each other. They differ only in contour integration over variable $p^{+}$ and $p'^{+}$.
 The calculation of the hadronic tensor for Fig.~\ref{fig:kernel-3_interference}(b) involves the contour integration for $p^{+}$ and is given as
\begin{equation}
    \begin{split}
    C_{1} & = \oint \frac{dp^{+}}{2\pi} \frac{e^{ip^{+}\left(x^{-}_{1} - z^{-}_{2}\right)}}{\left[\left(q+p\right)^{2}+i\epsilon\right]} 
     = \left( \frac{2\pi i}{2\pi}\right) \frac{\theta\left(x^{-}_{1} - z^{-}_{2}\right)}{2q^{-}}  e^{-i q^{+}\left( x^{-}_{1} - z^{-}_{2}\right) }  ,
    \end{split}
\end{equation}
and, the contour integration for $p'^{+}$ is given as
\begin{equation}
    \begin{split}
    C_{2} & = \oint \frac{dp'^{+}}{2\pi} \frac{e^{ip'^{+}\left(-x^{-}_{2} + z^{-}_{3}\right)}}{\left[\left(q+ p'\right)^2  -i\epsilon\right]\left[\left(q+ p' - p_{2}\right)^2-i\epsilon\right] } \\
    & = \left( \frac{-2\pi i}{2\pi}\right) \frac{\theta\left( x^{-}_{2} -z^{-}_{3}\right)}{4q^{-} \left(q^{-}-p^{-}_2\right)} e^{iq^{+}\left(x^{-}_{2} - z^{-}_{3}\right)}  \left[ \frac{-1 + e^{-i\mathcal{G}^{(p_2)}_0\left(x^{-}_{2} - z^{-}_{3}\right)}}{\mathcal{G}^{(p_2)}_0}  \right],
    \end{split}
\end{equation}
where $\mathcal{G}^{(p_2)}_0$ is given in Eq.~(\ref{eq:GP2M0_append}).

The final expression for the hadronic tensor [Fig.~\ref{fig:kernel-3_interference}(b)] yields
\begin{equation}
       \begin{split}
    W^{\mu\nu}_{3,c_2} & = \left[ C_{F}C_{A} \left(C_F - \frac{C_A}{2} \right) \right] \sum_f 2 \left[ -g^{\mu\nu}_{\perp \perp} \right] e^2_f g^4_s \int d (\Delta x^{-})   e^{iq^{+}(\Delta x^{-} )}    
  \left\langle P \left| \bar{\psi}_{_f}( \Delta x^{-}) \frac{\gamma^{+}}{4} \psi_{_f}(0)\right| P\right\rangle  \\
  & \times   \int  d \zeta^{-} d (\Delta z^{-})d^2 \Delta z_{\perp} \frac{dy}{2\pi} \frac{d^2 \ell_{2\perp}}{(2\pi)^2} \frac{d^2 k_{\perp}}{(2\pi)^2} e^{-i \mathcal{H}^{(\ell_2,p_2)}_{0} (\Delta z^{-})}     e^{i \pmb{k}_{\perp}\cdot \Delta\pmb{z}_{\perp}}  \theta( x^{-}_{1} -z^{-}_2) \theta( x^{-}_{2} -z^{-}_{3})  \\
  & \times \frac{1}{ yq^{-} } \left[ \frac{1-y+\eta y}{(1+\eta y)(1-\eta)}\right]  \frac{J_2}{ \left[ \pmb{\ell}_{2\perp} - \pmb{k}_\perp\right]^2 \left[ \left(1+\eta y\right)\pmb{\ell}_{2\perp} - y\pmb{k}_{\perp} \right]^2 }  \left[ -1 + e^{-i\mathcal{G}^{(p_2)}_{0}\left(x^{-}_{2}-z^-_3\right)} \right]  \\
  &
\times   \left\langle P_{A-1} \left| \bar{\psi}_{_{f}}(\zeta^{-},0) \frac{\gamma^+}{4}  \psi_{_{f}}(\zeta^{-},\Delta z^{-}, \Delta \pmb{z}_{\perp}) \right| P_{A-1} \right\rangle,
    \end{split} 
\label{eq:k3_b_intf_q_g_qqbar_qg_wmunu}
\end{equation}
where $\mathcal{G}^{(p_2)}_0$ is given in Eq.~(\ref{eq:GP2M0_append}) and $\mathcal{H}^{(\ell_2,p_2)}_{0}$ is given in Eq.~(\ref{eq:HL2P2M0_append}).

Adding the hadronic tensor associated with Figs.~\ref{fig:kernel-3_interference}(a) and \ref{fig:kernel-3_interference}(b), i.e., Eq.~(\ref{eq:k3_a_intf_q_gq_qqbar_g_q_wmunu}) and (\ref{eq:k3_b_intf_q_g_qqbar_qg_wmunu}), gives the following form of the hadronic tensor:
\begin{equation}
       \begin{split}
    W^{\mu\nu}_{3,c_1+c_2} & = -\left[ C_{F}C_{A} \left(C_F - \frac{C_A}{2} \right) \right] \sum_f 2 \left[ -g^{\mu\nu}_{\perp \perp} \right] e^2_f g^4_s \int d (\Delta x^{-})   e^{iq^{+}(\Delta x^{-} )}    
  \left\langle P \left| \bar{\psi}_{_f}( \Delta x^{-}) \frac{\gamma^{+}}{4} \psi_{_f}(0)\right| P\right\rangle  \\
  & \times   \int  d \zeta^{-} d (\Delta z^{-})d^2 \Delta z_{\perp} \frac{dy}{2\pi} \frac{d^2 \ell_{2\perp}}{(2\pi)^2} \frac{d^2 k_{\perp}}{(2\pi)^2} e^{-i \mathcal{H}^{(\ell_2,p_2)}_{0} (\Delta z^{-})}     e^{i \pmb{k}_{\perp}\cdot \Delta\pmb{z}_{\perp}}  \theta( \zeta^{-})  \\
  & \times \frac{1}{ yq^{-} } \left[ \frac{1-y+\eta y}{(1+\eta y)(1-\eta)}\right]  \frac{J_2}{ \left[ \pmb{\ell}_{2\perp} - \pmb{k}_\perp\right]^2 \left[ \left(1+\eta y\right)\pmb{\ell}_{2\perp} - y\pmb{k}_{\perp} \right]^2 }  \left[ 2 - 2\cos\left\{\mathcal{G}^{(p_2)}_{0}\zeta^{-}\right\} \right]  \\
  &
\times   \left\langle P_{A-1} \left| \bar{\psi}_{_{f}}(\zeta^{-},0) \frac{\gamma^+}{4}  \psi_{_{f}}(\zeta^{-},\Delta z^{-}, \Delta \pmb{z}_{\perp}) \right| P_{A-1} \right\rangle,
    \end{split} 
\label{eq:k3_ab_intf_q_g_qqbar_qg_wmunu_added}
\end{equation}
where $\mathcal{G}^{(p_2)}_0$ is given in Eq.~(\ref{eq:GP2M0_append}) and $\mathcal{H}^{(\ell_2,p_2)}_{0}$ is given in Eq.~(\ref{eq:HL2P2M0_append}).

%%%%%%%%%%%%%%%%%%%%%%%%%%%%%%%%%%%%%%%%%%%%%%%%%%%%%%%%%%%%%%%%%%%%%%%%%%%%%%%%%%%%%%%%%%%%%%%%%
\section{$\mbox{}$\!\!\!\!\!\!: Single-emission single-scattering kernel: Two quarks in the final state }
\label{append:kernel-4}
%%%%%%%%%%%%%%%%%%%%%%%%%%%%%%%%%%%%%%%%%%%%%%%%%%%%%%%%%%%%%%%%%%%%%%%%%%%%%%%%%%%%%%%%%%%%%%%%
%%%%%%%%%%%%%%%%%%%%%%%%%%%%%%%%%%%%%%%%%%%%%%%%%%%%%%%%%%%%%%%%%%%%%%%%%%%%%%%%%%%%%%%%%%%%%%%%%
In this section, we summarize the NLO calculation of all diagrams contributing to kernel-4 with two quarks in the final state. We analyze the singularity structure, contour integrations, and traces involved in the hadronic tensor. Since there are solely two diargams contributing herein, there will be no subsections in this appendix. 
\begin{figure}[!h]
    \centering 
    \begin{subfigure}[t]{0.47\textwidth}
        \centering        \includegraphics[height=1.45in]{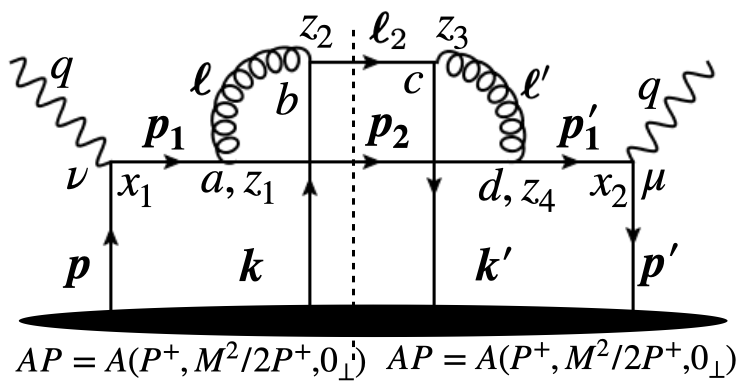}
        \caption{Quark quark scattering channel. }
    \end{subfigure}%
    \begin{subfigure}[t]{0.47\textwidth}
        \centering        \includegraphics[height=1.45in]{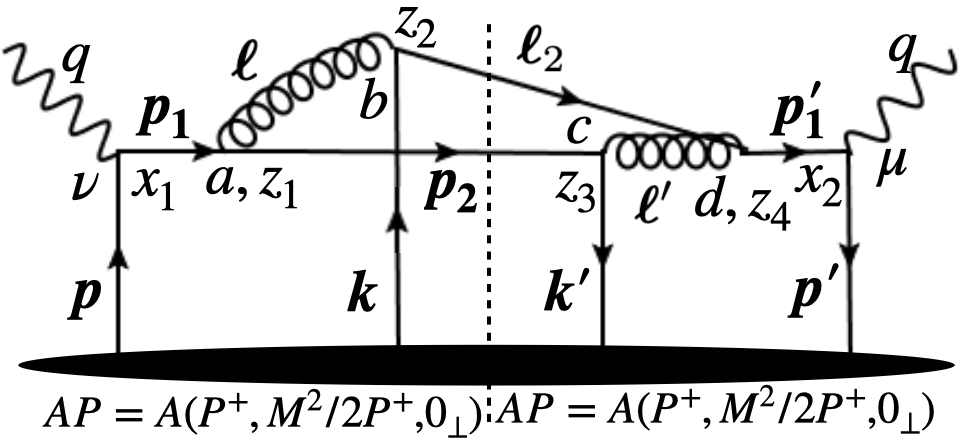}
        \caption{An interference diagram.}
    \end{subfigure}
\caption{A forward scattering diagram contributing to kernel-4.}
\label{fig:kernel-4_qq_scattering_interference}
\end{figure}

The hadronic tensor for Fig.~\ref{fig:kernel-4_qq_scattering_interference}(a) has the following form
\begin{equation}
\begin{split}
W^{\mu\nu}_{4,c} & =\sum_f \sum_{f'}  e^2_f  g^4_{s} \int d^4 x_1 d^4 x_2 d^4 z_{2} d^4 z_{3} \int \frac{d^4 p}{(2\pi)^4} \frac{d^4 p'}{(2\pi)^4}  \frac{d^4 \ell_{2}}{(2\pi)^4} \frac{d^4 p_{2}}{(2\pi)^4} e^{-ip'x_2} e^{ipx_1}\left\langle P \left| \bar{\psi}_{_f}(x_2) \frac{\gamma^{+}}{4} \psi_{_f}(x_1)\right| P\right\rangle  \\
  & \times e^{i\left(q+p'-p_{2}-\ell_{2}\right) z_{3}} e^{i(\ell_{2}+p_{2} - q - p)z_{2}} \left\langle P_{A-1} \left| \bar{\psi}_{_{f'}}(z_3)\frac{\gamma^+}{4}\psi_{_{f'}}(z_2)\right| P_{A-1} \right\rangle d^{(q+p-p_2)}_{\sigma_{1} \sigma_{2}} d^{(q+p'-p_2)}_{\sigma_{3} \sigma_{4}} (2 \pi) \delta\left(\ell_{2}^{2}\right) (2 \pi)  \delta\left(p_{2}^{2}-M^{2}\right) \\
& \times\frac{ {\rm Tr} \left[\gamma^- \gamma^{\mu} \left(\slashed{q}+\slashed{p}'+M\right) \gamma^{\sigma_4}\left(\slashed{p}_2+M\right)\gamma^{\sigma_{1}} \left( \slashed{q} + \slashed{p} + M\right) \gamma^{\nu}  \right] {\rm Tr} \left[ \gamma^- \gamma^{\sigma_3} \slashed{\ell}_2 \gamma^{\sigma_2} \right] 
}{  \left[\left(q+p'\right)^{2}-M^2-i\epsilon\right] \left[\left(q+p\right)^2 - M^2 + i\epsilon\right]  \left[\left(q+p'-p_2\right)^2 - i\epsilon\right] \left[\left(q+p -p_2\right)^2+i\epsilon\right]  } \delta^{ab}\delta^{cd} {\rm Tr}[t^{a}t^{d}]{\rm Tr}[t^{b}t^{c}],
\end{split} 
\label{eq:app_k4_qq_scat_wmnunu_1}
\end{equation}
where $M$ is the mass of the quark of flavor $f$. The above expression of the hadonic tensor has singularity when the denominator of the propagator for $p_1$, $\ell$, $\ell'$, and $p'_1$ becomes on-shell. It contains two simple poles for $p^+$ and $p'^+$. The contour integration for $p^+$ gives
\begin{equation}
\begin{split}
C_{1} & = \oint \frac{dp^{+}}{(2\pi)} \frac{e^{ip^{+}(x^{-}_{1}-z^{-}_{2})}}{\left[\left(q+p\right)^{2} - M^{2} + i\epsilon\right]\left[\left(q+p-p_{2}\right)^{2} + i\epsilon\right]} \\
      & = \frac{(2\pi i)}{2\pi} \frac{\theta\left(x^{-}_{1} - z^{-}_{2}\right)}{4q^{-}\left(q^{-}-p^{-}_{2}\right)}  e^{i\left[-q^{+} + \frac{M^2}{2q^{-}}\right](x^{-}_{1}-z^{-}_{2})}\left[ \frac{ -1 + e^{i \mathcal{G}^{(p_2)}_M(x^{-}_{1}-z^{-}_{2})} }{ \mathcal{G}^{(p_2)}_M}  \right],
\end{split}
%\label{eq:}
\end{equation}
where $\mathcal{G}^{(p_2)}_M$ is defined in Eq.~(\ref{eq:GMP2_append}). Similarly, the contour integration for $p'^+$ can be done
\begin{equation}
\begin{split}
C_{2} & = \oint \frac{dp'^{+}}{(2\pi)} \frac{e^{-ip'^{+}(x^{-}_{2}-z^{-}_{3})}}{\left[\left(q+p'\right)^{2} - M^{2} - i \epsilon\right]\left[\left(q+p'-p_{2}\right)^{2}- i\epsilon\right]} \\
      & = \frac{(-2\pi i)}{2\pi} \frac{\theta\left(x^{-}_{2} - z^{-}_{3}\right)}{4q^{-}\left(q^{-}-p^{-}_{2}\right)}  e^{i\left[q^{+} - \frac{M^2}{2q^{-}}\right](x^{-}_{2}-z^{-}_{3})}\left[ \frac{ -1 + e^{-i \mathcal{G}^{(p_2)}_M(x^{-}_{2}-z^{-}_{3})} }{ \mathcal{G}^{(p_2)}_M}  \right].
\end{split}
%\label{eq:}
\end{equation}
The trace in the numerator of the third line of Eq.~(\ref{eq:app_k4_qq_scat_wmnunu_1}) gives
\begin{equation}
    \begin{split}
       & {\rm Tr} \left[\gamma^- \gamma^{\mu} \left(\slashed{q}+\slashed{p}'+M\right) \gamma^{\sigma_4}\left(\slashed{p}_2+M\right)\gamma^{\sigma_{1}} \left( \slashed{q} + \slashed{p} + M\right) \gamma^{\nu}  \right] {\rm Tr} \left[ \gamma^- \gamma^{\sigma_3} \slashed{\ell}_2 \gamma^{\sigma_2} \right] d^{(q+p-p_2)}_{\sigma_{1} \sigma_{2}} d^{(q+p'-p_2)}_{\sigma_{3} \sigma_{4}} \\
       & = 32 [-g^{\mu\nu}_{\perp\perp}] (q^-)^2\left[ \frac{1 + (1-y)^2}{y\left(1-y+\eta y\right)} \right]  \left[ \left(\pmb{\ell}_{2\perp}-\pmb{k}_{\perp}\right)^2 + \kappa y^4 M^2 \right],
    \end{split}
\end{equation}
where $\kappa$ is defined in Eq.~(\ref{eq:kappa_append}). Finally, the hadronic tensor [Fig.~\ref{fig:kernel-4_qq_scattering_interference}(a)] reduces to the following form
\begin{equation}
\begin{split}
    W^{\mu\nu}_{4,c} & = \left[ \frac{C_F C_A}{2} \right] \sum_f \sum_{f'} 2 [-g^{\mu\nu}_{\perp \perp }] e^2_f g^4_s \int d (\Delta x^{-})   e^{iq^{+}(\Delta x^{-} )} 
    e^{-i[M^{2}/(2q^{-})](\Delta x^{-} )}
  \left\langle P \left| \bar{\psi}_{_f}( \Delta x^{-}) \frac{\gamma^{+}}{4} \psi_{_f}(0)\right| P\right\rangle  \\
  & \times    \int  d \zeta^{-} d (\Delta z^{-}) d^2 \Delta z_{\perp} \frac{dy}{2\pi} \frac{d^2 \ell_{2\perp}}{(2\pi)^2} \frac{d^2 k_{\perp}}{(2\pi)^2}  e^{-i (\Delta z^{-})\mathcal{H}^{(\ell_2 , p_2)}_{M} }     e^{i \pmb{k}_{\perp} \cdot \Delta\pmb{z}_{\perp}} \theta(\zeta^-)\\% \theta\left( x^{-}_{1}-z^{-}_2\right) \theta\left( x^{-}_{2}-z^{-}_3\right) \\
  & \times \left[ \frac{ 1 + \left(1-y\right)^2 }{y}\right]  \frac{1}{yq^-} \frac{ \left[\left(\pmb{\ell}_{2\perp} - \pmb{k}_{\perp}\right)^2 +\kappa y^4 M^2 \right]}{\left[  \left(\pmb{\ell}_{2\perp} -\pmb{k}_{\perp}\right)^2 + y^2\left(1-\eta\right)^2 M^2 \right]^2 }  \left[ 2 -2 \cos\left\{\mathcal{G}^{(p_2)}_{M}\zeta^-\right\} \right]\\
  &\times   \left\langle P_{A-1} \left| \bar{\psi}_{_{f'}}(\zeta^-,\Delta z^-, \Delta z_{\perp}) \frac{\gamma^+}{4} \psi_{_{f'}}(\zeta^-, 0)\right| P_{A-1} \right\rangle , 
\end{split}     \label{eq:K4_hq_light_q_scatt}
\end{equation}
where $\mathcal{G}^{(p_2)}_M$ is given in Eq.~(\ref{eq:GMP2_append}) and $\mathcal{H}^{(\ell_2,p_2)}_{M}$ is defined in Eq.~(\ref{eq:HL2P2M_append}).
%

%For identical particles
For the case when the final state [Fig.~\ref{fig:kernel-4_qq_scattering_interference}(a)] consists of identical quarks, we have three additional contributions. The first one stems from interchanging $\ell_2$ and $p_2$ in Fig.~\ref{fig:kernel-4_qq_scattering_interference}(a) giving the following hadronic tensor:
\begin{equation}
\begin{split}
    W^{\mu\nu}_{4,c} & = \left[ \frac{C_F C_A}{2} \right] \sum_f  2 [-g^{\mu\nu}_{\perp \perp }] e^2_f g^4_s \int d (\Delta x^{-})   e^{iq^{+}(\Delta x^{-} )} 
  \left\langle P \left| \bar{\psi}_{_f}( \Delta x^{-}) \frac{\gamma^{+}}{4} \psi_{_f}(0)\right| P\right\rangle  \\
  & \times    \int  d \zeta^{-} d (\Delta z^{-}) d^2 \Delta z_{\perp} \frac{dy}{2\pi} \frac{d^2 \ell_{2\perp}}{(2\pi)^2} \frac{d^2 k_{\perp}}{(2\pi)^2}  e^{-i (\Delta z^{-})\mathcal{H}^{(\ell_2 , p_2)}_{0} }     e^{i \pmb{k}_{\perp} \cdot \Delta\pmb{z}_{\perp}} \theta(\zeta^-)\\% \theta\left( x^{-}_{1}-z^{-}_2\right) \theta\left( x^{-}_{2}-z^{-}_3\right) \\
  & \times \left[ \frac{ 1 + y^2 }{1-y}\right] \frac{1}{\left(1-y+\eta y\right)q^-} \frac{1}{\pmb{\ell}^2_{2\perp}}  \left[ 2 -2 \cos\left\{\mathcal{G}^{(\ell_2)}_{0}\zeta^-\right\} \right]   \\
  &
\times \left\langle P_{A-1} \left| \bar{\psi}_{_{f}}(\zeta^-,\Delta z^-, \Delta z_{\perp}) \frac{\gamma^+}{4} \psi_{_{f}}(\zeta^-, 0)\right| P_{A-1} \right\rangle  , 
\end{split}     \label{eq:K4_light_heavy_flip}
\end{equation}
where $\mathcal{G}^{(\ell_2)}_0$ is given in Eq.~(\ref{eq:GL20_append}) and $\mathcal{H}^{(\ell_2,p_2)}_{0}$ is defined in Eq.~(\ref{eq:HL2P2M0_append}).
The second and third contribution comes from the process in Fig.~\ref{fig:kernel-4_qq_scattering_interference}(b), where either momenta $p_2$ and $\ell_2$ are as illustrated or they are interchanged. For the illustrated situation in Fig.~\ref{fig:kernel-4_qq_scattering_interference}(b), the hadronic tensor gives
\begin{equation}
\begin{split}
W^{\mu\nu}_{4,c_1} & = \sum_f e^2_f g^4_{s} \int d^4 x_1 d^4 x_2 d^4 z_{2} d^4 z_{3} \int \frac{d^4 p}{(2\pi)^4} \frac{d^4 p'}{(2\pi)^4}  \frac{d^4 \ell_{2}}{(2\pi)^4} \frac{d^4 p_{2}}{(2\pi)^4} e^{-ip'x_2} e^{ipx_1}\left\langle P \left| \bar{\psi}_{_f}(x_2) \frac{\gamma^{+}}{4} \psi_{_f}(x_1)\right| P\right\rangle  \\
  & \times e^{i\left(q+p'-p_{2}-\ell_{2}\right) z_{3}} e^{i(\ell_{2}+p_{2} - q - p)z_{2}} \left\langle P_{A-1} \left| \bar{\psi}_{_f}(z_3)\frac{\gamma^+}{4}\psi_{_f}(z_2)\right| P_{A-1} \right\rangle  (2 \pi) \delta\left(\ell_{2}^{2}\right) (2 \pi)  \delta\left(p_{2}^{2}\right) \delta^{ab} \delta^{cd} {\rm Tr}\left[t^{d}t^{b} t^{c} t^{a}\right] \\
& \times\frac{ {\rm Tr} \left[\gamma^- \gamma^{\mu} \left(\slashed{q}+\slashed{p}'\right) \gamma^{\sigma_4}\slashed{\ell}_2\gamma^{\sigma_{2}} \gamma^-\gamma^{\sigma_{3}}\slashed{p}_2\gamma^{\sigma_{1}}(\slashed{q}+\slashed{p})\gamma^{\nu}  \right]  
}{  \left[\left(q+p'\right)^{2}-i\epsilon\right] \left[\left(q+p\right)^2  + i\epsilon\right]  \left[\left(q+p'-\ell_2\right)^2 - i\epsilon\right] \left[\left(q+p -p_2\right)^2+i\epsilon\right]  } d^{(q+p-p_2)}_{\sigma_{1} \sigma_{2}} d^{(q+p'-\ell_2)}_{\sigma_{3} \sigma_{4}}.
\end{split} \label{eq:kernel4_intereference_wi_a}
\end{equation}
The above expression of the hadronic tensor has singularity when the denominator of the propagator for $p_1$, $\ell$, $\ell'$, and $p'_1$ becomes on-shell. It contains two simple poles for $p^+$ and $p'^+$. The contour integration for $p^+$ gives
\begin{equation}
\begin{split}
C_{1} & = \oint \frac{dp^{+}}{(2\pi)} \frac{e^{ip^{+}\left(x^{-}_{1}-z^{-}_{2}\right)}}{\left[\left(q+p\right)^{2}  + i\epsilon\right]\left[\left(q+p-p_{2}\right)^{2}+i\epsilon\right]} \\
      & = \frac{(2\pi i)}{2\pi} \frac{\theta\left(x^{-}_{1} - z^{-}_{2}\right)}{4q^{-}\left(q^{-}-p^{-}_{2}\right)} e^{-iq^{+}\left(x^{-}_{1}-z^{-}_{2}\right)}\left[ \frac{ -1 + e^{i \mathcal{G}^{(p_2)}_0(x^{-}_{1}-z^{-}_{2})} }{ \mathcal{G}^{(p_2)}_0}  \right],
\end{split}
\end{equation}
where $ \mathcal{G}^{(p_2)}_0$ is defined in Eq.~(\ref{eq:GP2M0_append}). Similarly, the contour integration for $p'^+$ gives
\begin{equation}
\begin{split}
C_{2} & = \oint \frac{dp'^{+}}{(2\pi)} \frac{e^{-ip'^{+}\left(x^{-}_{2}-z^{-}_{3}\right)}}{\left[\left(q+p'\right)^{2} - i\epsilon\right]\left[\left(q+p'-\ell_{2}\right)^{2} - i\epsilon\right]} \\
      & = \frac{(-2\pi i)}{2\pi} \frac{\theta\left(x^{-}_{2} - z^{-}_{3}\right)}{4q^{-}\left(q^{-}-\ell^{-}_{2}\right)}e^{iq^{+}\left(x^{-}_{2}-z^{-}_{3}\right)}\left[ \frac{ -1+e^{-i \mathcal{G}^{(\ell_2)}_0(x^{-}_{2}-z^{-}_{3})} }{ \mathcal{G}^{(\ell_2)}_0}  \right],
\end{split}
\end{equation}
where $\mathcal{G}^{(\ell_2)}_0$ is defined in Eq.~(\ref{eq:GL20_append}). The trace in the third line of Eq.~(\ref{eq:kernel4_intereference_wi_a}) simplifies to
\begin{equation}
    \begin{split}
         & {\rm Tr} \left[\gamma^- \gamma^{\mu} \left(\slashed{q}+\slashed{p}'\right) \gamma^{\sigma_4}\slashed{\ell}_2\gamma^{\sigma_{2}} \gamma^-\gamma^{\sigma_{3}}\slashed{p}_2\gamma^{\sigma_{1}}(\slashed{q}+\slashed{p})\gamma^{\nu}  \right] d^{(q+p-p_2)}_{\sigma_{2} \sigma_{1}} d^{(q+p'-\ell_2)}_{\sigma_{4} \sigma_{3}} \\
         & = \frac{32 (q^-)^2 \left[ -g^{\mu\nu}_{\perp\perp}\right]  }{(1-\eta )y(1-y)q^-} \left[ -\pmb{\ell}^{2}_{2\perp} +  \pmb{\ell}_{2\perp}\cdot \pmb{k}_{\perp} \right].
    \end{split}
\end{equation}
The final expression of the hadronic tensor for Fig.~\ref{fig:kernel-4_qq_scattering_interference}(b) is given by
\begin{equation}
\begin{split}
W^{\mu\nu}_{4,c_1} & =  \left[C_F C_A\left(C_F - \frac{C_A}{2} \right) \right]  \sum_f  2[-g^{\mu\nu}_{\perp \perp }] e^2_f  g^4_s \int d (\Delta x^{-})   e^{iq^{+}(\Delta x^{-} )}     
  \left\langle P \left| \bar{\psi}_{_f}( \Delta x^{-}) \frac{\gamma^{+}}{4} \psi_{_f}(0)\right| P\right\rangle  \\
  & \times  \int  d \zeta^{-} d (\Delta z^{-}) d^2 \Delta z_{\perp} \frac{dy}{2\pi} \frac{d^2 \ell_{2\perp}}{(2\pi)^2} \frac{d^2 k_{\perp}}{(2\pi)^2}   e^{-i (\Delta z^{-})\mathcal{H}^{(\ell_2 , p_2)}_{0} }     e^{i \pmb{k}_{\perp} \cdot \Delta\pmb{z}_{\perp}}  \theta\left(x^{-}_{1}-z^{-}_2\right) \theta\left(x^{-}_{2}-z^{-}_3\right) \\
  & \times \frac{1}{\left(1-\eta \right)(1-y)yq^-} \frac{\left[ -\pmb{\ell}^{2}_{2\perp} +  \pmb{\ell}_{2\perp}\cdot \pmb{k}_{\perp} \right]}{\left(\pmb{\ell}_{2\perp}-\pmb{k}_{\perp} \right)^2 \pmb{\ell}^2_{2\perp} } \left[ -1 + e^{i\mathcal{G}^{(p_2)}_{0}(x^{-}_{1} - z^{-}_{2})} \right] \left[ -1 + e^{-i\mathcal{G}^{(\ell_2)}_{0}(x^{-}_{2} - z^{-}_{3})} \right]  \\
  & \times \left\langle P_{A-1} \left| \bar{\psi}_{_f}(\zeta^-, \Delta z^-, \Delta z_{\perp}) \frac{\gamma^+}{4} \psi_{_f}(\zeta^-, 0)\right| P_{A-1} \right\rangle , 
\end{split}  
%\label{eq:}
\end{equation}
where $\mathcal{G}^{(\ell_2)}_0$ is defined in Eq.~(\ref{eq:GL20_append}), $\mathcal{G}^{(p_2)}_0$ is given in Eq.~(\ref{eq:GP2M0_append}), and $\mathcal{H}^{(\ell_2,p_2)}_{0}$ is given in Eq.~(\ref{eq:HL2P2M0_append}). After additional simplifications, one gets
\begin{equation}
\begin{split}
W^{\mu\nu}_{4,c_1} & = \left[C_F C_A\left(C_F - \frac{C_A}{2} \right) \right] \sum_f  2[-g^{\mu\nu}_{\perp \perp }] e^2_f g^4_s \int d (\Delta x^{-})   e^{iq^{+}(\Delta x^{-} )}     
  \left\langle P \left| \bar{\psi}_{_f}( \Delta x^{-}) \frac{\gamma^{+}}{4} \psi_{_f}(0)\right| P\right\rangle  \\
  & \times   \int   d\zeta^- d (\Delta z^{-}) d^2 \Delta z_{\perp} \frac{dy}{2\pi} \frac{d^2 \ell_{2\perp}}{(2\pi)^2} \frac{d^2 k_{\perp}}{(2\pi)^2}  e^{-i (\Delta z^{-})\mathcal{H}^{(\ell_2 , p_2)}_{0} }     e^{i \pmb{k}_{\perp} \cdot \Delta\pmb{z}_{\perp}} \\
  & \times  \frac{\theta\left(\zeta^-\right) }{\left(1-\eta \right)(1-y)yq^-} \frac{\left[ -\pmb{\ell}^{2}_{2\perp} +  \pmb{\ell}_{2\perp}\cdot \pmb{k}_{\perp} \right]}{\left(\pmb{\ell}_{2\perp}-\pmb{k}_{\perp} \right)^2 \pmb{\ell}^2_{2\perp} }  \left[ 1 - e^{i\mathcal{G}^{(p_2)}_{0}\zeta^-} - e^{-i\mathcal{G}^{(\ell_2)}_{0}\zeta^-}+e^{i\left\{\mathcal{G}^{(p_2)}_{0}-\mathcal{G}^{(\ell_2)}_{0}\right\}\zeta^-} \right]  \\
  & \times   \left\langle P_{A-1} \left| \bar{\psi}_{_f}(\zeta^-, \Delta z^-, \Delta z_{\perp}) \frac{\gamma^+}{4} \psi_{_f}(\zeta^-, 0)\right| P_{A-1} \right\rangle. 
\end{split}  
\label{eq:K4_b_interf_l2p2_as}
\end{equation}
Since the final state [Fig.~\ref{fig:kernel-4_qq_scattering_interference}(b)] consists of two identical quarks, the momentum of the two quarks could be interchanged. After performing $p_2\leftrightarrow\ell_2$, the resulting hadronic tensor is given as
\begin{equation}
\begin{split}
    W^{\mu\nu}_{4,c_2} & = \left[C_F C_A\left(C_F - \frac{C_A}{2} \right) \right] \sum_f  2[-g^{\mu\nu}_{\perp \perp }] e^2_f g^4_s \int d (\Delta x^{-})   e^{iq^{+}(\Delta x^{-} )}     
  \left\langle P \left| \bar{\psi}_{_f}( \Delta x^{-}) \frac{\gamma^{+}}{4} \psi_{_f}(0)\right| P\right\rangle  \\
  & \times   \int  d \zeta^{-} d (\Delta z^{-}) d^2 \Delta z_{\perp} \frac{dy}{2\pi} \frac{d^2 \ell_{2\perp}}{(2\pi)^2} \frac{d^2 k_{\perp}}{(2\pi)^2}  e^{-i (\Delta z^{-})\mathcal{H}^{(\ell_2 , p_2)}_{0} }     e^{i \pmb{k}_{\perp} \cdot \Delta\pmb{z}_{\perp}} \theta\left(x^{-}_{1}-z^{-}_2\right) \theta\left(x^{-}_{2}-z^{-}_3\right) \\
  & \times \frac{1}{\left(1-\eta\right)(1-y)yq^-} \frac{\left[ -\pmb{\ell}^{2}_{2\perp} +  \pmb{\ell}_{2\perp}\cdot \pmb{k}_{\perp} \right]}{\left(\pmb{\ell}_{2\perp}-\pmb{k}_{\perp}\right)^2 \pmb{\ell}^2_{2\perp} }  \left[ -1 + e^{i\mathcal{G}^{(\ell_2)}_{0}(x^{-}_{1} - z^{-}_{2})} \right] \left[ -1 + e^{-i\mathcal{G}^{(p_2)}_{0}(x^{-}_{2} - z^{-}_{3})} \right]   \\
  & \times  \left\langle P_{A-1} \left| \bar{\psi}_{_f}(\zeta^-, \Delta z^-, \Delta z_{\perp}) \frac{\gamma^+}{4} \psi_{_f}(\zeta^-, 0)\right| P_{A-1} \right\rangle.
\end{split}  
%\label{eq:}
\end{equation}
The above expression can be simplified further to give the following: 
\begin{equation}
\begin{split}
    W^{\mu\nu}_{4,c_2} & = \left[C_F C_A\left(C_F - \frac{C_A}{2} \right) \right] \sum_f  2[-g^{\mu\nu}_{\perp \perp }] e^2_f g^4_s \int d (\Delta x^{-})   e^{iq^{+}(\Delta x^{-} )}     
  \left\langle P \left| \bar{\psi}_{_f}( \Delta x^{-}) \frac{\gamma^{+}}{4} \psi_{_f}(0)\right| P\right\rangle  \\
  & \times   \int  d \zeta^{-} d (\Delta z^{-}) d^2 \Delta z_{\perp} \frac{dy}{2\pi} \frac{d^2 \ell_{2\perp}}{(2\pi)^2} \frac{d^2 k_{\perp}}{(2\pi)^2}   e^{-i (\Delta z^{-})\mathcal{H}^{(\ell_2 , p_2)}_{0} }     e^{i \pmb{k}_{\perp} \cdot \Delta\pmb{z}_{\perp}} \\
  & \times   \frac{\theta\left(\zeta^-\right)}{\left(1-\eta\right)(1-y)yq^-} \frac{\left[ -\pmb{\ell}^{2}_{2\perp} +  \pmb{\ell}_{2\perp}\cdot \pmb{k}_{\perp} \right]}{\left(\pmb{\ell}_{2\perp}-\pmb{k}_{\perp}\right)^2 \pmb{\ell}^2_{2\perp} } \left[1  - e^{-i\mathcal{G}^{(p_2)}_{0}\zeta^-} - e^{i\mathcal{G}^{(\ell_2)}_{0}\zeta^-} + e^{-i\left\{\mathcal{G}^{(p_2)}_{0} - \mathcal{G}^{(\ell_2)}_{0} \right\}\zeta^-}\right]   \\
  & \times   \left\langle P_{A-1} \left| \bar{\psi}_{_f}(\zeta^-, \Delta z^-, \Delta z_{\perp}) \frac{\gamma^+}{4} \psi_{_f}(\zeta^-, 0)\right| P_{A-1} \right\rangle .
\end{split}  
\label{eq:K4_b_interf_l2p2_interchanged}
\end{equation} 

Adding the two hadronic tensors associated with Fig.~\ref{fig:kernel-4_qq_scattering_interference}(b), i.e., Eq.~(\ref{eq:K4_b_interf_l2p2_as}) and (\ref{eq:K4_b_interf_l2p2_interchanged}), gives the following form of the hadronic tensor:
\begin{eqnarray}
\begin{split}
    W^{\mu\nu}_{4,c_1+c_2} & = \left[C_F C_A\left(C_F - \frac{C_A}{2} \right) \right] \sum_f  2[-g^{\mu\nu}_{\perp \perp }] e^2_f g^4_s \int d (\Delta x^{-})   e^{iq^{+}(\Delta x^{-} )}     
  \left\langle P \left| \bar{\psi}_{_f}( \Delta x^{-}) \frac{\gamma^{+}}{4} \psi_{_f}(0)\right| P\right\rangle  \\
  & \times   \int  d \zeta^{-} d (\Delta z^{-}) d^2 \Delta z_{\perp} \frac{dy}{2\pi} \frac{d^2 \ell_{2\perp}}{(2\pi)^2} \frac{d^2 k_{\perp}}{(2\pi)^2}   e^{-i (\Delta z^{-})\mathcal{H}^{(\ell_2 , p_2)}_{0} }     e^{i \pmb{k}_{\perp} \cdot \Delta\pmb{z}_{\perp}} \theta\left(\zeta^-\right) \\
  & \times   \left[2  - 2\cos\left\{\mathcal{G}^{(p_2)}_{0}\zeta^-\right\} - 2\cos\left\{\mathcal{G}^{(\ell_2)}_{0}\zeta^-\right\} + 2\cos\left\{\left(\mathcal{G}^{(p_2)}_{0} - \mathcal{G}^{(\ell_2)}_{0} \right)\zeta^-\right\}\right]   \\
  & \times   \frac{1}{\left(1-\eta\right)(1-y)yq^-}\frac{\left[ -\pmb{\ell}^{2}_{2\perp} +  \pmb{\ell}_{2\perp}\cdot \pmb{k}_{\perp} \right]}{\left(\pmb{\ell}_{2\perp}-\pmb{k}_{\perp}\right)^2 \pmb{\ell}^2_{2\perp} } \left\langle P_{A-1} \left| \bar{\psi}_{_f}(\zeta^-, \Delta z^-, \Delta z_{\perp}) \frac{\gamma^+}{4} \psi_{_f}(\zeta^-, 0)\right| P_{A-1} \right\rangle,
\end{split}  
\label{eq:K4_b_both_interf_wmunu_added}
\end{eqnarray} 
where $\mathcal{G}^{(\ell_2)}_0$ is defined in Eq.~(\ref{eq:GL20_append}), $\mathcal{G}^{(p_2)}_0$ is given in Eq.~(\ref{eq:GP2M0_append}), and $\mathcal{H}^{(\ell_2,p_2)}_{0}$ is given in Eq.~(\ref{eq:HL2P2M0_append}). 

\end{appendices}